\documentclass[12pt]{article}

\usepackage[T1]{fontenc} 
\usepackage{setspace}

\newcommand{\bodyStretch}{1.5}

\newcommand{\bodyspacing}{\setstretch{\bodyStretch}}

\usepackage{microtype} %
\usepackage[margin=.9in]{geometry} %
\usepackage[hang, small,labelfont=bf,up,textfont=it,up]{caption} %
\usepackage{booktabs} %
\usepackage{multirow}
\usepackage{siunitx}
\usepackage{float} %
\usepackage{lettrine} %
\usepackage{paralist} %
\usepackage[svgnames]{xcolor}
\usepackage{comment}
\usepackage[bottom]{footmisc} %
\usepackage{amsmath,amsfonts,amsthm, amssymb,textcomp, bm, bbm, harpoon} %
\usepackage[normalem]{ulem} %

\usepackage{algorithm, algorithmic} %

\usepackage{lineno}
\newcommand*\patchAmsMathEnvironmentForLineno[1]{%
  \expandafter\let\csname old#1\expandafter\endcsname\csname #1\endcsname
  \expandafter\let\csname oldend#1\expandafter\endcsname\csname end#1\endcsname
  \renewenvironment{#1}%
    {\linenomath\csname old#1\endcsname}%
    {\csname oldend#1\endcsname\endlinenomath}}%
\newcommand*\patchBothAmsMathEnvironmentsForLineno[1]{%
  \patchAmsMathEnvironmentForLineno{#1}%
  \patchAmsMathEnvironmentForLineno{#1*}}%
\AtBeginDocument{%
  \patchBothAmsMathEnvironmentsForLineno{equation}%
  \patchBothAmsMathEnvironmentsForLineno{align}%
  \patchBothAmsMathEnvironmentsForLineno{flalign}%
  \patchBothAmsMathEnvironmentsForLineno{alignat}%
  \patchBothAmsMathEnvironmentsForLineno{gather}%
  \patchBothAmsMathEnvironmentsForLineno{multline}%
}

\usepackage{abstract} %
\usepackage{titlesec} %
\usepackage{sidecap} %
\sidecaptionvpos{figure}{c} %
\usepackage[section]{placeins} %

\titleformat{\section}[block]
  {\large\bfseries}{\thesection.}{1em}{}

\titleformat{\subsection}[block]
  {\large}{\thesubsection.}{1em}{}

\titlespacing*{\section}
  {0pt}    %
  {0.5ex}  %
  {0.3ex}  %

\titlespacing*{\subsection}
  {0pt}
  {0.4ex}
  {0.2ex}

\usepackage{fancyhdr} %
\usepackage{graphicx}
\usepackage{soul} \soulregister\ref7 %
\usepackage{outlines}

\usepackage{listings} %
\usepackage{graphicx}
\usepackage{float}
\usepackage{booktabs, makecell} %

\usepackage[bottom]{footmisc}

\makeatletter
\renewcommand*\env@matrix[1][\arraystretch]{%
  \edef\arraystretch{#1}%
  \hskip -\arraycolsep
  \let\@ifnextchar\new@ifnextchar
  \array{*\c@MaxMatrixCols c}}
\makeatother

\newcolumntype{P}[1]{>{\centering\arraybackslash}p{#1}}

\definecolor{junglegreen}{rgb}{0.16, 0.67, 0.53}
\definecolor{lav}{rgb}{0.71, 0.49, 0.86}
\newcommand{\orange}[1]{\textcolor{orange}{#1}}

\newcommand{\rc}[1][]{\textsuperscript{\orange{(cite)}} }

\let\oldtextbf\textbf
\renewcommand{\textbf}[1]{\boldmath\oldtextbf{#1}\unboldmath}

\newif\ifshowSecrets

\ifshowSecrets

  \newcommand{\secretf}[1]{\footnote{\textcolor{junglegreen}{#1}}}
  \newcommand{\secreti}[1]{\textcolor{junglegreen}{#1}}

\else

  \newcommand{\secretf}[1]{}
  \newcommand{\secreti}[1]{}
  
  \excludecomment{secret}
\fi

\usepackage{pgfplotstable}
\pgfplotsset{compat=1.18}
\usepackage{csvsimple}
\usepackage{siunitx}

\usepackage{tikz}
\usetikzlibrary{shapes,decorations,arrows,calc,arrows.meta,fit,positioning,decorations.pathreplacing}
\tikzset{
    -Latex,auto,node distance =1 cm and 1 cm,semithick,
    state/.style ={ellipse, draw, minimum width = 0.7 cm},
    point/.style = {circle, draw, inner sep=0.04cm,fill,node contents={}},
    bidirected/.style={Latex-Latex,dashed},
    el/.style = {inner sep=2pt, align=left, sloped}
}

\usetikzlibrary{arrows,shapes.arrows,shapes.geometric,
shapes.multipart,decorations.pathmorphing,positioning,
swigs}

\usepackage{verbatim}

\newtheoremstyle{spaceythm}
  {\baselineskip}                          %
  {\dimexpr\baselineskip-\parskip\relax}   %
  {} %
  {} %
  {\bfseries} %
  {.} %
  {.5em} %
  {} %

\theoremstyle{spaceythm}
\newtheorem{theorem}{Theorem}
\newtheorem{assumption}{Assumption}

\newtheorem{proposition}{Proposition}

\newtheorem{corollary}{Corollary}
\newtheorem{definition}{Definition}

\newtheoremstyle{named}{}{}{\itshape}{}{\bfseries}{.}{.5em}{\thmnote{#3}}
\theoremstyle{named}

\usepackage{authblk}

\usepackage{sansmath} %
\usepackage{subcaption} %
\usepackage{hyperref}
\usepackage{xurl}

\usepackage[american]{babel}
\usepackage[style=american]{csquotes}
\usepackage[style=numeric-comp,autocite=superscript,sorting=none,sortcites=true]{biblatex}
\renewbibmacro{in:}{}
\DeclareNameAlias{default}{labelname}

\title{\vspace{-15mm}\fontsize{18pt}{18pt}\selectfont\textbf{Estimating conditional means under missingness-not-at-random with incomplete auxiliary variables}} %

\date{\vspace{-5ex}}

\author[1]{Maya B. Mathur\thanks{Correspondence to: Maya B. Mathur (\href{mailto:mmathur@stanford.edu}{mmathur@stanford.edu}), Quantitative Sciences Unit, 3180 Porter Drive, Palo Alto, CA, 94304.}}

\author[2]{Shaun Seaman}

\author[3,4]{Willi Zhang}

\author[5,6,7]{Sean McGrath}

\author[8]{Ilya Shpitser}

\affil[1]{Quantitative Sciences Unit and Department of Pediatrics, School of Medicine, Stanford University, USA}

\affil[2]{MRC Biostatistics Unit, University of Cambridge, England}

\affil[3]{Centre for Epidemiology and Community Medicine, Region Stockholm, Stockholm, Sweden}

\affil[4]{Department of Global Public Health, Karolinska Institutet, Stockholm, Sweden}

\affil[5]{Department of Statistics and Data Science, Washington University in St.\ Louis, USA}
\affil[6]{Bursky School of Public Health, Washington University in St.\ Louis, USA}
\affil[7]{Department of Biostatistics, School of Public Health, Yale University, USA}

\affil[8]{Department of Computer Science, Whiting School of Engineering, Johns Hopkins University, USA}

\def\ind{\perp\!\!\!\perp}
\newcommand{\notind}{\not\!\perp\!\!\!\perp}

\newcommand{\OmegaMAR}{\Omega_{\text{MAR}}}

\newcommand{\supp}{\operatorname{support}}

\newcommand{\Xvec}{\vec{X}}

\newcommand{\xvec}{\vec{x}}

\newcommand{\xvect}{\xvec^\top}
\newcommand{\Xvect}{\Xvec^\top}

\newcommand{\KMAR}{K_\text{MAR}}

\newcommand{\Vobs}[2]{V_{#1}^{\text{obs}}(#2)}
\newcommand{\Wobs}[2]{W_{#1}^{\text{obs}}(#2)}

\newcommand{\Lobs}{L^{\text{obs}}}

\newcommand{\Lmis}{L^{\text{mis}}}

\newcommand{\Wrc}{\overline{W}_R}

\newcommand{\muMIA}{\mu_{\text{MIA}}}
\newcommand{\muMIAx}{\mu_{\text{MIA}}(x)}

\newcommand{\muMAR}{\mu_{\text{MAR}}}

\newcommand{\muWCC}{\mu_{\text{WCC}}}

\newcommand{\muhatMIA}{\widehat \mu_{\text{MIA}}}

\newcommand{\muhatMIAice}{\widehat \mu_{\text{MIA-ICE}}}
\newcommand{\muhatMIAipw}{\widehat \mu_{\text{MIA-IPW}}}
\newcommand{\muhatMIAaipw}{\widehat \mu_{\text{MIA-AIPW}}}
\newcommand{\muhatMIAtmle}{\widehat \mu_{\text{MIA-TMLE}}}
\newcommand{\muhatMAR}{\widehat \mu_{\text{MAR}}}

\newcommand{\pMAR}{p_{\text{MAR}}}
\newcommand{\WRdef}{\big\{ W_j \in W : W_j \notind R(X \cup W) \mid X \big\}}

\usepackage{pifont}%
\newcommand{\cmark}{\textcolor{junglegreen}{\ding{51}} }%

\newcommand{\xmark}{\textcolor{red}{\ding{55}} }

\newcommand{\qmark}{\ensuremath{\boldsymbol{\mathcal{?}}} }

\newcommand{\PMNAR}{\mathcal P_{\text{MNAR}}}

\newcommand{\Cov}{\mathrm{Cov}}

\usepackage{xr-hyper}   %
\usepackage[commandnameprefix=ifneeded]{changes} %
\input{xr_supp_labels.tex}

\newif\ifredact
\redactfalse

\ifredact
\DeclareSourcemap{
  \maps[datatype=bibtex]{
    \map[overwrite=true]{
      \step[fieldsource=keywords, match=\regexp{redactme}, final]
      \step[fieldset=author, fieldvalue={{[REDACTED-FOR-BLINDED-REVIEW]}}]
      \step[fieldsource=url,  null]
      \step[fieldsource=doi,  null]
      \step[fieldsource=note, null]
    }
  }
}

  \makeatletter
  \renewcommand\AB@authlist{}%
  \renewcommand\AB@affillist{}%
  \renewcommand\AB@authors{}%
  \makeatother
  \author{[Authors redacted for blinded review]}

\fi

\begin{document}

\singlespacing

\maketitle %

\thispagestyle{fancy} %

\singlespacing

\renewcommand{\abstractname}{Abstract}

\begin{abstract}
\noindent Estimators assuming missingness at random (MAR) can fail under missingness not at random (MNAR). Introducing complete auxiliary variables sometimes restores MAR by breaking dependence between analysis variables and missingness. However, if the auxiliaries are themselves incomplete, MAR typically remains violated. 

\noindent We derive necessary and sufficient, weaker-than-MAR conditions under which the MAR functional $\muMAR$ for a conditional mean $E[Y \mid X]$ is correct.  We also propose an alternative identifying functional for $E[Y \mid X]$, called $\muMIA$ for ``Marginalization over Incomplete Auxiliaries'', along with sufficient conditions for $\muMIA$ to be correct. 

\noindent For MNAR distributions in which all missingness patterns can occur and certain context-specific independences do not occur, there are two straightforward cases where $\muMIA$ is correct for a strict superset of the MNAR distributions for which $\muMAR$ is correct: (1) $X$ is complete and $Y$ is incomplete; or (2) all auxiliary variables are incomplete. In case (1), $\muMAR$ is always incorrect, barring knife-edge cases. Determining whether either case holds requires no knowledge of the MNAR mechanism. 

\noindent We propose estimators of $\muMIA$, including g-computation and doubly robust estimators. Our R package \texttt{miapack} implements a semiparametric iterated-conditional-expectations estimator. We illustrate with real data and present simulations comparing estimators of $\muMIA$ and $\muMAR$.
\\*
\end{abstract}

\vspace{0.5cm}

\pagebreak

\bodyspacing

\section{Introduction}

\newcommand{\nonparamIDFootnote}{An estimand is \textit{nonparametrically identified} if it is uniquely determined by the observed-data distribution under the assumed independence restrictions alone, without additional parametric assumptions. However, parametric or semiparametric restrictions are often involved in estimation even when they are not required for identification.}

Estimators that assume missingness at random (MAR) are widely used to mitigate bias and efficiency loss due to missing data \autocite{li2015multiple,white2010bias,perkins2018principled}. Such estimators include multiple imputation \autocite{rubin2004multiple}, inverse-probability of missingness weighting (IPW) \autocite{wooldridge2007inverse,robins1995analysis,sun2018inverse}, and full-information maximum likelihood \autocite{carpenter2021missing}. These are all estimators of the \textit{MAR functional} for the joint distribution, meaning the identifying functional constructed by assuming a MAR data-generating mechanism. An \textit{identifying functional} is a population-level mapping from the observed-data distribution to the possible space of the target estimand, typically constructed by assuming a certain data-generating model (DGM). An identifying functional is correct for a given class of DGMs if, throughout that class, it is equal to the target estimand.

Informally, MAR states that for any given missingness pattern, the probability that the pattern occurs for a given individual is independent of the variables that are missing for that pattern given the variables that are observed for that pattern.\footnote{The term \textit{missingness at random} is used inconsistently in the literature, denoting more than one model. We adopt a widely-used definition of MAR that is sometimes called \textit{everywhere-MAR}. It is this version of MAR that justifies repeated-sampling properties such as unbiasedness and consistency \autocite{seaman2013meant}.} MAR is often hard to justify on substantive grounds except under certain forms of monotone missingness.\autocite{robins1997non,nabi2022causal} When MAR is violated, which is probably often the case in practice, \textit{missingness not at random (MNAR)} is said to hold. For some MNAR data-generating models (DGMs), the joint distribution of the variables of interest is identified, so it is in principle possible to construct the correct identifying functional for the joint distribution of all variables, or for a specific estimand.\autocite{nabi2020full,mathur2025pitfalls,phung2025recursive} However, regardless of whether MAR or MNAR holds, the correct functional typically depends on the precise DGM \autocite{nabi2022causal}, which is rarely known in practice.\footnote{Interestingly, under certain positivity assumptions, Nabi et al.\autocite{nabi2020full} show that if the joint distribution of the variables of interest \textit{and} the missingness indicators (the \textit{full law}) is nonparametrically identified, there exists a single identifying functional that applies regardless of the DGM. This result does not apply, though, if a given estimand of interest is identified but the full law is not.} Some work has therefore sought weak conditions -- requiring only partial knowledge of the DGM -- under which specific functionals are correct.\autocite{correa2018generalized,correa2019identification,robins1995analysis,iwn,nmar2,van2023missing} These developments parallel earlier work in causal inference on identification with unmeasured confounding.\autocite{shpitser2006identification,pearl2009causality}

When a specific estimand is of interest, a distinction can be made between measured variables that are involved in the estimand (here called \textit{analysis variables}) and other measured variables used to mitigate bias due to missing data (called \textit{auxiliary variables}). For example, in IPW, auxiliary variables are often included in the model for the missingness mechanism. When data on the analysis variables are MNAR, introducing appropriately chosen auxiliary variables can sometimes restore MAR -- that is, the joint data comprising the analysis and auxiliary variables are MAR.\autocite{iwn,madley2025analyses} However, often the auxiliary variables themselves are incomplete (i.e., missing for some analysis units).\autocite{madley2025analyses} In these cases, MNAR will generally still hold in the joint data comprising the analysis variables and the auxiliary variables. That is, in most cases, incomplete auxiliary variables do not restore MAR if it did not already hold \autocite{madley2025analyses}. With sufficiently detailed knowledge of the MNAR mechanism, constructing a valid functional for the target estimand may be possible.\autocite{nabi2020full,zhang2026use} However, since such knowledge is rarely available, a widespread practice is to simply treat the joint data on all analysis and auxiliary variables as if they were MAR. Alternatively, there are auxiliary-based methods that do not assume the auxiliaries restore MAR, but these methods typically still require the auxiliaries to be complete.\autocite{miao2018identification}

Under MNAR, the MAR functional for the joint distribution is incorrect. However, identifying the full joint distribution is not necessary because the auxiliary variables are not involved in the estimand. Indeed, for many common estimands, it is not even necessary to estimate the entire joint distribution of the analysis variables. We focus on estimating marginal and conditional expectations, $E[Y \mid X]$, given their relevance to common estimands such as population and subgroup prevalences, average treatment effects, and conditional average treatment effects. Finding sufficient conditions such that MNAR holds, yet the MAR functional for the specific estimand $E[Y \mid X]$ is correct, is an open problem.  Simulation studies have assessed performance of various MAR-based estimators under a variety of MNAR DGMs \autocite{white2010bias, marshall2010comparison, lee2012recovery, beunckens2008simulation, carpenter2006comparison, ccvmi}. 

This paper focuses on misspecification arising from performing estimation under an incorrect identifying functional for the missingness mechanism, in particular performing estimation under MAR when the true mechanism is MNAR. We do not consider misspecification caused by introducing incorrect parametric or semiparametric restrictions on $p(y \mid x)$ or on nuisance distributions (e.g., assuming variables are related linearly when the true relationships are nonlinear). Stated otherwise, we assume that given any assumed identifying functional, a consistent estimator \textit{of that functional} is available, but the functional itself might be incorrect due to a misspecified missingness mechanism. We therefore study a single representation of the MAR functional, namely IPW.

This paper makes two main contributions. First, we provide sufficient conditions, which cover some MNAR cases, under which the MAR functional $\muMAR$ for $E[Y \mid X]$ is correct. Within a broad class of MNAR models, the sufficient conditions for $\muMAR$ to be correct are also necessary. Second, we propose an alternative identifying functional for $E[Y \mid X]$, called $\muMIA$ for ``Marginalization over Incomplete Auxiliaries'' (MIA). The functional comprises two components: (1) the expectation of $Y$ among complete cases; and
(2) the distribution of the auxiliary variables $W$ given $X$ among all individuals with $X$ and $W$ observed. The functional $\muMIA$ is correct for many MNAR cases in which $\muMAR$ is incorrect. We provide sufficient conditions for $\muMIA$ to be correct, and hence for consistent estimators of $\muMIA$ to be consistent for $E[Y \mid X]$. Among MNAR distributions in which: (1) all missingness patterns on the incomplete variables can occur; and (2) certain restrictions on context-specific independences hold, there are two key cases in which $\muMIA$ is correct for a strict superset of the distributions for which $\muMAR$ is correct. The first case occurs if $X$ is complete but the outcome is incomplete. The second occurs if all auxiliary variables are incomplete. In fact, in the first case, $\muMAR$ is incorrect regardless of what auxiliary variables, if any, are included. Determining whether each of these cases holds requires only identifying which variables are incomplete, based on the study design or observed data, without additional untestable assumptions.

This paper is structured as follows. Section \ref{sec:setting} establishes notation. Given our focus on the IPW representation of $\muMAR$, Section \ref{sec:wcc_review} reviews general weighted complete-case (WCC) estimators and IPW estimators as a special case. Section \ref{sec:main_results} develops the main results regarding MAR and MIA and discusses some practical considerations. Section \ref{sec:estimation_and_software} discusses estimation of MIA and its implementation in our R package, \texttt{miapack}. Section \ref{sec:applied_example} applies MIA and MAR estimators to a real longitudinal study regarding individual changes in sexual identity over time. Section \ref{sec:sims} presents a simulation study comparing MIA to MAR estimators in a small number of illustrative DGMs. Section \ref{sec:discussion} concludes. 

\section{Preliminaries}
\label{sec:setting}

\subsection{Notation}

Table \ref{tb:notation} summarizes all key notation. Letters in italicized uppercase (e.g., $L$) denote sets of variables or singleton variables. When discussing regression, we will introduce notation such as $\Xvec$ to denote vectors. When necessary for clarity, we use subscripts ($L_j$) for members in a set, and superscripts ($L^i$) to denote variables for a given individual. (Given the special meaning of ``observed'' in the context of missing data, we refer to ``individuals'' rather than ``observations''.) We use $p(\cdot)$ generically for densities and probability mass functions. Its arguments are values, as in $p(w \mid x)$; in the conditioning, a capital letter denotes conditioning on the random variable itself, so that, for example, $p(w \mid X)$ is a random variable that is a function of $X$, whereas $p(w \mid x)$ is a fixed value. We assume data are independent and identically distributed (iid). 

Let $Y$ be a singleton outcome and $X$ a set of predictor variables. The estimand of interest is $E \big[ Y \mid X \big]$, so $V := X \cup \{Y\}$ is the set of analysis variables. In the case of estimating a marginal mean, $X := \emptyset$. When discussing special cases involving a discrete $X$, we write $E \big[ Y \mid X = x \big]$. Let $W$ be a set of auxiliary variables, meaning variables that are not in $V$ but that will be used in identifying functionals for $E[Y \mid X]$. For any subset $L$ of $V \cup W$, $L^+$ and $L^-$ are, respectively, the sets of complete and incomplete variables in $L$. The set of missingness indicators for $L$ is $R(L) := \{ R_{L_j} \}$ for all $j$ such that $L_j \in L^-$, where $R_{L_j} := 1$ if $L_j$ is observed and $R_{L_j} := 0$ otherwise. Define $M(L)$ as a categorical variable on the positive integers, which denotes patterns of missingness. In particular, $M(L)=1$ denotes the complete-case pattern (i.e., $R_{L_j} = 1$ for all $L_j \in L^-$) and other values ($M(L) > 1$) denote other patterns. The sets of variables in $L$ that are observed and missing, respectively, for pattern $M(L)=m$ are $L^{\text{obs}}(m)$ and $L^{\text{mis}}(m)$. We write $R(L) = 1$ to mean that all members of $R(L)$ are equal to 1.
For shorthand, we write $R$ for $R( V \cup W )$, and similarly for $M$. We assume all variables in $L$ have well-defined values for all individuals and that missing data occur when, for whatever reason, the analyst does not observe those values. (This excludes cases in which ``missing'' data represent undefined quantities, such as quality of life among dead individuals.) 

\singlespacing

\begin{table}[h]
\resizebox{0.97\columnwidth}{!}{%
\begin{tabular}{@{}llc@{}}
\toprule
\textbf{Notation} & \textbf{Definition} & \textbf{Section} \\ \midrule

\multicolumn{3}{@{}l}{\textbf{Analysis and auxiliary variables}} \\
$L$ & Generic set of variables comprising complete members $L^+$ and incomplete members $L^-$ & \ref{sec:setting} \\
$Y$ & Scalar outcome variable in estimand $E[Y \mid X]$ & \ref{sec:setting} \\
$X$ & Predictor variables in estimand $E[Y \mid X]$ & \ref{sec:setting} \\
$V := X \cup \{Y\}$ & Analysis variables & \ref{sec:setting} \\
$W:= W^+ \cup W^-$ & Auxiliary variables (disjoint from $V$), where $W^+$ are complete and $W^-$ are incomplete
& \ref{sec:setting} \\

\midrule

\multicolumn{3}{@{}l}{\textbf{Missingness indicators and patterns}} \\

$R(L)$ & Missingness indicators for incomplete variables in $L$; shorthand $R := R(V \cup W)$ & \ref{sec:setting} \\
$M(L)$ & Categorical missingness pattern for $L$; shorthand $M := M( V \cup W )$ & \ref{sec:setting} \\

$V_{\text{obs}}(m), W_{\text{obs}}(m)$ & Subsets of $V$ and $W$, respectively, that are observed for pattern $M = m$ & \ref{sec:setting} \\

$V_{\text{mis}}(m), W_{\text{mis}}(m)$ & Subsets of $V$ and $W$, respectively, that are missing for pattern $M = m$ & \ref{sec:setting} \\

$R_{XW}$ & $1\{ R(X \cup W) = \mathbf{1} \}$ & \ref{sec:af4} \\

\midrule

\multicolumn{3}{@{}l}{\textbf{Notation for WCC}} \\

$K$ & Generic function of $V \cup W$ used as weights in WCC & \ref{sec:wcc_review} \\

$\muWCC(K)$ & WCC functional for $E[Y \mid X]$ for given weight function $K$ & \ref{sec:wcc_conditions} \\

$\muMAR$ & WCC functional for $E[Y \mid X]$ with $K$ based on MAR($V \cup W$) & \ref{sec:ipmw_setup} \\

$\Omega$ & $K \cdot p(M = 1 \mid X, Y, W)$ & \ref{sec:wcc_conditions} \\

\midrule

\multicolumn{3}{@{}l}{\textbf{Notation for MIA}} \\
$\muMIA$ & Functional for $E[Y \mid X]$ defined as $\int_w E[Y \mid X, W, M=1] \, p(w \mid X, R_{XW}=1) \, dw$ & \ref{sec:af4} \\
$W_R$, $\Wrc$ & $W_R := \WRdef$; $\Wrc := W \setminus W_R$ & \ref{sec:af4} \\

\midrule

\multicolumn{3}{@{}l}{\textbf{Notation for domination results}} \\
$\PMNAR$ & Distributions such that MNAR($V \cup W^+$), \ref{ass:positivity}, and \ref{ass:no_csi} all hold (implies MNAR($V \cup W$) also holds) & \ref{sec:domination} \\
\bottomrule
\end{tabular}
}
\caption{\label{tb:notation} Glossary of key notation. ``Section'' indicates where the notation is introduced. WCC = weighted complete-case; MIA = Marginalization over Incomplete Auxiliaries.}
\end{table}

\bodyspacing

\subsection{Assumptions on the data-generating model}
\label{sec:DGM_assumptions}

Throughout, we will make some basic regularity assumptions on $p(v,w,r)$, which rule out distributions with degeneracies such as knife-edge perfect cancellations (Supplement, Section \ref{supp-sec:review_assumptions}). Certain necessity results will also invoke two assumptions of substantive importance; when this happens, it will be stated explicitly. Those substantive assumptions will not be required to establish any subsequent sufficiency results. In particular, those assumptions are irrelevant to whether the MIA functional is correct; they are relevant to MIA only when defining distribution classes within which the MIA functional dominates the MAR functional.

\begin{assumption}[Positivity] \label{ass:positivity}
Conditional on any value $(v, w) \in \supp(V \cup W)$, every missingness pattern on the incomplete variables can occur, i.e., $p(r \mid v, w)>0$ for all $r \in \{ 0, 1\}^{|R|}$ and $(v,w) \in \supp(V \cup W).$
\end{assumption}

\begin{assumption}[Restrictions on context-specific independence (CSI)]
\label{ass:no_csi}
For every indicator $R^* \in R$, $Z \in (V \cup W)$, and $C \subseteq ( V \cup W \cup R ) \setminus \{ Z, R^* \}$ such that $Z \notind R^* \mid C$, we have $Z \notind R^* \mid c$ for every $c$ such that $p(c)>0$. That is, there are no context-specific independences such that $Z \notind R^* \mid C$ but $Z \ind R^* \mid c'$ for some $c' \in \supp(C)$.
\end{assumption}

Assumption \ref{ass:positivity} excludes forms of monotone missingness that are caused by determinism among missingness indicators. Since complete variables do not have missingness indicators, their existence does not violate Assumption \ref{ass:positivity}. Under Assumption \ref{ass:no_csi}, CSI between members of $V \cup W$ is permitted. As an example of CSI that violates Assumption \ref{ass:no_csi}, consider a hypothetical study in which colonoscopies measuring indicators of colon health are performed for all patients aged $\ge$50 years old and younger patients who have certain risk factors ($Z$). Suppose $R^*$ is the missingness indicator for colon health and $C$ is age in years. Conditional on $C < 50$, $Z$ is associated with $R^*$. However, conditional on $C \ge 50$, $R^* = 1$ deterministically, so $Z$ and $R^*$ are independent. (The determinism involving $R^*$ also violates Assumption \ref{ass:positivity}.) 

\subsection{Missingness models}
\label{sec:mar_mnar}

The results given in this paper do not assume the DGM obeys MAR. However, we will derive conditions under which the MAR functional for $E[Y \mid X]$ is correct even though the DGM is MNAR. The MAR and MNAR missingness mechanisms are as follows.
\begin{definition}[MAR and MNAR]
\label{def:eMAR}
Given a set of variables $L := L^+ \cup L^-$, \textit{missingness at random} (MAR) holds wrt $L$, denoted MAR($L$), if for every missingness pattern $m$ such that $p(M(L) = m)>0$, we have:
\[ 1\{ M(L) = m \} \ind \Lmis(m) \mid \Lobs(m). \]
Otherwise, \textit{missing not at random} (MNAR) holds wrt $L$, denoted MNAR($L$).
\end{definition}
\noindent Several definitions of MAR appear in the literature; this one is sometimes called \textit{everywhere MAR} or \textit{missing always at random} \autocite{doretti2018missing,rubin1976inference,mealli2015clarifying}.
\newcommand{\LobsFootnote}{\footnote{The classical literature on MAR typically writes $\Lobs$, suppressing the pattern $m$, to refer to all data, across all individuals, that are observed in the realized data. We avoid this notation, concurring with others that it can be misleading.\autocite{seaman2013meant}}}

\noindent We will later need the following result. All proofs appear in the Supplement.

\begin{proposition}%
\label{prop:wneg_cant_fix_mnar}
Suppose Assumptions \ref{ass:positivity} and \ref{ass:no_csi} hold. If MNAR($V \cup W^+$) holds, then MNAR($V \cup W$) also holds.
\end{proposition}

\section{Weighted complete-case functionals and estimators}
\label{sec:wcc_review}

To fix ideas, we first review a basic WCC estimator for a conditional mean $E[Y \mid X]$, where $X$ is discrete. First suppose we choose a weight function $K(V \cup W)$ that is correctly specified for the inverse-probability of being a complete case, i.e., $K := p(M = 1 \mid V, W)^{-1}$. For brevity, we simply write $K$. Given such a weight function, a basic Hájek-style estimator for the conditional expectation at $X=x$ is:
\begingroup
\addtolength{\jot}{0.5em}
\begin{align*}
\widehat E_{\text{Hájek}} \big[ Y \mid X = x \big] &= \frac{ \sum_{i=1}^n 1\{ X^i = x, M^i=1 \} \cdot K^i Y^i }{ \sum_{i=1}^n 1\{ X^i = x, M^i = 1 \} \cdot K^i }, \notag 
\intertext{where $X^i$ denotes the value of $X$ for the $i^{th}$ individual, and similarly for other variables. By standard arguments:}
\widehat E_{\text{Hájek}} \big[ Y \mid X=x \big] &\xrightarrow{p} \frac{E \big[ p(M = 1 \mid Y, X=x, W ) \cdot KY \mid X=x \big] }{E \big[ p(M = 1 \mid Y, X=x, W ) \cdot K \mid X=x \big] }. 
\end{align*}
\endgroup
For a given set of variables $L$, we call $p(M(L) = 1 \mid L)$ the \textit{response model}. Typically, the response model (here $p(M = 1 \mid V, W)$) is not known and must be estimated. For example, if MAR holds, the response model is identified and can be estimated as described below in Section \ref{sec:ipmw_setup}. This would yield a consistent estimator $\widehat E_{\text{Hájek}}[Y \mid X=x]$ \autocite{robins1994estimation}. 

More general semiparametric regression estimators, such as quasi-likelihood estimators, can accommodate continuous or higher-dimensional $X$, as well as categorical or binary $Y$. We next define a corresponding class of WCC functionals that includes those based on the quasi-likelihood or on some GLMs, in which there are user-specified weights $K$ as well as an indicator $1\{ M = 1 \}$ that restricts to complete cases. The weight function $K$ may be any strictly positive, measurable function of $V \cup W$, which may be constant in any subset of $V \cup W$. This generalization will allow consideration of arbitrarily misspecified weights that are fixed conditional on $V \cup W$. 

\begin{definition}[Weighted complete-case functional]
\label{def:wcc}

Let $\lambda(X; \beta)$ be an assumed, correctly specified model for $E[Y \mid X]$, where $\beta$ is a $d \times 1$ vector of parameters and $\beta_0$ is the true value of $\beta$. Define the \textit{full-data estimating
function}:
\[ U_t(\beta; q) := q(X; \beta, \phi) \, \epsilon(Y, X; \beta), \]
where $q(X; \beta, \phi)$ returns a $d \times 1$ vector and may depend on nuisance parameters $\phi$, and $\epsilon(Y, X; \beta) := Y - \lambda(X; \beta)$ returns a scalar. Because $\lambda$ is correctly specified, for every bounded, measurable $q$ we have $E \big[ U_t(\beta_0; q) \big] = 0$. Thus, $E_n \big[ U_t(\beta; q) \big] = 0$, where $E_n$ denotes the sample average, is an unbiased estimating equation. Due to missing data, $U_t(\beta; q)$ is not observable for all individuals. Multiplying by $1\{M=1\} K$ yields the \textit{weighted complete-case estimating function}:
\[ U_w(\beta; q) := 1\{M=1\} \, K \, U_t(\beta; q), \]
which depends only on the observed components of $V \cup W$. If $\beta_q^*$ is a solution to $E \big[ U_w(\beta; q) \big] = 0$, then $\beta_q^*$ is called a \textit{weighted complete-case (WCC) functional} for $\beta_0$, and $\lambda(X; \beta_q^*)$ is the corresponding WCC functional for $E[Y \mid X]$.
\end{definition}

For example, letting $\Xvec$ denote a column vector of variables in $X$, the weighted least squares (WLS) estimating equations $E_n \big[ U_w(\beta; q) \big] = 0$ arise from setting $\lambda(X; \beta) := \Xvect \beta$ and $q(X; \beta, \phi) := \Xvec$. Any solution $\beta_q^*$ of $E \big[ U_w(\beta; q) \big] = 0$ is a WCC functional for $\beta_0$, whose closed form is $\beta_q^* = \big( E[ 1\{M=1\} K \, \Xvec \Xvect ] \big)^{-1} E[ 1\{M=1\} K \, \Xvec Y ].$ More generally, consider the population quasi-likelihood equations\autocite{wedderburn1974quasi} under the canonical link, with $h(\cdot)$ defined as corresponding inverse-link function, so that $\lambda(X; \beta) := h( \Xvect \beta )$ and $q(X; \beta, \phi) := \Xvec$. If $\beta_q^*$ is a solution to $E\big[ 1\{M=1\} K \, \Xvec \,(Y - h( \Xvect \beta )) \big] = 0$, then $\beta^*_q$ is a WCC functional for $\beta_0$. The corresponding WCC functional for the conditional mean is $\mu_\mathrm{WCC}(\beta_q^*) = \lambda(X; \beta_q^*) = h( \xvect \beta_q^* )$.

If $E \big[ U_w (\beta_0; q) \big] = 0$ and $E \big[ U_w (\beta; q) \big] \neq 0$ for all $\beta \neq \beta_0$, then $\beta_q^* = \beta_0$ and $\mu_{\rm WCC} (q)$ is a correct identifying functional for $E[Y \mid X]$. Henceforth, however, we will say that $\mu_{\rm WCC} (q)$ is correct if and only if $\mu_{\rm WCC} (q')$ is correct for all bounded, measurable functions $q'$. In this case, $\beta_q^*$ and $\mu_{\rm WCC} (q)$ are the same for all $q$ and we can write them as $\beta^*$ and $\mu_{\rm WCC}$. That is, correctness of the WCC functional requires the same property -- namely unbiasedness of the estimating function across all bounded, measurable $q$ -- as is fulfilled by the corresponding functional of the full-data estimating function. Additionally, requiring correctness across all $q$ rules out knife-edge cases in which a specific choice of $q(x)$, but not other choices, introduces perfect algebraic cancellations in the weighted estimating function, where those cancellations do not occur in the full-data estimating function (proof of Theorem \ref{thm:wcc_conditions}, Supplement).

If $K$ is correctly specified as $p(M = 1 \mid V, W)^{-1}$, then it is easy to show by iterated expectations that $E[U_w(\beta_0; q)] = E[U_t(\beta_0; q)] = 0$, so the WCC functionals $\beta_q^*$ and $\lambda(X; \beta_q^*)$ are correct. In Section \ref{sec:main_results}, we will establish weaker sufficient conditions for a given WCC functional $E[Y \mid X]$ to be correct, accommodating a potentially misspecified $K$.

\subsection{Inverse-probability weighting (IPW)}
\label{sec:ipmw_setup}

The IPW functional is a special case of WCC with the weight function $K := \pMAR( M = 1 \mid V, W )^{-1}$. As a simple starting point, under MAR with monotone missingness, it is convenient to write the response model as:
\begingroup
\small
\addtolength{\jot}{0.5em}
\begin{align*}
\pMAR \left( M = 1 \mid V, W \right) &= \prod_{m = 2}^{J} 
p \left(M \leq m-1 \mid M \leq m, \Vobs{}{m}, \Wobs{}{m} \right),
\end{align*}
\endgroup
\noindent where the patterns are ordered from most complete ($M=1$) to least complete ($M = J$). This expression can be estimated either using a collection of patternwise estimators, or by a single joint model \autocite{robins1994estimation,seaman2013review}. For nonmonotone MAR, the patterns do not have a natural ordering, but Sun \& Tchetgen Tchetgen (STT) \autocite{sun2018inverse} proposed estimators using the alternative representation:
\begingroup
\addtolength{\jot}{0.5em}
\begin{align}
\label{eq:MAR_CCmodel_functional}
\pMAR(M = 1 \mid V, W) &= 1 - \sum_{m > 1} p \left( M = m \mid \Vobs{}{m}, \Wobs{}{m} \right).
\end{align}
\endgroup
Under Assumption \ref{ass:positivity}, the parameters of the pattern-specific models contributing to the sum are estimable by joint maximum likelihood \autocite{sun2018inverse,ross2023accounting}. However, in finite samples the resulting estimates for the parameters of the response model may fail to converge or yield negative probabilities. To mitigate these issues, STT introduced a constrained Bayesian estimator that samples from the posterior distribution under an unconstrained model but discards samples that yield response probabilities less than a specified constant \autocite{sun2018inverse,ross2023accounting}. 

\section{Main results}
\label{sec:main_results}

\subsection{Necessary and sufficient conditions for WCC}
\label{sec:wcc_conditions}

For common subclasses of WCC functionals, such as WLS, it is well-known that if the weight function $K$ is the correctly-specified inverse of the response model, i.e., $p(M = 1 \mid V, W)^{-1}$, then the WCC functional is correct, so the corresponding functional $\muWCC(K)$ for $E[Y \mid X]$ is also correct. As noted in the Introduction, for some MNAR models, the response model is identified. However, in those cases, the identifying functional for the joint distribution $p(v, w)$ is by definition different from its counterpart under MAR. Nevertheless, even when $K$ is arbitrarily misspecified -- e.g., if it is specified under an incorrect MAR assumption -- there exist cases in which the WCC functional for $E[Y \mid X]$ is correct. 

\begin{theorem}%

\label{thm:wcc_conditions}

Let $K$ be a strictly positive, measurable function, $\muWCC(K)$ a corresponding WCC functional for $E[Y \mid X]$, and $ \Omega := K \cdot p(M = 1 \mid X, Y, W) $. Then $\muWCC(K)$ is correct if and only if $\Cov(\Omega, Y \mid X) = 0$. This condition is termed (C-WCC).
\end{theorem}

\begin{corollary}[Sufficient conditions for (C-WCC)]
\label{coro:wcc_suff_conditions}

Let $H$ be the minimal subset of $V \cup W$ on which $K$ depends. (This definition is formalized in the Supplement, Section \ref{supp-sec:H}.) Each of the following conditions is sufficient for (C-WCC) to hold:
\begin{align*}
& \text{(C1-WCC) } \quad \Omega = 1\text{, i.e., } K = p( M = 1 \mid V, W)^{-1}; \text{ or } \\
& \text{(C2-WCC) } \quad E[ Y \mid X, H, M = 1 ] = E[ Y \mid X ].
\end{align*}
\end{corollary}
\noindent(C1-WCC) is the standard condition stating that the weight function is correctly specified. A special case occurs if the weight function assumes the MAR response model and MAR indeed holds. However, (C1-WCC) can also hold under MNAR. If (C2-WCC) holds, then regardless of whether the DGM exhibits MAR or MNAR, a correct weight function is not necessary to have $\muWCC(K) = E[Y \mid X]$. In fact, the functional would be correct even with constant weights, corresponding to unweighted complete-case analysis. We return to this connection in the Discussion.

Let $\KMAR$ be the weight function under an assumed MAR($V \cup W$) model, and let $\muMAR := \muWCC(\KMAR)$. Proposition \ref{prop:wneg_cant_fix_mnar} established that under Assumptions \ref{ass:positivity} and \ref{ass:no_csi}, MNAR($V \cup W^+$) implies MNAR($V \cup W$). Additionally, if MNAR($V \cup W^+$) holds, $X$ is complete, and $Y$ is incomplete, then the resulting form of MNAR($V \cup W$) is such that $\muMAR$ is incorrect:

\begin{theorem}%
\label{thm:IPW_with_incomplete_Y}
Suppose Assumptions \ref{ass:positivity} and \ref{ass:no_csi} hold. If $V^- = \{Y\}$ and MNAR($V \cup W^+$) holds, then $\muMAR$ is incorrect, excluding edge cases in which $\OmegaMAR$ is functionally dependent on $Y$, yet $\Cov(\OmegaMAR, Y \mid X) = 0$.
\end{theorem}
For example, consider the DGM depicted in Figure \ref{fig:sim_dags}, graph (a). (This figure is specifically a directed acyclic graph [DAG], which depicts variables' generative structure via directed edges representing their causal relationships.\autocite{mohan2021graphical, nabi2020full, nabi2022causal, bhattacharya2020identification}) This graph implies that $R_{W_1}$ is independent of all other variables, and $R_Y$ is independent of all other variables conditional on $W_1$. Therefore, the true response model is:
\begingroup
\addtolength{\jot}{0.5em}
\begin{align*}
p(M = 1 \mid V, W) & := p(R_{W_1} = R_Y = 1 \mid Y, X_1, X_2, W) = p(R_{W_1} = 1) \; p(R_Y = 1 \mid W) \\
&= p(R_{W_1} = 1) \; p(R_Y = 1 \mid W, R_{W_1} = 1).
\end{align*}
\endgroup
This function does not depend on $Y$. In contrast, the response model under an incorrect MAR assumption is:
\begingroup
\small
\addtolength{\jot}{0.5em}
\begin{align*}
\pMAR(M = 1 \mid V, W) &= 1 - \big[ p( M = 2 \mid Y, X_1, X_2) + p( M = 3 \mid X_1, X_2, W) + p( M = 4 \mid X_1, X_2) \big].
\end{align*}
\endgroup
where $M =2$ denotes the pattern $(R_{W_1} = 0, R_Y = 1)$; $M =3$ denotes $(R_{W_1} = 1, R_Y = 0)$; and $M =4$ denotes $(R_{W_1} = 0, R_Y = 0)$.
Since $R \notind Y \mid \{ X_1, X_2 \}$, the function $p(M=2 \mid Y, X_1, X_2  )$ depends on $Y$ under Assumptions \ref{ass:positivity} and \ref{ass:no_csi}. This violates the sufficient condition (C2-WCC). In fact, Theorem \ref{thm:IPW_with_incomplete_Y} could be applied immediately to see that $\muMAR$ is incorrect. Later simulations illustrate the inconsistency of corresponding estimators (Section \ref{sec:sims}). 

\subsection{Marginalization over incomplete auxiliaries (MIA)}
\label{sec:af4}

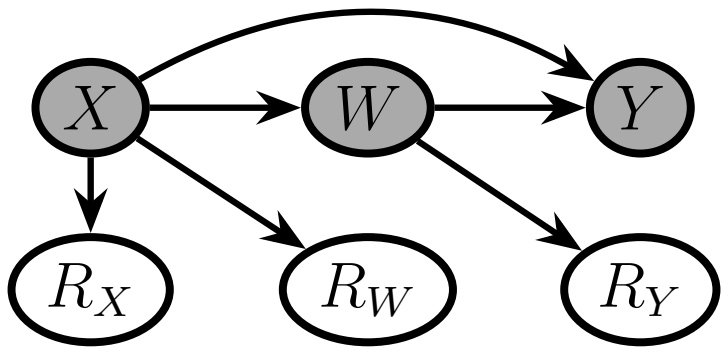
\begin{figure}[!ht]
\centering
\begin{tikzpicture}
\tikzset{
  ell/.style={draw,fill=white, inner sep=2pt, line width=1.5pt},
  swig vsplit={gap=3pt, line color right=red},
  every path/.style={>=Stealth, line width=1.2pt, >={Stealth[scale=1.0]}}
}
\node[name=X1, ell, shape=ellipse, fill={rgb:black,1;white,2}]{$X$};
\node[name=W, ell, shape=ellipse, right=10mm of X1, fill={rgb:black,1;white,2}]{$W$};
\node[name=Y, ell, shape=ellipse, right=10mm of W, fill={rgb:black,1;white,2}]{$Y$};
\node[name=RW, ell, shape=ellipse, below=5mm of W]{$R_{W}$};
\node[name=RY, ell, shape=ellipse, below=5mm of Y]{$R_Y$};
\node[name=RX, ell, shape=ellipse, below=5mm of X1]{$R_X$};

\draw[->,line width=1.2pt,>=stealth]
(X1) edge (W)
(X1) edge (RX)
(X1) edge (RW)
(W) edge (RY)
(W) edge (Y)
(X1) edge[out=30,in=150] (Y);
\end{tikzpicture}
\caption{\label{fig:mediation_dag} A DGM in which estimating the total effect of $X$ on $Y$ requires marginalizing over $W$ rather than conditioning on it.}
\end{figure}

Our proposed identifying functional for $E[Y \mid X]$ is:
\begingroup
\addtolength{\jot}{0.5em}
\begin{align}
 \muMIA  &:= \int_{w} E \big[ Y \mid X, W = w, M = 1 \big] \; p( w \mid X, R_{XW} = 1 ) \; dw \label{eq:mia_sp} \\
&= E \Big[ E \big[ Y \mid X, W, M = 1 \big] \mid X, R_{XW} = 1 \Big]. \label{eq:mia_ice}
\end{align}
\endgroup
\noindent where $R_{XW} := 1\{ R(X \cup W) = \mathbf{1} \}$. As in IPW, the $X := \emptyset$ case yields the marginal distribution and expectation of $Y$. The $\muMIA$ functional can be viewed as a marginalization over $p(w \mid X, R_{XW} = 1)$ of the unweighted complete-case (uCC) functional $E[Y \mid X, W, M = 1]$. MIA's marginalization is useful when $W$ cannot simply be included in the conditioning set $X$, because the estimand is specifically $E[Y \mid X]$ and not $E[Y \mid X, W]$. For example, consider the simple model in Figure \ref{fig:mediation_dag}. Suppose the estimand is the average treatment effect $E[Y(x=1)] - E[Y(x=0)]$, where $Y(x)$ denotes a potential outcome. In Figure \ref{fig:mediation_dag}, this is equal to the full-data estimand $E[Y \mid X]$, but generally is \textit{not} equal to $E[Y \mid X, W]$ because $W$ is a mediator. In this case, marginalization over $W$ rather than conditioning is necessary. We further consider connections between MIA and uCC in the Discussion.

\noindent The following sufficiency conditions are immediate.
\begin{theorem}%
\label{thm:mia_conditions}
Let $ W_R := \WRdef $ and $ \Wrc := W \setminus W_R $. The following conditions suffice for $\muMIA$ to be correct:
\begingroup
\addtolength{\jot}{0.5em}
\begin{align*}
& \text{(C1-MIA) } \; E[ Y \mid X, W, M = 1 ] = E[ Y \mid X, W ] \text{; and }\\
& \text{(C2-MIA) } \; E[ Y \mid X, W ] = E[ Y \mid X, \Wrc ].
\end{align*}
\end{theorem}
\endgroup
A special case of (C2-MIA) occurs if $W_R = \emptyset$. If (C1-MIA) and (C2-MIA) are strengthened to require full independence $Y \ind 1\{ M = 1 \} \mid X \cup W$ and $Y \ind W_R \mid (X \cup \Wrc)$ respectively, the strengthened conditions also suffice for the entire conditional distribution $p(y \mid x)$ to be identified by $\int_{w} p( y \mid x, w, M =1 ) \; p( w \mid x, R_{XW} = 1 ) \; dw$. We discuss estimation of $\muMIA$ in Section \ref{sec:estimation_and_software} below. 

\subsection{Key cases in which $\muMIA$ dominates $\muMAR$ over MNAR distributions}
\label{sec:domination}

Given distinct functionals for an estimand of interest, it is natural to compare the permissiveness of their respective identification assumptions. To this end, the next definition 
describes, for a specified class of distributions, an identifying functional that is correct for a strict superset of those distributions for which a second functional is correct:
\begin{definition}[Dominating functional]
\label{def:dominating_functional}
Let $\theta_A$ and $\theta_B$ be two distinct identifying functionals intended to represent a parameter $\theta$, and let $\mathcal P$ be a given class of data-generating distributions on a set of variables $L$. Let $\mathcal P_A = \{ p \in \mathcal P : \theta_A = \theta \text{ in } p \}$ and $\mathcal P_B = \{ p \in \mathcal P : \theta_B = \theta \text{ in } p \}$.  We say that $\theta_A$ \textit{dominates} $\theta_B$ if $\mathcal P_B \subsetneq \mathcal P_A$.
\end{definition}

For example, familiar doubly robust functionals can be motivated through domination. In the simple case of estimating an average treatment effect in a nonrandomized study with no missing data, an augmented inverse propensity score weighting functional dominates the corresponding g-computation and IPW functionals. This is because the augmented functional is correct when either the g-computation functional or the IPW functional is correct, whereas the converse is not true. Trivially, $\muMAR$ dominates $\muMIA$ over the class of all MAR distributions: $\muMAR$ is correct for every MAR distribution, but there exist MAR distributions for which $\muMIA$ is incorrect. However, our focus is on MNAR, and there are important cases in which $\muMIA$ dominates $\muMAR$ over the class of MNAR distributions. To this end, the next result states that if MNAR holds, all auxiliary variables are incomplete, and the sufficient conditions for $\muMIA$ to be correct are violated, then $\muMAR$ is incorrect.

\begin{theorem}%
\label{thm:IPW_with_Wneg}
Suppose Assumptions \ref{ass:positivity} and
\ref{ass:no_csi} hold and $W = W^-$. If
(C1-MIA) or (C2-MIA) is violated, then $\muMAR$ is incorrect, barring the edge cases described in Theorem \ref{thm:IPW_with_incomplete_Y}.
\end{theorem}
\noindent The next result follows from Theorems \ref{thm:IPW_with_incomplete_Y} and \ref{thm:IPW_with_Wneg} and is illustrated in Figure \ref{fig:mia_domination}.
\begin{theorem}%
\label{thm:dominating_cases_af4}
Fix $V \cup W$ and let $\PMNAR$ be the class of distributions for which MNAR($V \cup W^+$) holds, and for which Assumptions \ref{ass:positivity} and \ref{ass:no_csi} hold. (By Proposition \ref{prop:wneg_cant_fix_mnar}, distributions in $\PMNAR$ also exhibit MNAR($V \cup W$).) If $W = W^-$ or $V^- = \{Y\}$, then $\muMIA$ dominates $\muMAR$ over $\PMNAR$, barring the edge cases described in Theorem \ref{thm:IPW_with_incomplete_Y}. We therefore refer to the $W = W^-$ or $V^- = \{Y\}$ cases as \textit{domination cases}.
\end{theorem}

\newcommand{\thisFigScale}{0.6}
\begin{figure}
\centering
\scalebox{\thisFigScale}{%
\begin{tikzpicture}[
  x=11cm, y=6.4cm, even odd rule,
  ellab/.style={font=\fontsize{14}{14}\selectfont, inner sep=2pt},
  annot/.style={font=\fontsize{14}{14}\selectfont, align=center, inner sep=1pt},
  arr/.style  ={-Latex, line width=0.7pt, shorten >=1pt}]
\def\Acx{0.30}\def\Acy{0.44}\def\Arx{0.29}\def\Ary{0.32}
\def\Bcx{0.44}\def\Bcy{0.44}\def\Brx{0.18}\def\Bry{0.26}
\def\MIAlabx{0.20}\def\MARlabx{0.58}
\def\annotxL{-0.115}   %
\def\annotxR{0.84}     %
\def\greentipx{0.13}   %
\pgfmathsetmacro{\MIAtip}{\Acy+\Ary*sqrt(1-((\MIAlabx-\Acx)/\Arx)^2)}
\pgfmathsetmacro{\MARtip}{\Bcy+\Bry*sqrt(1-((\MARlabx-\Bcx)/\Brx)^2)}
\pgfmathsetmacro{\slivermid}{0.5*(\Acx+\Arx+\Bcx+\Brx)}
\draw[line width=1.2pt] (-0.06,0.00) rectangle (0.785,1.08);
\node[anchor=west, font=\fontsize{18}{14}\selectfont] at (-0.035,1.00)
      {MNAR wrt $V \cup W^{+}$};
\begin{scope}
  \clip (\Acx,\Acy) ellipse [x radius=\Arx, y radius=\Ary];
  \fill[green!45!black!35] (-1,-1) rectangle (2,2)
        (\Bcx,\Bcy) ellipse [x radius=\Brx, y radius=\Bry];
\end{scope}
\begin{scope}
  \clip (\Bcx,\Bcy) ellipse [x radius=\Brx, y radius=\Bry];
  \fill[red!30] (-1,-1) rectangle (2,2)
        (\Acx,\Acy) ellipse [x radius=\Arx, y radius=\Ary];
\end{scope}
\draw[line width=1.4pt] (\Acx,\Acy) ellipse [x radius=\Arx, y radius=\Ary];
\draw[line width=1.4pt] (\Bcx,\Bcy) ellipse [x radius=\Brx, y radius=\Bry];
\node[ellab] (MIAlabel) at (\MIAlabx,0.87) {$\muMIA$ is correct};
\node[ellab] (MARlabel) at (\MARlabx,0.87) {$\muMAR$ is correct};
\draw[arr] (MIAlabel.south) -- (\MIAlabx,\MIAtip);
\draw[arr] (MARlabel.south) -- (\MARlabx,\MARtip);
\node[annot, anchor=east, text=green!25!black] (greentxt) at (\annotxL,\Acy)
      {MIA sufficient conditions hold \\ but (C-WCC) is violated};
\draw[arr] (greentxt.east) -- (\greentipx,\Acy);
\node[annot, anchor=west, text=red!55!black] (redtxt) at (\annotxR,\Acy)
      {Edge cases: The function $\OmegaMAR(V,W)$ \\ depends on $Y$, but $\Cov(\OmegaMAR, Y \mid X) = 0$.};
\draw[arr] (redtxt.west) -- (\slivermid,\Acy);
\end{tikzpicture}%
}
\caption{\label{fig:mia_domination} Regions of the MNAR($V \cup W^+$) space in which $\muMIA$ and $\muMAR$, respectively, are correct for the $V^- = \{Y\}$ and $W = W^-$ cases.}
\end{figure}
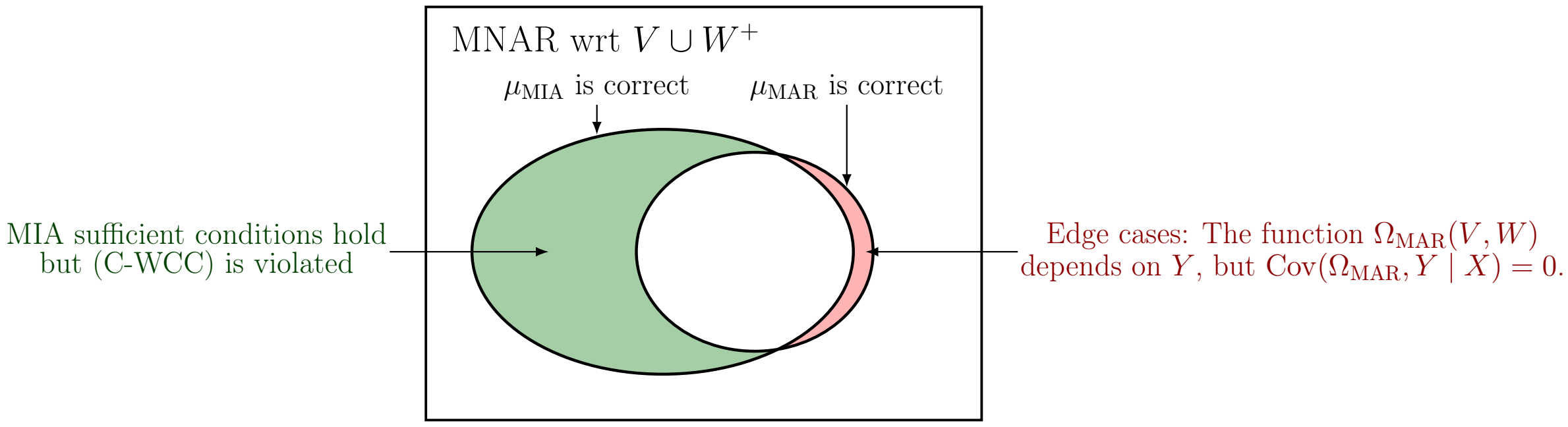

\subsection{Practical considerations regarding MIA}
\label{sec:practical}

 \newcommand{\mohanFootnote}{ }
 
 Even within MIA's domination cases, it is certainly possible for both $\muMAR$ and $\muMIA$ to be incorrect. Thus, analysts should still consider on substantive grounds whether $W$ can be chosen so as to render MIA's sufficient conditions plausible. Figure \ref{fig:dgm_flowchart} provides a taxonomy of DGMs that may serve as a useful heuristic when choosing among analyses for a given study. In terms of causal structures underlying DGMs, mean-dependence that violates (C1-MIA) could occur if (but not only if): (1) $Y$ shares unmeasured common causes with $R$, or $Y$ itself affects $R$; and (2) at least one of the resulting $Y$--$R$ causal paths is not blocked by $X \cup W$. (See the Supplement, Section \ref{supp-sec:DAGs} for a primer on causal paths and structures that block them.) Critically, if causal paths in (1) are specifically between $Y$ and $R_Y$, and there is no subset $Z$ of measured variables that blocks all such paths, then the situation is, in a sense, hopeless because $E[Y \mid X]$ is not nonparametrically identified (box 12 in Figure \ref{fig:dgm_flowchart}).\autocite{mohan2021graphical}\mohanFootnote Assessing (C2-MIA) involves considering if, conditional on $X$, $Y$ may be mean-dependent on members of $W$ that are associated with $R(X \cup W)$ conditional on $X$. (C2-MIA) requires case-by-case assessment on substantive grounds and can be demanding, especially if $X$ or $W$ is high-dimensional. (If $X$ is complete, as in the $V^- = \{Y\}$ case, then this requirement becomes somewhat less stringent, pertaining only to $R(W)$.) 

\begingroup\setstretch{1}

\DeclareRobustCommand{\flowchip}[1]{%
  \raisebox{-0.2ex}{\mbox{\setlength{\fboxsep}{0pt}\setlength{\fboxrule}{0.6pt}%
    \fcolorbox{black}{#1}{\rule{0pt}{6pt}\rule{6pt}{0pt}}}}}

\begin{figure}[!ht]
\centering

\colorlet{fillYes}{red!30}         %
\colorlet{fillMaybe}{orange!25}    %
\colorlet{fillNo}{black!8}         %
\newcommand{\flowhead}[1]{#1}
\newcommand{\flowsub}[1]{\\[3pt]{\fontsize{10.5}{12}\selectfont #1}}
\newcommand{\floweg}[2]{\\[4pt]{\fontsize{10}{11.5}\selectfont\color{black!60}%
  e.g.\ $#1$\\ or $#2$}}
\newcommand{\boxnum}[2]{\node[badge] at (#1.north west) {#2};}

\resizebox{\textwidth}{!}{%
\begin{tikzpicture}[
  auto=false, node distance=0pt, every path/.style={},
  box/.style={draw, line width=1.0pt, text width=5.2cm, align=center,
              inner sep=5pt, font=\fontsize{13}{15.5}\selectfont,
              execute at begin node={\parindent=0pt \parskip=0pt
                \hyphenpenalty=10000 \exhyphenpenalty=0}},
  quest/.style={box, fill=white, minimum height=3.0cm},
  questmnar/.style={box, fill=fillYes, minimum height=3.0cm},
  concl/.style={box, minimum height=2.4cm},
  mnar/.style   ={concl, fill=fillYes},
  maybe/.style  ={concl, fill=fillMaybe},
  notmnar/.style={concl, fill=fillNo},
  arr/.style={draw, -Latex, line width=0.7pt, shorten >=1pt,
              rounded corners=5pt},
  plainline/.style={draw, -, line width=0.7pt},
  ellab/.style={font=\fontsize{11}{13}\selectfont, inner sep=2pt},
  badge/.style={circle, draw, line width=0.9pt, fill=white, inner sep=0pt,
                minimum size=7.0mm, font=\fontsize{11.5}{11.5}\selectfont},
]

\def\cA{0}   \def\cB{6.4}  \def\cC{12.8} \def\cD{19.2}
\def\yT{0}   \def\yM{-4.2} \def\yB{-8.4} \def\yBB{-12.6} \def\yBBB{-17.2}
\def\yFork{-14.8}

\node[quest] (qA) at (\cA, \yT)
  {Does $E[Y \mid X, M {=} 1] = E[Y \mid X]$ hold?
   };

\node[quest] (qB) at (\cB, \yT)
  {Does $V^- \ind R(V) \mid (V^+ \cup W^+)$ hold?
   };

\node[mnar] (gate) at (\cC, \yT)
  {\flowhead{$\mathrm{MNAR}(V \cup W^{+})$ and $\mathrm{MNAR}(V \cup W)$} hold
  \flowsub{(under \ref{ass:positivity}, \ref{ass:no_csi})$^\dagger$}
  };

\node[maybe] (aA) at (\cA, \yM)
  {\flowhead{uCC functional omitting $W$ is correct}
  \flowsub{(by definition)}};

\node[notmnar] (aB) at (\cB, \yM)
  {\flowhead{$\mathrm{MAR}(V \cup W^{+})$ holds; corresponding functional is correct}
  };

\node[questmnar] (qC) at (\cC, \yM)
  {Does $E[Y \mid X, W, M {=} 1] = E[Y \mid X, W]$ hold?
   };

\node[mnar] (aCno) at (\cD, \yM)
  {\flowhead{(C1-MIA) is violated}};

\node[mnar] (aC) at (\cC, \yB)
  {\flowhead{(C1-MIA) holds}};

\node[questmnar] (qE) at (\cC, \yBB)
  {Does (C2-MIA) also hold?};

\node[mnar] (aF) at (\cB, \yBB)
  {\flowhead{$\muMIA$ is correct}};

\node[questmnar] (qD) at (\cD, \yBB)
  {Is $Y$ associated with $R_Y$ via causal path(s) not blocked by any measured variables?$\ddagger$};

\node[mnar] (aD) at (\cC, \yBBB)
  {\flowhead{$E[Y \mid X]$ is not nonparametrically identified}
  \flowsub{(under \ref{ass:positivity}, \ref{ass:no_csi})$^\dagger$}
  };

\node[mnar] (aE) at (\cD, \yBBB)
{
  \flowhead{$E[Y \mid X]$ might be nonparametrically identified by a bespoke functional}
  }
  ;

\draw[arr] (qA.south) -- node[ellab, midway, right=1pt] {Yes} (aA.north);
\draw[arr] (qB.south) -- node[ellab, midway, right=1pt] {Yes} (aB.north);

\draw[arr] (qA.east) -- node[ellab, midway, above=1pt] {No} (qB.west);
\draw[arr] (qB.east) -- node[ellab, midway, above=1pt] {No} (gate.west);

\draw[arr] (gate.south) -- (qC.north);

\draw[arr] (qC.south) -- node[ellab, midway, right=1pt] {Yes} (aC.north);
\draw[arr] (qC.east) -- node[ellab, midway, above=1pt] {No} (aCno.west);

\draw[arr] (aC.south) -- (qE.north);
\draw[arr] (qE.west) -- node[ellab, midway, above=1pt] {Yes} (aF.east);
\draw[arr] (qE.east) -- node[ellab, midway, above=1pt] {No} (qD.west);

\draw[arr] (aCno.south) -- (qD.north);

\coordinate (fork) at (\cD, \yFork);
\draw[plainline] (qD.south) -- (fork);
\draw[arr] (fork) -| node[ellab, pos=0.3, above=1pt] {Yes} (aD.north east);
\draw[arr] (fork) -- node[ellab, midway, right=1pt] {No} (aE.north);

\boxnum{qA}{1}   \boxnum{aA}{2}
\boxnum{qB}{3}   \boxnum{aB}{4}
\boxnum{gate}{5}
\boxnum{qC}{6}   \boxnum{aC}{7}
\boxnum{qE}{8}   \boxnum{aF}{9}
\boxnum{aCno}{10} \boxnum{qD}{11}
\boxnum{aD}{12}  \boxnum{aE}{13}

\draw[draw, -, line width=0.7pt, decorate,
      decoration={brace, mirror, amplitude=6pt, raise=4pt}]
  (aA.south west) -- (aB.south east)
  node[midway, below=14pt, text width=11.0cm, align=center,
       font=\fontsize{11.5}{13.5}\selectfont,
       execute at begin node={\parindent=0pt \parskip=0pt
         \hyphenpenalty=10000 \exhyphenpenalty=10000}]
  {$W^-$ unnecessary for point identification};
\end{tikzpicture}%
}

\caption{Shading indicates whether
  MNAR($V \cup W^+$) holds, which defines the cases we focus on: \flowchip{fillNo}~no;
  \flowchip{fillMaybe}~maybe; \flowchip{fillYes}~yes. $\dagger$: Assumptions \ref{ass:positivity} and \ref{ass:no_csi} together imply that when a member of $V \cup W$ is associated with $R$, it is also associated with membership in every missingness pattern. A formalization is given in Definition \ref{supp-def:pattern_regularity} (Supplement). $\ddagger$: Implies the first sufficient condition for non-identifiability given by Theorem 3 in Mohan \& Pearl (2021) \autocite{mohan2021graphical}.}\label{fig:dgm_flowchart}
\end{figure}
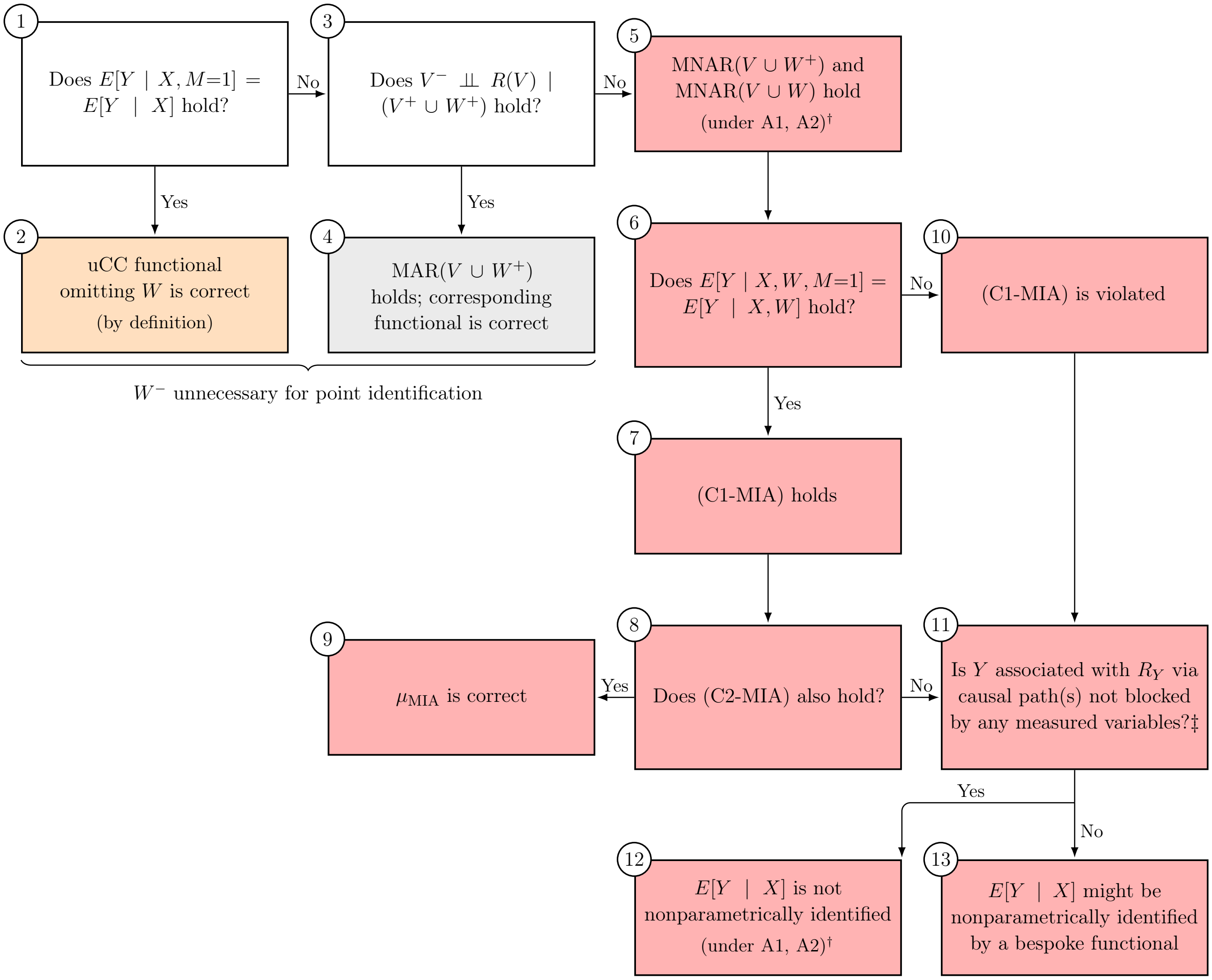

\endgroup

Since the composition of $W$ is at the analyst's discretion, $W$ can be chosen strategically with the MIA conditions in mind. Regarding (C1-MIA), in terms of causal structures underlying DGMs, one should generally include in $W$ any variables that are plausibly common causes of $Y$ and $R(V)$ or intermediates on paths from $Y$ to $R(V)$; if any such variables are not in $X$, then they typically must be in $W$ for (C1-MIA) to be fulfilled. In prospective studies, such variables can sometimes be collected intentionally, analogously to the planned collection of confounders in prospective observational studies. Variables whose inclusion in $W$ is \textit{necessary} to ensure (C1-MIA) are themselves connected to $Y$ via a causal path of some kind; depending on the structure of such path(s), these $W$ members may be associated with $Y$ conditional on $X \cup \Wrc$. If so, and if these necessary $W$ members are also associated with $R(X \cup W)$ conditional on $X$, they risk violating (C2-MIA). It can be helpful to separately consider potential problematic associations with $R(X)$ and with $R(W)$, as follows. Let ${W_{R(X)}} := \{ W_j \in W \, : \, W_j \notind R(X) \mid X \}$ and ${W_{R(W)}} := \{ W_j \in W \, : \, W_j \notind R(W) \mid X \}$, so that $W_R = W_{R(X)} \cup W_{R(W)}$. 

We first consider how to minimize the risk of (C2-MIA)-violating associations between $R(W)$ and variables whose inclusion in $W$ is necessary to fulfill (C1-MIA). Of course, it would be ideal to collect complete versions of all such variables. If this is infeasible, another option is to collect versions of $W$ members whose missingness is random within strata of $X$, so that $W_{R(W)}$ is empty. In prospective studies, intentional random sampling may be possible, as is often done for expensive measurements, such as biomarker assays, microbiome sequencing, or imaging. In retrospective data or when re-analyzing prospective data, missingness in $W$ that renders $W_{R(W)}$ empty may occur if $W$ is recorded only during certain time periods due to intermittent abstraction or auditing. For example, $W$ might indicate quality of care (e.g., protocol adherence, healthcare team experience, classroom instructional quality), which improves both outcomes ($Y$) and thoroughness of data recording ($R(V)$); if quality of care is measured only for randomly audited time periods or individuals, then $W_{R(W)}$ is empty. As a caveat, auditing initiated when poor care is suspected would likely violate (C2-MIA) (as well as MAR($V \cup W$)). Physical limitations of data collection devices can also cause $R(W)$ to be isolated. If $W$ is recorded by a wearable device, missing data may occur when users remove the device or the device fails to record. For example, for stick-on heart monitors, removal and recording failures may have causes entirely unrelated to $V$ and $W$, such as skin irritation or frequency of exposure to water. These arguments require that the quantity being measured ($W$) does not itself affect whether the device is removed or fails to record. 

We now consider (C2-MIA)-violating associations between $R(X)$ and variables whose inclusion in $W$ is necessary. In one of the two key use cases for MIA, namely when $V^- = \{ Y \}$, $R(X)$ is empty, so $W_{R(X)}$ is automatically empty. The pessimistic scenario is that $X$ is incomplete and one has no control over its missingness mechanism. In this case, one should consider on substantive grounds whether $W$ plausibly affects or shares common causes with $R(X)$, in which case (C2-MIA) will likely be violated, along with MAR($V \cup W$). 

Assumptions~\ref{ass:positivity} and~\ref{ass:no_csi} merit some additional discussion. Both can fail in real-world studies; for example, monotone dropout violates Assumption~\ref{ass:positivity}, and variables that are measured only within certain demographic or clinical strata could violate Assumption~\ref{ass:no_csi}. We reiterate that these assumptions are not needed for any of our sufficiency results or for establishing the WCC conditions (Theorem \ref{thm:wcc_conditions}). The assumptions are used to establish the two key cases in which $\muMAR$ is incorrect (Theorems \ref{thm:IPW_with_incomplete_Y} and \ref{thm:IPW_with_Wneg}), and downstream domination results (Theorem \ref{thm:dominating_cases_af4}). Although counterexamples to these domination results can be constructed by relaxing Assumption \ref{ass:positivity} or \ref{ass:no_csi}, such counterexamples are far from generic, requiring quite specific forms of non-positivity or CSI. (Some examples are given in the Supplement, Section~\ref{supp-sec:sims_assumptions_violated}).

\section{Estimation and software}
\label{sec:estimation_and_software}

We now describe several classes of estimators of $\muMIA$, which use iterated conditional expectations (ICE), IPW, augmented IPW (AIPW), or targeted maximum likelihood estimation (TMLE). The ICE estimator is implemented in our R package \texttt{miapack}, \href{https://cran.r-project.org/web/packages/miapack/index.html}{available on CRAN}.\autocite{miapack} To ground concepts, we begin with the case where all members of $X$ are discrete; as we show, existing causal inference software can then easily be repurposed to estimate $\muMIAx$ by subsetting the data to retain only individuals with $X=x$ and $R_{XW}=1$. The Supplement (Section \ref{supp-sec:estimation_general_case}) details a generalized AIPW estimator of $\muMIAx$ that does not require subsetting to $X=x$, thus accommodating cases in which: (1) $X$ contains continuous variables; or (2) $X$ only contains discrete variables but is high-dimensional.

For the discrete-$X$ case, we index the MIA functional and its estimators by the value of $X$ at which they are evaluated, e.g., $\muMIAx$. An ICE estimator for $\muMIAx$ can be obtained in a manner equivalent to g-computation in causal inference:
\begingroup
\addtolength{\jot}{0.5em}
\begin{align}
\label{eq:muhatMIAice}
\muhatMIAice(x) &= \frac{ \sum_{i=1}^n 1 \{ X^i = x \} \, R^i_{XW} \, \widehat g(X^i, W^i) }{ \sum_{i=1}^n 1 \{ X^i = x \} \, R^i_{XW}}.
\end{align}
\endgroup
where superscript $i$ indexes the individual in the dataset and $\widehat g(x, w)$ is an estimator of $g(x, w) := E[ Y \mid X = x, W = w, M = 1 ]$. That is, $\muhatMIAice(x)$ is equivalent to a g-computation estimator for the expectation of $Y$ among the subset of individuals for whom $X=x$ and $R_{XW} = 1$, where $W$ plays the role of adjustment covariates in g-computation. A variance estimator can be obtained via nonparametric bootstrapping.

As with g-computation, a noniterative conditional expectations estimator (NICE) is also possible, in which $p(w \mid x, R_{XW} = 1)$ is estimated semiparametrically or parametrically. This estimator will often be more prone to misspecification than is the ICE estimator, so is relegated to the Supplement (Section \ref{supp-sec:nice_estimator}). Alternatively, an IPW estimator is: 
\begingroup
\addtolength{\jot}{0.5em}
\begin{align}
\label{eq:muhatMIAipw}
\muhatMIAipw(x) &= \frac{ \sum_{i=1}^n 1 \{ X^i = x \} \, R^i_{XW}  \, \frac{R_Y^i}{\widehat \pi(X^i, W^i)} Y^i }{ \sum_{i=1}^n 1 \{ X^i = x \} \, R^i_{XW}  },
\end{align}
\endgroup
where $\widehat \pi(x, w)$ is an estimator of $\pi(x, w) := p(R_Y = 1 \mid X=x, W=w, R_{XW} = 1)$. Finally, the AIPW estimator solves the estimating equation given by the nonparametric efficient influence function of $\muMIAx$. This estimator is:
\begingroup
\addtolength{\jot}{0.5em}
\begin{align}
\label{eq:muhatMIAaipw}
\muhatMIAaipw(x) &= \frac{ \sum_{i=1}^n 1 \{ X^i = x \} \, R^i_{XW}  \, \Big[ \widehat g(X^i, W^i) + \frac{R_Y^i}{\widehat \pi(X^i, W^i)} \left( Y^i - \widehat g(X^i, W^i) \right) \Big] }{ \sum_{i=1}^n 1 \{ X^i = x \} \, R^i_{XW} }.
\end{align}
\endgroup
This estimator is nonparametrically efficient and doubly robust, meaning that it is consistent if $\widehat g(x,w)$ or $\widehat \pi(x,w)$ (but not necessarily both) is consistent. The TMLE estimator $\muhatMIAtmle(x)$ is asymptotically equivalent to $\muhatMIAaipw(x)$ and is described in the Supplement, Section \ref{supp-sec:discrete_X_aipw_tmle}.

Estimators $\widehat g (x, w)$ and $\widehat \pi (x, w)$ can be obtained by fitting parametric models for $g(x, w)$ and $\pi (x, w)$. Alternatively, if AIPW or TMLE is used, $g (x, w)$ and $\pi (x, w)$ can be estimated using data-adaptive procedures, such as a super learner.\autocite{vanderlaan2007superlearner}
Estimators $\muhatMIAaipw(x)$ and $\muhatMIAtmle(x)$ are consistent and asymptotically normally distributed if $\widehat g (x, w)$ and $\widehat \pi (x, w)$ are consistent. When data-adaptive procedures are used, the product of the convergence rates of $\widehat g (x, w)$ and $\widehat \pi (x, w)$ is faster than $1/\sqrt{n}$ (Supplement, Section \ref{supp-sec:estimation_general_case}). The asymptotic variance of $\sqrt{n} \{ \muhatMIAaipw (x) - \muMIAx \}$ and $\sqrt{n} \{ \muhatMIAtmle (x) - \muMIAx \}$ has a closed form arising from the efficient influence function (Supplement, Section \ref{supp-sec:discrete_X_aipw_tmle}).

The AIPW and TMLE estimators can be implemented using existing software (e.g.,\ R packages \texttt{gcomputation}\autocite{gcomputation} or \texttt{tmle}\autocite{tmle}) for estimating the average causal effect of a non-randomized binary treatment on an outcome via adjustment for measured covariates. Simply define $W$ to be the covariates, $R_Y$ to be the exposure, and $Y$ to be the outcome. For individuals with $R_Y = 0$, set $Y$ equal to zero (this ensures that the average causal effect estimated by the software is equal to the mean potential outcome when exposed). Restrict the analysis to individuals with $R_{XW} = 1$ and $X=x$. The standard AIPW and TMLE estimators of the average causal effect are then equivalent to $\muhatMIAaipw(x)$ and $\muhatMIAtmle(x)$, respectively.

\section{Applied example}
\label{sec:applied_example}

Building on prior work led by author WZ,\autocite{zhang2024population} we examined whether individuals who initially identified as non-heterosexual were more or less likely to experience changes in sexual identity over time, compared to those who initially identified as heterosexual. Methods are detailed in the Supplement (Section \ref{supp-sec:applied}); here we briefly summarize. We used data from the 2014 subcohort of the Stockholm Public Health Cohort (SPHC);\autocite{ces_regionstockholm_sphc_data, svensson2013cohort} see the section Data Availability for information on accessing the data. At baseline in 2014, stratified random samples were drawn from the population of Stockholm County using the Total Population Register. Participants were followed up through a survey in 2021. The analysis was restricted to individuals aged 25--60 years at baseline.

The analysis and auxiliary variables were as follows. The predictor of interest, $X$, was sexual identity in 2014, coded as non-heterosexual [$X=1$] vs. heterosexual [$X=0$]. The outcome, $Y$, was change in sexual identity from 2014 to 2021, coded as yes [$Y=1$] vs. no [$Y=0$]. Changes in sexual identity may reflect, for example, changes in subjective experiences of attraction or changes in social self-identification. There were three complete auxiliary variables: age, sex, and country of birth. There was also an incomplete auxiliary variable: occupation ($W_1$). This variable was binned into three categories: manual \& field trades, service \& support, and expertise \& leadership. We believe, on substantive grounds, that occupation is an important auxiliary variable: workplaces can function as sources of social support or of stigmatization, both of which may shape individuals' self-identification ($Y$).\autocite{devine2007sexual} Additionally, occupation may affect attrition from the cohort ($R_Y$). Although $X$ was complete, $Y$ had 30.6\% missing data primarily due to attrition, and $W_1$ had 30.2\% missing data. Among individuals working in manual and field trades occupations, 35.9\% had missing data on $Y$, compared with 31.5\% among those in service and support and 25.5\% among those in expertise and leadership.

\begin{figure}[!ht]
\centering
\begin{minipage}[c]{0.30\linewidth}
  \centering
  \resizebox{\linewidth}{!}{%
  \begin{tikzpicture}
  \tikzset{
    ell/.style={draw,fill=white, inner sep=2pt, line width=1.5pt},
    swig vsplit={gap=3pt, line color right=red},
    every path/.style={>=Stealth, line width=1.2pt, >={Stealth[scale=1.0]}}
  }
     \node[name=X, ell, shape=ellipse] at (0,0) {$X$};

     \node[name=Y, ell, shape=ellipse, fill={rgb:black,1;white,2},
        right=50mm of X] {$Y$};

     \node[name=RY, ell, shape=ellipse, below=10mm of Y] {$R_Y$};
     \path let \p1 = (X), \p2 = (Y) in
     coordinate (MidAB) at ($(\p1)!.5!(\p2)$); %
     \node[name=W, ell, shape=ellipse, fill={rgb:black,1;white,2}] at (MidAB) {$W_1$};
     \node[name=RW, ell, shape=ellipse, below=10mm of W] {$R_{W_1}$};

      \draw[->, line width=1.2pt, >=stealth] (X) edge (W);
      \draw[->, line width=1.2pt, >=stealth] (X) edge (RW);
      \draw[->, line width=1.2pt, >=stealth] (X) edge (RY);
      \draw[->, line width=1.2pt, >=stealth] (X)[out=30, in=150] edge (Y);
      \draw[->, line width=1.2pt, >=stealth] (W) edge (Y);
      \draw[->, line width=1.2pt, >=stealth] (W) edge (RY);

  \end{tikzpicture}%
  }
\end{minipage}%
\hfill
\begin{minipage}[c]{0.68\linewidth}
  \centering
  \includegraphics[width=\linewidth]{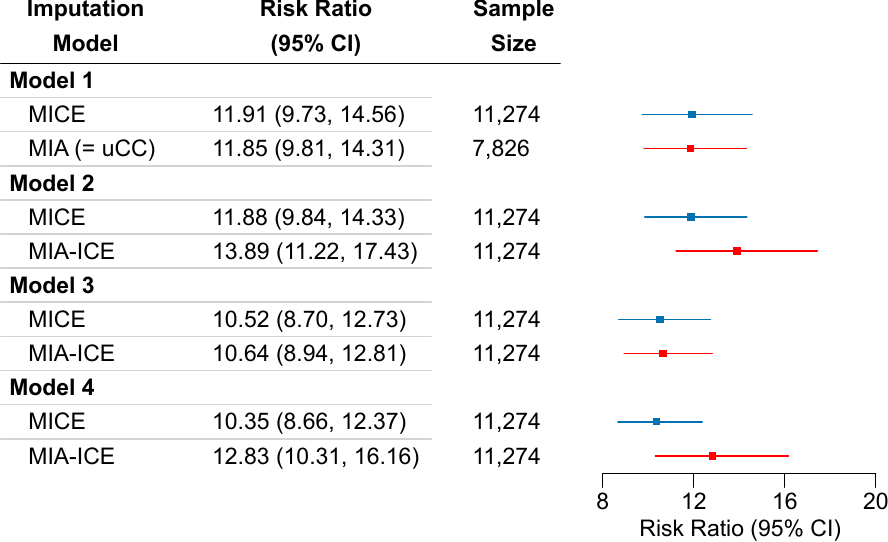}
\end{minipage}
\caption{\textbf{Left:} A possible causal structure representing the applied example.
$X$: Sexual identity in 2014 (non-heterosexual vs.\ heterosexual). $Y$: Change in
sexual identity from 2014 to 2021 (yes vs.\ no). $W_1$: Occupation (an incomplete
auxiliary variable). The three complete auxiliary variables are omitted for
simplicity. Gray nodes represent incomplete variables.
\textbf{Right:} Effect estimates by model, where risk ratios above 1 indicate a
positive association between initial non-heterosexual identity and subsequent
identity change. Model 1 included no auxiliary variables. Model 3 included age, sex,
and country of birth. Models 2 and 4 had the same specifications as Models 1 and 3,
respectively, but additionally included occupation. Abbreviations: uCC, unweighted complete-case
analysis; CI, confidence interval; MICE, multiple imputation by chained equations.
\label{fig:applied_example_dag}\label{fig:applied_forest_plot}}
\end{figure}
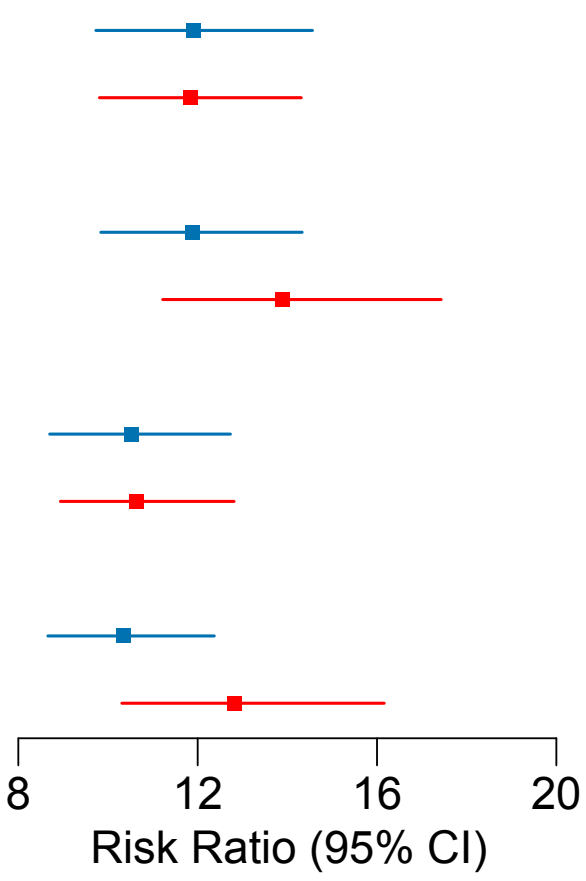

Figure \ref{fig:applied_example_dag} illustrates one plausible DGM, where $X$ affects $Y$, partially mediated by $W_1$. Additionally, $X$ affects both $R_{W_1}$ and $R_Y$. The three complete auxiliary variables, omitted from the figure for simplicity, are presumed common causes of $R_Y$ and $Y$. Critically, the incomplete auxiliary variable $W_1$ is also a presumed common cause of $R_Y$ and $Y$. Thus, we believe MNAR($V \cup W^+$) likely holds; by Proposition \ref{prop:wneg_cant_fix_mnar}, this implies MNAR($V \cup W$) also holds. If MNAR does hold, and Assumptions \ref{ass:positivity} and \ref{ass:no_csi} hold, then $\muMAR$ is incorrect. On the other hand, if the DGM depicted in Figure \ref{fig:applied_example_dag} is correct, then the sufficient conditions for MIA hold. (In particular, $W_R = \emptyset$, so (C2-MIA) holds trivially.) However, because $V^- = \{Y\}$ here, it is not necessary to assume the specific MNAR mechanism in Figure \ref{fig:applied_example_dag} to conclude that $\muMIA$ dominates $\muMAR$ over the class of MNAR distributions. Therefore, the only untestable assumptions on the DGM needed to conclude (from Theorem \ref{thm:dominating_cases_af4}) that $\muMIA$ is the ``safer'' choice for this problem are that any form of MNAR holds and that Assumptions \ref{ass:positivity} and \ref{ass:no_csi} hold. 

We obtained $\muhatMAR$ and $\muhatMIAice$ using four different sets of auxiliary variables: \textit{Model 1} used no auxiliary variables; \textit{Model 2} used occupation; \textit{Model 3} used age, sex, and country of birth; and \textit{Model 4} used occupation, age, sex, and country of birth. We used the ICE estimator due to its ease of implementation and desirable finite-sample properties in our simulation study (Section \ref{sec:sims} below).
To estimate $\muMAR$, we used multiple imputation by chained equations (MICE) with 50 imputations.\autocite{van2011mice} Fully conditional imputation was conducted using predictive mean matching, using $X$, $Y$, and the auxiliary variables as predictors in the mean model for each variable to be imputed. (We chose MICE instead of IPW because no general software is available to estimate IPW under nonmonotone MAR, and for our purposes, programming an implementation general enough to handle this dataset would be prohibitively complicated [B. Sun, personal communication, April 4, 2025].) 

Since $\muMIA$ coincides with the unweighted complete-case functional $E[Y \mid X=x, M=1]$ when there are no auxiliary variables, we refer to MIA as uCC (unweighted complete-case analysis) for Model 1. For MICE and uCC, the outcome model was a Poisson regression of $Y$ on $X$ alone, with robust variance estimation. Because $X$ is binary, the outcome model is saturated, yielding a risk ratio that does not depend on the choice of link function. For MIA estimation via ICE, the conditional outcome model $g(x,w) := E[Y \mid X=x, W=w, M=1]$ was a logistic regression model including all main effects and two-way interactions among members of $X \cup W$. Per Eq. (\ref{eq:muhatMIAice}), $\muhatMIAice$ was obtained as the sample average of $\widehat g(x,w)$ within the $(X=x, R_{XW} = 1)$ stratum of the dataset. Because $X$ is binary, this is equivalent to modeling the outer expectation in $\muMIA$ via a saturated regression model. 
 
Figure \ref{fig:applied_forest_plot} presents risk ratios with confidence intervals for each model. When all auxiliary variables were complete (Models 1 and 3), MICE and MIA/uCC yielded quite similar estimates. However, in Models 2 and 4, which introduced the incomplete auxiliary variable, MIA estimates suggested a stronger association of initial sexual identity with change in sexual identity than did MICE. Diagnostic plots did not indicate model-fit or convergence problems for either MIA or MICE (Supplement, Section \ref{supp-sec:applied}).

\section{Simulation studies}
\label{sec:sims}

We conducted simulation studies comparing the performance of estimators of $\muMAR$ and two estimators of $\muMIA$ (using ICE and TMLE, respectively) under various forms of MNAR. In Study 1, all DGMs violated (C-WCC) but some did fulfill (C1-MIA) and (C2-MIA), and Assumptions \ref{ass:positivity} and \ref{ass:no_csi} held. In Study 2, we explored performance when $W$ was high-dimensional. In Study 3, we introduced violations and near-violations of Assumptions \ref{ass:positivity} and \ref{ass:no_csi}, but (C-WCC) sometimes held. 

Here we briefly summarize the methods and results, focusing only on some key scenarios for brevity. Detailed methods and comprehensive results for all scenarios are provided in the Supplement (Section \ref{supp-sec:sim_study}). Comprehensive results can also be browsed via interactive tables available at \url{https://www.mayamathur.com/mia-simulations} and in the supplementary file \texttt{sim\_results\_table.html}. The interactive tables also include results for MAR-assuming multiple imputation and additional performance metrics.

\subsection{Study 1: Univariate, incomplete $W$}

\newcommand{\scaleCoef}{.8}  %
\newcommand{\textSize}{\footnotesize} %
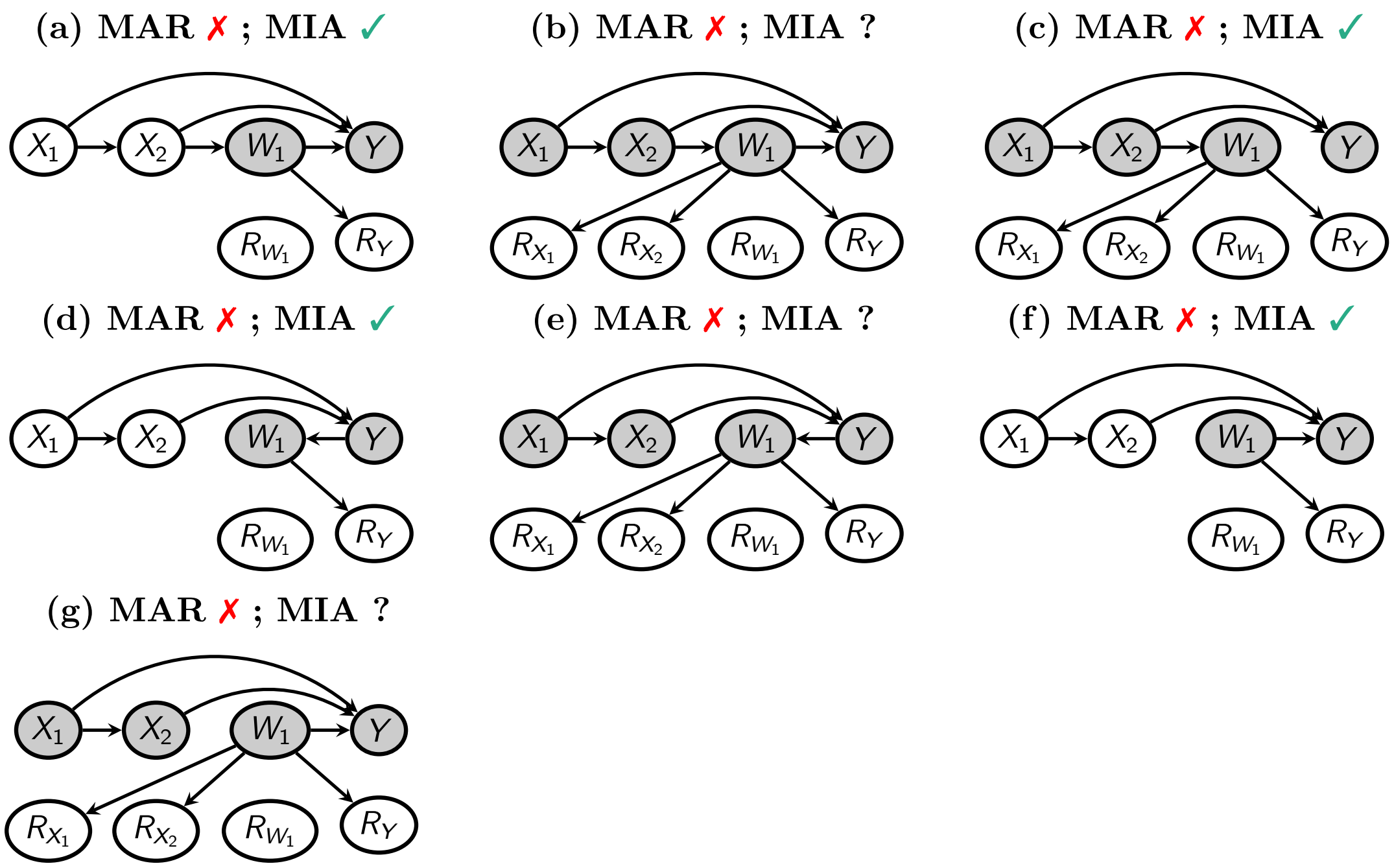
\begin{figure}[!ht]
\centering
\setlength{\tabcolsep}{2mm}
\renewcommand{\arraystretch}{1} %
\begin{tabular}{ccc}
\textbf{\textSize (a) MAR \xmark; MIA \cmark} & \textbf{ \textSize (b) MAR \xmark; MIA \qmark} & \textbf{ \textSize (c) MAR \xmark; MIA \cmark} \\
\scalebox{\scaleCoef}{ \begin{tikzpicture}
\sansmath
\tikzset{
  line width=1.2pt,
  >=stealth,
  ell/.style={draw, fill=white, ellipse, inner sep=2pt, line width=1.5pt},
  shaded/.style={draw, fill=gray!40, ellipse, inner sep=2pt, line width=1.5pt}
}

\node[name=X1, ell]                         {$X_1$};
\node[name=X2, ell, right=5mm of X1]       {$X_2$};
\node[name=W,  shaded, right=5mm of X2]    {$W_1$};
\node[name=Y,  shaded, right=5mm of W]     {$Y$};
\node[name=RW, ell, below=5mm of W]        {$R_{W_1}$};
\node[name=RY, ell, below=5mm of Y]        {$R_Y$};

\draw[->,line width=1.2pt,>=stealth]
(X1) edge (X2)
(X2) edge (W)
(W) edge (Y)
(W) edge (RY);
\draw[->,line width=1.2pt,>=stealth] (X2) to[out=30,in=150] (Y);
\draw[->,line width=1.2pt,>=stealth] (X1) to[out=40,in=140] (Y);

\end{tikzpicture} } &
\scalebox{\scaleCoef}{ \begin{tikzpicture}
\sansmath
\tikzset{
  line width=1.2pt,
  >=stealth,
  ell/.style={draw, fill=white, ellipse, inner sep=2pt, line width=1.5pt},
  shaded/.style={draw, fill=gray!40, ellipse, inner sep=2pt, line width=1.5pt}
}

\node[name=X1, shaded]                         {$X_1$};
\node[name=X2, shaded, right=5mm of X1]       {$X_2$};
\node[name=W,  shaded, right=5mm of X2]    {$W_1$};
\node[name=Y,  shaded, right=5mm of W]     {$Y$};
\node[name=RX1, ell, below=5mm of X1]        {$R_{X_1}$};
\node[name=RX2, ell, below=5mm of X2]        {$R_{X_2}$};
\node[name=RW, ell, below=5mm of W]        {$R_{W_1}$};
\node[name=RY, ell, below=5mm of Y]        {$R_Y$};

\draw[->,line width=1.2pt,>=stealth]
(X1) edge (X2)
(X2) edge (W)
(W) edge (Y)
(W) edge (RX1)
(W) edge (RX2)
(W) edge (RY);
\draw[->,line width=1.2pt,>=stealth] (X2) to[out=30,in=150] (Y);
\draw[->,line width=1.2pt,>=stealth] (X1) to[out=40,in=140] (Y);

\end{tikzpicture} } &
\scalebox{\scaleCoef}{ \begin{tikzpicture}
\sansmath
\tikzset{
  line width=1.2pt,
  >=stealth,
  ell/.style={draw, fill=white, ellipse, inner sep=2pt, line width=1.5pt},
  shaded/.style={draw, fill=gray!40, ellipse, inner sep=2pt, line width=1.5pt}
}

\node[name=X1, shaded]                         {$X_1$};
\node[name=X2, shaded, right=5mm of X1]       {$X_2$};
\node[name=W,  shaded, right=5mm of X2]    {$W_1$};
\node[name=Y,  shaded, right=5mm of W]     {$Y$};
\node[name=RX1, ell, below=5mm of X1]        {$R_{X_1}$};
\node[name=RX2, ell, below=5mm of X2]        {$R_{X_2}$};
\node[name=RW, ell, below=5mm of W]        {$R_{W_1}$};
\node[name=RY, ell, below=5mm of Y]        {$R_Y$};

\draw[->,line width=1.2pt,>=stealth]
(X1) edge (X2)
(X2) edge (W)
(W) edge (RX1)
(W) edge (RX2)
(W) edge (RY);
\draw[->,line width=1.2pt,>=stealth] (X2) to[out=30,in=150] (Y);
\draw[->,line width=1.2pt,>=stealth] (X1) to[out=40,in=140] (Y);

\end{tikzpicture} } \\
\textbf{ \textSize (d) MAR \xmark; MIA \cmark} & \textbf{ \textSize (e) MAR \xmark; MIA \qmark} & \textbf{\textSize (f) MAR \xmark; MIA \cmark} \\
\scalebox{\scaleCoef}{ \begin{tikzpicture}
\sansmath
\tikzset{
  line width=1.2pt,
  >=stealth,
  ell/.style={draw, fill=white, ellipse, inner sep=2pt, line width=1.5pt},
  shaded/.style={draw, fill=gray!40, ellipse, inner sep=2pt, line width=1.5pt}
}

\node[name=X1, ell]                         {$X_1$};
\node[name=X2, ell, right=5mm of X1]       {$X_2$};
\node[name=W,  shaded, right=5mm of X2]    {$W_1$};
\node[name=Y,  shaded, right=5mm of W]     {$Y$};
\node[name=RW, ell, below=5mm of W]        {$R_{W_1}$};
\node[name=RY, ell, below=5mm of Y]        {$R_Y$};

\draw[->,line width=1.2pt,>=stealth]
(X1) edge (X2)
(Y) edge (W)
(W) edge (RY);
\draw[->,line width=1.2pt,>=stealth] (X2) to[out=30,in=150] (Y);
\draw[->,line width=1.2pt,>=stealth] (X1) to[out=40,in=140] (Y);

\end{tikzpicture} } &
\scalebox{\scaleCoef}{ \begin{tikzpicture}
\sansmath
\tikzset{
  line width=1.2pt,
  >=stealth,
  ell/.style={draw, fill=white, ellipse, inner sep=2pt, line width=1.5pt},
  shaded/.style={draw, fill=gray!40, ellipse, inner sep=2pt, line width=1.5pt}
}

\node[name=X1, shaded]                         {$X_1$};
\node[name=X2, shaded, right=5mm of X1]       {$X_2$};
\node[name=W,  shaded, right=5mm of X2]    {$W_1$};
\node[name=Y,  shaded, right=5mm of W]     {$Y$};
\node[name=RX1, ell, below=5mm of X1]        {$R_{X_1}$};
\node[name=RX2, ell, below=5mm of X2]        {$R_{X_2}$};
\node[name=RW, ell, below=5mm of W]        {$R_{W_1}$};
\node[name=RY, ell, below=5mm of Y]        {$R_Y$};

\draw[->,line width=1.2pt,>=stealth]
(X1) edge (X2)
(Y) edge (W)
(W) edge (RX1)
(W) edge (RX2)
(W) edge (RY);
\draw[->,line width=1.2pt,>=stealth] (X2) to[out=30,in=150] (Y);
\draw[->,line width=1.2pt,>=stealth] (X1) to[out=40,in=140] (Y);

\end{tikzpicture} } &
\scalebox{\scaleCoef}{ \begin{tikzpicture}
\sansmath
\tikzset{
  line width=1.2pt,
  >=stealth,
  ell/.style={draw, fill=white, ellipse, inner sep=2pt, line width=1.5pt},
  shaded/.style={draw, fill=gray!40, ellipse, inner sep=2pt, line width=1.5pt}
}

\node[name=X1, ell]                         {$X_1$};
\node[name=X2, ell, right=5mm of X1]       {$X_2$};
\node[name=W,  shaded, right=5mm of X2]    {$W_1$};
\node[name=Y,  shaded, right=5mm of W]     {$Y$};
\node[name=RW, ell, below=5mm of W]        {$R_{W_1}$};
\node[name=RY, ell, below=5mm of Y]        {$R_Y$};

\draw[->,line width=1.2pt,>=stealth]
(X1) edge (X2)
(W) edge (Y)
(W) edge (RY);
\draw[->,line width=1.2pt,>=stealth] (X2) to[out=30,in=150] (Y);
\draw[->,line width=1.2pt,>=stealth] (X1) to[out=40,in=140] (Y);

\end{tikzpicture} } \\
\textbf{ \textSize (g) MAR \xmark; MIA \qmark} &  &  \\
\scalebox{\scaleCoef}{ \begin{tikzpicture}
\sansmath
\tikzset{
  line width=1.2pt,
  >=stealth,
  ell/.style={draw, fill=white, ellipse, inner sep=2pt, line width=1.5pt},
  shaded/.style={draw, fill=gray!40, ellipse, inner sep=2pt, line width=1.5pt}
}

\node[name=X1, shaded]                         {$X_1$};
\node[name=X2, shaded, right=5mm of X1]       {$X_2$};
\node[name=W,  shaded, right=5mm of X2]    {$W_1$};
\node[name=Y,  shaded, right=5mm of W]     {$Y$};
\node[name=RX1, ell, below=5mm of X1]        {$R_{X_1}$};
\node[name=RX2, ell, below=5mm of X2]        {$R_{X_2}$};
\node[name=RW, ell, below=5mm of W]        {$R_{W_1}$};
\node[name=RY, ell, below=5mm of Y]        {$R_Y$};

\draw[->,line width=1.2pt,>=stealth]
(X1) edge (X2)
(W) edge (Y)
(W) edge (RX1)
(W) edge (RX2)
(W) edge (RY);
\draw[->,line width=1.2pt,>=stealth] (X2) to[out=30,in=150] (Y);
\draw[->,line width=1.2pt,>=stealth] (X1) to[out=40,in=140] (Y);

\end{tikzpicture} } &
&
\end{tabular}
\caption{DGMs used in the simulation study. Gray nodes are incomplete variables. \cmark indicates that the sufficient (and, in the case of $\muMAR$, necessary) conditions for the correctness of $\muMAR$ and $\muMIA$, respectively, hold. \xmark indicates that the necessary and sufficient conditions for $\muMAR$ are violated, so the corresponding estimators will be inconsistent. \qmark indicates that the sufficient (but not necessary) conditions for $\muMIA$ to be correct are violated.}
\label{fig:sim_dags}
\end{figure}

\subsubsection{Data generation}

We simulated from DGMs depicted in Figure \ref{fig:sim_dags} and used six different sample sizes ranging from $N = 200$ to $N=10{,}000$. We ran 1,000 simulation replicates for each unique combination of simulation parameters. All variables were binary except $Y$, which was generated from a linear regression model that included main effects of all parents of $Y$ (i.e., its direct causes) and all 2-way interactions between these parents. We considered two estimands:
\begingroup
\addtolength{\jot}{0.5em}
\begin{align*}
\beta_0 &:= E[Y \mid X_1 = 0, X_2 = 0]\\
\beta_{X_2} &:= E[Y \mid X_1 = 0, X_2 = 1] - E[Y \mid X_1 = 0, X_2 = 0].
\end{align*}
\endgroup

Under appropriate causal assumptions, $\beta_{X_2}$ represents a conditional average treatment effect of $X_2$, holding constant the confounder $X_1 = 0$. The effect $\beta_{X_2}$ was 3.22 for DGMs (a) and (b), in which $X_2$ also affects $Y$ through $W_1$, and 2 for the remaining DGMs. The intercept $\beta_0$ was $0.54$ for DGMs (a) and (b); $0$ for (c), (d), and (e); and $1$ for (f) and (g). 
In all cases, $W = W^-$ constituted a single variable $W_1$ (as shown in Figure \ref{fig:sim_dags}). Since $W = W^-$, by Theorem \ref{thm:dominating_cases_af4}, $\muMIA$ dominates $\muMAR$ over the class of MNAR distributions satisfying Assumptions \ref{ass:positivity} and \ref{ass:no_csi}. Each missingness indicator $R_j \in \{ R_{X_1}, R_{X_2}, R_{W_1}, R_{Y} \}$ either had no parents or had $W_1$ as its sole parent. These variables were simulated as:
$$ R_j \sim
\begin{cases}
\mathrm{Bern}(0.5), & \text{if $R_j$ has no parents}, \\
\mathrm{Bern}\!\left( \operatorname{expit}(-1 + 3 W_1) \right), & \text{if $W_1$ is the sole parent of $R_j$.}
\end{cases} $$

\subsubsection{Estimation}

To estimate $\muMAR$, we implemented the Sun \& Tchetgen Tchetgen (STT) estimator using custom R code. In our DGMs, the densities $p(y \mid x)$ implied by the models for $p(y \mid x, w)$ and $p(w \mid x)$ were typically mixture distributions without a tractable parametric form. Accordingly, we used Huber-White (also known as HC0) heteroskedasticity-consistent standard errors\autocite{huber1967behavior} for the outcome model. To obtain $\muhatMIAice$, we used our R package \texttt{miapack}, using linear regression to estimate $E[Y \mid X, W, M=1]$. Confidence intervals were obtained by bootstrapping. To estimate $\muhatMIAtmle$, we used the R package \texttt{tmle}, using logistic regression to estimate the response model $\pi(x,w)$.

\subsubsection{Theoretical predictions}

Regarding $\muMAR$, (C-WCC) is violated in all DGMs. In DGMs (a), (d), and (f), this follows immediately from Theorem \ref{thm:IPW_with_incomplete_Y} because $V^- = \{Y\}$. However, in those DGMs, $\muMIA$ is correct because (C1-MIA) and (C2-MIA) hold. In the remaining DGMs, all analysis variables are incomplete ($V^- = \{ Y, X_1, X_2 \}$), so Theorem \ref{thm:IPW_with_incomplete_Y} does not apply. Nevertheless, in DGM (c), (C1-MIA) holds, and so does (C2-MIA) because $Y \ind W_1 \mid \{ X_1, X_2 \}$. In the remaining DGMs (b), (e), and (g), the sufficient conditions for MIA are violated: in particular, the edges $W_1 \to R_{X_1}$ and $W_1 \to {R_{X_2}}$ and $Y \to W_1$ or $W_1 \to Y$ violate (C2-MIA). To summarize, theory indicates that $\muhatMIA$ will be consistent (and hence asymptotically unbiased) for at least DGMs (a), (c), (d), and (f); and that $\muhatMAR$ will be inconsistent for all DGMs. For brevity, when discussing consistency and unbiasedness, we refer to the parameter vector $(\beta_0, \beta_{X_2})$ unless specifically referring to one component.

\subsubsection{Results}

Table \ref{tb:sim_n1000_wdim1} presents bias, RMSE, and CI coverage results for a midrange sample size of $N=1{,}000$. 

\vspace{-0.5cm} \paragraph{Performance of MIA in DGMs fulfilling its sufficient conditions.}

In DGMs (a), (c), (d), and (f), at sample sizes $N \ge 2{,}000$, simulation results demonstrated the asymptotics guaranteed from theory: $\muhatMIAtmle$ and $\muhatMIAice$ were both approximately unbiased (bias at most 0.01 in absolute value), and CI coverage was at least 93\%. At the smallest sample size ($N=200$), $\muhatMIAice$ remained approximately unbiased, whereas $\muhatMIAtmle$ was substantially biased for $\beta_{X_2}$ in DGMs (a) and (f) (0.12 and 0.22, respectively). CIs for $\muhatMIAice$ exhibited mild undercoverage (as low as 90\%) for most $N=200$ scenarios, but CIs for $\muhatMIAtmle$ exhibited more severe undercoverage (as low as 72\%). The relatively better small-sample performance of $\muhatMIAice$ over $\muhatMIAtmle$ may reflect instability in the estimated TMLE weights at small sample sizes, whereas $\muhatMIAice$ does not use weights. At $N=200$, the MIA estimators and $\muhatMAR$ had comparable RMSE; as the sample size increased, the advantage of the MIA estimators generally grew, since the bias of $\muhatMAR$ did not diminish. At $N=10{,}000$, the RMSE of each MIA estimator was at most that of $\muhatMAR$ for both estimands in all four DGMs.

\vspace{-0.5cm} \paragraph{Performance of MAR.}

Since all DGMs in Study 1 violated (C-WCC), $\muhatMAR$ was usually biased. The exceptions concerned $\beta_{X_2}$: at $N \ge 5{,}000$, the bias of $\muhatMAR$ for $\beta_{X_2}$ was at most 0.03 in absolute value in DGMs (c), (d), (f), and (g). However, $\muhatMAR$ remained biased for $\beta_0$ in all DGMs except (c), where its bias was small ($-0.03$).

\vspace{-0.5cm} \paragraph{Relative performance of MIA and MAR in DGMs violating conditions for both.}

In two such DGMs, $\muhatMAR$ and its CIs either substantially or modestly outperformed the MIA estimators in terms of coverage (DGMs (b) and (e) respectively). In DGM (g), $\muhatMAR$ outperformed the MIA estimators for estimation of $\beta_{0}$, but the opposite was true for $\beta_{X_2}$. In these scenarios, $\muhatMAR$ usually had better RMSE than the MIA estimators. 

\newcommand{\colwidth}{1.42cm}   %
\newcommand{\dagwidth}{0.70cm}
\newcommand{\coefwidth}{0.95cm}

\begin{table}[h]
\centering
\scriptsize
\setlength{\tabcolsep}{3pt}         %
\renewcommand{\arraystretch}{1.15}  %
\begin{tabular}{
    >{\centering\arraybackslash}p{\dagwidth}
    >{\centering\arraybackslash}p{\coefwidth}
    >{\centering\arraybackslash}p{\colwidth} >{\centering\arraybackslash}p{\colwidth} >{\centering\arraybackslash}p{\colwidth}
    >{\centering\arraybackslash}p{\colwidth} >{\centering\arraybackslash}p{\colwidth} >{\centering\arraybackslash}p{\colwidth}
    >{\centering\arraybackslash}p{\colwidth} >{\centering\arraybackslash}p{\colwidth} >{\centering\arraybackslash}p{\colwidth}
}
\toprule
\textbf{DGM} & \textbf{} &
\multicolumn{3}{c}{\textbf{Bias}} &
\multicolumn{3}{c}{\textbf{RMSE}} &
\multicolumn{3}{c}{\textbf{CI coverage}} \\
\cmidrule(lr){3-5} \cmidrule(lr){6-8} \cmidrule(lr){9-11}
 & & MAR (IPW) & MIA-T & MIA-I & MAR (IPW) & MIA-T & MIA-I & MAR (IPW) & MIA-T & MIA-I \\
\midrule
\multirow{2}{*}{(a)} & $\widehat \beta_{X_2}$
    & \xmark / -0.06 & \cmark / \textbf{-0.01} & \cmark / \textbf{-0.01} & 0.24 & 0.24 & 0.22 & \textbf{0.95} & \textbf{0.93} & \textbf{0.95} \\
 & $\widehat \beta_0$
    & \xmark / 0.20 & \cmark / \textbf{0.00} & \cmark / \textbf{0.00} & 0.24 & 0.14 & 0.14 & 0.79 & \textbf{0.95} & \textbf{0.95} \\
\midrule
\multirow{2}{*}{(b)} & $\widehat \beta_{X_2}$
    & \xmark / -0.38 & \qmark / -0.83 & \qmark / -0.84 & 0.69 & 0.89 & 0.89 & 0.66 & 0.20 & 0.21 \\
 & $\widehat \beta_0$
    & \xmark / 0.68 & \qmark / 1.06 & \qmark / 1.06 & 0.87 & 1.09 & 1.08 & 0.55 & 0.03 & 0.04 \\
\midrule
\multirow{2}{*}{(c)} & $\widehat \beta_{X_2}$
    & \xmark / \textbf{0.00} & \cmark / \textbf{-0.01} & \cmark / \textbf{0.00} & 0.35 & 0.26 & 0.25 & 0.89 & 0.92 & \textbf{0.94} \\
 & $\widehat \beta_0$
    & \xmark / \textbf{-0.02} & \cmark / \textbf{0.01} & \cmark / \textbf{0.00} & 0.26 & 0.22 & 0.20 & 0.90 & 0.88 & 0.92 \\
\midrule
\multirow{2}{*}{(d)} & $\widehat \beta_{X_2}$
    & \xmark / \textbf{0.00} & \cmark / \textbf{0.00} & \cmark / \textbf{0.00} & 0.23 & 0.24 & 0.24 & \textbf{0.93} & \textbf{0.93} & \textbf{0.94} \\
 & $\widehat \beta_0$
    & \xmark / 0.05 & \cmark / \textbf{0.00} & \cmark / \textbf{0.00} & 0.13 & 0.13 & 0.13 & \textbf{0.94} & \textbf{0.94} & \textbf{0.94} \\
\midrule
\multirow{2}{*}{(e)} & $\widehat \beta_{X_2}$
    & \xmark / -0.45 & \qmark / -0.54 & \qmark / -0.56 & 0.55 & 0.58 & 0.60 & 0.40 & 0.23 & 0.22 \\
 & $\widehat \beta_0$
    & \xmark / 0.60 & \qmark / 0.67 & \qmark / 0.69 & 0.66 & 0.69 & 0.70 & 0.24 & 0.02 & 0.02 \\
\midrule
\multirow{2}{*}{(f)} & $\widehat \beta_{X_2}$
    & \xmark / \textbf{-0.01} & \cmark / \textbf{0.00} & \cmark / \textbf{0.00} & 0.29 & 0.27 & 0.27 & \textbf{0.95} & \textbf{0.94} & \textbf{0.94} \\
 & $\widehat \beta_0$
    & \xmark / 0.27 & \cmark / \textbf{0.00} & \cmark / \textbf{0.00} & 0.30 & 0.14 & 0.14 & 0.57 & \textbf{0.95} & \textbf{0.95} \\
\midrule
\multirow{2}{*}{(g)} & $\widehat \beta_{X_2}$
    & \xmark / 0.11 & \qmark / 0.05 & \qmark / \textbf{0.00} & 0.69 & 0.33 & 0.29 & 0.82 & 0.89 & \textbf{0.93} \\
 & $\widehat \beta_0$
    & \xmark / 0.68 & \qmark / 0.83 & \qmark / 0.83 & 0.76 & 0.85 & 0.84 & 0.49 & 0.02 & 0.01 \\
\bottomrule
\end{tabular}
\caption{\label{tb:sim_n1000_wdim1} Simulation results for $N=1{,}000$, $|W|=1$ scenarios. $\beta_{X_2}=3.22$ for DGMs (a) and (b) and $\beta_{X_2}=2$ otherwise. MAR (IPW): Sun \& Tchetgen Tchetgen inverse-probability weighting estimator. MIA-I: ICE estimator. MIA-T: TMLE estimator. \cmark indicates fulfillment of the sufficient (and in the case of $\muMAR$, necessary) conditions for $\muMAR$ and $\muMIA$, respectively. \xmark indicates that those conditions are violated. \qmark indicates that the sufficient (but not necessary) conditions for $\muMIA$ are violated. Boldface indicates approximately correct performance.}
\end{table}

\subsection{Study 2: Multivariate $W$ including both complete and incomplete members}
\label{sec:sims_study2}

Study 2 was identical to Study 1 except that the univariate $W$ was replaced with 10 correlated, mixed-type variables (5 complete and 5 incomplete), each with the same edges as those depicted for $W_1$ in Figure \ref{fig:sim_dags}, and with interactions among members of $W$ on all child nodes. In DGMs (a) and (b), $\beta_{X_2}$ was therefore approximately 3.13 and $\beta_0$ approximately 0.60; in the remaining DGMs, the estimands were as in Study 1. Since $W$ now included complete members, domination results for $W = W^-$ cases no longer apply, but the fulfillment or violation of MIA's sufficient conditions is not affected. It was infeasible to run the Sun \& Tchetgen Tchetgen (STT) estimator because, with 10 auxiliary variables, there were at least $2^{6}$ missingness patterns, leading to a combinatorial explosion of parameters to be estimated. We instead used multiple imputation by chained equations (MICE), using predictive mean matching for continuous variables and logistic regression for binary variables, and generating 50 imputed datasets.\autocite{van2011mice}
We used sample sizes ranging from $N=1{,}000$ to $N=10{,}000$ throughout Study 2. 

For DGMs fulfilling MIA's sufficient conditions, results resembled those of Study 1: at $N \ge 2{,}000$, both MIA estimators were approximately unbiased (bias at most 0.01 in absolute value), with CI coverage of at least 93\%. At $N=1{,}000$, coverage was closer to nominal in Study 2 than in Study 1: at least 95\% versus 92\% for $\muhatMIAice$, and at least 90\% versus 88\% for $\muhatMIAtmle$. In DGMs (b), (e), and (g), which violate MIA's sufficient conditions, the MIA estimators remained biased, though generally less so than in Study 1; at $N=10{,}000$, their bias for $\beta_0$ was 0.53, 0.11, and 0.71, respectively, versus 1.06, 0.67, and 0.83 in Study 1. MICE was approximately unbiased for $\beta_0$ in all DGMs (bias at most 0.02 in absolute value at $N = 10{,}000$), but was biased for $\beta_{X_2}$ in DGMs (b) and (g) ($-0.12$ and $-0.17$) and slightly biased in DGMs (a) and (f) ($-0.04$ and $-0.05$). This comparatively good performance arises because the incomplete members of $W$ were missing only 10\% of the time and were correlated with other members of $W$, including complete members, so MAR($V \cup W$) nearly held.

\subsection{Study 3: Violations and near-violations of Assumptions \ref{ass:positivity} and \ref{ass:no_csi}}
\label{sec:sim_study_3}

Although the conditions for $\muMIA$ and $\muMAR$ rely only on weak regularity assumptions, the domination results also rely on Assumptions \ref{ass:positivity} and \ref{ass:no_csi}. When either is violated, it is possible for MNAR($V \cup W^+$) to hold, yet for $\muMAR$ to be correct and $\muMIA$ to be incorrect. However, examples are far from generic, requiring quite specific forms of non-positivity or CSI, as Study 3 will illustrate. Heuristically, this is because the violation of either (C1-MIA) or (C2-MIA) usually involves some kind of association between $Y$ and $R$, potentially via a member of $W_R$ (if (C2-MIA) is violated). Thus, in the WCC functional $\muMAR$, the MAR-assuming response model will often depend on $Y$. In turn, since $\Omega$ is a function of the MAR-assuming response model, typically $\Cov(\Omega, Y \mid X) \ne 0$, which violates (C-WCC).

In Study 3, we accordingly constructed counterexamples to Theorems \ref{thm:IPW_with_incomplete_Y} ($\muMAR$ with $V^- = \{Y\}$) and \ref{thm:IPW_with_Wneg} ($\muMAR$ with $W = W^-$) by introducing specific violations of Assumptions \ref{ass:positivity} or \ref{ass:no_csi}. We chose counterexamples in which $\muMIA$ is incorrect, so they also serve as counterexamples to Theorem \ref{thm:dominating_cases_af4} (domination over MNAR distributions). Thus, we used DGMs in which: (1) MNAR($V \cup W^+$) held, which is a premise of Theorem \ref{thm:dominating_cases_af4}; (2) MAR($V \cup W$) held, which requires introducing violations of Assumption \ref{ass:positivity} or \ref{ass:no_csi}; and (3) (C2-MIA) was violated. We also probed the effect of near-violations of Assumptions \ref{ass:positivity} or \ref{ass:no_csi} by constructing a parallel set of DGMs in which a weak form of MNAR($V \cup W$) held. As expected, when MAR($V \cup W$) held, $\muhatMAR$ was approximately unbiased for $(\beta_0, \beta_{X_2})$ at larger sample sizes ($N \ge 5{,}000$); when weak MNAR($V \cup W$) held, $\muhatMAR$ was modestly biased (at most 0.14 at $N = 10{,}000$). Also by design, $\muhatMIA$ was biased in all DGMs, with bias for the intercept of approximately 0.63 and 0.53 for the two DAGs; for the DAG with $V^- = \{Y\}$, it was nonetheless approximately unbiased for $\beta_{X_2}$ (Supplement, Section \ref{supp-sec:sims_assumptions_violated}).

\section{Discussion}
\label{sec:discussion}

\subsection{Summary and implications for analyzing MNAR data}

This paper has made two contributions regarding conditional means under unknown MNAR mechanisms. First, we derived conditions weaker than MAR, which are necessary as well as sufficient within a broad class, under which the MAR-assuming WCC functional is correct. A sufficient condition for $\muMAR$ to be correct is that, conditional on the predictors $X$, the outcome $Y$ is mean-independent of the complete-case indicator and of the variables upon which the assumed response model depends (C2-WCC). Second, we proposed the alternative functional $\muMIA$, which dominates $\muMAR$ over the class of MNAR distributions satisfying Assumptions \ref{ass:positivity} and \ref{ass:no_csi} in two cases that are common in practice: when $X$ is complete but the outcome is incomplete ($V^- = \{Y\}$), and when all auxiliary variables are incomplete ($W = W^-$).

Faced with suspected MNAR that cannot be resolved by introducing complete auxiliaries, a commonly used approach is to include incomplete auxiliary variables and apply a MAR estimator. However, as Proposition \ref{prop:wneg_cant_fix_mnar} shows, this strategy usually fails. The purpose of MIA is to improve on this widespread approach if $W = W^{-}$ or $V^{-} = \{ Y \}$, commonly occurring cases in practice. In those domination cases, $\muMIA$ dominates $\muMAR$ over the class of MNAR($V \cup W^+$) distributions satisfying Assumptions \ref{ass:positivity} and \ref{ass:no_csi}. Determining whether each case applies only requires identifying which variables are incomplete, by considering the study design or observed data. Thus, in the domination cases, $\muMIA$ is arguably a ``safer'' point identification method than $\muMAR$, though it is certainly possible that both functionals are incorrect. In the special case where $V^- = \{Y\}$, $\muMAR$ is always incorrect under MNAR (again under Assumptions \ref{ass:positivity} and \ref{ass:no_csi}), whereas $\muMIA$ may be correct. As a simple functional that identifies only $E[Y \mid X]$, MIA permits straightforward ICE and doubly robust estimators that can, in some contexts, reduce the risk of misspecifying nuisance models. 

Thus, in the presence of suspected MNAR that cannot be resolved by introducing complete auxiliaries, and within the domination cases, MIA-based estimation could reasonably serve as the primary analysis. To address the possibility that, unexpectedly, MAR does hold, MAR-based estimation could be used as a sensitivity analysis. Outside the two domination cases, neither functional dominates the other over the class of MNAR distributions, so treating them as co-primary may be reasonable. As noted in Section \ref{sec:af4}, the unweighted complete-case (uCC) functional $E[Y \mid X, W, M=1]$ identifies $E[Y \mid X, W]$ under (C1-MIA) alone.\autocite{nmar2} It may therefore be informative to report a uCC estimate of $E[Y \mid X, W]$ alongside the MIA and MAR estimates of $E[Y \mid X]$.\autocite{nmar2,ccvmi}  Of course, when substantive knowledge renders implausible the assumptions underlying any of the aforementioned methods, it may be reasonable to omit that method. Reporting these three estimates is only a minimal recommendation; for example, bounds obtained under worst-case forms of MNAR can be informative,\autocite{carpenter2021missing,mvalue} as can sensitivity analyses under a variety of specific MNAR mechanisms.\autocite{nabi2020full,carpenter2021missing,white2011strategy,little2012prevention,bell2014handling,phung2025recursive} 

Beyond identification, MIA estimators can offer practical advantages over MAR estimators, arising from MIA's less ambitious estimation target: the conditional mean alone, rather than the entire joint distribution. Because MIA involves fewer nuisance models, MIA estimators may be at less risk of misspecification. The ICE estimator of $\muMIA$ requires only two models: an outcome model and a nuisance outer regression, which is saturated when $X$ is binary. In contrast, the STT estimator of $\muMAR$ requires not only an outcome model but also a nuisance model for each missingness pattern. AIPW and TMLE estimators of $\muMIA$ can be used to further reduce the risk of misspecification. However, in small samples, those estimators sometimes performed worse than ICE in terms of both point estimation and CI coverage.

\subsection{Connections to previous literature}

Identification results exist for certain subclasses of MNAR mechanisms, but those results generally require strong untestable assumptions about the specific MNAR mechanism. Like some previous work,\autocite{van2023missing,nmar2,robins1995analysis} our results accommodate limited knowledge of the MNAR mechanism by targeting a specific estimand rather than the full joint distribution. If $V^- = \{Y\}$ or $W = W^-$, then no assumptions beyond Assumptions \ref{ass:positivity} and \ref{ass:no_csi} are needed to conclude that MIA estimators are a ``safer'' choice than MAR estimators for MNAR distributions, at least if one is primarily concerned with consistency. As noted above, we consider MIA estimators as complements rather than replacements for MAR estimators and the many other principled methods for missing data. The choice of which analysis to designate as primary should be informed by substantive consideration of whether MAR or MNAR is more likely, and whether $V^- = \{Y\}$ or $W = W^-$ holds. 

Previous work has addressed nonparametric identification of $E[Y \mid X, W]$ via the uCC functional $E[Y \mid X, W, M=1]$, for which $1\{ M = 1 \} \ind Y \mid (X \cup W)$ clearly suffices.\autocite{glynn1986regression,bartlett2014improving,white2010bias} This condition is sometimes called \textit{outcome-independent missingness}.\autocite{yang2019causal} (C1-MIA) is exactly the identification $E[Y \mid X, W, M=1] = E[Y \mid X, W]$ itself. Hence, uCC identifies the \textit{conditional} estimand $E[Y \mid X, W]$ under weaker conditions than those under which MIA identifies the \textit{marginal} estimand $E[Y \mid X]$. This is because uCC allows certain violations of (C2-MIA). However, when the identification target is instead $E[Y \mid X]$, as in the present paper, MIA accommodates different forms of MNAR than does the uCC functional that conditions on both $X$ and $W$, i.e., $E[Y \mid X, W, M=1]$. Whereas Theorem \ref{thm:dominating_cases_af4} provided a domination result relating the classes of MNAR distributions for which MIA and MAR, respectively, are correct, a practically useful analog does not appear to exist for comparing MIA or MAR to uCC.

As noted in Section \ref{sec:wcc_conditions}, uCC is the special case of WCC in which the weights $K := 1$. Thus $H = \emptyset$, and one of the sufficient conditions (C2-WCC) for WCC reduces to the classical condition, $1\{ M=1 \} \ind Y \mid (X \cup W)$. However, if $H \ne \emptyset$, then (C2-WCC) restricts the dependence of $Y$ on not only the \textit{true} response indicator via $M$, but also on the variables involved in the \textit{assumed} response model, via $H$.  

\subsection{Limitations}

Some limitations were discussed in Section \ref{sec:practical}. Additionally, while AIPW and TMLE estimators of $\muMIAx$ are efficient in the nonparametric model, this does not indicator whether they are more efficient than MAR estimators within the smaller model class in which both $\muMIA$ and $\muMAR$ are correct. An estimator exploiting the restrictions of that model class might be more efficient than either the MIA estimators or the MAR estimator we consider. Furthermore, efficiency is an asymptotic property; at small sample sizes, $\muhatMIAice$ sometimes outperformed $\muhatMIAtmle$. In our simulations, a MIA estimator usually had better RMSE than the MAR estimator whenever the MIA estimates were unbiased, but our scenarios were not exhaustive. Characterizing efficient estimation when both MIA and MAR are correct is a direction for future work. Second, we have treated the choice of auxiliary variables $W$ as fixed, but in practice, sometimes the analyst can choose from among a large set of candidate auxiliary variables. In principle, one could derive an algoritthm which, given such a set of candidates, indicates whether there exists a subset $W$ that would suffice for a given identifying functional to be correct (e.g., $\muMIA$ or $\muMAR$). However, it appears that such an algorithm would require detailed knowledge of the data-generating MNAR mechanism, so its utility seems limited. 

\singlespacing

\vspace{0.5cm}

\small
\paragraph{Data availability.}\label{sec:data-availability} Public repositories contain all code required to reproduce the simulation study (\url{https://github.com/mayamathur/iai}) and applied example (\url{https://github.com/willizhang/incomplete-auxiliary-variable-in-imputation}). The dataset on sexual identity can be accessed after approved ethics application to the Steering Committee of the Stockholm Public Health Cohort (\url{https://www.ces.regionstockholm.se/projekt-och-uppdrag/halsa-stockholm/SPHC-data}). Code for the R package \texttt{miapack} is publicly available (\url{https://github.com/stmcg/miapack}).

\vspace{-0.7cm}

\paragraph{Funding.} MBM was supported by National Institutes of Health grants R01 LM013866, UL1TR003142, P30CA124435, and P30DK116074. IS was supported by ONR N00014-21-1-2820, NSF 2040804, NSF CAREER 1942239, and NIH R01 AI127271-01A1. WZ was supported by the Swedish Research Council for Health, Working Life and Welfare (Grant nr 2022-00159 to Dr. Kyriaki Kosidou) and Region Stockholm. The funders had no role in the design, conduct, or reporting of this research.

\vspace{-0.7cm}

\paragraph{Ethics approval.} The analyses for the applied example were conducted under existing ethical approvals (DNR: 2022-02626-01, 2023-06610-02, 2024-02155-02).

\vspace{-0.7cm}

\paragraph{Acknowledgments.} Dr. Kyriaki Kosidou coordinated and acquired funding for the analyses of the Stockholm Public Health Cohort.

\vspace{-0.7cm}

\paragraph{Conflicts of interest.} None.

\vspace{-0.7cm}

\paragraph{Generative AI.} Claude Opus assisted with searching for related articles, drafting visuals and R code, and endless proofreading. All AI-drafted code was reviewed line-by-line and tested by one or more authors.

\pagebreak

\begin{singlespace}
\printbibliography
\end{singlespace}

\end{document}


\date{\vspace{-5ex}}  %
\maketitle %

\thispagestyle{empty}

\setcounter{tocdepth}{4} %

\singlespacing

\begingroup
\hypersetup{
  linkcolor=black,
  urlcolor=black,
  citecolor=black
}

\singlespacing

\setlength{\parskip}{0pt}  %

\tableofcontents 
\endgroup  %

\begin{table}[H]
\centering
\singlespacing
\footnotesize
\setlength{\tabcolsep}{4pt}
\renewcommand{\arraystretch}{1.1}
\begin{tabular}{@{}p{2.3cm}p{8.5cm}p{3.8cm}@{}}
\toprule
\textbf{Result} & \textbf{Description} & \textbf{Relies on} \\
\midrule
 
\multicolumn{3}{@{}l}{\textbf{Assumptions}} \\[1pt]
 
Ass.\ \ref{ass:positivity}
& Positivity
& --- \\
 
Ass.\ \ref{ass:no_csi}
& Restrictions on CSI
& --- \\
 
\midrule
\multicolumn{3}{@{}l}{\textbf{Definitions}} \\[1pt]
 
Def.\ \ref{def:perfect_cancellations_sums_fns}
& Perfect cancellations of dependences
& --- \\
 
Def.\ \ref{def:pattern_regularity}
& Pattern regularity
& --- \\
 
Def.\ \ref{def:functionally_determining_set}
& Functionally determining sets
& --- \\
 
\midrule
\multicolumn{3}{@{}l}{\textbf{Remarks}} \\[1pt]
 
Rmk.\ \ref{rm:Omega_function_vs_rv}
& $\OmegaMAR$ as a function and as a random variable
& Lem.\ \ref{lm:unique_mfds} \\
 
Rmk.\ \ref{rm:averaging_cancellations}
& On the exclusion of perfect cancellations
& Lem.\ \ref{lm:Y_R_plus_MNAR_muMAR_incorrect} \\
 
\midrule
\multicolumn{3}{@{}l}{\textbf{Supporting results}} \\[1pt]
 
Lem.\ \ref{lm:pattern_regularity}
& Sufficient conditions for pattern regularity
& Ass.\ \ref{ass:positivity}, \ref{ass:no_csi}; Def.\ \ref{def:perfect_cancellations_sums_fns}, \ref{def:pattern_regularity} \\
 
Prop.\ \ref{prop:wneg_cant_fix_mnar}
& Incomplete auxiliaries cannot turn MNAR into MAR
& Ass.\ \ref{ass:positivity}, \ref{ass:no_csi}; Def.\ \ref{def:pattern_regularity}; Lem.\ \ref{lm:pattern_regularity} \\
 
Lem.\ \ref{lm:unique_mfds}
& Uniqueness of MFDS
& Def.\ \ref{def:functionally_determining_set} \\
 
Coro.\ \ref{coro:mfds_sums}
& MFDS of a sum
& Def.\ \ref{def:perfect_cancellations_sums_fns}, \ref{def:functionally_determining_set}; Lem.\ \ref{lm:unique_mfds} \\
 
Prop.\ \ref{prop:H_characterization}
& Characterization of $\HMAR$
& Ass.\ \ref{ass:positivity}, \ref{ass:no_csi}; Def.\ \ref{def:perfect_cancellations_sums_fns}, \ref{def:functionally_determining_set}; Lem.\ \ref{lm:pattern_regularity}, \ref{lm:unique_mfds}; Coro.\ \ref{coro:mfds_sums} \\
 
Lem.\ \ref{lm:empty_interior_solution_set}
& Knife-edge nature of expectation constraint
& --- \\
 
Lem.\ \ref{lm:C1_MIA_numerator}
& $\neg$ (C1-MIA) $\Rightarrow p(M = 1 \mid V, W)$ is functionally dependent on $Y$
& --- \\
 
Lem.\ \ref{lm:C2_MIA_Y_R_path}
& $\neg$ (C2-MIA) $\wedge$ ($W = W^-$) $\Rightarrow$ unblocked $Y$--$R$ path
& --- \\
 
Lem.\ \ref{lm:Omega_FD_on_Y}
& Functional dependence of $\OmegaMAR$ on $Y$
& Rmk.\ \ref{rm:Omega_function_vs_rv} \\
 
Lem.\ \ref{lm:Y_R_plus_MNAR_muMAR_incorrect}
& Unblocked $Y$--$R$ path $+$ MNAR $\Rightarrow \muMAR$ incorrect
& Ass.\ \ref{ass:positivity}, \ref{ass:no_csi}; Lem.\ \ref{lm:pattern_regularity}, \ref{lm:Omega_FD_on_Y}; Coro.\ \ref{coro:mfds_sums}; Thm.\ \ref{thm:wcc_conditions} \\
 
\midrule
\multicolumn{3}{@{}l}{\textbf{Main results}} \\[1pt]
 
Thm.\ \ref{thm:wcc_conditions}
& Necessary and sufficient conditions for $\muWCC(K)$
& --- \\
 
Coro.\ \ref{coro:wcc_suff_conditions}
& Sufficient conditions for (C-WCC)
& Def.\ \ref{def:functionally_determining_set}; Thm.\ \ref{thm:wcc_conditions} \\
 
Thm.\ \ref{thm:IPW_with_incomplete_Y}
& $\muMAR$ with complete $X$ and incomplete $Y$
& Ass.\ \ref{ass:positivity}, \ref{ass:no_csi}; Prop.\ \ref{prop:wneg_cant_fix_mnar}; Lem.\ \ref{lm:Y_R_plus_MNAR_muMAR_incorrect} \\
 
Thm.\ \ref{thm:mia_conditions}
& Sufficient conditions for $\muMIA$ 
& --- \\
 
Thm.\ \ref{thm:IPW_with_Wneg}
& $\muMAR$ with all auxiliary variables incomplete
& Ass.\ \ref{ass:positivity}, \ref{ass:no_csi}; Lem.\ \ref{lm:C1_MIA_numerator}, \ref{lm:C2_MIA_Y_R_path}, \ref{lm:Omega_FD_on_Y}, \ref{lm:Y_R_plus_MNAR_muMAR_incorrect}; Thm.\ \ref{thm:wcc_conditions} \\
 
Thm.\ \ref{thm:dominating_cases_af4}
& Domination over MNAR distributions
& Ass.\ \ref{ass:positivity}, \ref{ass:no_csi}; Lem.\ \ref{lm:Y_R_plus_MNAR_muMAR_incorrect}; Prop.\ \ref{prop:wneg_cant_fix_mnar}; Thm.\ \ref{thm:IPW_with_incomplete_Y}, \ref{thm:mia_conditions}, \ref{thm:IPW_with_Wneg} \\
 
\bottomrule
\end{tabular}
\caption{\label{tb:results_index} Index of all theoretical results and their dependencies, i.e., other results or assumptions used to establish the result in the first column.
}
\end{table}

\ifx\flowchip\undefined
\DeclareRobustCommand{\flowchip}[1]{%
  \raisebox{-0.2ex}{\mbox{\setlength{\fboxsep}{0pt}\setlength{\fboxrule}{0.6pt}%
    \fcolorbox{black}{#1}{\rule{0pt}{6pt}\rule{6pt}{0pt}}}}}
\fi

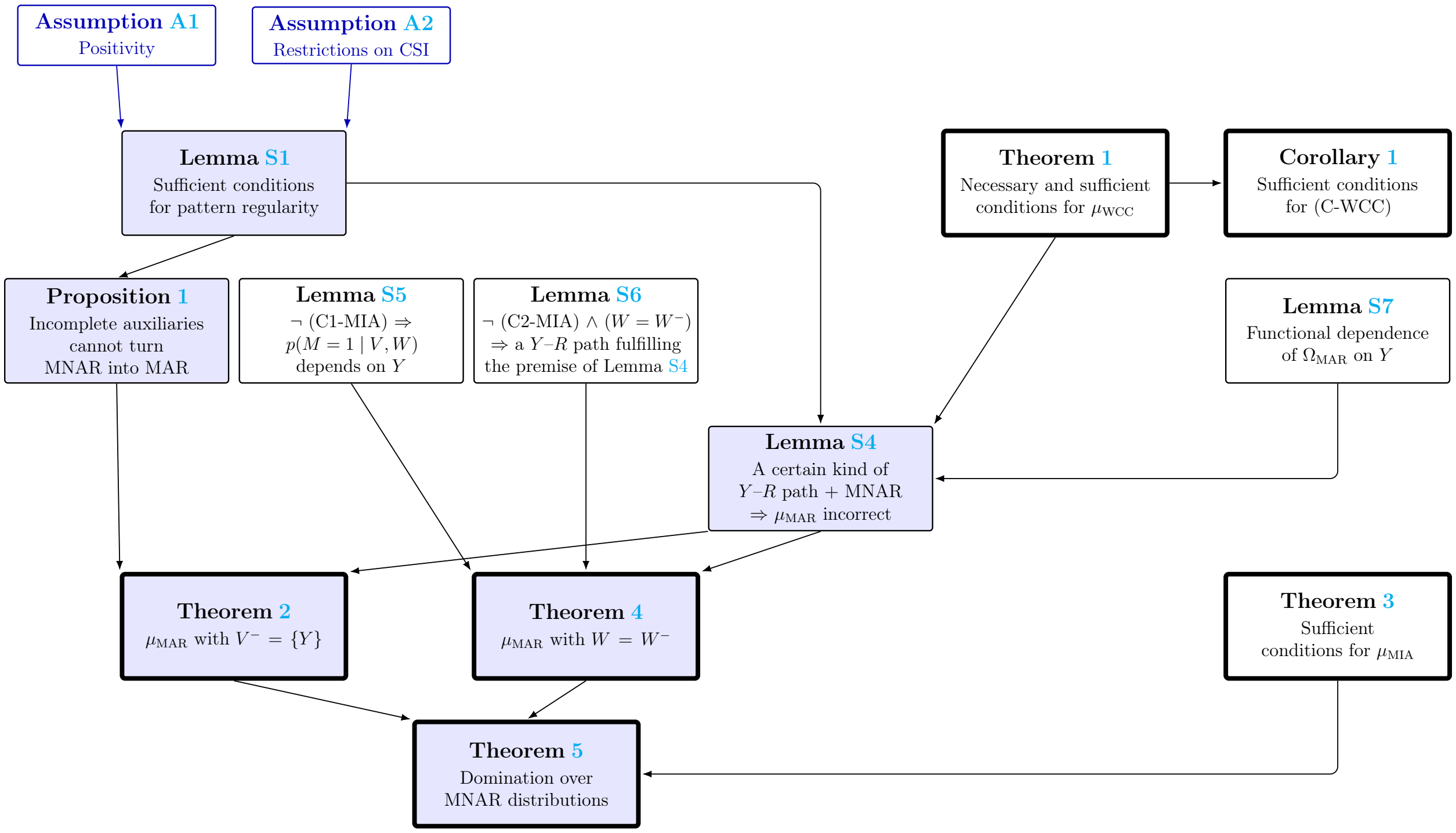
\begin{figure}[!ht]
\centering

\colorlet{assText}{blue!70!black}   %
\colorlet{assFill}{blue!10}         %
\newcommand{\deplab}[1]{{\fontsize{13.5}{16}\selectfont\textbf{#1}}}
\newcommand{\depsub}[1]{\\[2.5pt]{\fontsize{12}{14}\selectfont #1}}

\resizebox{\textwidth}{!}{%
\begin{tikzpicture}[
  auto=false, node distance=0pt, every path/.style={},
  box/.style={draw, rounded corners=2pt, text width=4.8cm, align=center,
              inner sep=5pt, minimum height=2.4cm,
              execute at begin node={\parindent=0pt \parskip=0pt
                \hyphenpenalty=10000 \exhyphenpenalty=0}},
  supp/.style={box, line width=0.9pt, fill=white},          %
  main/.style={box, line width=3pt, fill=white},            %
  blueS/.style={supp, fill=assFill},                        %
  blueM/.style={main, fill=assFill},
  ass/.style={box, line width=0.9pt, fill=white, draw=assText, text=assText,
              text width=4.2cm, minimum height=1.3cm},
  arr/.style={draw, -Latex, line width=0.7pt, shorten >=1pt, rounded corners=5pt},
  plainline/.style={draw, -, line width=0.7pt},
]

\def\xA{0}    \def\xB{5.4}  \def\xC{10.8} \def\xD{16.2}
\def\xE{21.6} \def\xF{28.1}
\def\yz{0}    \def\yo{-3.4}  \def\yt{-6.8} \def\yth{-10.2}
\def\yfo{-13.6} \def\yfi{-17.0}
\def\xAB{2.7} \def\xTD{9.43}

\node[ass] (A2) at (\xA,\yz)
  {\deplab{Assumption \ref{ass:positivity}}\depsub{Positivity}};

\node[ass] (A3) at (\xB,\yz)
  {\deplab{Assumption \ref{ass:no_csi}}\depsub{Restrictions on CSI}};

\node[blueS] (L1) at (\xAB,\yo)
  {\deplab{Lemma \ref{lm:pattern_regularity}}%
   \depsub{Sufficient conditions for pattern regularity}};

\node[blueS] (CO1) at (\xA,\yt)
  {\deplab{Proposition \ref{prop:wneg_cant_fix_mnar}}%
   \depsub{Incomplete auxiliaries cannot turn MNAR into MAR}};

\node[supp] (LC1) at (\xB,\yt)
  {\deplab{Lemma \ref{lm:C1_MIA_numerator}}%
   \depsub{$\neg$ (C1-MIA) $\Rightarrow$ \mbox{$p(M = 1 \mid V, W)$} depends on $Y$}};

\node[supp] (LC2) at (\xC,\yt)
  {\deplab{Lemma \ref{lm:C2_MIA_Y_R_path}}%
   \depsub{$\neg$ (C2-MIA) $\wedge$ \mbox{($W = W^-$)} $\Rightarrow$ a \mbox{$Y$--$R$} path fulfilling the premise of Lemma \ref{lm:Y_R_plus_MNAR_muMAR_incorrect}}};

\node[main] (TW) at (\xE,\yo)
  {\deplab{Theorem \ref{thm:wcc_conditions}}%
   \depsub{Necessary and sufficient conditions for $\muWCC$}};

\node[main] (COW) at (\xF,\yo)
  {\deplab{Corollary \ref{coro:wcc_suff_conditions}}%
   \depsub{Sufficient conditions for (C-WCC)}};

\node[supp] (LOM) at (\xF,\yt)
  {\deplab{Lemma \ref{lm:Omega_FD_on_Y}}%
   \depsub{Functional dependence of $\OmegaMAR$ on $Y$}};

\node[blueS] (LCON) at (\xD,\yth)
  {\deplab{Lemma \ref{lm:Y_R_plus_MNAR_muMAR_incorrect}}%
   \depsub{A certain kind of \mbox{$Y$--$R$} path $+$ MNAR $\Rightarrow$ \mbox{$\muMAR$ incorrect}}};

\node[blueM] (TY) at (\xAB,\yfo)
  {\deplab{Theorem \ref{thm:IPW_with_incomplete_Y}}%
   \depsub{$\muMAR$ with $V^- = \{ Y \}$}};

\node[blueM] (TWN) at (\xC,\yfo)
  {\deplab{Theorem \ref{thm:IPW_with_Wneg}}%
   \depsub{$\muMAR$ with $W = W^-$}};

\node[main] (TM) at (\xF,\yfo)
  {\deplab{Theorem \ref{thm:mia_conditions}}%
   \depsub{Sufficient conditions for $\muMIA$}};

\node[blueM] (TD) at (\xTD,\yfi)
  {\deplab{Theorem \ref{thm:dominating_cases_af4}}%
   \depsub{Domination over MNAR distributions}};

\draw[arr, draw=assText] (A2.south) -- (L1.north west);
\draw[arr, draw=assText] (A3.south) -- (L1.north east);

\draw[arr] (L1.south) -- (CO1.north);
\draw[arr] (L1.east) -- (\xD,\yo) -- (LCON.north);

\draw[arr] (TW.south)  -- (LCON.north east);
\draw[arr] (LOM.south) |- (LCON.east);

\draw[arr] (CO1.south)       -- (TY.north west);
\draw[arr] (LCON.south west) -- (TY.north east);

\draw[arr] (LC1.south)  -- (TWN.north west);
\draw[arr] (LC2.south)  -- (TWN.north);
\draw[arr] (LCON.south) -- (TWN.north east);

\draw[arr] (TW.east) -- (COW.west);

\draw[arr] (TY.south)  -- (TD.north west);
\draw[arr] (TWN.south) -- (TD.north);
\draw[arr] (TM.south) |- (TD.east);

\end{tikzpicture}%
}

\caption[Dependency structure of the results]{Dependency structure of key theoretical results. Each edge indicates that the child relies on the parent; only direct dependences are drawn, so a result also relies on all of its ancestors. Thick borders mark results stated in the main text; thin borders mark those appearing only in the Supplement. \flowchip{assFill}~: reliance on Assumptions \ref{ass:positivity} (positivity) and \ref{ass:no_csi} (restrictions on CSI), which are used for Lemma \ref{lm:pattern_regularity} and then inherited by its descendants. Lemmas \ref{lm:C1_MIA_numerator}, \ref{lm:C2_MIA_Y_R_path} and \ref{lm:Omega_FD_on_Y} require only the regularity assumptions, so they appear as roots.}\label{fig:proof_dependency_flowchart}
\end{figure}

\newpage

\pagebreak

\pagebreak
\section{Theoretical preliminaries}

Supporting definitions and results that appear only in the Online Supplement have numbers prefaced with ``S'' (e.g., Lemma S1). 

\subsection{Directed acyclic graphs}
\label{sec:DAGs}

\newcommand{\exhaustiveFootnote}{}

Some proofs below use directed acyclic graphs (DAGs)\autocite{pearl2009causality,nabi2022causal,bhattacharya2020identification} to encode conditional independences and dependences in a joint distribution. We briefly summarize a standard formalization of DAGs; further details are provided in Section 1.2 of reference \cite{pearl2009causality}. In an arbitrary DAG $G_L$ comprising nodes $L$, a \textit{path} between two members $L_i$ and $L_j$ of $L$ is an unbroken sequence of nodes and edges of any orientation between $L_i$ and $L_j$. A given path between two disjoint sets $A \subset L$ and $B \subset L$ is said to be \textit{d-separated} or \textit{blocked} by a set $C \subseteq ( L \setminus (A \cup B) )$ if, for some set of variables $\left( L_i, L_j, L_k \right)$: (i) the path includes a chain $L_i \rightarrow L_j \rightarrow L_k$ or a fork $L_i \leftarrow L_j \rightarrow L_k$, where $L_j \in C$; or (ii) the path includes a collider $L_i \rightarrow L_j \leftarrow L_k$, where neither $L_j$ nor any of its descendants is in $C$. A set $C$ is said to \textit{d-separate} $A$ from $B$ in a graph $G_L$ if and only if $C$ blocks every path in $G_L$ from $A$ to $B$. D-separation between $A$ and $B$ conditional on $C$ is denoted $A \dsep{G_L} B \mid C$.  

A distribution $p$ over $L$ is \textit{Markov wrt $G_L$} if, for all disjoint sets of nodes $A$, $B$, and $C$ such that $C$ d-separates $A$ from $B$, the corresponding statistical independence $A \ind B \mid C$ holds in $p$. Conversely, $p$ is \textit{faithful} to $G_L$ if, for all disjoint sets of nodes $A$, $B$, and $C$ such that $A \ind B \mid C$ holds in $p$, the corresponding d-separation holds in $G_L$. Without restrictions on $p$, trivially, there exists DAG(s) such that $p$ is Markov wrt the DAG(s). However, if $p$ exhibits (for example)\exhaustiveFootnote certain forms of determinism or perfect cancellations of association, then there may not exist a DAG to which $p$ is faithful.

Informally, a DAG is \textit{causal} if each edge $A \to B$ represents a potential causal relationship of $A$ on $B$.\autocite{pearl2009causality} While our theoretical results do not require a causal interpretation of the DAG, it is difficult to envision DGMs that do not reflect an underlying causal process.\autocite{robins1997non} We therefore use causal language when discussing the practical implications of our results, since doing so facilitates assessing the assumptions' plausibility (see Section \ref{main-sec:practical} in the main text). 

In our context, let $G$ be a DAG whose node set is $(V \cup W \cup R)$, along with unmeasured variables $U$ as needed. As noted above, we restrict consideration to distributions that do not involve perfect cancellations of associations (\ref{ass:no_cancellations}). Formally, we consider distributions satisfying regularity Assumption \ref{ass:partial_faithfulness}, under which d-separation in the graph and the corresponding statistical independence in the distribution coincide for the sets to which that assumption applies. We nonetheless keep the two distinct notationally, writing $L_1 \ind L_2 \mid L_3$ for the statistical independence and $L_1 \dsep{G} L_2 \mid L_3$ for the d-separation. 

Regarding context-specific independences (CSI), given an edge between any two nodes $L_i$ and $L_j$ and a conditioning set $L_c$, we refer to the edge as ``inactive conditional on $L_c = l_c$'' if the corresponding causal relationship ceases to hold conditional on $L_c = l_c$. (If using a non-causal interpretation of the DAG, ``causal relationship'' can simply be replaced by ``conditional dependence''.) More generally, a path between $L_i$ and $L_j$ is inactive conditional on $L_c = l_c$ if and only if at least one of its edges is inactive conditional on $L_c = l_c$.

Given a missing-data DAG, $G$, a missingness pattern $m$ is compatible with $G$ (or \textit{graph-compatible} for shorthand) if there exists an $r \in \{ 0, 1 \}^{|R|}$ such that:
$$1\{ M = m \} := 1\{ R = r \}.$$
That is, a graph-compatible pattern represents any assignment of values in $\{ 0, 1 \}$ to members of $R$. Importantly, in a given distribution that is Markov wrt the graph, it is possible that any given graph-compatible pattern never occurs, due to determinism. Therefore, we will use phrasing like ``there exists a graph-compatible pattern'' instead of the ambiguous ``there exists a pattern''. 

\subsection{Review of assumptions}
\label{sec:review_assumptions}

\noindent For brevity, we will not restate the regularity assumptions in each result, and invoke them only when they play a specific technical role.

\newcommand{\faithFootnote}{\footnote{As noted in Section \ref{sec:DAGs}, the existence of a DAG fulfilling the ``$\Leftarrow$'' direction (i.e., $A \dsep{G} B \mid C \Rightarrow A \ind B \mid C$) is not restrictive; this holds for any distribution over a finite set of variables. The ``$\Rightarrow$'' direction is weaker than assuming $p(v,w,r,u)$ is faithful to $G$: $A \ind B \mid C \Rightarrow A \dsep{G} B \mid C$ is only required when the conditioning set and at least one of the two independent sets consists of members of $(V \cup W \cup U)$, excluding $R$. Assumption \ref{ass:partial_faithfulness} thus permits determinism among the missingness indicators: for example, if $R_1 = R_2$, then $R_2$ is independent of every $A \subseteq (V \cup W \cup R \cup U) \setminus \{ R_1, R_2 \}$ conditional on $R_1$, even where $R_2$ and $A$ are conditionally d-connected. Non-determinism among missingness indicators is assumed only in results invoking Assumption \ref{ass:positivity}. That assumption is substantive, so is presented separately from the regularity assumptions.}} 

\setcounter{assumption}{-1} %
\begin{assumption}[Regularity assumptions] \label{ass:boring_assumptions}
\hfill\\ %
\vspace{-1cm}

\begingroup
\renewcommand{\theenumi}{(\roman{enumi})}
\renewcommand{\labelenumi}{\theenumi}
\makeatletter
\renewcommand{\p@enumi}{\theassumption}
\makeatother
\begin{enumerate}

\item \label{ass:partial_faithfulness} (Partial faithfulness.) There exists a DAG, $G$, with node set $(V \cup W \cup R \cup U)$, where $U$ is a potentially empty set of unmeasured variables, such that for all
disjoint sets of nodes $A \subset (V \cup W \cup R \cup U)$, $B \subset (V \cup W \cup U)$, and
$C \subset (V \cup W \cup U)$:\faithFootnote
$$ A \ind B \mid C \quad \Leftrightarrow \quad A \dsep{G} B \mid C.$$

\item \label{ass:no_determinism} (No determinism of $V \cup W$.) There are no deterministic constraints among members of $V \cup W$, i.e., $\supp(V \cup W) = \prod_{i=1}^{|V|} \supp(V_i) \prod_{j=1}^{|W|} \supp(W_j)$. 

\item \label{ass:cc_positivity} (Complete-case pattern positivity.) The complete-case pattern occurs with positive probability conditional on any $(v,w) \in \supp(V \cup W)$, i.e., $p(M = 1 \mid v, w) > 0$.

\item \label{ass:no_cancellations} (No perfect cancellations.) The DGM does not involve perfect cancellations of associations of the kind formalized in Definition \ref{def:perfect_cancellations_sums_fns} below.
\end{enumerate}
\endgroup
\end{assumption}

The next two assumptions are reproduced from the main text for convenience.

\begin{assumption}[Positivity] \label{ass:positivity}
Conditional on any value $(v, w) \in \supp(V \cup W)$, every missingness pattern on the incomplete variables can occur:
$$p(r \mid v, w)>0 \quad \forall \, r \, \in \{ 0, 1\}^{|R|} \text{ and } (v,w) \in \supp(V \cup W).$$
\end{assumption}

\begin{assumption}[Restrictions on context-specific independence (CSI)]
\label{ass:no_csi}
For every indicator $R^* \in R$, $Z \in (V \cup W)$, and $C \subseteq ( V \cup W \cup R ) \setminus \{ Z, R^* \}$ such that $Z \notind R^* \mid C$, we have $Z \notind R^* \mid c$ for every $c$ such that $p(c)>0$. That is, there are no context-specific independences such that $Z \notind R^* \mid C$ but $Z \ind R^* \mid c'$ for some $c' \in \supp(C)$.
\end{assumption}

The next definition formalizes Assumption \ref{ass:no_cancellations} (no perfect cancellations). Heuristically, in a sum of functions, perfect cancellations are said to occur if a sum is constant wrt a given random variable, but at least one function is not constant wrt that variable. 

\newcommand{\RMMFootnote}{}

\begin{sdefinition}[Perfect cancellations of dependences: Sums of functions]
\label{def:perfect_cancellations_sums_fns}

Let $$\mathcal{G} = \{ g_1(L), g_2(L), \cdots g_{J}(L) \}$$
be a set of functions of a set of variables $L$. Suppose there exist no
deterministic constraints among members of $L$, in the sense that for any $L_i \in L$,
there exists no $L' \subseteq ( L \setminus \{ L_i \} )$ such that $L_i$ is fixed
conditional on $L'$. Suppose the functions in $\mathcal{G}$ are such that, for some
$L^* \in L$, $\sum_m g_m(L)$ is constant in $L^*$, but there exists a
$g_m(L) \in \mathcal{G}$ that is not constant in $L^*$. Then we say that there are
perfect cancellations of dependences in the sum $\sum_m g_m(L)$.\RMMFootnote
\end{sdefinition}

\subsection{Pattern regularity}

Lemma \ref{lm:pattern_regularity} below establishes that Assumptions \ref{ass:positivity} and \ref{ass:no_csi} together imply a property we call \textit{pattern regularity}, which essentially states that when an incomplete variable is associated with a missingness indicator, it is also associated with membership in every pattern (Definition \ref{def:pattern_regularity} below). Pattern regularity pertains to missingness indicators, which may exhibit determinism, so is not implied by the regularity assumptions alone. 

\newcommand{\patternfaithFootnote}{}

\begin{sdefinition}[Pattern regularity]
\label{def:pattern_regularity}
Let $L := V \cup W$, $L^* \in L$, and $R^*\in R$. Pattern regularity holds if, for every choice of conditioning set $C \subseteq ( ( V \cup W ) \setminus \{ L^* \} )$ (including $C = \emptyset$),\patternfaithFootnote
$$ L^* \notind R^* \mid C \; \Rightarrow \; L^* \notind 1\{M=m\} \mid C \text{ for every graph-compatible pattern } m.$$
\end{sdefinition}

\begin{slemma}[Sufficient conditions for pattern regularity]
\label{lm:pattern_regularity}
If a given distribution $p(v,w,r,u)$ satisfies Assumptions \ref{ass:positivity}
(positivity) and \ref{ass:no_csi}, and exhibits no perfect
cancellations of dependences per Definition \ref{def:perfect_cancellations_sums_fns},
then pattern regularity holds.

\begin{proof}
Let $L := V \cup W$. Arguing by contradiction, suppose pattern regularity is violated. Then there exist $L^* \in L$, $R^* \in R$, $C \subseteq ( L \setminus \{ L^* \} )$, and a graph-compatible pattern $\mt$ such that $L^* \notind R^* \mid C$ but $1\{ M = \mt \} \ind L^* \mid C$. The latter
implies that for every $c \in \supp(C)$,
$$ p(\mt \mid l^*, c) = p(\mt \mid c),$$
i.e. $p(\mt \mid l^*, c)$ is constant in $l^*$. Letting $R^c := R \setminus \{ R^* \}$ and
writing $r_A(m)$ for the value that $R_A \subseteq R$ takes when $M = m$, we have
\begingroup
\addtolength{\jot}{0.5em}
\begin{align}
\label{eq:interm2}
p(\mt \mid l^*, c) &= p(r^*(\mt) \mid l^*, c) \; p(r^c(\mt) \mid l^*, c, r^*(\mt)).
\end{align}
\endgroup
By Assumption \ref{ass:positivity}, both factors on the RHS are
strictly positive, so we can take logarithms to obtain a sum:
$$ \log p(\mt \mid l^*, c) = \log p(r^*(\mt) \mid l^*, c)
   + \log p(r^c(\mt) \mid l^*, c, r^*(\mt)).$$
By Assumption \ref{ass:no_csi}, the dependence
$L^* \notind R^* \mid C$ holds at every $c \in \supp(C)$. Since $R^*$ is binary,
this implies that both $p(R^* = 1 \mid l^*, c)$ and $p(R^* = 0 \mid l^*, c)$ are not constant in
$l^*$. Therefore, regardless of whether $r^*(\mt)=1$ or $r^*(\mt)=0$, the first term on the RHS, namely $p(r^*(\mt) \mid l^*, c)$, is not constant in $l^*$. However, the LHS is constant in $l^*$. Since there are no deterministic constraints among members of $V \cup W$ (Assumption \ref{ass:no_determinism}),
this is a perfect cancellation of dependences per Definition
\ref{def:perfect_cancellations_sums_fns}, contradicting Assumption
\ref{ass:no_cancellations}. \qedhere
\end{proof}
\end{slemma}

\begin{proposition}[Incomplete auxiliary variables cannot turn MNAR into MAR]
\label{prop:wneg_cant_fix_mnar}
Suppose Assumptions \ref{ass:positivity} and \ref{ass:no_csi} hold. If MNAR($V \cup W^+$) holds, then MNAR($V \cup W$) also holds.

\begin{proof} 
By Lemma \ref{lm:pattern_regularity}, Assumptions \ref{ass:positivity} and \ref{ass:no_csi} imply pattern regularity. In turn, pattern regularity implies that, for any set of variables $L$, MAR($L$) is equivalent to $R(L) \ind L^- \mid L^+$. Suppose MNAR($V \cup W^+$) holds:
$$ 1\{ M = m \} \notind \Vmis{}{m} \mid ( \Vobs{}{m} \cup W^+ ).$$
By pattern regularity, this implies:
$$ R(V) \notind V^- \mid ( V^+ \cup W^+ ).$$
By Assumption \ref{ass:partial_faithfulness} (partial faithfulness), there exists a path $\eta$ between a member $V^* \in V^-$ and $R^* \in R(V)$ that is unblocked conditional on $( V^+ \cup W^+ )$. Arguing by contradiction, suppose MAR($V \cup W$) holds, so by pattern regularity:
$$ R(V \cup W) \ind V^- \mid ( V^+ \cup W ).$$
By Assumption \ref{ass:partial_faithfulness}, this implies that $\eta$ is blocked conditional on $( V^+ \cup W )$. Since $\eta$ is unblocked conditional on the smaller set $( V^+ \cup W^+ )$, at least one member $W^* \in W^-$ must be a noncollider on $\eta$. The subpath of $\eta$ between $W^*$ and $R^*$ is unblocked conditional on $( V^+ \cup W^+ )$, and since $W^* \in W^-$, this means $R^* \ndsep{G} W^- \mid ( V^+ \cup W^+ )$. By Assumption \ref{ass:partial_faithfulness} and pattern regularity, this immediately implies that MNAR($V \cup W$) holds. This is a contradiction.
\end{proof}
\end{proposition}

\subsection{Notation and properties of $\OmegaMAR$}

\begin{sremark}[$\OmegaMAR$ as a function and as a random variable]
\label{rm:Omega_function_vs_rv}
We write $\OmegaMAR(v, w)$ for the deterministic function
\begingroup
\addtolength{\jot}{0.5em}
\begin{align*}
\OmegaMAR(v, w) := \frac{ p( M = 1 \mid V=v, W = w) } { \pMAR( M = 1 \mid V=v, W = w) },
\end{align*}
\endgroup
and $\OmegaMAR$ for the random variable $\OmegaMAR(V, W)$ that it induces, exactly as for any random variable generated by a structural equation. The phrase ``functionally dependent on $Y$'' means that $Y$ belongs to the MFDS of $\OmegaMAR$. By Assumption \ref{ass:no_determinism} and Lemma \ref{lm:unique_mfds}, that MFDS exists and is unique, so the phrase is well defined. Statements of statistical properties, such as $\Cov(\OmegaMAR, Y \mid X) =0$, refer to the random variable. 

Note that both the numerator and denominator of $\OmegaMAR(V,W)$ are strictly positive: the numerator by Assumption \ref{ass:cc_positivity}, and the denominator by the definition of $\KMAR$. Therefore, we can write:
\begingroup
\addtolength{\jot}{0.5em}
\begin{align*}
\log \OmegaMAR = \log p( M = 1 \mid V, W) - \log \pMAR( M = 1 \mid V, W).
\end{align*}
\endgroup
This will be used in subsequent results. 
\end{sremark}

\section{Proofs of theoretical results}

\subsection{Definition and properties of $H$}
\label{sec:H}

We will need the following definitions. We use $\subsetneq$ to denote a strict subset. 

\begin{sdefinition}[Functionally determining sets]
\label{def:functionally_determining_set}
Let $g(L)$ be a function of a set of variables $L$. Consider a set $L' \subseteq L$. We say $L'$ is a \textit{functionally determining set} (FDS) for $g(L)$ if $g(l) = g(\lt)$ for every pair of values $l \in \text{support}(L)$ and $\lt \in \text{support}(L)$ such that $l$ and $\lt$ have the same value for every member of $L'$. (If $g(L)$ is constant in $L$, then trivially, the empty set is an FDS.) We say $L'$ is a \textit{minimal functionally determining set} (MFDS) for $g(L)$ if there exists no strict subset of $L'$ that is an FDS.
If there exists a unique MFDS for $g(L)$, called $Z^*$, then we say $g(L)$ is \textit{functionally dependent} on every $L_i \in Z^*$ and is \textit{functionally independent} of every $L_j \notin Z^*$.  
\end{sdefinition}

\begin{slemma}[Uniqueness of MFDS]
\label{lm:unique_mfds}
Let $g(L)$ be a function of a set of variables $L$. Suppose there are no deterministic constraints among members of $L$. Then an MFDS for $g(L)$ exists and is unique.

\begin{proof}
Trivially, $L$ itself is an FDS for $g(L)$. Therefore, since $L$ is finite, an MFDS exists. If $L$ itself is an MFDS, then the claim holds trivially. Suppose instead that there exists an $S \subsetneq L$ that is an MFDS for $g(L)$. Pick an arbitrary member $S_i \in S$, and consider any other subset $T \subsetneq L$ such that $S_i \notin T$. We claim that $T$ is not an FDS, and hence also not an MFDS. Let $\lt_1$ and $\lt_2$ be two values of $L$ that agree for all variables except $S_i$. Choose the values of $S_i$ in $\lt_1$ and $\lt_2$ such that $g(\lt_1) \ne g(\lt_2)$. Such a choice is possible because $S$ is an MFDS, and since there are no deterministic constraints among members of $L$, both $\lt_1$ and $\lt_2$ are supported. Since $\lt_1$ and $\lt_2$ agree on all members of $T$ and yet $g(\lt_1) \ne g(\lt_2)$, $T$ cannot be an FDS. Therefore, every FDS for $g(L)$ must contain every member of $S$. Since $S$ itself is minimal, it follows that $S$ is the unique MFDS.
\end{proof}
\end{slemma}

For example, suppose $L := \{ L_1, L_2, L_3 \}$ where $L_2 = 2 L_3$, and there are no deterministic relationships between $L_1$ and $\{ L_2, L_3 \}$. Consider a function $g(L)$ that can be written as $g(L) = L_1 + L_2$. Thus, $\{ L_1, L_2 \}$ is an FDS for $g(L)$, and it is minimal because $g(L)$ cannot be written as a function of either $L_1$ or $L_2$ alone. However, the MFDS is not unique: $g(L)$ can alternatively be written as $g(L) := L_1 + 2 L_3$, so $\{ L_1, L_3 \}$ is also an MFDS. If there are no deterministic constraints among members of $L$, then Lemma \ref{lm:unique_mfds} guarantees that the MFDS is unique. 

\noindent In the context of WCC, recall that $K$ is the weight function. Since we assume there are no deterministic constraints among members of $V \cup W$ (Assumption \ref{ass:no_determinism}), Lemma \ref{lm:unique_mfds} implies that there exists a unique MFDS for any choice of $K$. This MFDS is called $H$. If $\KMAR := \pMAR(M = 1 \mid V, W)^{-1}$, we denote the MFDS as $\HMAR \subseteq V \cup W$. As shown in Proposition \ref{prop:H_characterization} below, under Assumptions \ref{ass:positivity} and \ref{ass:no_csi}, $\HMAR$ comprises those members of $V \cup W$ that are dependent on $R$ conditional on the complete variables along with some strict subset of the incomplete variables. 

\begin{scorollary}[MFDS of a sum]
\label{coro:mfds_sums}
Let $L$ be a set of random variables and $\mathcal G$ a set of functions $\{ g_1(L), \cdots, g_{J}(L) \}$. Assume there are no deterministic constraints among members of $L$, and that $g(L) := \sum_m g_m(L)$ exhibits no perfect cancellations per Definition \ref{def:perfect_cancellations_sums_fns}. Suppose some $L^* \in L$ is in the MFDS of some function $g_m(L) \in \mathcal G$. Then $L^*$ is also in the MFDS for $g(L)$.

\begin{proof}
Since $g_m(L)$ is not constant in $L^*$, the assumption of no perfect cancellations in the sum immediately implies that $g(L)$ is not constant in $L^*$, so every FDS for $g(L)$ must contain $L^*$. Since there are no deterministic constraints among members of $L$, Lemma \ref{lm:unique_mfds} establishes that $L^*$ is also in the unique MFDS. 
\end{proof}
\end{scorollary}

\noindent We next give a graphical characterization of $\HMAR$.

\begin{sproposition}[Characterization of $\HMAR$]
\label{prop:H_characterization}

Suppose Assumptions \ref{ass:positivity} and \ref{ass:no_csi} hold. Assume that the sum
$$\sum_{m > 1} p \left( M = m \mid \Vobs{}{m}, \Wobs{}{m} \right)$$
exhibits no perfect cancellations of dependences per Definition \ref{def:perfect_cancellations_sums_fns}. 

Then $\HMAR$ can be expressed as:
\[ \HMAR = \Bigl\{
Z \in (V \cup W) :
\, \, \exists \, \, Q_Z \subsetneq (V^- \cup W^-) \setminus \{ Z \} \; \text{s.t.} \;
Z \notind R \mid \left( (V^+ \cup W^+ \cup Q_Z) \setminus \{ Z \} \right)
\Bigr\}. \]
\normalsize
(Note that $Q_Z$ may be the empty set and need not be the same set for every $Z \in \HMAR$.) 

\begin{proof}
In graphical terms (see Section \ref{sec:DAGs} above), the claim is that any given $Z \in (V \cup W)$ is in $\HMAR$ if and only if there exists a path $\lamZR$ between $Z$ and a member $R^*$ of $R$, such that there exists a set $Q_Z \subsetneq (V^- \cup W^-) \setminus \{ Z \}$ such that $\lamZR$ is unblocked conditional on $(V^+ \cup W^+ \cup Q_Z ) \setminus \{ Z \}$.

\textbf{Claim 1: Path $\lamZR$ exists $\Rightarrow Z \in \HMAR$.}
For a given graph, there exists one graph-compatible pattern $\mt$ such that $Z$ is observed and: 
$$\left( V^-_{\text{obs}}(\mt) \cup W^-_{\text{obs}}(\mt) \right) \setminus \{ Z \} = Q_Z.$$
That is, the set of incomplete variables that are observed for pattern $\mt$ is such that $\lamZR$ is unblocked conditional on that set along with the complete variables.  The incomplete variables that are \textit{missing} for pattern $\mt$ are exactly $(V^- \cup W^-) \setminus ( \{ Z \} \cup Q_Z )$, which is nonempty by assumption. Therefore, $\mt$ is well defined and is not the complete-case pattern. 

By Assumption \ref{ass:positivity}, $p(\mt \mid v, w) > 0$ for all supported $(v,w)$. By Assumptions \ref{ass:positivity} and \ref{ass:no_csi}, pattern regularity holds (Lemma \ref{lm:pattern_regularity}). Therefore, the statistical dependence implied by $\lamZR$ (by Assumption \ref{ass:partial_faithfulness}) propagates to the pattern indicators:
\begingroup
\addtolength{\jot}{0.5em}
\begin{align*}
& \Mequals{\mt} \notind Z \mid \left( (\Vobs{}{\mt} \cup \Wobs{}{\mt})  \setminus \{ Z \} \right).
\end{align*}
\endgroup
Therefore, $p \left( M = \mt \mid Z, \Vobs{}{\mt}, \Wobs{}{\mt} \right)$ is not constant in $Z$.
Since there are no deterministic constraints among members of $V \cup W$, Lemma \ref{lm:unique_mfds} states that an MFDS exists for $p \left( M = \mt \mid Z, \Vobs{}{\mt}, \Wobs{}{\mt} \right)$. Therefore, $Z$ is in the MFDS for:
\begingroup
\addtolength{\jot}{0.5em}
\begin{align}
&p \left( M = \mt \mid Z, \Vobs{}{\mt}, \Wobs{}{\mt} \right). \label{eq:barbara}
\end{align}
\endgroup
The response model is:
\begingroup
\addtolength{\jot}{0.5em}
\begin{align}
p_{\text{MAR}}(M = 1 \mid V, W) &= 1 - \sum_{m > 1} p \left( M = m \mid \Vobs{}{m}, \Wobs{}{m} \right). \label{eq:jack}
\end{align}
\endgroup
 As noted above, $\mt$ is not the complete-case pattern, so expression (\ref{eq:barbara}) is a summand in expression (\ref{eq:jack}). By the assumption that the sum exhibits no perfect cancellations of dependences, Corollary \ref{coro:mfds_sums} establishes that $Z$ is also in the MFDS for $p_{\text{MAR}}(M = 1 \mid V, W)$, and hence for its inverse, $\KMAR$. Therefore, $Z \in \HMAR$.

\textbf{Claim 2: $Z \in \HMAR \Rightarrow$ path $\lamZR$ exists.} Since $Z \in \HMAR$,  at least one summand in the RHS of Eq. (\ref{eq:jack}) is functionally dependent on $Z$. Thus, there exists some pattern $\mt$ such that 
$$p(M = \mt \mid \Vobs{}{\mt}, \Wobs{}{\mt} )$$
is not constant in $Z$,  and $\mt$ is not the complete-case pattern because the sum runs over $m>1$ only. Since $1\{ M = \mt \}$ is a deterministic function of $R$, the soundness of d-separation implies that there exists a path $\lamZR$ between $Z$ and a member of $R$, such that $\lamZR$ is unblocked conditional on
$$( \Vobs{}{\mt} \cup \Wobs{}{\mt} ) \setminus \{ Z \},$$
Setting $Q_Z := ( V^-_{\text{obs}}(\mt) \cup W^-_{\text{obs}}(\mt) ) \setminus \{ Z \}$ (i.e., $Q_Z$ is the set of incomplete variables other than $Z$ that are observed for pattern $\mt$), we have:
$$ Z \notind R \mid \left( ( V^+ \cup W^+ \cup Q_Z ) \setminus \{ Z \} \right), $$
as desired.  Additionally, since $\mt$ is not the complete-case pattern, at least one incomplete variable is missing for $\mt$. By construction, the incomplete variables missing for $\mt$ are exactly $(V^- \cup W^-) \setminus ( \{ Z \} \cup Q_Z )$, so that set is nonempty, completing the proof. \qedhere
\end{proof}
\end{sproposition}

\subsection{Theorem \ref{thm:wcc_conditions}: Necessary and sufficient conditions for $\muWCC(K)$ to be correct}
\label{sec:wcc_conditions}

We first establish a lemma formalizing the concept of a knife-edge constraint, which will be invoked in later results. 

\begin{slemma}[Knife-edge nature of expectation constraint]
\label{lm:empty_interior_solution_set}

Let $X$ be a random variable and let $g_1(X)$ be a bounded, integrable function such that $p( g_1(X) \ne 0 ) > 0$. Suppose there exists another bounded, integrable, nonnegative function $g_2(X)$ such that:
$$ E[ g_1(X) \, g_2(X) ] = 0.$$
Then for any $\epsilon > 0$, there exists a perturbed version of $g_2(X)$, called $\widetilde g_2(X)$, such that
$$ E\!\left[\,| g_2(X) - \widetilde g_2(X) |\,\right] < \epsilon,$$
but
$$ E[ g_1(X)\, \widetilde g_2(X) ] \neq 0.$$

An equivalent, measure-theoretic statement is as follows. Let $\mathbbm P$ be the probability measure governing the distribution of $X$ and
$$ L^1_+(\mathbbm P) := \{ g(X): E [ \, |g(X)| \, ] < \infty \text{ and } g(X) \ge 0 \}, $$
i.e., nonnegative functions with finite $L^1$-norm. The solution set to the expectation constraint is:
$$ \mathcal S := \left\{ g_2(X) \in L^1_+(\mathbbm P) : E[ g_2(X)\, g_1(X) ] = 0 \right\}.$$
Then, wrt the $L^1$ norm, $\mathcal S$ has empty interior in
$L^1_+(\mathbbm P)$.

\begin{proof}
Suppose wlog that $p( g_1(X) > 0 ) > 0$. (The proof is symmetric for the case where $p( g_1(X) < 0 ) > 0$.) Therefore, letting
$$\mathcal X^* := \{ x \, : \, g_1(x) > 0 \},$$
we have $p(X \in \mathcal X^*) > 0$. Fix any $g_2 \in \mathcal S$ and $\delta > 0$, and define
$$ \widetilde g_2(x) := g_2(x) + \delta \, 1\{ x \in \mathcal X^* \}.$$
Then $p( \widetilde g_2(x) \ge 0 ) = 1$, and
$$ E \left[ \, |\widetilde g_2(X) - g_2(X)| \, \right] = \delta \, p( X \in \mathcal X^* ), $$
which can be made arbitrarily small (i.e., smaller than any tolerance $\epsilon$) by choosing $\delta$ sufficiently small. However,
\begin{align*}
&E[ g_1(X)\, \widetilde g_2(X) ] \, = \, E[ g_1(X)\, g_2(X) ]
+ \delta \int_{\mathcal X^*} g_1(x)\, d\mathbbm P \, = \, \delta \int_{\mathcal X^*} g_1(x)\, d\mathbbm P \; > \; 0.
\end{align*}
This shows that every $L^1$-neighborhood of $g_2(X)$ contains functions outside $\mathcal S$, so $\mathcal S$ has empty interior.
\end{proof}
\end{slemma}

\noindent For brevity, we will use ``knife-edge cases'' to describe those in which $K$, despite being misspecified, is coincidentally chosen to create perfect cancellations with a certain bias term. That is, we will ultimately use Lemma \ref{lm:empty_interior_solution_set} to show that $K$ must be chosen from an empty-interior set of candidate functions, so there exists an arbitrarily small perturbation to $K$ such that the perfect cancellation ceases to hold.

\newcommand{\qXFootnote}{ }

\begin{theorem}[Necessary and sufficient conditions for $\muWCC(K)$ to be correct]
\label{thm:wcc_conditions}

Let $K$ be a strictly positive, measurable function. Given $K$, let $\muWCC(K)$ be a corresponding WCC functional for $E[Y \mid X]$. Let $ \Omega := K \cdot p(M = 1 \mid X, Y, W) $. Then $\muWCC(K)$ is correct\qXFootnote if and only if:
\begingroup
\addtolength{\jot}{0.5em}
\begin{align*}
& \text{(C-WCC) } \quad \quad  \Cov(\Omega, Y \mid X) = 0.
\end{align*}
\endgroup

\begin{proof}

Recall from Definition \ref{main-def:wcc} that, for a given $q$, the weighted
complete-case estimating equation is
\begingroup
\addtolength{\jot}{0.5em}
\begin{align}
\label{eq:wtd_est_eqs}
E \big[ U_w(\betabold; q) \big] := E \big[ 1\{M=1\} \, K \, q(X; \betabold, \phibold)
  \, \epsilon(Y, X; \betabold) \big] &= 0,
\end{align}
\endgroup
where $\lambda(X; \betabold)$ is an assumed, correctly specified model for $E[Y \mid X]$; $q(X; \betabold, \phibold)$ returns a $d \times 1$ vector and may depend on nuisance parameters $\phibold$; and $\epsilon(Y, X ; \betabold) := Y - \lambda(X; \betabold)$ returns a scalar. For brevity, we subsequently drop the indexing of $q$ and $\epsilon$ by the parameters.

For a given $q(X)$, let $\betaWCC$ denote the solution to Eq. (\ref{eq:wtd_est_eqs}), which is assumed unique. Then $\muWCC(K)$ is correct if and only if $\betaWCC = \betabold_0$ for every bounded, measurable $q(X)$. In turn, that holds if and only if Eq. (\ref{eq:wtd_est_eqs}) holds when evaluated at $\betabold_0$ for every such $q(X)$. Iterating expectations over $(X, Y, W)$:
$$ E \Big\{ K \, q(X) \, \epsilon(Y, X) \, E  \big[ 1\{M=1\}  \mid X, Y, W \big] \Big\} = 0 $$
$$ E \big[ \Omega \cdot q(X) \, \epsilon(Y, X) \big] = 0. $$
Iterating expectations over $X$:
\begin{align}
\label{eq:38309}  %
E \Big\{  q(X) \, E \big[ \Omega \cdot \epsilon(Y, X) \mid X \big] \Big\} = 0.
\end{align}
This holds for all bounded, measurable $q(X)$ if and only if
\begingroup
\addtolength{\jot}{0.5em}
\begin{align*}
E \big[ \Omega \cdot \epsilon(Y, X) \mid X \big] &= 0. \label{eq:carol}
\end{align*}
\endgroup
(To see the ``only if'' direction, let
$\psi(X) := E \big[ \Omega \, \epsilon(Y, X) \mid X \big]$ and let every entry of $q(X)$ be equal to $\text{sign}( \psi(X) )$, which is bounded and
measurable. Then Eq. (\ref{eq:38309}) above becomes $E \big| \psi(X) \big| = 0$, so $\psi(X) = 0$ almost surely.) We have:
\begingroup
\addtolength{\jot}{0.5em}
\begin{align*}
E \big[ \Omega \cdot \epsilon(Y, X) \mid X \big] &= E\big[ \Omega \, (Y - E[Y \mid X]) \mid X \big] \\
&= E\big[ \Omega \cdot Y  \mid X \big] - E\big[ \Omega \cdot E[ Y \mid X ] \mid X \big] \\
&= \Cov(\Omega, Y \mid X). \qedhere
\end{align*}
\endgroup

\end{proof}
\end{theorem}

\begin{corollary}[Sufficient conditions for (C-WCC)]
\label{coro:wcc_suff_conditions}

Each of the following conditions is sufficient for (C-WCC) to hold:
\begingroup
\addtolength{\jot}{0.5em}
\begin{align*}
& \text{(C1-WCC) } \; \Omega = 1\text{, i.e., } K = p( M = 1 \mid V, W)^{-1} \\
& \text{(C2-WCC) } \; E[ Y \mid H, X, M = 1 ] = E[ Y \mid X ]. \\
\end{align*}
\endgroup

\begin{proof}

\noindent \textbf{Sufficiency of (C1-WCC).} 
 As established in the proof of Theorem \ref{thm:wcc_conditions},
$\Cov(\Omega, Y \mid X) = E \big[ \Omega \cdot \epsilon(Y, X) \mid X \big]$. Under (C1-WCC):

$$ E \big[ \Omega \cdot \epsilon(Y, X) \mid X \big] = E \big[ \epsilon(Y, X) \mid X \big]. $$
Under a correct mean model specification,
$$ E \big[ \epsilon(Y, X) \mid X \big] = E \big[ Y - E[Y \mid X] \mid X \big] = 0, $$
so (C-WCC) holds.

\noindent \textbf{Sufficiency of (C2-WCC).}

In the proof of Theorem \ref{thm:wcc_conditions}, it was established that (C-WCC) is equivalent to 
$$ E  \big[ 1\{M=1\} \, K \, q(X) \, \epsilon(Y, X) \big] = 0. $$
This can be written as:
\begingroup
\addtolength{\jot}{0.5em}
\begin{align*}
E \Big\{ E \big[ 1\{M=1\} \, K \, q(X) \, \epsilon(Y, X) \mid X, H, M=1 \big] \Big\} &= 0 \\ 
\intertext{By the definition of $H$, $K$ is fixed conditional on $H$:}
E \Big\{ 1\{M=1\} \, K \, q(X) \, E \big[ \epsilon(Y, X) \mid X, H, M=1 \big] \Big\} &= 0 \\
E \Big\{ 1\{M=1\} \, K \, q(X) \, E \big[ Y - E[Y \mid X] \mid X, H, M=1 \big] \Big\} &= 0 \\
E \Big\{ 1\{M=1\} \, K \, q(X) \, \left( E \big[ Y \mid X, H, M=1 \big] - E[Y \mid X] \right) \Big\} &= 0.
\end{align*}
\endgroup
Clearly, (C2-WCC) suffices for this to hold. 
\end{proof}
\end{corollary}

\subsection{Theorem \ref{thm:IPW_with_incomplete_Y}: $\muMAR$ with complete $X$ and incomplete $Y$}

\begin{slemma}[Implications of unblocked $Y$--$R$ path and MNAR]
\label{lm:Y_R_plus_MNAR_muMAR_incorrect}
Suppose Assumptions \ref{ass:positivity} and \ref{ass:no_csi} hold, and suppose that MNAR($V \cup W$) holds. If there exists: (1) a path $\eta$ between $Y$ and a member $R^* \in R$; and (2) a set $Q \subseteq (X \cup W)$ such that $(X^+ \cup W^+) \subseteq Q$ and $\eta$ is unblocked conditional on $Q$, then $\muMAR$ is incorrect, barring perfect cancellations such that $\OmegaMAR$ is functionally dependent on $Y$, yet $\Cov(\OmegaMAR, Y \mid X) = 0$.
\begin{proof}
Let $m^*$ be the missingness pattern in which exactly the members of $\{ Y \} \cup Q$ are observed. This pattern is graph-compatible: every member of $V \cup W$ outside $\{ Y \} \cup Q$ is incomplete, because $Q \cup \{ Y \}$ contains all of $V^+ \cup W^+$. By Assumption \ref{ass:positivity}, $m^*$ is realized with positive probability, and by construction $( \Vobs{}{m^*} \cup \Wobs{}{m^*} ) \setminus \{ Y \} = Q$.
 
Since $\eta$ is unblocked conditional on $Q$, Assumption \ref{ass:partial_faithfulness} gives $Y \notind R^* \mid Q$. Pattern regularity holds by Assumptions \ref{ass:positivity} and \ref{ass:no_csi} (Lemma \ref{lm:pattern_regularity}), so:
\begin{equation}
\label{eq:pattern_dependence}
Y \notind 1 \{ M = m^* \} \mid Q.
\end{equation}
Because $Y$ is observed in $m^*$, expression (\ref{eq:pattern_dependence}) says that $p \big( M = m^* \mid \Vobs{}{m^*}, \Wobs{}{m^*} \big)$ is functionally dependent on $Y$. There are two cases.
 
\medskip
 
\begin{quote}
\noindent \textbf{Case 1: $m^*$ is the complete-case pattern.} Then $Q = (V \cup W) \setminus \{ Y \}$, so expression (\ref{eq:pattern_dependence}) becomes $Y \notind 1 \{ M = 1 \} \mid \big( ( V \cup W ) \setminus \{ Y \} \big)$. Because $1 \{ M = 1 \}$ is binary, this says that the true response model $p( M = 1 \mid V, W)$ is functionally dependent on $Y$.
 
\medskip
 
\noindent \textbf{Case 2: $m^*$ is not the complete-case pattern.} Then $m^* > 1$, so $p \big( M = m^* \mid \Vobs{}{m^*}, \Wobs{}{m^*} \big)$ is one of the summands in:
\begin{equation}
\label{eq:pmar_sum}
\pMAR( M = 1 \mid V, W ) = 1 - \sum_{m > 1} p \big( M = m \mid \Vobs{}{m}, \Wobs{}{m} \big).
\end{equation}
Since that summand is functionally dependent on $Y$, Corollary \ref{coro:mfds_sums} (which invokes Assumption \ref{ass:no_cancellations} prohibiting perfect cancellations of dependences) establishes that the sum, and hence $\pMAR( M = 1 \mid V, W )$, is functionally dependent on $Y$.
\end{quote}
 
\medskip
 
\noindent In either case, at least one factor of $\OmegaMAR$ is functionally dependent on $Y$. By Lemma \ref{lm:Omega_FD_on_Y}, which applies because MNAR($V \cup W$) holds, $\OmegaMAR$ is itself functionally dependent on $Y$. Barring perfect cancellations, $\Cov(\OmegaMAR, Y \mid X) \ne 0$, so by Theorem \ref{thm:wcc_conditions}, $\muMAR$ is incorrect.
\end{proof}
\end{slemma}

\begin{sremark}[On the exclusion of perfect cancellations]
\label{rm:averaging_cancellations}
The exclusion of cases in which $\OmegaMAR$ is functionally dependent on $Y$, yet $\Cov(\OmegaMAR, Y \mid X) = 0$ is stated as a single condition, but can also be expressed as two more primitive conditions. Since
\[ \Cov(\OmegaMAR, Y \mid X) = \Cov \big( E[ \OmegaMAR \mid X, Y], \, Y \mid X \big), \]
it would suffice to assume that: (1) when marginalizing $\OmegaMAR(Y, X, w)$ over $p( w \mid Y, X)$, the dependences of $\OmegaMAR(Y, X, w)$ on $Y$ do not cancel out; and (2) the resulting mean dependence $E[\OmegaMAR \mid Y, X] \ne E[\OmegaMAR \mid X]$ is not an uncorrelated-but-dependent configuration in which a function that depends on $Y$ nevertheless outputs a random variable that is \textit{uncorrelated} with $Y$. For brevity, we state (1) and (2) as a single exclusion in Lemma \ref{lm:Y_R_plus_MNAR_muMAR_incorrect} and in the theorems that follow from it.
\end{sremark}

\begin{theorem}[$\muMAR$ with complete $X$ and incomplete $Y$]
\label{thm:IPW_with_incomplete_Y}
Suppose Assumptions \ref{ass:positivity} and \ref{ass:no_csi} hold. If $V^- = \{Y\}$ and MNAR($V \cup W^+$) holds, then $\muMAR$ is incorrect, barring edge cases in which $\OmegaMAR$ is functionally dependent on $Y$, yet $\Cov(\OmegaMAR, Y \mid X) = 0$.

\begin{proof}
Since $V^- = \{ Y \}$ and every member of $W^+$ is complete, $R(V \cup W^+) = \{ R_Y \}$, As established in the proof of Proposition \ref{prop:wneg_cant_fix_mnar}, pattern regularity implies that MNAR($V \cup W^+$) is equivalent to its variable-level form, here $Y \notind R_Y \mid ( X \cup W^+ )$. By Assumption \ref{ass:partial_faithfulness}, there is therefore a path $\eta$ between $Y$ and $R_Y$ that is unblocked conditional on $( X \cup W^+ )$. Take $Q := X \cup W^+$, which is a subset of $X \cup W$ and contains $X^+ \cup W^+$. By Proposition \ref{prop:wneg_cant_fix_mnar}, MNAR($V \cup W^+$) implies MNAR($V \cup W$), so Lemma \ref{lm:Y_R_plus_MNAR_muMAR_incorrect} applies and yields the conclusion.
\end{proof}
\end{theorem}

\subsection{Theorem \ref{thm:mia_conditions}: Sufficient conditions for $\muMIA$ to be correct}

\begin{theorem}[Sufficient conditions for $\muMIA$ to be correct]
\label{thm:mia_conditions}

Let $R_{XW} := 1\{ R(X \cup W) = \mathbf{1} \}$ and:
$$ W_R := \WRdef $$
and $ \Wrc := W \setminus W_R $. The following conditions suffice for $\muMIA$ to be correct:
\begingroup
\addtolength{\jot}{0.5em}
\begin{align*}
& \text{(C1-MIA) } \quad E[ Y \mid X, W, M = 1 ] = E[ Y \mid X, W ] \text{; and }\\
& \text{(C2-MIA) } \quad E[ Y \mid X, W ] = E[ Y \mid X, \Wrc ].
\end{align*}
\endgroup
A special case of (C2-MIA) occurs if $W_R = \emptyset$.
\begin{proof}
The MIA functional can be written as:
    \begingroup
    \addtolength{\jot}{0.5em}
    \begin{align}
        \muMIAx &:= \int_{w_R, \wrc} \, E[ Y \mid X, w_R, \wrc, M = 1 ] \, p( w_R, \wrc \mid X, R_{XW} = 1) d w_R \, d \wrc \notag \\
        \intertext{By (C1-MIA), the conditioning on $M=1$ can be dropped in the first term:}
        &= \int_{w_R, \wrc} E[ Y \mid X, w_R, \wrc ] \, p( w_R, \wrc \mid X, R_{XW} = 1) \, d w_R \, d \wrc \\
        \intertext{By (C2-MIA), the conditioning on $W_R$ can now also be dropped in the first term:}
        &= \int_{w_R, \wrc} E[ Y \mid X, \wrc ] \, p( w_R, \wrc \mid X, R_{XW} = 1) \, d w_R \, d \wrc \\
        \intertext{By regularity Assumption \ref{ass:partial_faithfulness}, the componentwise definition of $\Wrc$ gives $W_j \dsep{G} R(X \cup W) \mid X$ for every $W_j \in \Wrc$. Since any path between $\Wrc$ and $R(X \cup W)$ is a path between some $W_j \in \Wrc$ and $R(X \cup W)$, it follows that $\Wrc \dsep{G} R(X \cup W) \mid X$. The Markov property then yields $\Wrc \ind R(X \cup W) \mid X$, and hence $\Wrc \ind R_{XW} \mid X$ because $R_{XW}$ is a function of $R(X \cup W)$:}
        &= \int_{\wrc} \Bigg\{ E[ Y \mid X, \wrc ] \, p( \wrc \mid X) \int_{w_R} p( w_R \mid \wrc, X, R_{XW} = 1) \, d w_R \Bigg\} \, d \wrc. \notag \\
        \intertext{The inner integral is 1, yielding $E[Y \mid X]$ as desired.} \notag 
    \end{align}
    \endgroup
\end{proof}

\end{theorem}

\subsection{Theorem \ref{thm:IPW_with_Wneg}: $\muMAR$ with all auxiliary variables incomplete}

To prove Theorem \ref{thm:IPW_with_Wneg}, we will need the following lemmas. 

\begin{slemma}[Implications of $\neg$ (C1-MIA)]
\label{lm:C1_MIA_numerator}
If (C1-MIA) is violated, then the true response model $p( M = 1 \mid V, W)$ is functionally dependent on $Y$. (Note: Because $1\{ M = 1 \}$ is binary, this conclusion is equivalent to $Y \notind 1 \{ M = 1 \} \mid (X \cup W)$.)
 
\begin{proof}
Since (C1-MIA) is violated, $E[ Y \mid X, W, M = 1] \ne E[ Y \mid X, W]$, so $p( y \mid X, W, M = 1) \ne p( y \mid X, W)$ on a set of positive probability. By Bayes' rule:
\begingroup
\addtolength{\jot}{0.5em}
\begin{align*}
p( M = 1 \mid X, W, y) = p( M = 1 \mid X, W) \cdot
  \frac{ p( y \mid X, W, M = 1) }{ p( y \mid X, W) },
\end{align*}
\endgroup
and the ratio is not identically one, so $p( M = 1 \mid V, W)$ is functionally dependent on $Y$.
\end{proof}
\end{slemma}

\newcommand{\RAncestryFootnote}{ }

\begin{slemma}[Implications of $\neg \text{ (C2-MIA) } \wedge (W = W^-)$]
\label{lm:C2_MIA_Y_R_path}
Suppose $W = W^-$. If (C2-MIA) is violated, then  there exists a path $\eta$ between $Y$ and a member $R^* \in R$, and a set $Q \subseteq (X \cup W)$ such that $X^+ \subseteq Q$ and $\eta$ is unblocked conditional on $Q$.

\begin{proof}
Throughout, we use the basic fact that if a given path is unblocked conditional on a given set, then every subpath of the path is also unblocked conditional on the same conditioning set.

\noindent \textbf{Step 1: Constructing two constituent paths.} Since (C2-MIA) is violated, $E[ Y \mid X, W ] \ne E[ Y \mid X, \Wrc ]$, so there exists a path $\alpha$ between $Y$ and a member $W^* \in W_R$, such that $\alpha$ is unblocked conditional on $( X \cup \Wrc )$. By the definition of $W_R$, there also exists a path between $W^*$ and a member of $R(X \cup W)$ such that the path is unblocked conditional on $X$; truncate it at the member of $R$ closest to $W^*$ and call the resulting path $\gamma$, with terminal node $R^* \in R$. Thus $\gamma$ is unblocked conditional on $X$, and $\gamma$ has no members of $R$ as nonterminal nodes.

\noindent \textbf{Step 2: Handling the case in which a member $\Rt \in R$ is a nonterminal node on $\alpha$.}\RAncestryFootnote  For this case, let $\eta$ denote the $Y$--$\widetilde R$ subpath of $\alpha$. Since $\alpha$ is unblocked conditional on $Q := ( X \cup \Wrc )$, so is $\eta$. Since $Q \subseteq (X \cup W)$ and $X^+ \subseteq Q$, this completes the proof for the cases defining Step 2.

Henceforth, we assume for the rest of the proof that no member of $R$ lies on $\alpha$.

\noindent \textbf{Step 3: Showing that $\gamma$ is unblocked conditional on $( X \cup \Wrc )$.} Since $\gamma$ is unblocked conditional on $X$ alone, it suffices to show that no members of $\Wrc$ are nonterminal nodes on $\gamma$. Suppose that a member $\widetilde W \in W$ is a nonterminal node on $\gamma$. We claim $\widetilde W \notin \Wrc$. Since $\gamma$ is unblocked conditional on $X$, so is its $\widetilde W$--$R^*$ subpath. Under Assumption \ref{ass:partial_faithfulness}, this establishes that $\widetilde W \notind R^* \mid X$. Since $R^* \in R(X \cup W)$, this means $\widetilde W \in W_R$, as claimed. Since $\gamma$ is unblocked conditional on $X$ and no members of $\Wrc$ are nonterminal on $\gamma$, adding $\Wrc$ to the conditioning set cannot block $\gamma$; hence $\gamma$ is unblocked conditional on $( X \cup \Wrc )$.

\noindent \textbf{Step 4: Splicing the two paths into a composite path $\eta$.} Figure \ref{fig:witnessing_path_Wneg_figure}(a) illustrates the following arguments. Sequentially denote the nodes of $\gamma$ as $(\gamma_1, \gamma_2, \ldots)$ with $\gamma_1 := W^*$, and let $k$ be the highest index such that the node $\gamma_k$ also lies on $\alpha$. Call this node $B$; it is well-defined because $\gamma_1 = W^*$ lies on $\alpha$, so $B$ may be $W^*$ itself. Thus, no node of $\gamma$ after $B$ lies on $\alpha$. Let $\alpha'$ be the $Y$--$B$ subpath of $\alpha$ and $\gamma'$ the $B$--$R^*$ subpath of $\gamma$. Every node of $\gamma'$ other than $B$ lies after $B$ on $\gamma$ and hence not on $\alpha$, so $\alpha'$ and $\gamma'$ share only $B$. Thus, the concatenation of $\alpha'$ and $\gamma'$, called $\eta$, is a path between $Y$ and $R^*$. Note that $W^*$ lies on $\eta$ only if $B$ is $W^*$. Also, by Steps 2 and 3, $R^*$ is the only member of $R$ on $\eta$, and no member of $\Wrc$ lies on $\gamma'$.

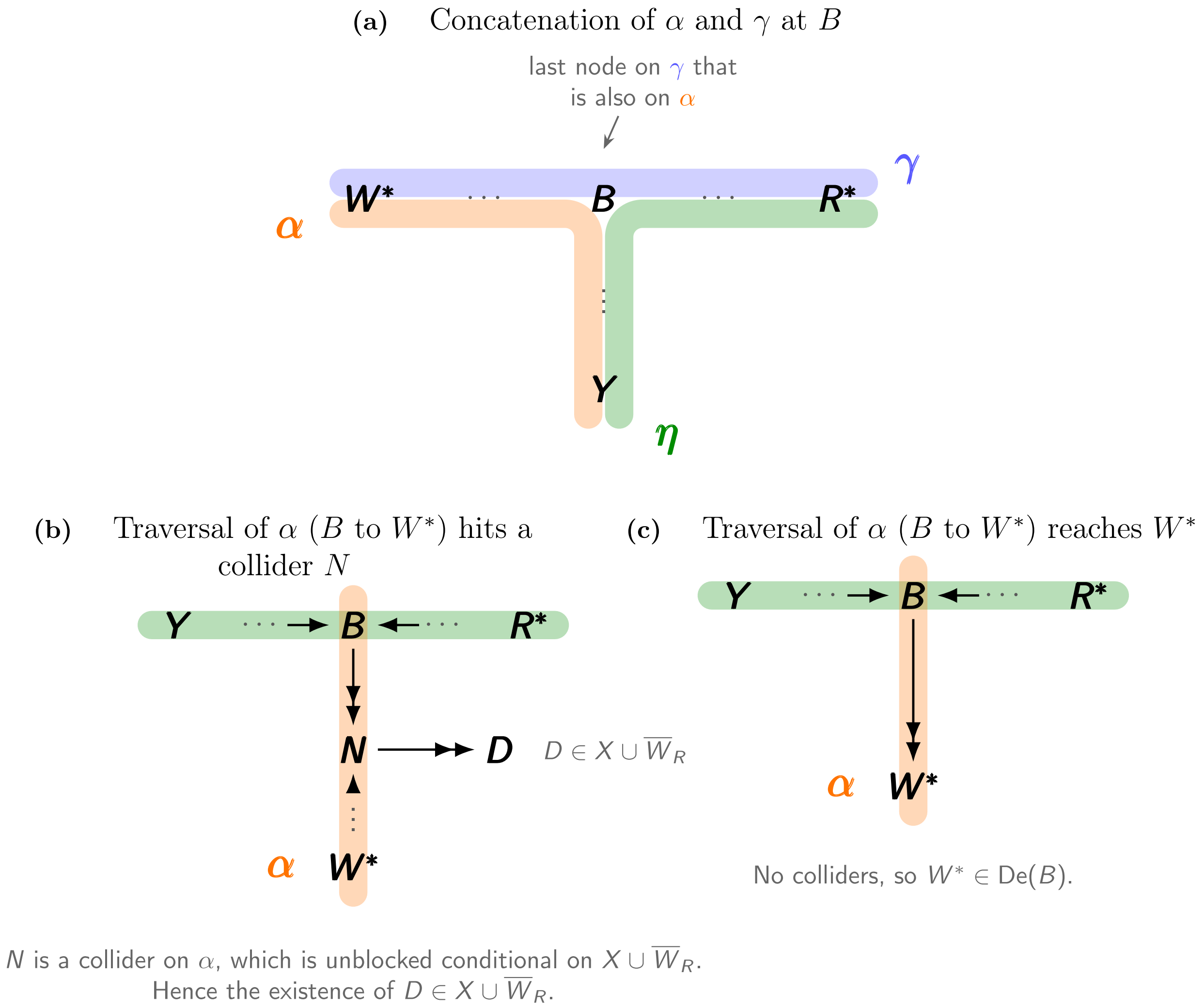
\begin{figure}[!ht]
\centering
\colorlet{cgam}{blue!65}
\colorlet{calp}{orange!90!red}
\colorlet{clam}{green!55!black}
\newcommand{\nlab}[1]{\pmb{#1}}
\newcommand{\pathchip}[1]{\textcolor{#1!28}{\rule[0.25ex]{1.1em}{5pt}}}
\newcommand{\panellab}[1]{{\footnotesize\textbf{(#1)}}}
\tikzset{
  pnl/.style={
    nd/.style={inner sep=2pt, outer sep=2pt, font=\large},
    dts/.style={inner sep=1pt, outer sep=1pt, font=\normalsize, color=black!65},
    pl/.style={inner sep=1pt, outer sep=1pt, font=\Large, text=##1},
    ann/.style={color=black!60, font=\footnotesize, align=center},
    hl/.style={-, draw=##1, line width=11pt, opacity=0.28,
               line cap=round, line join=round, rounded corners=9pt},
    seg/.style={-{Latex[scale=1.1]}, line width=0.9pt, color=black},
    dseg/.style={-{Latex[scale=1.1]Latex[scale=1.1]}, line width=0.9pt, color=black},
  }
}
\def\ro{6pt}   %
\def\rc{10pt}  %

\panellab{a} \quad Concatenation of $\alpha$ and $\gamma$ at $B$

\vspace{1mm}
\begin{tikzpicture}[pnl]
\sansmath \sffamily
\coordinate (aW) at (0,0);
\coordinate (aB) at (3.2,0);
\coordinate (aR) at (6.4,0);
\coordinate (aY) at (3.2,-2.6);
\draw[hl=cgam] ([xshift=-\rc,yshift=\ro]aW) -- ([xshift=\rc,yshift=\ro]aR);
\draw[hl=calp] ([xshift=-\rc,yshift=-\ro]aW) -- ([xshift=-\ro,yshift=-\ro]aB)
                                             -- ([xshift=-\ro,yshift=-\rc]aY);
\draw[hl=clam] ([xshift=\ro,yshift=-\rc]aY) -- ([xshift=\ro,yshift=-\ro]aB)
                                            -- ([xshift=\rc,yshift=-\ro]aR);
\node[nd] (aWn) at (aW) {$\nlab{W^*}$};
\node[nd] (aBn) at (aB) {$\nlab{B}$};
\node[nd] (aRn) at (aR) {$\nlab{R^*}$};
\node[nd] (aYn) at (aY) {$\nlab{Y}$};
\node[dts] at (1.6,0)    {$\cdots$};
\node[dts] at (4.8,0)    {$\cdots$};
\node[dts] at (3.2,-1.3) {$\vdots$};
\node[pl=cgam, right=3mm of aRn, yshift=4mm]   {$\nlab{\gamma}$};
\node[pl=calp, left=3mm of aWn, yshift=-4mm]   {$\nlab{\alpha}$};
\node[pl=clam, below right=1mm and 3mm of aYn] {$\nlab{\eta}$};
\node[ann, above=8mm of aBn, xshift=4mm] (alab)
      {last node on \textcolor{cgam}{$\gamma$} that\\ is also on \textcolor{calp}{$\alpha$}};
\draw[color=black!60, -{Stealth[scale=0.9]}, line width=0.7pt]
      ([xshift=-2mm]alab.south) -- ([yshift=3.5mm]aBn.north);
\end{tikzpicture}

\vspace{6mm}

\begin{minipage}[t]{0.48\textwidth}
\centering
\panellab{b} \quad Traversal of $\alpha$ ($B$ to $W^*$) hits a collider $N$

\vspace{1mm}
\begin{tikzpicture}[pnl]
\sansmath \sffamily
\coordinate (bY) at (0,0);
\coordinate (bB) at (2.4,0);
\coordinate (bR) at (4.8,0);
\coordinate (bN) at (2.4,-1.7);
\coordinate (bW) at (2.4,-3.3);
\coordinate (bD) at (4.4,-1.7);
\draw[hl=clam] ([xshift=-\rc]bY) -- ([xshift=\rc]bR);
\draw[hl=calp] ([yshift=\rc]bB)  -- ([yshift=-\rc]bW);
\node[nd] (bYn) at (bY) {$\nlab{Y}$};
\node[nd] (bBn) at (bB) {$\nlab{B}$};
\node[nd] (bRn) at (bR) {$\nlab{R^*}$};
\node[nd] (bNn) at (bN) {$\nlab{N}$};
\node[nd] (bWn) at (bW) {$\nlab{W^*}$};
\node[nd] (bDn) at (bD) {$\nlab{D}$};
\node[dts] (bE1) at (1.15,0)     {$\cdots$};
\node[dts] (bE2) at (3.65,0)     {$\cdots$};
\node[dts] (bE3) at (2.4,-2.55)  {$\vdots$};
\draw[seg]  (bE1) -- (bBn);
\draw[seg]  (bE2) -- (bBn);
\draw[seg]  (bE3) -- (bNn);
\draw[dseg] (bBn) -- (bNn);
\draw[dseg] (bNn) -- (bDn);
\node[pl=calp, left=2mm of bWn] {$\nlab{\alpha}$};
\node[ann, right=1mm of bDn, align=left] {$D \in X \cup \Wrc$};
\node[ann, below=6mm of bWn]
  {$N$ is a collider on $\alpha$, which is unblocked conditional on $X \cup \Wrc$. \\ Hence the existence of $D \in X \cup \Wrc$.};
\end{tikzpicture}
\end{minipage}
\hfill
\begin{minipage}[t]{0.48\textwidth}
\centering
\panellab{c} \quad Traversal of $\alpha$ ($B$ to $W^*$) reaches $W^*$

\vspace{1mm}
\begin{tikzpicture}[pnl]
\sansmath \sffamily
\coordinate (cY) at (0,0);
\coordinate (cB) at (2.4,0);
\coordinate (cR) at (4.8,0);
\coordinate (cW) at (2.4,-2.6);
\draw[hl=clam] ([xshift=-\rc]cY) -- ([xshift=\rc]cR);
\draw[hl=calp] ([yshift=\rc]cB)  -- ([yshift=-\rc]cW);
\node[nd] (cYn) at (cY) {$\nlab{Y}$};
\node[nd] (cBn) at (cB) {$\nlab{B}$};
\node[nd] (cRn) at (cR) {$\nlab{R^*}$};
\node[nd] (cWn) at (cW) {$\nlab{W^*}$};
\node[dts] (cE1) at (1.15,0) {$\cdots$};
\node[dts] (cE2) at (3.65,0) {$\cdots$};
\draw[seg]  (cE1) -- (cBn);
\draw[seg]  (cE2) -- (cBn);
\draw[dseg] (cBn) -- (cWn);
\node[pl=calp, left=2mm of cWn] {$\nlab{\alpha}$};
\node[ann, below=6mm of cWn]
  {No colliders, so $W^* \in \De(B)$.};
\end{tikzpicture}
\end{minipage}

\caption{\label{fig:witnessing_path_Wneg_figure}Schematic of key steps in
Lemma \ref{lm:C2_MIA_Y_R_path}. Colored strips trace the paths
\protect\pathchip{calp}~$\alpha$, \protect\pathchip{cgam}~$\gamma$, and $\eta$. Ellipses denote any number of intermediate
nodes and edges. A double-headed arrow ($\twoheadrightarrow$) denotes a
directed path of any length. Panel (a) shows Step 4: $\eta$ is formed by splicing the $Y$--$B$ subpath of
$\alpha$ to the $B$--$R^*$ subpath of $\gamma$. Panels (b) and (c) show the two ways the traversal of Step 7, Case 2.2 can terminate. The traversal runs along $\alpha$, starting at $B$ and heading toward $W^*$. Either way, $B$ is shown to have a descendant that is in $X \cup \Wrc \cup \{ W^* \}$.}
\end{figure}

\noindent \textbf{Step 5: Showing that if $B \in X$, then $B$ is a collider on $\eta$.} The terminal nodes of $\alpha$ are $Y$ and $W^*$, and those of $\gamma$ are $W^*$ and $R^*$; none is in $X$. So $B \in X$ implies $B$ is nonterminal on both paths. Since $\alpha$ is unblocked conditional on $( X \cup \Wrc )$ and $\gamma$ conditional on $X$, any members of $X$ on either path must be colliders. Therefore, $B$ is a collider on both paths. The two edges of $\eta$ adjacent to $B$ are an edge of $\alpha$ and an edge of $\gamma$, so both point into $B$, making $B$ a collider on $\eta$.

\newenvironment{caselevel}[1]
  {\par\begingroup\advance\leftskip by #1\relax\parindent=0pt\ignorespaces}
  {\par\endgroup}

\noindent \textbf{Step 6: Showing that no member of $X$ is a noncollider on $\eta$.} A nonterminal node of $\eta$ other than $B$ is nonterminal, with the same collider status, on $\alpha$ or on $\gamma$; since $\alpha$ is unblocked conditional on $( X \cup \Wrc )$ and $\gamma$ conditional on $X$, no member of $X$ is a noncollider on either. The node $B$ is covered by Step 5.

\noindent \textbf{Step 7: Showing that every collider on $\eta$ is either in $D := X \cup \Wrc \cup \{ W^* \}$, or has a descendant in $D$.}

\begin{caselevel}{1.5em}
\noindent \textbf{Step 7, Case 1: A collider on $\eta$ that is not $B$ itself.} Let $N$ be such a collider, so $N$ is also a collider on $\alpha$ or on $\gamma$. Each of those paths is unblocked conditional on $( X \cup \Wrc )$ by Steps 1 and 3, so $N$ is in $X \cup \Wrc$ or has a descendant in that set.

\medskip

\noindent \textbf{Step 7, Case 2: $B$ itself is a collider on $\eta$.}

\begin{caselevel}{1.5em}
\noindent \textbf{Step 7, Case 2.1: $B$ is also a collider on $\alpha$ or on $\gamma$.} Since both $\alpha$ and $\gamma$ are unblocked conditional on $( X \cup \Wrc )$ (Steps 1 and 3), the argument of Case 1 applies unchanged, and $B$ has a descendant in $X \cup \Wrc$.
\end{caselevel}
\begin{caselevel}{1.5em}
\noindent \textbf{Step 7, Case 2.2: $B$ is a collider on neither $\alpha$ nor $\gamma$.} If $B$ is $W^*$, then $B \in D$ and the claim holds. Suppose instead that $B$ is not $W^*$. Since $B$ is a collider on $\eta$, it is nonterminal on $\eta$, so $B$ is not $Y$; and the terminal nodes of $\alpha$ are $Y$ and $W^*$, so $B$ is nonterminal on $\alpha$. Both edges of $\eta$ at $B$ point into $B$, so in particular the edge of $\alpha$ at $B$ on the $Y$ side points into $B$; since, by assumption for Case 2.2, $B$ is not a collider on $\alpha$, the remaining edge of $\alpha$ at $B$ (on the $W^*$ side) points out of $B$. Traverse $\alpha$ from $B$ toward $W^*$. Each node reached by an incoming edge is a descendant of $B$, and is either a collider on $\alpha$ -- in which case it has a descendant in $X \cup \Wrc$, and so does $B$ -- or a noncollider whose remaining edge points onward. The traversal either ends at such a collider or reaches the terminal node $W^*$, in which case $W^*$ is a descendant of $B$. This establishes the claim.
\end{caselevel}
\end{caselevel}

\noindent \textbf{Step 8:  Constructing the conditioning set $Q$. } Let $\Noncoll(\eta)$ denote the set of noncolliders on $\eta$, and define:
\[ Q := (X \cup W) \setminus \Noncoll(\eta). \]
To show that $\eta$ is unblocked conditional on $Q$, it suffices to show that every collider on $\eta$ has a descendant in $Q$. By Step 7, every collider on $\eta$ is in $D = X \cup \Wrc \cup \{ W^* \}$ or has a descendant in $D$, so it suffices to show $D \subseteq Q$. The definition of $D$ implies that $D \subseteq (X \cup W)$, so it remains to show that no member of $D$ is a noncollider on $\eta$:
\begin{outline}[enumerate]
\1 For members of $X$, this is Step 6.
\1 For members of $\Wrc$: a node of $\gamma'$ is never in $\Wrc$ by Step 3; a noncollider of $\alpha$ is never in $\Wrc$ because $\alpha$ is unblocked conditional on $( X \cup \Wrc )$; and $B \notin \Wrc$ by Step 3.
\1 For $W^*$: by Step 4, $W^*$ lies on $\eta$ only if $B = W^*$, so if $B \ne W^*$ the claim is vacuous. If $B = W^*$, then $W^*$ is invoked in Step 7 only in Case 2, in which $B$ is a collider on $\eta$.
\end{outline}

Hence $\eta$ is unblocked conditional on $Q$.  

Additionally, by definition, $Q \subseteq (X \cup W)$. It remains to show that $X^+ \subseteq Q$, i.e., that no members of $X^+$ are in $\Noncoll(\eta)$, since the latter set is excluded from $Q$. Again, this was established in Step 6. Thus, $Q$ satisfies the requirements in the lemma statement. 
\end{proof}
\end{slemma}

\begin{slemma}[Functional dependence of $\OmegaMAR$ on $Y$]
\label{lm:Omega_FD_on_Y}
Suppose MNAR($V \cup W$) holds. If at least one of $p( M = 1 \mid V, W)$ and $\pMAR( M = 1 \mid V, W)$ is functionally dependent on $Y$, then $\OmegaMAR$ is functionally dependent on $Y$.
 
\begin{proof}
By Remark \ref{rm:Omega_function_vs_rv}, we can write:
\[ \OmegaMAR = \log p( M = 1 \mid V, W) - \log \pMAR( M = 1 \mid V, W). \]
\noindent \textbf{Case 1: Exactly one of the two summands is functionally dependent on $Y$.} Then, trivially, the difference is also functionally dependent on $Y$, establishing the claim. 

\noindent \textbf{Case 2: Both summands are functionally dependent on $Y$.} Arguing by contradiction, suppose that $\OmegaMAR$ is functionally independent of $Y$. This requires the two summands' dependences on $Y$ to exactly cancel. There are two ways this could arise. First, it could arise structurally, because $p( M = 1 \mid V, W) = \pMAR( M = 1 \mid V, W)$, but this is ruled out by MNAR($V \cup W$).\footnote{If $p( M = 1 \mid V, W) = \pMAR( M = 1 \mid V, W)$, then $\sum_{m > 1} \big[ p \big( M = m \mid \Vobs{}{m}, \Wobs{}{m} \big) - p( M = m \mid V, W) \big] = 0$. Since $p( \cdot \mid V, W)$ and $\pMAR( \cdot \mid V, W)$ are both probability mass functions over the patterns, a failure of MAR($V \cup W$) implies that at least one summand with $m > 1$ is nonzero. The displayed sum would then itself be a perfect cancellation of dependences, prohibited by Assumption \ref{ass:no_cancellations}.} Otherwise the cancellation is a non-generic cancellation between two distinct functions, which violates regularity Assumption \ref{ass:no_cancellations}. Thus, $\OmegaMAR$ is functionally dependent on $Y$.
\end{proof}
\end{slemma}

\begin{theorem}[$\muMAR$ with all auxiliary variables incomplete]
\label{thm:IPW_with_Wneg}
Suppose Assumptions \ref{ass:positivity} and \ref{ass:no_csi} hold. Suppose $W = W^-$ and MNAR($V \cup W$) holds. If (C1-MIA) or (C2-MIA) is violated, then $\muMAR$ is incorrect, barring edge cases in which $\OmegaMAR$ is functionally dependent on $Y$, yet $\Cov(\OmegaMAR, Y \mid X) = 0$.
 
\begin{proof}
Suppose first that (C1-MIA) is violated. By Lemma \ref{lm:C1_MIA_numerator}, the true response model $p( M = 1 \mid V, W)$ is functionally dependent on $Y$, so by Lemma \ref{lm:Omega_FD_on_Y}, so is $\OmegaMAR$. With the exclusion of edge cases, $\Cov(\OmegaMAR, Y \mid X) \ne 0$, so by Theorem \ref{thm:wcc_conditions}, $\muMAR$ is incorrect.\footnote{This branch invokes neither Assumption \ref{ass:positivity} nor \ref{ass:no_csi}.} Now suppose that (C2-MIA) is violated. By Lemma \ref{lm:C2_MIA_Y_R_path}, there exists a path $\eta$ between $Y$ and a member of $R$, and a set $Q \subseteq (X \cup W)$ such that $X^+ \subseteq Q$ and $\eta$ is unblocked conditional on $Q$. Lemma \ref{lm:Y_R_plus_MNAR_muMAR_incorrect} then yields the conclusion.
\end{proof}
\end{theorem}

\subsection{Theorem \ref{thm:dominating_cases_af4}: Domination over MNAR distributions}
\label{sec:domination_thm}

\begin{theorem}[Domination over MNAR distributions]
\label{thm:dominating_cases_af4}
 
\newcommand{\tempFootnote}{Q: Do we need to also stipulate MAR($V$)? A: No. The only route by which $\muMAR$ could be correct for a distribution in $\PMNAR$ is an averaging cancellation, which the present theorem excludes; no separate condition on MAR($V$) is needed.}
 
Fix $V \cup W$ and let $\PMNAR$ be the class of distributions such that MNAR($V \cup W^+$) holds, and such that Assumptions \ref{ass:positivity} and \ref{ass:no_csi} hold. (By Proposition \ref{prop:wneg_cant_fix_mnar}, distributions in $\PMNAR$ also exhibit MNAR($V \cup W$).) If $W = W^-$ or $V^- = \{Y\}$, termed \textit{domination cases}, then $\muMIA$ dominates $\muMAR$ over $\PMNAR$, barring edge cases in which $\OmegaMAR$ is functionally dependent on $Y$, yet $\Cov(\OmegaMAR, Y \mid X) = 0$.
 
\begin{proof}
In the $V^- = \{Y\}$ case, Theorem \ref{thm:IPW_with_incomplete_Y} establishes that $\muMAR$ is incorrect for every $p \in \PMNAR$. In the $W = W^-$ case, the contrapositive of Theorem \ref{thm:IPW_with_Wneg} establishes that if $\muMAR$ is correct for a given $p \in \PMNAR$, then (C1-MIA) and (C2-MIA) both hold, so by Theorem \ref{thm:mia_conditions}, $\muMIA$ is also correct for that $p$. In either case, the distributions in $\PMNAR$ for which $\muMAR$ is correct are a subset of those for which $\muMIA$ is correct. It therefore suffices to show that the subset is strict, i.e., to find for each case a distribution in $\PMNAR$ such that (C1-MIA) and (C2-MIA) hold but $\muMAR$ is incorrect. For both the $V^- = \{Y\}$ and the $W = W^-$ case, examples appear in Figure \ref{main-fig:sim_dags} in the main text. \qedhere
\end{proof}
\end{theorem}

To provide intuition regarding the exclusion of edge cases in which $\OmegaMAR(V,W)$ depends on $Y$, yet $\Cov(\OmegaMAR, Y \mid X) = 0$, consider the following analogous case in the simple context of estimating an average potential outcome $E[ Y(x) ]$ in a nonrandomized study. Suppose it is known that at least the following edges are present in the data-generating DAG: $X \gets C_1 \to Y \to C_2 \gets X \to Y$. Thus, conditioning $C_1$ blocks a noncausal path between $X$ and $Y$, but conditioning $C_2$ unblocks a different noncausal path. There may be additional latent variables in the DAG, so it is not known whether $C_1$ blocks all noncausal paths. Consider two functionals intended to represent the expected potential outcome $E[ Y(x) ]$ via adjustment for $C_1$ and $C_2$ respectively: 
$$ \theta_A := \int_{c_1} \, E[ Y \mid x, c_1] \, p(c_1) \, dc_1 $$
$$ \theta_B := \int_{c_2} \, E[ Y \mid x, c_2] \, p(c_2) \, dc_2$$
It is clear that $\theta_A$ is the preferable functional: it blocks at least one noncausal path without unblocking any others, whereas $\theta_B$ unblocks one noncausal path without blocking any others. However, it cannot be said that $\theta_A$ dominates $\theta_B$ over the class $\mathcal P$ of all distributions $p(x,y,c_1,c_2)$ unless certain knife-edge cases are excluded. In particular, if there is a latent confounder $U$ with the structure $X \gets U \to Y$, then it is trivial to find cases in which $\theta_A$ is incorrect, and among these, there exist knife-edge cases in which $\theta_B$ is correct. The knife-edge cases arise when the associations produced by the three distinct noncausal paths that are unblocked conditional on $C_2$ (i.e., $X \gets U \to Y$, $X \gets C_1 \to Y$, and $X \to C_2 \gets Y$) cancel perfectly.  

If such a perfect cancellation does not occur, $\theta_B$ can also be correct if there is an uncorrelated-but-dependent configuration: namely, the noncausal component of the dependence between $X$ and $Y$ conditional on $C_2$ does not affect the expectation of $Y$ but rather only higher moments. However, barring these two kinds of cases, it can easily be shown that $\theta_A$ dominates $\theta_B$ over $\mathcal P$. This statement is analogous to Theorem \ref{thm:dominating_cases_af4}.

\section{Estimation via AIPW or TMLE: Discrete $X$}
\label{sec:discrete_X_aipw_tmle}

If $X$ is discrete, $\muMIAx$ can be viewed in terms of an average treatment effect in a subgroup.
That is, if we interpret $Y$ as an outcome, $R_Y$ as a binary treatment indicator, and $\{  X, W, R_{XW} \}$ as a sufficient set of confounders, then $\muMIAx$ becomes the expectation of the potential outcome under treatment (often denoted $Y^{(1)}$) in the subpopulation of individuals with $R_{XW} = 1$ and $X=x$.
It follows from this that the efficient influence function (EIF) of $\muMIAx$ is\autocite{wei2023efficient}
\begingroup
\small
\addtolength{\jot}{0.5em}
\begin{align}
\label{eq:mia_eif}
 \phi(O) &:= \frac{ R_{XW} \, 1 \{ X = x \} }{ p(R_{XW} = 1) \, p(X = x \mid R_{XW} = 1) } \times \left( g(X, W) +
      \frac{ R_Y }{ \pi (X, W) }
      ( Y - g (X, W) ) - \muMIAx
      \right).
\end{align}
\endgroup
Two asymptotically efficient (and asymptotically equivalent) estimators $\muhatMIAaipw(x)$ and $\muhatMIAtmle(x)$ of $\muMIAx$ are now described.
The first is the solution to a set of unbiased estimating equations.
These estimating equations are obtained by taking the sample average of the EIF, and replacing the unknown functions $g(X,W)$ and $\pi(X,W)$ with their estimates, and the unknown probabilities $p(R_{XW}=1)$ and $p(X=x \mid R_{XW} = 1)$ with their corresponding sample proportions.
This leads to estimating equations:
\begingroup
\addtolength{\jot}{0.5em}
\begin{align}
\label{eq:mia_eif_est_eqs}
\sum_{i=1}^n 1 \{ X^i = x \} \, R^i_{XW}  \, \Big[ \widehat g(X^i, W^i) + \frac{R_Y^i}{\widehat \pi(X^i, W^i)} \left( Y^i - \widehat g(X^i, W^i) \right) - \muhatMIA(x)  \Big] &= 0.
\end{align}
\endgroup
The solution to these estimating equations is the AIPW estimator: 
\begingroup
\addtolength{\jot}{0.5em}
\begin{align}
\label{eq:muhatMIAaipw}
\muhatMIAaipw(x) &= \frac{ \sum_{i=1}^n 1 \{ X^i = x \} \, R^i_{XW}  \, \Big[ \widehat g(X^i, W^i) + \frac{R_Y^i}{\widehat \pi(X^i, W^i)} \left( Y^i - \widehat g(X^i, W^i) \right) \Big] }{ \sum_{i=1}^n 1 \{ X^i = x \} \, R^i_{XW} }.
\end{align}
\endgroup
The second asymptotically efficient estimator of $\muMIAx$ is the TMLE $\muhatMIAtmle(x)$.
This is obtained by using the standard TMLE procedure for estimating an expected potential outcome under treatment, using $Y$ as the outcome, $R_Y$ as the treatment indicator, and $W$ as the confounders, and restricting the data set to those individuals with $R_{XW}=1$ and $X=x$.\autocite{gruber2009targeted}

If the estimators of $g(X,W)$ and $\pi(X,W)$ converge to the corresponding true functions and the product of their convergence rates is faster than $n^{-1/2}$, then
\begin{align*}
\sqrt{n}\left\{\muhatMIA(x) - \mu_{\text{MIA}}(x)\right\}
\; \xrightarrow{d} \;
N\!\left(0,\ \mathrm{Var}\{\phi(O)\}\right),
\end{align*}
where $\muhatMIA(x)$ denotes $\muhatMIAaipw$ or $\muhatMIAtmle$ and $ \xrightarrow{d}$ denotes convergence in distribution.
Therefore,
$$\mathrm{Var}\{\muhatMIA(x)\} \approx \mathrm{Var}\{\phi(O)\}/n. $$ 

This variance $\mathrm{Var}\{\phi(O)\}/n$ can be estimated as $n^{-1}$ times the sample variance of expression~(\ref{eq:mia_eif}).
As $g(x,w)$, $\pi(x,w)$ and $\mu_{\text{MIA}}(x)$ are unknown, they are replaced by their estimates.
Likewise, the unknown probability $p(R_{XW}=1)\,p(X = x \mid R_{XW}=1)$ is replaced by the corresponding sample proportion $\frac{1}{n} \sum_{i=1}^{n} R^i_{XW}\,\mathbb{1}\{X^i = x\}$.
This leads to the variance estimator:
\begingroup
\small
\addtolength{\jot}{0.5em}
\begin{align*}
\widehat{\mathrm{Var}}\{ \muhatMIA(x) \}
\;=\; \frac{\sum_{i=1}^{n} R^i_{XW}\,\mathbb{1}\{X^i = x\}
      \left[
        \widehat g(X^i,W^i)
        + \frac{R^i_Y}{\widehat\pi(X^i,W^i)}
          \{Y^i - \widehat g(X^i,W^i)\}
        - \muhatMIA(x)
      \right]^{2}}
     { \left( \sum_{i=1}^{n} R^i_{XW}\,\mathbb{1}\{X^i = x\} \right)^{2} }
\end{align*}
\endgroup
for $\muhatMIA(x)$.
A $(1-\alpha)$ Wald confidence interval for $\mu_{\text{MIA}}(x)$ is given by
$$\muhatMIA(x) \pm z_{1-\alpha/2}
\sqrt{\widehat{\mathrm{Var}}\{\muhatMIA(x)\}}.$$

\section{Estimation via AIPW: Arbitrary variable types in $X$}
\label{sec:estimation_general_case}

The task of estimating $\muMIA$ is closely related to the task of estimating conditional average treatment effects (CATEs) studied by Morzywolek, Decruyenaere and Vansteelandt\autocite{morzywolek2026weighted} (henceforth MDV).
We shall first describe the latter task and some of MDV's results concerning it.
We shall then explain how estimation of $\muMIA$ can be seen as a special case of that task.

Throughout this section only, we adopt MDV's notation, which conflicts with ours in
several places. Within this section: $V$ denotes the CATE conditioning subvector rather
than the analysis variables $X \cup \{ Y \}$; $Z$ denotes the observed data vector rather
than a generic member of $V \cup W$; $A$ denotes a binary treatment rather than a generic
set of nodes; $g$ and $\mathcal{G}$ denote the CATE function and its model class rather
than a generic function and set of functions; $\pi$, $\psi$, and $Q$ denote a propensity
score, a pseudo-outcome, and an outcome regression, respectively. Elsewhere in this
Supplement, all of these symbols carry their usual meanings.

\subsection{An existing AIPW estimator of a CATE}

MDV assume that $n$ iid observations of $Z = (Y, A, X)$ are available, where $X$ is a vector of covariates, $A$ is a binary indicator of treatment, and $Y$ is an outcome. The potential outcomes when treated and untreated are denoted as $Y^1$ and $Y^0$, respectively. The CATE of interest is defined as $E (Y^1 - Y^0 \mid V)$, where $V$ is either $X$ or a subvector of $X$.
Define $\pi (X) = P(A=1 \mid X)$. MDV's Assumptions 1--3 are conditional exchangeability (i.e.\ $Y^a \ind A \mid X$ for $a \in \{0,1\}$), positivity (i.e.\ $0 < \pi(x) < 1$ for all possible values $x$ of $X$), and consistency (i.e.\ $Y^a = Y$ when $A=a$). These assumptions ensure that $E(Y^a \mid X) = E(Y \mid X, A=a)$, and hence the CATE is identified as
\begin{eqnarray}
  E (Y^1 - Y^0 \mid V)
  & = &
  E \{ E (Y^1 - Y^0 \mid X) \mid V \}
  \nonumber \\
  & = &
  E \{ E(Y \mid X, A=1) - E(Y \mid X, A=0) \mid V \}.
  \label{eq:CATE.definition.equivalent}
\end{eqnarray}

MDV consider a number of weighting functions $\omega(X)$, but for simplicity we consider only $\omega(X)=1$.
MDV define the loss function
\begin{equation}
  \mathcal{L} (g) = E[ \{ Y^1 - Y^0 - g(V) \}^2 ],
  \label{eq:CATE.loss}
\end{equation}
where $g = g(V)$ is some function of $V$.
The true value of $g$ is defined as the function that minimizes $\mathcal{L} (g)$ over a specified class $\mathcal{G}$ of functions $g$.
MDV note that if $\mathcal{G}$ includes the true CATE function, then the true value of $g (V)$ equals the CATE.
Otherwise, the true value of $g$ is the closest function to the CATE in the class $\mathcal{G}$ in the sense of an expected squared error.

MDV also define $Q^a = E(Y \mid X, A=a)$.
In Section~\ref{sect:Shaunproof1} we show that the true value of $g$ is also the function that minimizes the loss function
\begin{equation}
  E[ \{ E(Y \mid X, A=1) - E(Y \mid X, A=0) - g(V) \}^2 ]
\label{eq:equivalent.loss.fn}
\end{equation}
over the class $\mathcal{G}$.

MDV's Lemma 4 states that the true value of $g$ is also the function $g$ that minimizes the loss function
\begin{equation}
E [ \{ \psi (X, A, Y) - g \}^2 ]
\label{eq:AIPW1}
\end{equation}
where
\begin{equation}
\psi (X, A, Y) = \frac{A}{ \pi(X) } \{ Y - Q^1 (X) \} + Q^1 (X)
  - \frac{1 - A}{ 1 - \pi(X) } \{ Y - Q^0 (X) \} - Q^0 (X).
\label{eq:AIPW2}
\end{equation}

\subsection{Repurposing the AIPW estimator for $\muMIA$}

We now show how estimation of $\muMIA$ can be reframed as a special case of estimation of the true value of $g$.
This involves making three modifications to the problem of estimating the CATE that we have just discussed.
First, define $Y^0 = 0$.
This implies that $E(Y \mid X, A=0) = 0$, so that the definition of CATE given by expression~(\ref{eq:CATE.definition.equivalent}) reduces to
$
E \{ E(Y \mid X, A=1) \mid V \}
$.
Second, redefine the notation, replacing $A$ by $R_Y$, $X$ by $(X, W, R_{XW})$ and $V$ by $(X, R_{XW})$, so that the CATE is now written as
$
E \{ E(Y \mid X, W, R_{XW}, R_Y=1) \mid X, R_{XW} \}
$.
Third, define $Y = 0$ when $R_{XW}=0$, so that the CATE can be written as
\begin{equation}
E \{ E(Y \mid X, W, R_{XW}=1, R_Y=1) \mid X, R_{XW}=1 \}
\label{eq:recognise}
\end{equation}
when $R_{XW}=1$ and zero when $R_{XW}=0$.
Note that Eq.~(\ref{eq:recognise}) is $\muMIA$.

Having made these three modifications, the loss function defined by Eq.~(\ref{eq:equivalent.loss.fn}) becomes
\begin{equation*}
  E[ \{ E(Y \mid X, W, R_{XW} = R_Y=1) - g(X) \}^2 ].
\end{equation*}
and expressions~(\ref{eq:AIPW1}) and~(\ref{eq:AIPW2}) become
\begin{equation*}
E [ \{ \psi (X, W, Y, R_{XW}, R_Y) - g \}^2 ]
\end{equation*}
and
\begin{equation*}
\psi (X, W, Y, R_{XW}, R_Y) = R_{XW} \left[ \frac{ R_Y }{ \pi(X, W) } \{ Y - Q (X, W) \} + Q (X, W) \right].
\end{equation*}
with $\pi(X, W)$ and $Q(X, W)$ defined as $\pi (X, W) = P(R_Y=1 \mid X, W, R_{XW}=1)$ and $Q (X, W) = E(Y \mid X, W, R_{XW} = R_Y = 1)$.

The random variable $\psi (X, W, Y, R_{XW}, R_Y)$ is called the \textit{pseudo-outcome}.
In Section~\ref{sect:Shaunproof2} we show that
\begin{equation}
  E \{ \psi (X, W, Y, R_{XW}, R_Y) \mid X, R_{XW} = 1 \}
  = \muMIA,
  \label{eq:Shaunequation}
\end{equation}
which partly justifies the use of the pseudo-outcomes in the estimation procedure described below.

Having framed estimation of $\muMIA$ as a special case of CATE estimation, we can estimate $g$ using the estimation procedure described by MDV in their Section 3.3.
First, the $n$ individuals are partitioned into $J$ folds.
Typically, $J$ is taken to be 2, 5 or 10.
Now, estimate the nuisance functions $\pi(X, W)$ and $Q(X, W)$ using the data on the individuals with $R_{XW}=1$ in $J-1$ of the $J$ folds. This can be done using parametric models or data-adaptive methods. We call this Step A.
Step B involves calculating the pseudo-outcome for each individual in the remaining fold, replacing the unknown functions $\pi(X, W)$ and $Q(X, W)$ by their estimates, and then regressing the pseudo-outcomes of the individuals with $R_{XW} = 1$ in the remaining fold on the covariates $X$.
This regression can be done using quasi-maximum likelihood estimation of the parameters in a working parametric model for $\muMIA = E \{ \psi (X, W, Y, R_{XW}, R_Y) \mid X, R_{XW} = 1 \}$ (e.g.\ a linear regression model if $Y$ is continuous).
We call this a working model, because only the model for the conditional {\it expectation} of $\psi (X, W, Y, R_{XW}, R_Y)$ given $X$ and $R_{XW} = 1$ needs to be correctly specified.
Alternatively, a data-adaptive method (e.g.\ a super learner\autocite{vanderlaan2007superlearner}) can be used.
In either case, the objective function to be minimized is
\[
\sum_{i \in \mathcal{S}} R_{XW_i} \; \{ \psi (X_i, W_i, Y_i, R_{XW_i}, R_{Yi}) - g(X_i) \}^2
+ \Delta (g),
\]
where $\mathcal{S}$ denote the set of indices of the individuals in the remaining fold, and $\Delta (g)$ denotes a penalization term that will usually be present when employing a data-adaptive algorithm to avoid the resulting estimate of $g$ being an overly ``complex'' function (for a parametric model, $\Delta (g) = 0$).
This provides an estimate of $g (X)$.
Steps A and B are repeated a total of $J$ times, with Step A using a different $J-1$ of the $J$ folds each time and Step B using the fold not used in Step A. 
Finally, the $J$ estimates of $g$ thus obtained are averaged to give one overall estimate $\hat{g}$ of $g(X)$.

MDV's Section 3.4 describes conditions under which $\hat{g}$ has ``oracle behavior'', i.e.\ is asymptotically equivalent to the corresponding estimator that uses the true unknown functions $\pi (X, W)$ and $Q (X, W)$.
This is useful because it means that, provided these conditions are satisfied, the asymptotic behavior of $\hat{g}$ does not depend on the behavior of the estimators of $\pi (X, W)$ and $Q (X, W)$.
One benefit of oracle behavior is that it may simplify variance estimation.
In particular, when Step B uses a parametric working model, $\hat{g}$ has oracle behavior when the product of the convergence rates of the estimators of $\pi (X, W)$ and $Q(X, W)$ used in Step A is $o_p (n^{-1/2})$.
This allows, for example, the estimators of $\pi (X, W)$ and $Q(X, W)$ to each converge at the $n^{1/4}$ rate, which is slower than the parametric $n^{1/2}$ rate.
The repeated-sampling variance of the parameters in the working parametric model for $g$ can then be estimated using a standard robust estimator, e.g.\ a sandwich estimator.
When Step B instead uses a data-adaptive procedure, $\hat{g}$ has oracle behavior under even weaker conditions on the estimators of $\pi (X, W)$ and $Q(X, W)$ (see MDV's Section 3.4.1).

\subsection{Proof that true value of $g$ minimizes (\ref{eq:equivalent.loss.fn})}

\label{sect:Shaunproof1}

Define $\epsilon^a = Y^a - E(Y \mid X, A=a)$.
Now $\mathcal{L} (g)$ can be rewritten as
\begin{eqnarray}
  \mathcal{L} (g)
  & = &
  E[ \{ Y^1 - Y^0 - g(V) \}^2 ]
  \nonumber \\
  & = &
  E ( E[ \{ Y^1 - Y^0 - g(V) \}^2 \mid X ] )
  \nonumber \\
  & = &
  E ( E[ \{ E(Y \mid X, A=1) + \epsilon^1 - E(Y \mid X, A=0) - \epsilon^0 - g(V) \}^2 \mid X ] )
  \nonumber \\
  & = &
  E ( E[ \{ E(Y \mid X, A=1) - E(Y \mid X, A=0) - g(V) \}^2 \mid X ] )
  \nonumber \\
  && +
  2 \; E ( E[ \{ E(Y \mid X, A=1) - E(Y \mid X, A=0) - g(V) \} (\epsilon^1 - \epsilon^0) \mid X ] )
  \nonumber \\
  && +
  E [ E\{ (\epsilon^1 - \epsilon^0)^2 \mid X \} ]
  \nonumber \\
  & = &
  E[ \{ E(Y \mid X, A=1) - E(Y \mid X, A=0) - g(V) \}^2 ]
  \nonumber \\
  && +
  2 \;   E [ \{ E(Y \mid X, A=1) - E(Y \mid X, A=0) - g(V) \} \times E (\epsilon^1 - \epsilon^0 \mid X) ]
  \nonumber \\
  && +
  E\{ (\epsilon^1 - \epsilon^0)^2 \}.
  \nonumber \\ 
  & = &
  E[ \{ E(Y \mid X, A=1) - E(Y \mid X, A=0) - g(V) \}^2 ]
  \nonumber \\
  && +
  E\{ (\epsilon^1 - \epsilon^0)^2 \}.
  \label{eq:expand.loss}
\end{eqnarray}
Note that line~(\ref{eq:expand.loss}) follows because $E( \epsilon^a \mid X ) = E(Y^a \mid X) - E(Y \mid X, A=a) = E(Y^a \mid X, A=a) - E(Y \mid X, A=a) = E(Y \mid X, A=a) - E(Y \mid X, A=a) = 0$.
The second term in expression~(\ref{eq:expand.loss}) does not depend on $g$.
Hence, the function $g$ that minimizes $\mathcal{L} (g)$ is indeed the function that minimizes expression~(\ref{eq:equivalent.loss.fn}).

\subsection{Proof of equation~(\ref{eq:Shaunequation})}

\label{sect:Shaunproof2}

\begingroup
\addtolength{\jot}{0.5em}
\footnotesize
\begin{align*}
  & E \{ \psi (X, W, Y, R_{XW}, R_Y) \mid X, R_{XW} = 1 \} \\
  &\quad = E \left[ \frac{ R_Y }{ \pi(X, W) } \{ Y - Q (X, W) \} + Q (X, W) \mid X, R_{XW} = 1 \right] \\
  &\quad = E \{ Q(X, W) \mid X, R_{XW} = 1 \} + E \left[ \frac{ R_Y }{ \pi(X, W) } \{ Y - Q (X, W) \} \mid X, R_{XW} = 1 \right] \\
  &\quad = E \{ E (Y \mid X, W, R_{XW} = R_Y = 1) \mid X, R_{XW} = 1 \} \\
  &\qquad\quad + E \left( E \left[ \frac{ R_Y }{ \pi(X, W) } \{ Y - Q (X, W) \} \mid X, W, R_{XW} = 1 \right] \mid X, R_{XW} = 1 \right) \\
  &\quad = \muMIA + E \left( P (R_Y = 1 \mid X, W, R_{XW} = 1) \, E \left[ \frac{ Y - Q (X, W) }{ \pi(X, W) } \mid X, W, R_{XW} = R_Y = 1 \right] \mid X, R_{XW} = 1 \right) \\
  &\quad = \muMIA + E \left( E \left[ Y - Q (X, W) \mid X, W, R_{XW} = R_Y = 1 \right] \mid X, R_{XW} = 1 \right) \\
  &\quad = \muMIA + E (0 \mid X, R_{XW} = 1) \\
  &\quad = \muMIA.
\end{align*}
\endgroup

\section{Estimation via NICE: Arbitrary variable types in $X$}

\label{sec:nice_estimator}

Analogously to estimation of the g-formula, an alternative to the ICE estimator is to estimate $p( w \mid x, R_{XW} = 1 )$ using any consistent nonparametric or sufficiently smooth semiparametric or parametric estimators, such as moments or maximum likelihood estimators. In turn, an estimator $\muhatMIAnice(x)$ of $\muMIAx$ can be obtained using analytic or numerical integration over $W$. Our R package \texttt{miapack} implements this estimator when $Y$ is continuous or binary and $W$ consists of normal, binary, or categorical variables. Briefly, the package uses the following computational algorithm, which accommodates both discrete and continuous $X$.

First, the outcome model $E [ Y \mid X=x, W, M = 1 ]$ is estimated using a generalized linear model (GLM) with the identity link if $Y$ is continuous, or the logit link if $Y$ is binary. The nuisance function $p( w \mid x, R_{XW} = 1 )$ is obtained using the factorization:
\begin{equation*}
    p( w \mid x, R_{XW} = 1 ) = \prod_{j} p( w_j \mid x, w_1, \dots, w_{j-1}, R_{XW} = 1 )
\end{equation*}
for any user-specified, arbitrary ordering of the members of $W$. The distribution of each $W_j \in W$ conditional on $X, W_1, \dots, W_{j-1}, R_{XW} = 1$ is estimated using an appropriate GLM (linear models for continuous variables, logistic models for binary variables, and multinomial log-linear models for categorical variables). After estimating the component models, the package obtains $\muhatMIAnice(x)$ using Algorithm \ref{alg:MIA}.

Confidence intervals for $\muhatMIAnice(x)$ are constructed using the nonparametric bootstrap \autocite{efron1994introduction}. This procedure relies on standard regularity conditions that hold for common parametric and smooth semiparametric estimators. Bootstrap replicates are obtained in a standard manner, as follows. Letting $N$ denote the sample size, each bootstrap replicate is obtained by: (1) sampling $N$ individuals independently with equal probability and with replacement from the original dataset; (2) obtaining $\muhatMIAnice(x)$ in resampled data. Using the bootstrap replicates, the package constructs standard types of confidence intervals for $\muhatMIAnice(x)$, including percentile intervals and bias-corrected and accelerated (BCa) intervals. 

\begin{algorithm}[H]
\caption{Monte Carlo NICE estimator of $\muMIAx$}
\label{alg:MIA}
\begin{algorithmic}[1]
\REQUIRE Data $(Y^i, X^i, W^i, R^i)$ for units of analysis $(i \in \{1, \cdots, N\})$, evaluation point $x$, number of Monte Carlo iterations $B$
\ENSURE $\muhatMIAnice$

\STATE Estimate $\widehat{y}(x, w) := E[Y \mid X, W, M = 1]$ using complete cases

\FOR{$j = 1,\dots, |W|$}
        \STATE Estimate $\widehat p(w_j \mid x, w_{<j}, R_{XW} = 1)$ using all individuals with all variables in $X \cup W$ observed
\ENDFOR

\FOR{$i = 1,\dots, B$}

    \FOR{$j = 1,\dots, |W|$}
        \STATE Simulate a value $\wt_j^{(i)}$ of $W_j$ from $\widehat{p}(w_j \mid x, w_{<j}, R_{XW} = 1)$
    \ENDFOR
    \STATE Compute $u^{(i)} = \widehat{y}(x, \wt^{(i)})$
\ENDFOR
\STATE \textbf{return} $\displaystyle \muhatMIAnice = \frac{1}{B} \sum_{i=1}^B u^{(i)}$
\end{algorithmic}
\end{algorithm}

\section{Applied example: Changes in sexual identity}
\label{sec:applied}

The Stockholm Public Health Cohort (SPHC) is a prospective, population-based cohort in Stockholm County, Sweden.\autocite{svensson2013cohort, ces_regionstockholm_sphc_data} To date, five subcohorts have been established in the years 2002, 2006, 2010, 2014, and 2021. For this study, we used data from the 2014 subcohort (i.e., SPHC 2014). 

At baseline (i.e., in 2014), area-stratified random samples were drawn from the population of Stockholm County aged 16 years and older. The sampling frame was selected from the Total Population Register (TPR) on 2014-08-31 and comprised approximately 1.8 million individuals. Participants were followed up through a survey in 2021. In total, 53,037 individuals were sampled in SPHC 2014, of whom 22,250 (42.0\%) participated in the 2014 survey and 14,704 (27.7\%) participated in the 2021 follow-up.

Sexual identity was measured in 2014 and 2021 with the question, “How do you define your sexual orientation?” with four response options: “heterosexual,” “homosexual,” “bisexual,” and “none of the above.” For simplicity, participants were categorized as heterosexual or non-heterosexual. Among participants, 20,249 (91.0\%) reported their sexual identity in 2014 and 13,885 (62.4\%) did so in 2021. Missingness in 2021 was primarily due to loss to follow-up.

In Sweden, each individual registered as a resident is assigned a unique personal identity number, which enables linkage to the national and regional registers. Age, sex, and country of birth were obtained from the Total Population Register. Occupation was measured in the 2014 survey with the question “What is/was your occupation or work tasks?”

\begin{figure}[H]
  \centering
  \begin{minipage}{0.35\textwidth}
    \includegraphics[width=\linewidth]{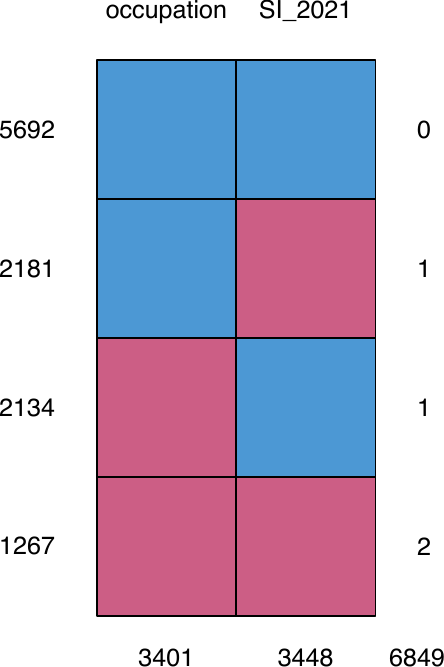}
  \end{minipage}%
  \begin{minipage}{0.45\textwidth}
    \captionof{figure}{Missing data pattern. The plot shows the pattern of missing and observed data across occupation and SI\_2021 (sexual identity in 2021). Blue cells indicate observed data, and red cells indicate missing data. Each row represents a unique missing data pattern. Numbers on the right show the number of variables missing in that pattern, numbers on the left indicate the count of observations for each pattern, and numbers along the bottom show total missing cases per variable and overall.}
    \label{fig:missing_pattern}
  \end{minipage}
\vspace{1.5em}
\end{figure}

 In this study, the analysis was restricted to individuals aged 25–60 years at baseline. This age range was chosen because many younger individuals had not yet entered the workforce, whereas most older individuals were likely to be retired; in both cases, participants often reported missing data on occupation. In Sweden, people generally retire after age 60 years, although retirement age varies between individuals.\autocite{informationsverige_pension_support} The definitions of self-reported and registry-based variables are available in Table \ref{tb:applied_var_defns}. The characteristics of the selected participants are summarized in Table \ref{tb:baseline_characteristics_2}. The missing data pattern is illustrated in Figure \ref{fig:missing_pattern}. Specifically, among participants with missing data on occupation, 37.3\% also had missing data on sexual identity in 2021; conversely, among those with missing sexual identity data, 36.7\% had missing occupation data.

 For each MIA estimator, we inspected residual diagnostic plots for the conditional outcome model $E[Y \mid X, W, M=1]$ using the DHARMa (Diagnostics for HierArchical Regression Models) package\autocite{Hartig2026DHARMa} (Figures \ref{fig:applied_dharma_mia_ice_model_2}--\ref{fig:applied_dharma_mia_ice_model_4}). For arbitrary GLMs, the package creates residual plots that can be interpreted similarly to residual plots from ordinary least squares.\autocite{dunn1996randomized} This facilitates inspection of residuals from GLMs with non-identity links. We also inspected MICE convergence plots (Figures \ref{fig:applied_mice_model_1_convergence}--\ref{fig:applied_mice_model_4_convergence}) and bar charts of imputed values for incomplete categorical variables (Figures \ref{fig:applied_mice_model_1_distribution_id}--\ref{fig:applied_mice_model_4_distribution_occ}). For both estimators, visual inspection and diagnostic tests did not indicate substantial model misspecification or convergence problems.

For comparison to the ICE estimator, we also included results from the NICE estimator (see Section \ref{sec:nice_estimator} and Figure \ref{fig:applied_forest_plot_check}) in which $p(w \mid x, R_{XW} = 1)$ is estimated using a factorization of $W$ in an arbitrary ordering, with each component estimated via a GLM. This estimator, called ``MIA-NICE'', will often be more prone to misspecification than is the ICE estimator, so was included only for comparison.

\vspace{0.3em}

\begin{table}[H]
\centering
\resizebox{0.95\columnwidth}{!}{%
\begin{tabular}{@{}lll@{}}
\toprule
\textbf{Variable} & \textbf{Definition} & \textbf{Source} \\
\midrule

\makecell[l]{Sexual identity} 
& Heterosexual or non-heterosexual 
& Self-report \\ \hline

Age (years) 
& \makecell[l]{Age in 2014} 
& TPR \\ \hline

Sex 
& \makecell[l]{Female or male, as assigned at birth or legally registered} 
& TPR \\ \hline

Country of birth 
& \makecell[l]{Sweden, Europe (excl. Sweden), or outside Europe} 
& TPR \\ \hline 

Occupation 
& \makecell[l]{Manual and field trades, service and support, or expertise\\ and leadership} 
& Self-report \\

\bottomrule
\end{tabular}
}
\vspace{0.3em}

\begin{minipage}{0.95\columnwidth}
\caption{\label{tb:applied_var_defns}Variable definitions and sources. Abbreviations: TPR, Total Population Register.}
\end{minipage}

\end{table}

\begin{table}[H]
\centering
\small
\resizebox{0.92\textwidth}{!}{%
\begin{tabular}{l c c c c}
\toprule
& \multicolumn{3}{c}{\textbf{Identity change}} & \\
\cmidrule(lr){2-4}
\textbf{Characteristics} & \textbf{No (\%)} & \textbf{Yes (\%)} & \textbf{Missing (\%)} & \textbf{Total N (\%)} \\
\midrule

Total N & 7,468 (100) & 358 (100) & 3,448 (100) & 11,274 (100) \\

Age (years), mean (SD) & 44.9 (9.7) & 43.7 (9.9) & 41.8 (9.9) & 43.9 (9.8) \\

Sex & & & & \\
\quad Male & 3,276 (43.9) & 135 (37.7) & 1,603 (46.5) & 5,014 (44.5) \\
\quad Female & 4,192 (56.1) & 223 (62.3) & 1,845 (53.5) & 6,260 (55.5) \\

Country of birth & & & & \\
\quad Sweden & 6,378 (85.4) & 228 (63.7) & 2,477 (71.8) & 9,083 (80.6) \\
\quad Europe (excl. Sweden) & 599 (8.0) & 45 (12.6) & 423 (12.3) & 1,067 (9.5) \\
\quad Outside Europe & 491 (6.6) & 85 (23.7) & 548 (15.9) & 1,124 (10.0) \\

Occupation & & & & \\
\quad Manual and field trades & 349 (4.7) & 26 (7.3) & 210 (6.1) & 585 (5.2) \\
\quad Service and support & 1,218 (16.3) & 77 (21.5) & 595 (17.3) & 1,890 (16.8) \\
\quad Expertise and leadership & 3,899 (52.2) & 123 (34.4) & 1,376 (39.9) & 5,398 (47.9) \\
\quad Missing & 2,002 (26.8) & 132 (36.9) & 1,267 (36.7) & 3,401 (30.2) \\

Sexual identity in 2014 & & & & \\
\quad Heterosexual & 7,083 (94.8) & 188 (52.5) & 3,043 (88.3) & 10,314 (91.5) \\
\quad Non-heterosexual & 385 (5.2) & 170 (47.5) & 405 (11.7) & 960 (8.5) \\

\bottomrule
\end{tabular}
}

\vspace{0.3em}

\begin{minipage}{0.92\columnwidth}
\caption{\label{tb:baseline_characteristics_2}Characteristics of participants aged 25–60 years at baseline in SPHC 2014. Variable definitions are available in Table \ref{tb:applied_var_defns}. Abbreviations: SD, standard deviation; SPHC, Stockholm Public Health Cohort.}
\end{minipage}

\end{table}

\newcommand{\appliedPlotCoef}{0.75}

\begin{figure}[h]
\centering
\includegraphics[width=\appliedPlotCoef\linewidth]{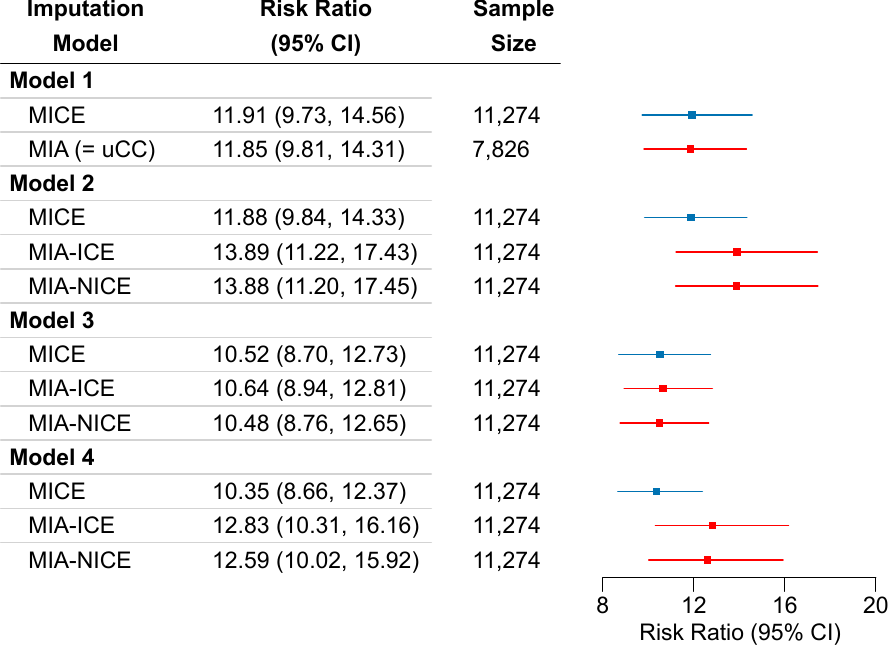}
\caption{Comparison of MIA-NICE and MIA-ICE estimates. For MIA-NICE Models 3 and 4, auxiliary variables were factorized in the arbitrary order sex, age, country of birth, and (for Model 4) occupation. Estimates from MICE and uCC are also shown for comparison. Abbreviations: uCC, unweighted complete-case analysis; CI, confidence interval; MICE, multiple imputation by chained equations.}
\label{fig:applied_forest_plot_check}
\end{figure}

\begin{figure}[h]
\centering
\includegraphics[width=\appliedPlotCoef\linewidth]{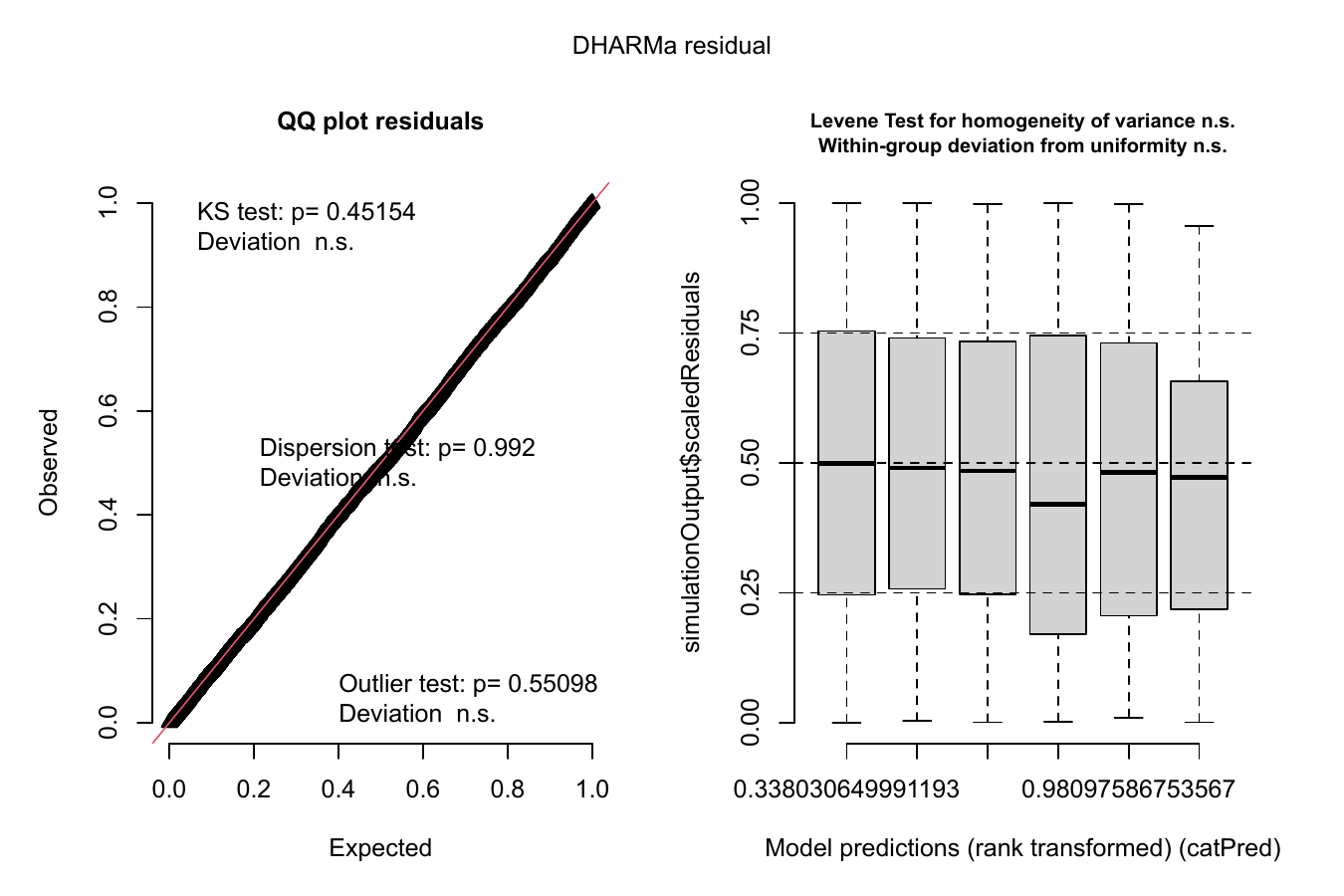}
\caption{\black{DHARMa residual diagnostics for the inner regression of MIA-ICE Model 2. The left panel shows the quantile-quantile plot of simulated scaled residuals, with tests for correct distribution, dispersion, and outliers. The right panel shows simulated scaled residuals against model predictions, with tests for within-group uniformity and homogeneity of variance.}}
\label{fig:applied_dharma_mia_ice_model_2}
\end{figure}

\begin{figure}[h]
\centering
\includegraphics[width=\appliedPlotCoef\linewidth]{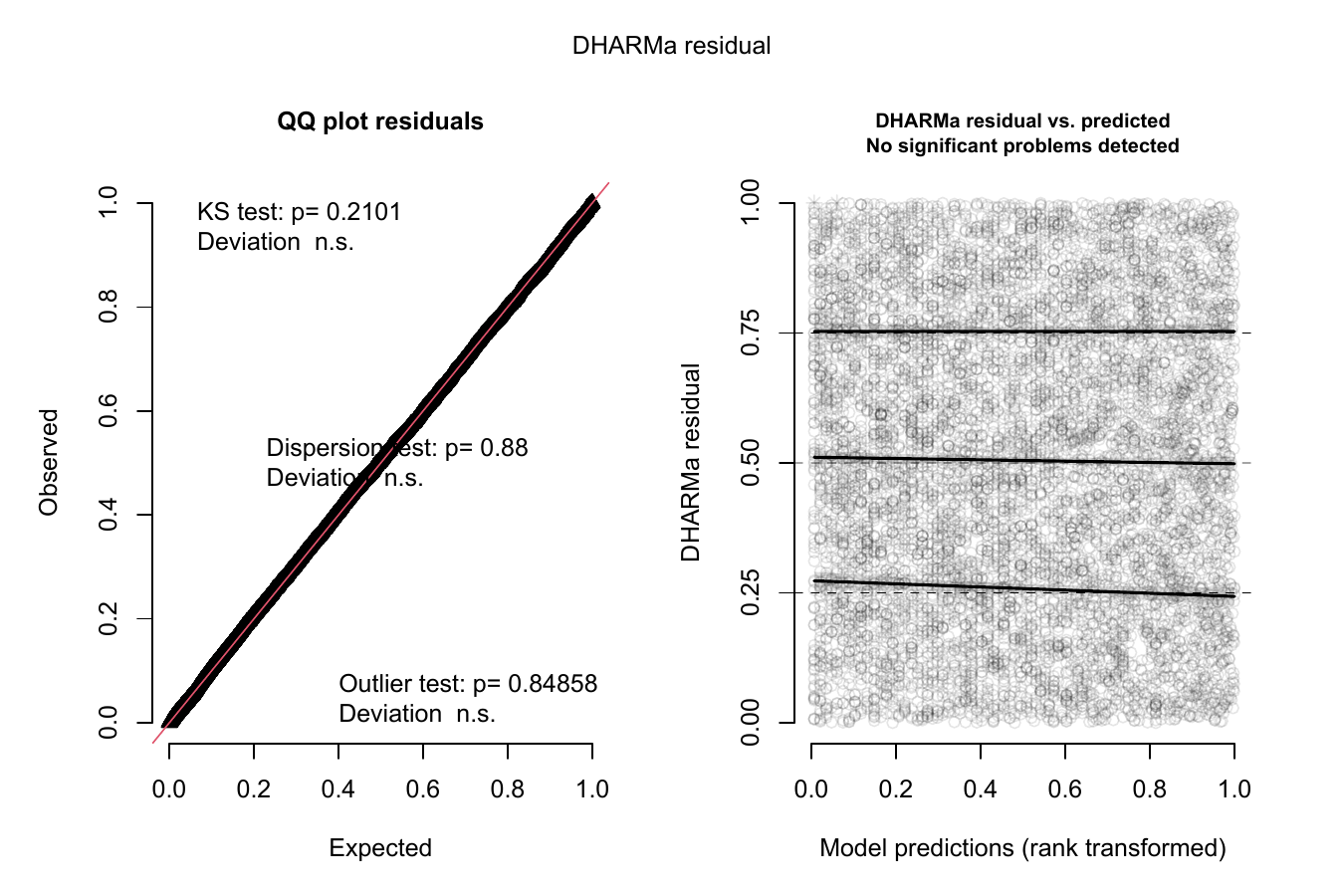}
\caption{\black{DHARMa residual diagnostics for the inner regression of MIA-ICE Model 3. The left panel shows the quantile-quantile plot of simulated scaled residuals, with tests for correct distribution, dispersion, and outliers. The right panel shows simulated scaled residuals against model predictions, with tests for within-group uniformity and homogeneity of variance.}}
\label{fig:applied_dharma_mia_ice_model_3}
\end{figure}

\begin{figure}[h]
\centering
\includegraphics[width=\appliedPlotCoef\linewidth]{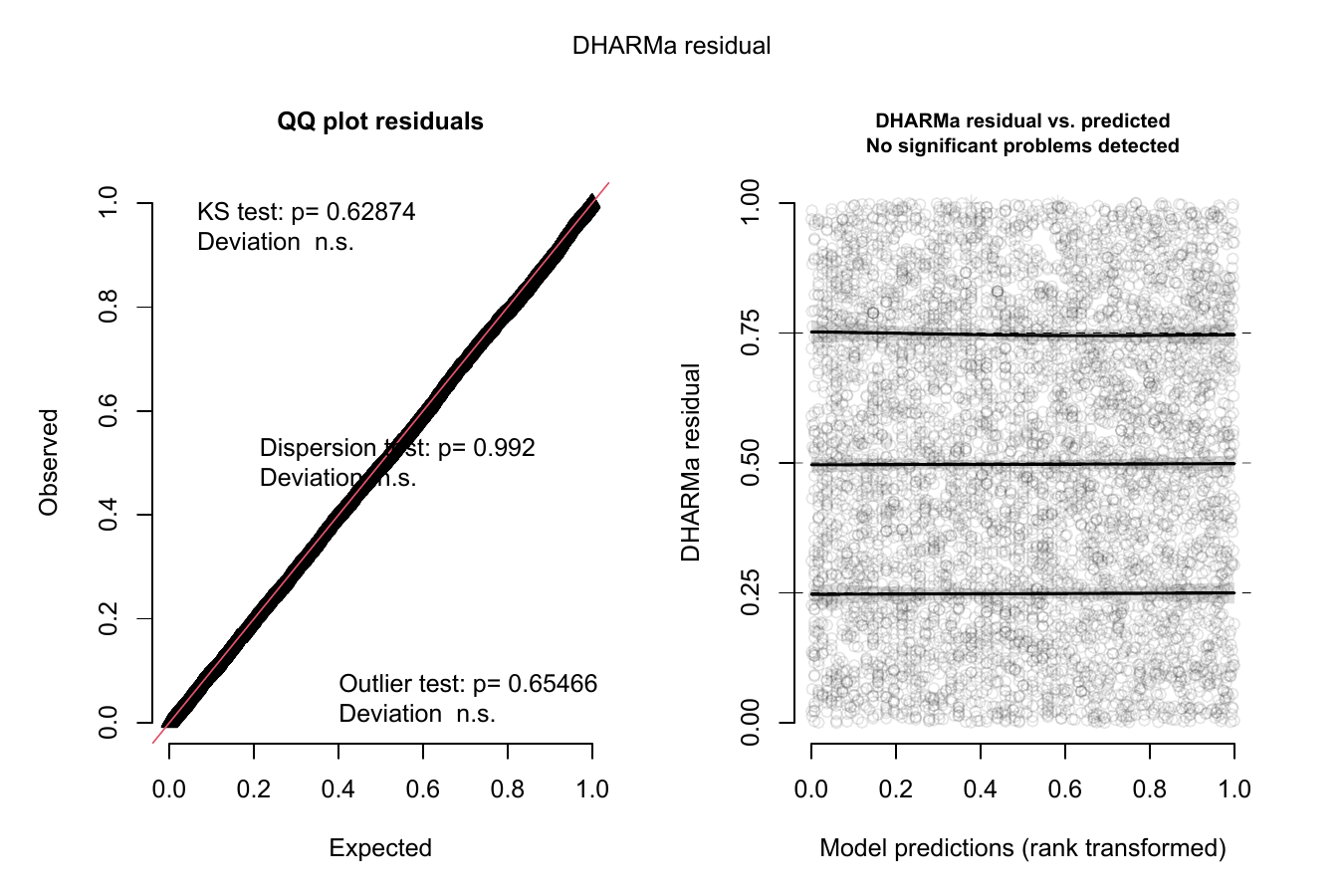}
\caption{\black{DHARMa residual diagnostics for the inner regression of MIA-ICE Model 4. The left panel shows the quantile-quantile plot of simulated scaled residuals, with tests for correct distribution, dispersion, and outliers. The right panel shows simulated scaled residuals against model predictions, with tests for within-group uniformity and homogeneity of variance.}}
\label{fig:applied_dharma_mia_ice_model_4}
\end{figure}

\begin{figure}[h]
\centering
\includegraphics[width=\appliedPlotCoef\linewidth]{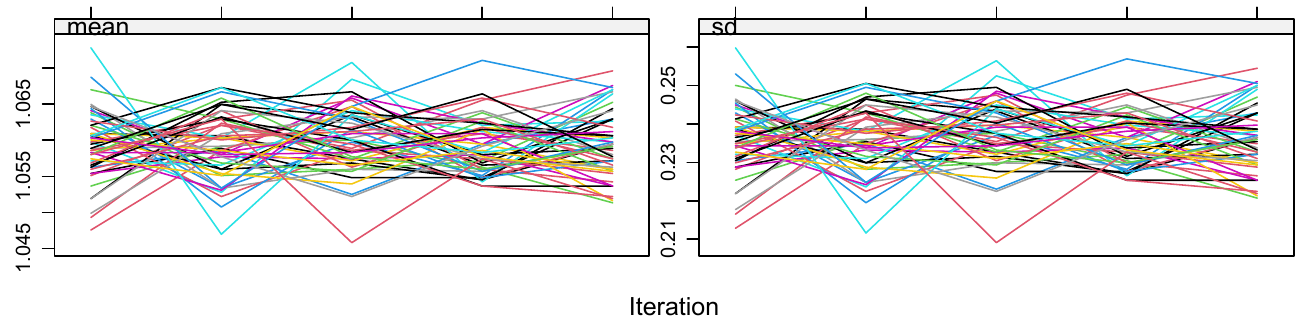}
\caption{\black{MICE convergence plot, Model 1.}}
\label{fig:applied_mice_model_1_convergence}
\end{figure}

\begin{figure}[h]
\centering
\includegraphics[width=\appliedPlotCoef\linewidth]{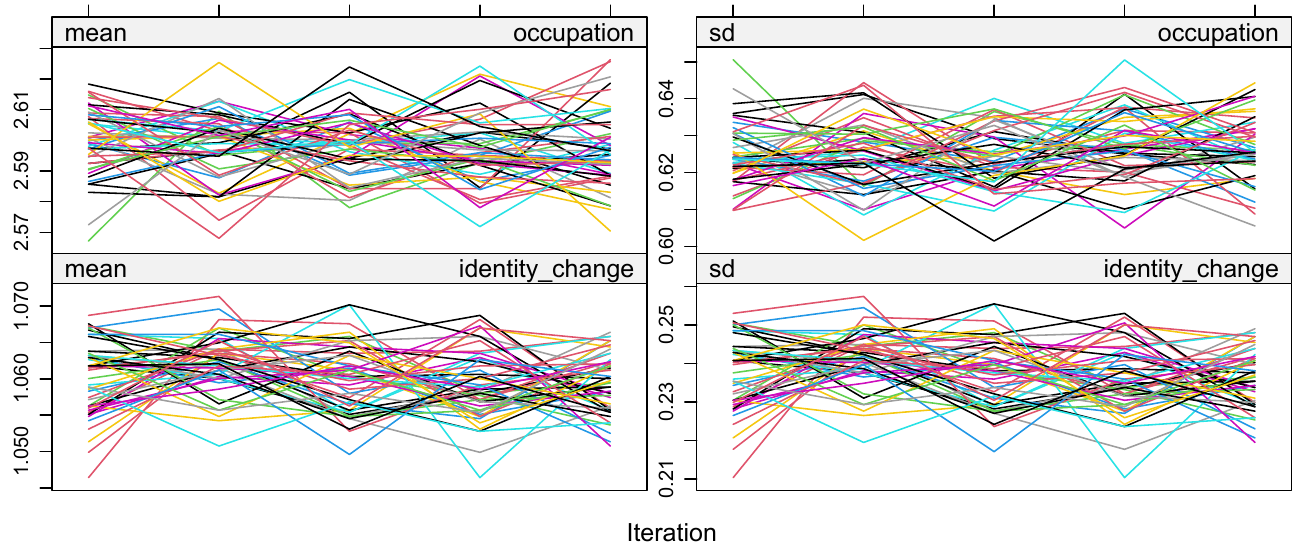}
\caption{\black{MICE convergence plot, Model 2.}}
\label{fig:applied_mice_model_2_convergence}
\end{figure}

\begin{figure}[h]
\centering
\includegraphics[width=\appliedPlotCoef\linewidth]{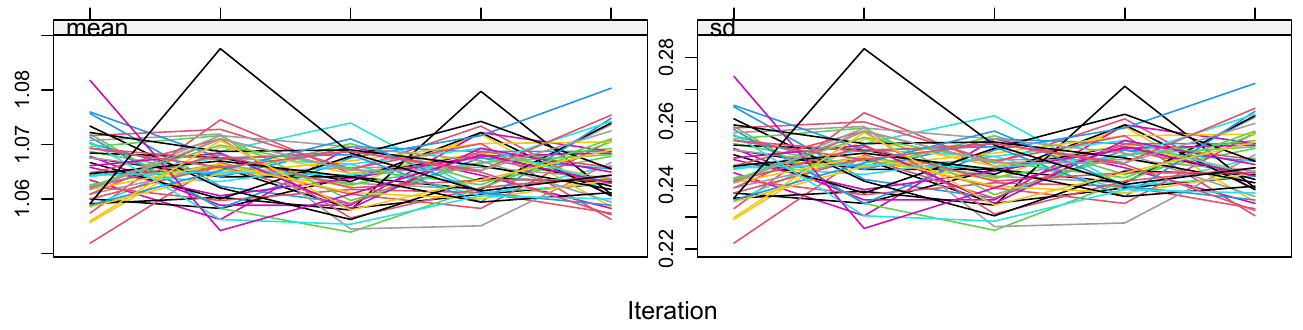}
\caption{\black{MICE convergence plot, Model 3.}}
\label{fig:applied_mice_model_3_convergence}
\end{figure}

\begin{figure}[h]
\centering
\includegraphics[width=\appliedPlotCoef\linewidth]{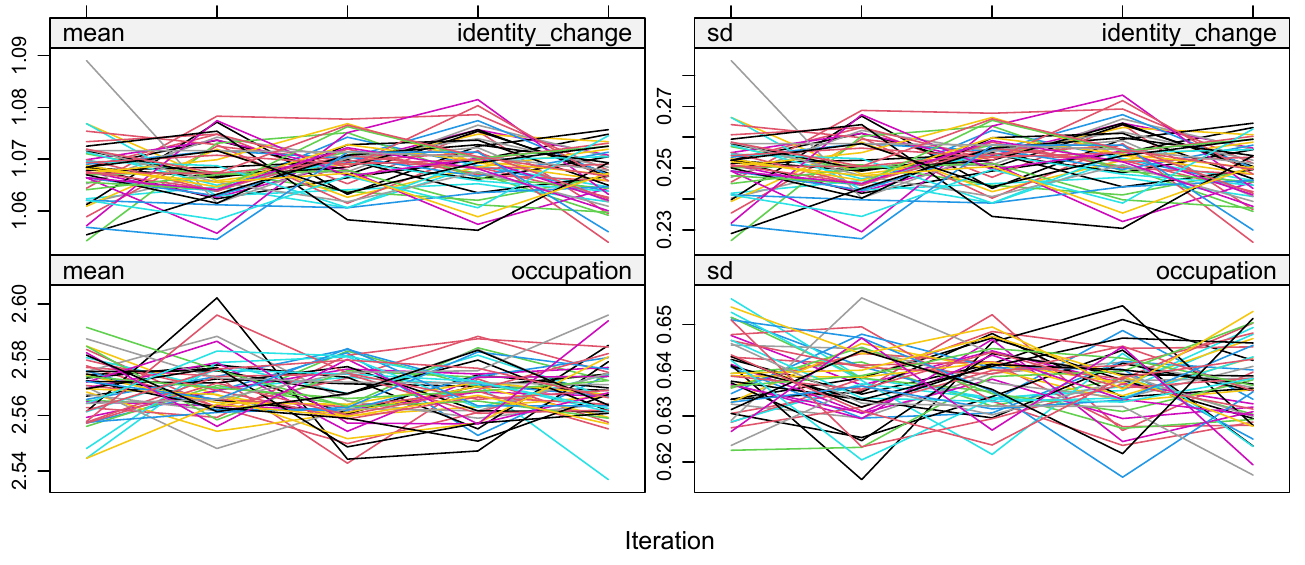}
\caption{\black{MICE convergence plot, Model 4.}}
\label{fig:applied_mice_model_4_convergence}
\end{figure}

\begin{figure}[h]
\centering
\includegraphics[width=\appliedPlotCoef\linewidth]{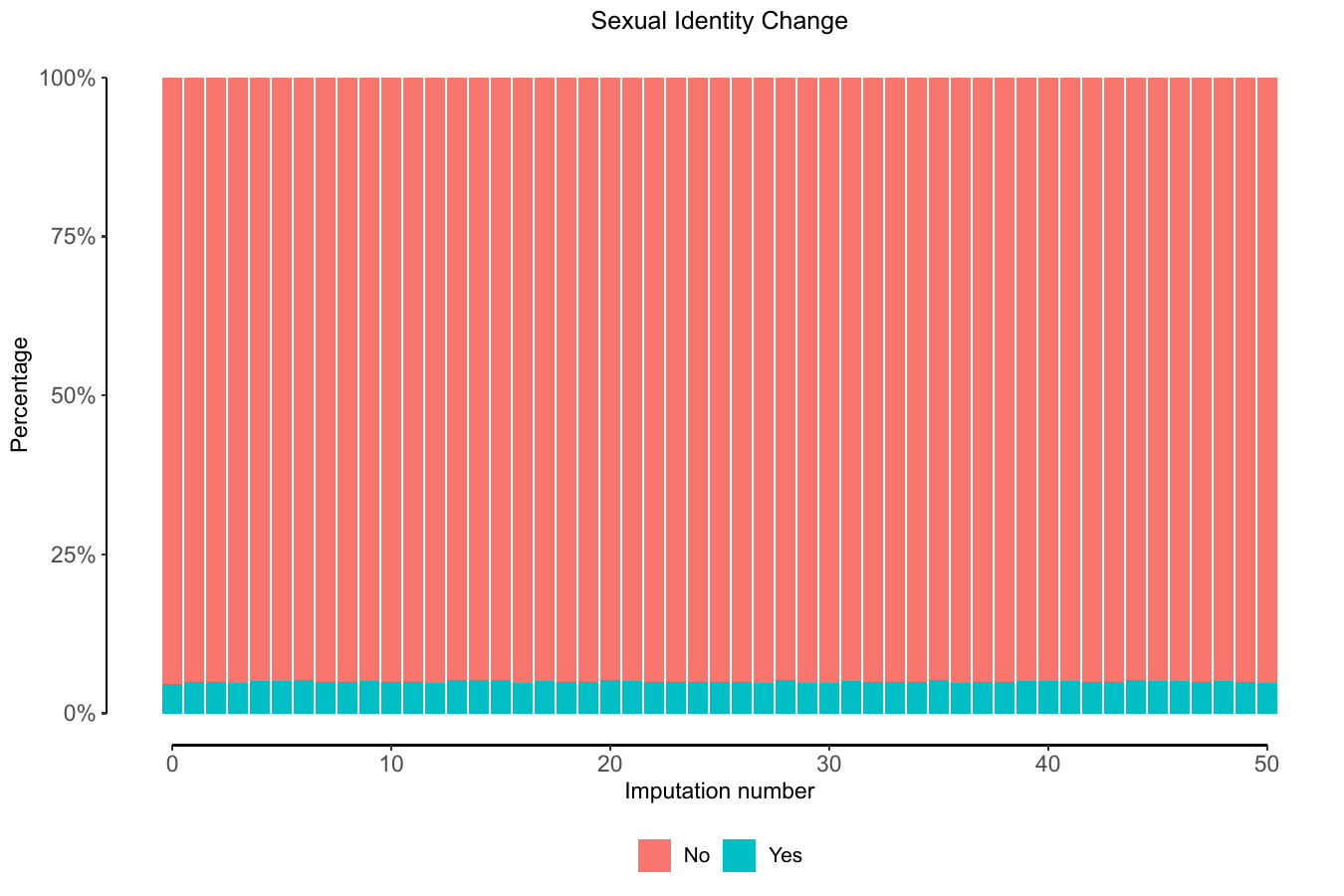}
\caption{\black{Distribution of observed and imputed values for sexual identity change, Model 1. Imputation 0 represents the original incomplete dataset; imputations 1–50 represent the imputed datasets.}}
\label{fig:applied_mice_model_1_distribution_id}
\end{figure}

\begin{figure}[h]
\centering
\includegraphics[width=\appliedPlotCoef\linewidth]{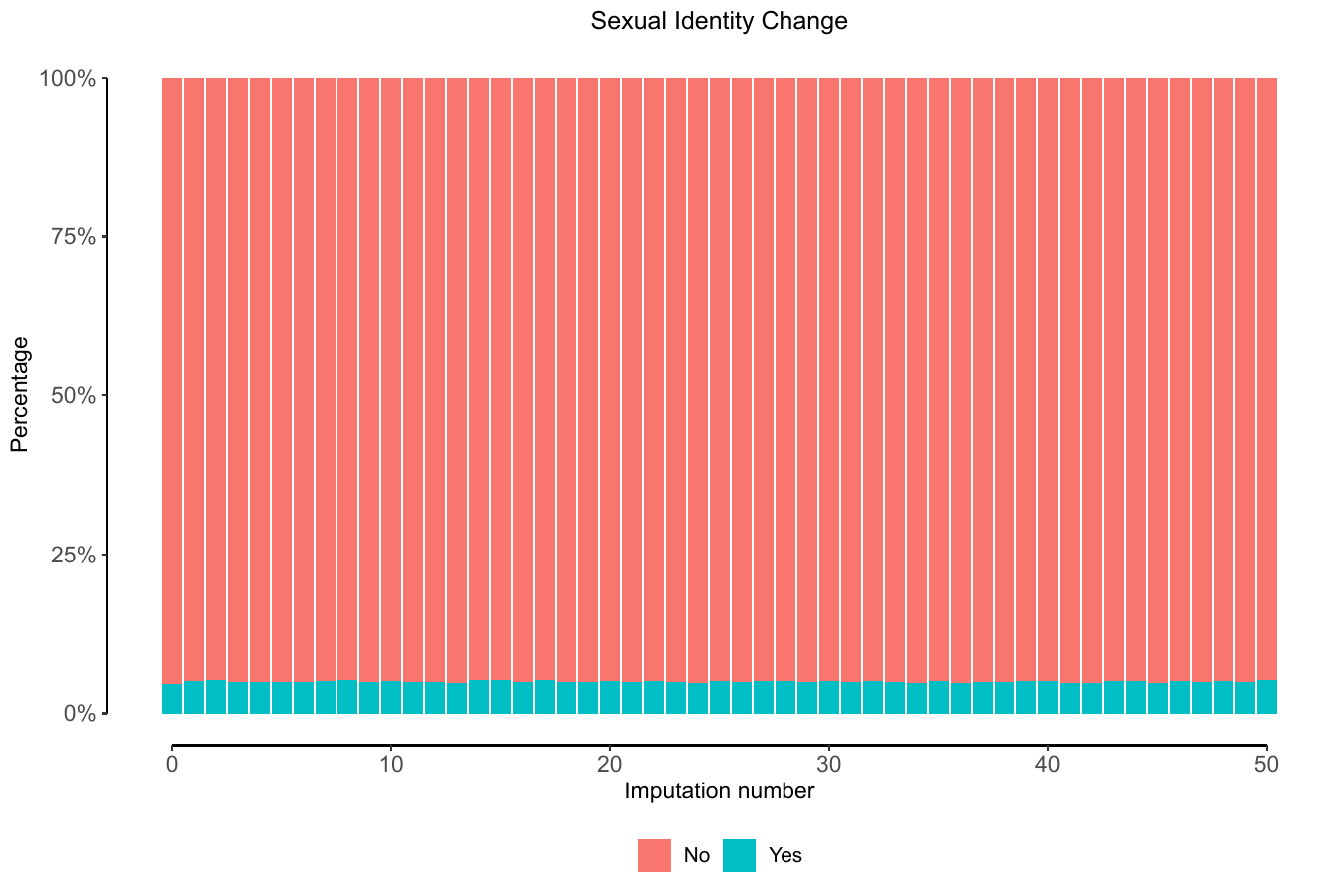}
\caption{\black{Distribution of observed and imputed values for sexual identity change, Model 2. Imputation 0 represents the original incomplete dataset; imputations 1–50 represent the imputed datasets.}}
\label{fig:applied_mice_model_2_distribution_id}
\end{figure}

\begin{figure}[h]
\centering
\includegraphics[width=\appliedPlotCoef\linewidth]{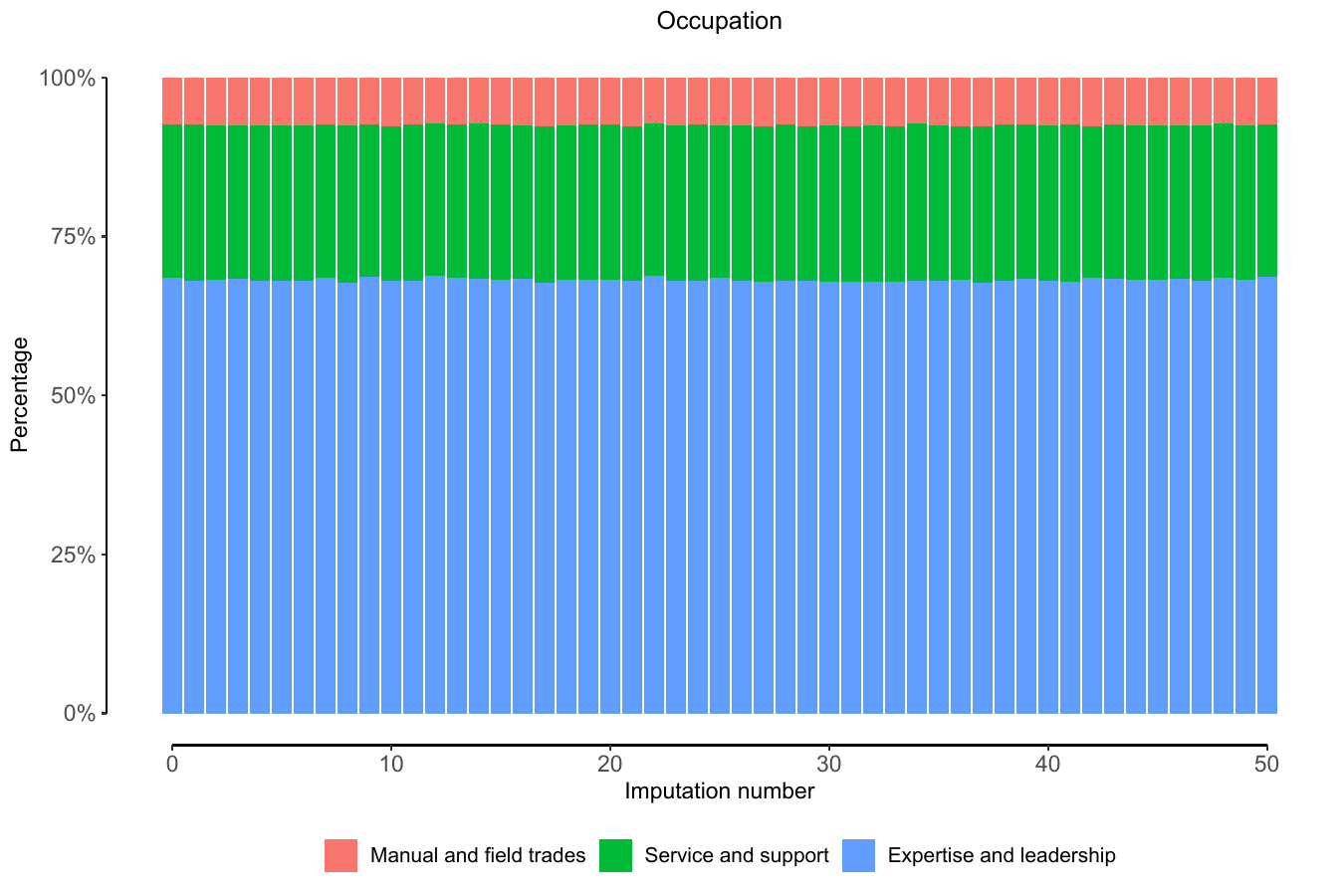}
\caption{\black{Distribution of observed and imputed values for occupation, Model 2. Imputation 0 represents the original incomplete dataset; imputations 1–50 represent the imputed datasets.}}
\label{fig:applied_mice_model_2_distribution_occ}
\end{figure}

\begin{figure}[h]
\centering
\includegraphics[width=\appliedPlotCoef\linewidth]{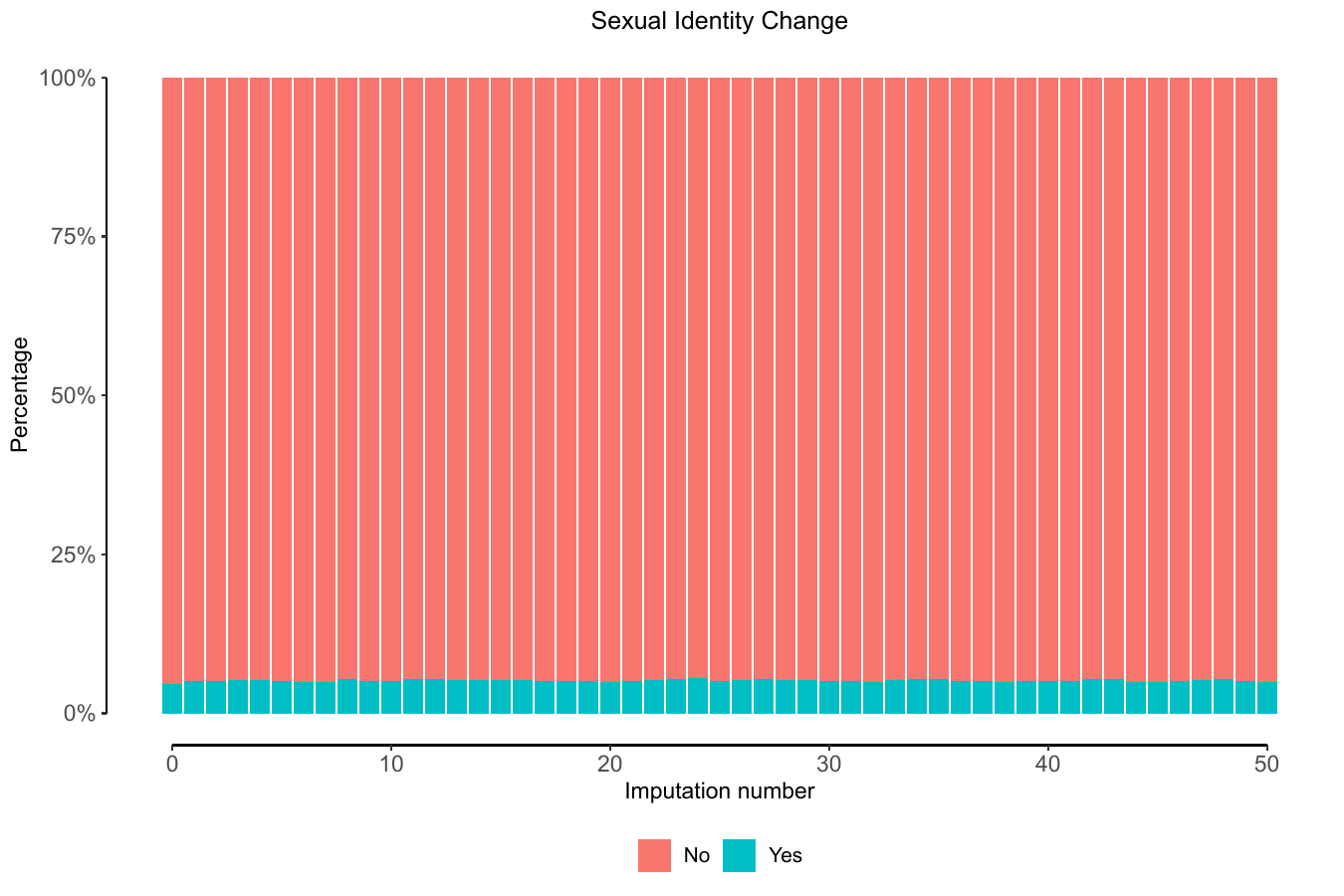}
\caption{\black{Distribution of observed and imputed values for sexual identity change, Model 3. Imputation 0 represents the original incomplete dataset; imputations 1–50 represent the imputed datasets.}}
\label{fig:applied_mice_model_3_distribution_id}
\end{figure}

\begin{figure}[h]
\centering
\includegraphics[width=\appliedPlotCoef\linewidth]{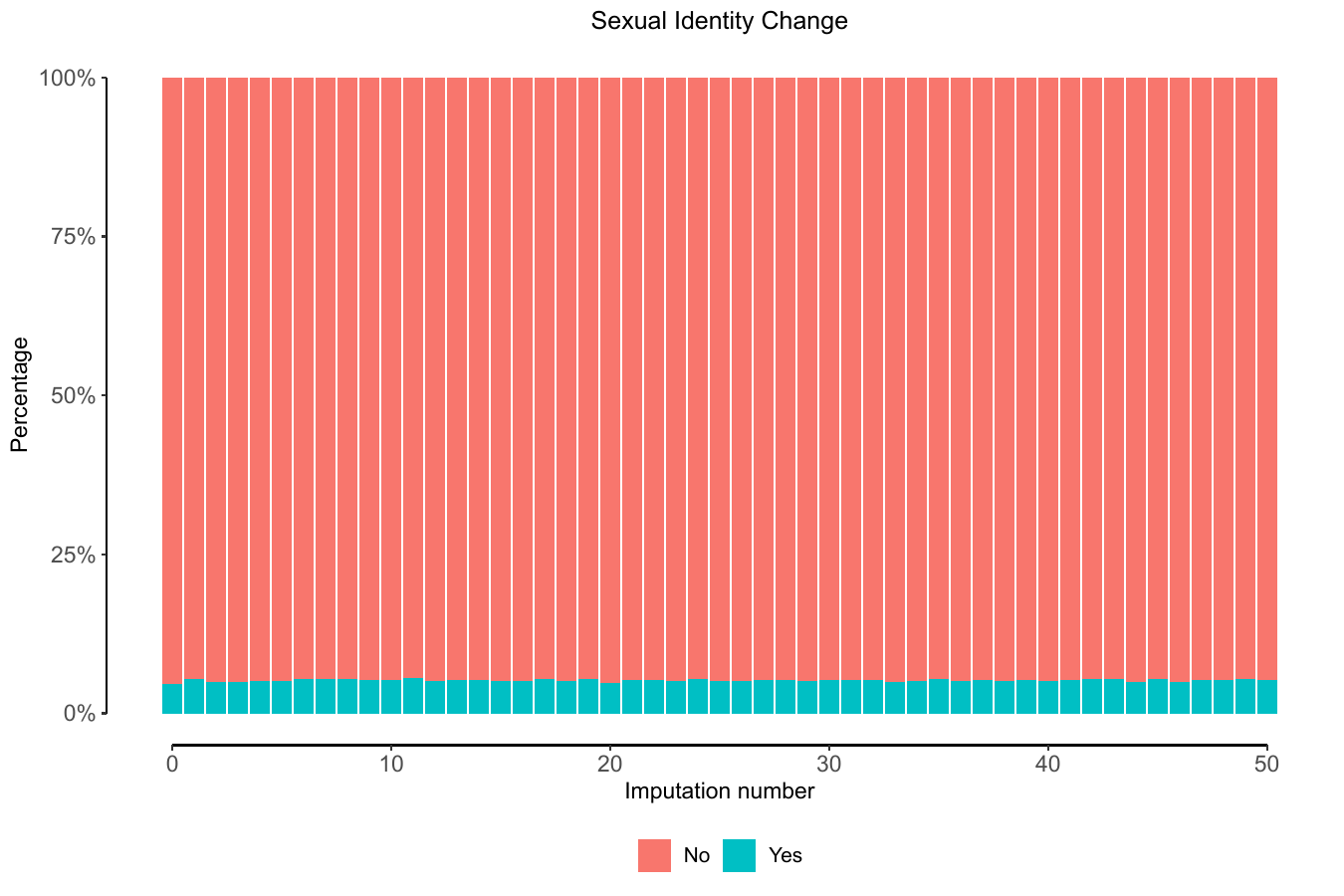}
\caption{\black{Distribution of observed and imputed values for sexual identity change, Model 4. Imputation 0 represents the original incomplete dataset; imputations 1–50 represent the imputed datasets.}}
\label{fig:applied_mice_model_4_distribution_id}
\end{figure}

\begin{figure}[h]
\centering
\includegraphics[width=\appliedPlotCoef\linewidth]{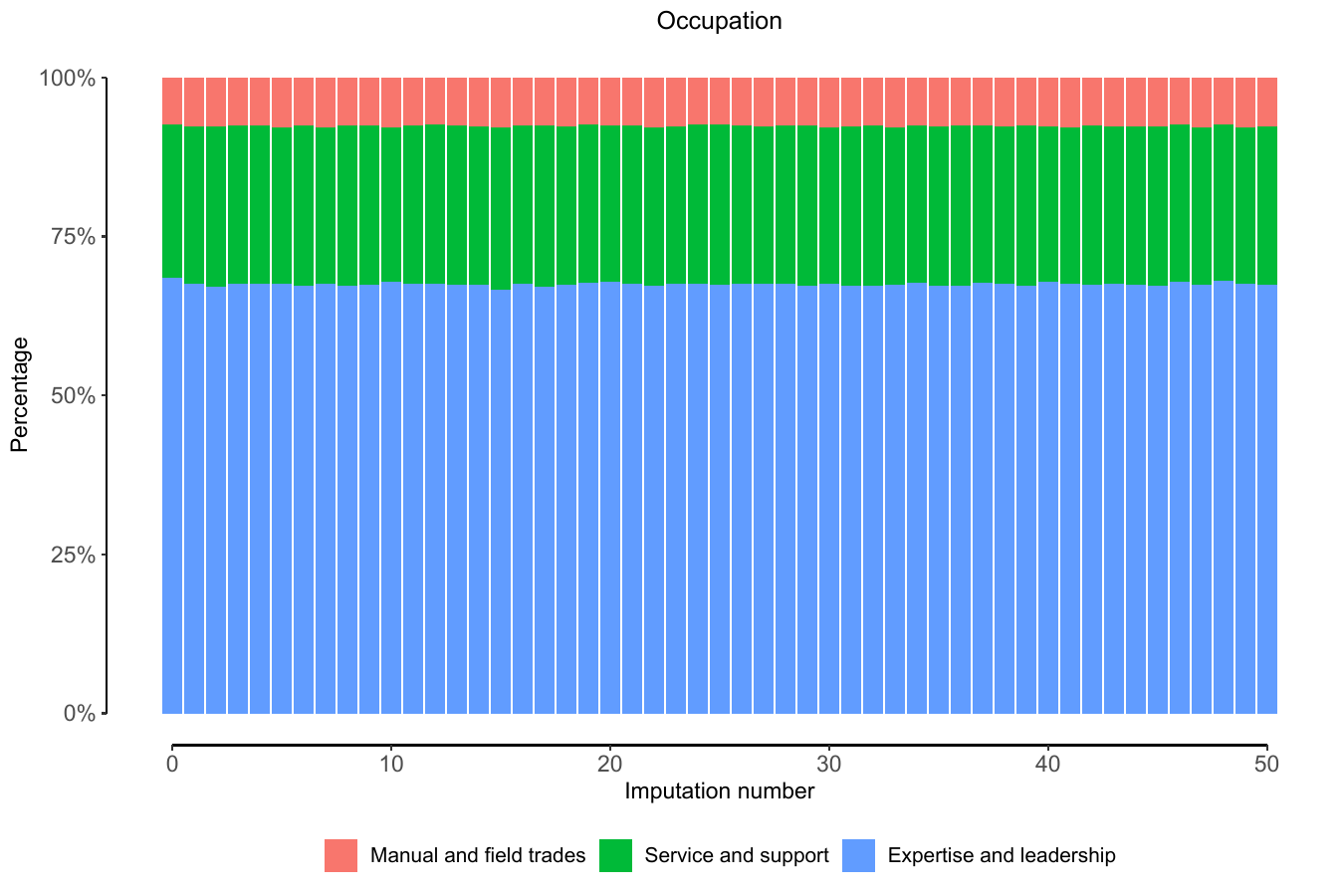}
\caption{\black{Distribution of observed and imputed values for occupation, Model 4. Imputation 0 represents the original incomplete dataset; imputations 1–50 represent the imputed datasets.}}
\label{fig:applied_mice_model_4_distribution_occ}
\end{figure}

\FloatBarrier

\pagebreak 

\section{Simulation study: Detailed methods and results}

\label{sec:sim_study}

\subsection{Study 1: $|W|=1$; Assumptions \ref{ass:positivity} and \ref{ass:no_csi} hold}

\label{sec:sims_univariate_W}

\subsubsection{Data generation}

Data-generating processes for each DAG are given below.

\subsubsection*{DGM (a)}
\begin{figure}[H]
\centering
\noindent
\parbox[t]{0.58\textwidth}{
\[
\begin{aligned}
X_1 &\sim \text{Bern}(0.5) \\
X_2 &\sim \text{Bern}\!\left(\text{expit}(-1 + 3X_1)\right) \\
W_1 &\sim \text{Bern}\!\left(\text{expit}(-1 + 3X_2)\right) \\
Y &\sim \text{N}\!\left(2.6X_1 + 2X_2 + 2W_1 + X_1W_1 + X_1X_2,\ 1\right) \\
R_{W_1} &\sim \text{Bern}(0.5) \\
R_Y &\sim \text{Bern}\!\left(\text{expit}(-1 + 3W_1)\right)
\end{aligned}
\]
}
\hfill
\parbox[t]{0.38\textwidth}{
\vspace{-0.5em}
\centering
\scalebox{1.15}{%
  \begin{tikzpicture}
\sansmath
\tikzset{
  line width=1.2pt,
  >=stealth,
  ell/.style={draw, fill=white, ellipse, inner sep=2pt, line width=1.5pt},
  shaded/.style={draw, fill=gray!40, ellipse, inner sep=2pt, line width=1.5pt}
}

\node[name=X1, ell]                         {$X_1$};
\node[name=X2, ell, right=5mm of X1]       {$X_2$};
\node[name=W,  shaded, right=5mm of X2]    {$W_1$};
\node[name=Y,  shaded, right=5mm of W]     {$Y$};
\node[name=RW, ell, below=5mm of W]        {$R_{W_1}$};
\node[name=RY, ell, below=5mm of Y]        {$R_Y$};

\draw[->,line width=1.2pt,>=stealth]
(X1) edge (X2)
(X2) edge (W)
(W) edge (Y)
(W) edge (RY);
\draw[->,line width=1.2pt,>=stealth] (X2) to[out=30,in=150] (Y);
\draw[->,line width=1.2pt,>=stealth] (X1) to[out=40,in=140] (Y);

\end{tikzpicture}
}
}
\end{figure}

\subsubsection*{DGM (b)}

\begin{figure}[H]
\centering
\noindent
\parbox[t]{0.58\textwidth}{
\[
\begin{aligned}
X_1 &\sim \text{Bern}(0.5) \\
X_2 &\sim \text{Bern}\!\left(\text{expit}(-1 + 3X_1)\right) \\
W_1 &\sim \text{Bern}\!\left(\text{expit}(-1 + 3X_2)\right) \\
Y &\sim \text{N}\!\left(2.6X_1 + 2X_2 + 2W_1 + X_1 W_1 + X_1 X_2,\ 1\right) \\
R_{X_2} &\sim \text{Bern}\!\left(\text{expit}(-1 + 3W_1)\right) \\
R_{W_1} &\sim \text{Bern}(0.5) \\
R_{X_1} &\sim \text{Bern}\!\left(\text{expit}(-1 + 3W_1)\right) \\
R_Y &\sim \text{Bern}\!\left(\text{expit}(-1 + 3W_1)\right)
\end{aligned}
\]
}
\hfill
\parbox[t]{0.38\textwidth}{
\vspace{-0.5em}
\centering
\scalebox{1.15}{%
  \begin{tikzpicture}
\sansmath
\tikzset{
  line width=1.2pt,
  >=stealth,
  ell/.style={draw, fill=white, ellipse, inner sep=2pt, line width=1.5pt},
  shaded/.style={draw, fill=gray!40, ellipse, inner sep=2pt, line width=1.5pt}
}

\node[name=X1, shaded]                         {$X_1$};
\node[name=X2, shaded, right=5mm of X1]       {$X_2$};
\node[name=W,  shaded, right=5mm of X2]    {$W_1$};
\node[name=Y,  shaded, right=5mm of W]     {$Y$};
\node[name=RX1, ell, below=5mm of X1]        {$R_{X_1}$};
\node[name=RX2, ell, below=5mm of X2]        {$R_{X_2}$};
\node[name=RW, ell, below=5mm of W]        {$R_{W_1}$};
\node[name=RY, ell, below=5mm of Y]        {$R_Y$};

\draw[->,line width=1.2pt,>=stealth]
(X1) edge (X2)
(X2) edge (W)
(W) edge (Y)
(W) edge (RX1)
(W) edge (RX2)
(W) edge (RY);
\draw[->,line width=1.2pt,>=stealth] (X2) to[out=30,in=150] (Y);
\draw[->,line width=1.2pt,>=stealth] (X1) to[out=40,in=140] (Y);

\end{tikzpicture}
}
}
\end{figure}

\subsubsection*{DGM (c)}

\begin{figure}[H]
\centering
\noindent
\parbox[t]{0.58\textwidth}{
\[
\begin{aligned}
X_1 &\sim \text{Bern}(0.5) \\
X_2 &\sim \text{Bern}\!\left(\text{expit}(-1 + 3X_1)\right) \\
W_1 &\sim \text{Bern}\!\left(\text{expit}(-1 + 3X_2)\right) \\
Y &\sim \text{N}\!\left(2.6X_1 + 2X_2 + X_1 X_2,\ 1\right) \\
R_{X_2} &\sim \text{Bern}\!\left(\text{expit}(-1 + 3W_1)\right) \\
R_{W_1} &\sim \text{Bern}(0.5) \\
R_{X_1} &\sim \text{Bern}\!\left(\text{expit}(-1 + 3W_1)\right) \\
R_Y &\sim \text{Bern}\!\left(\text{expit}(-1 + 3W_1)\right)
\end{aligned}
\]
}
\hfill
\parbox[t]{0.38\textwidth}{
\vspace{-0.5em}
\centering
\scalebox{1.15}{%
  \begin{tikzpicture}
\sansmath
\tikzset{
  line width=1.2pt,
  >=stealth,
  ell/.style={draw, fill=white, ellipse, inner sep=2pt, line width=1.5pt},
  shaded/.style={draw, fill=gray!40, ellipse, inner sep=2pt, line width=1.5pt}
}

\node[name=X1, shaded]                         {$X_1$};
\node[name=X2, shaded, right=5mm of X1]       {$X_2$};
\node[name=W,  shaded, right=5mm of X2]    {$W_1$};
\node[name=Y,  shaded, right=5mm of W]     {$Y$};
\node[name=RX1, ell, below=5mm of X1]        {$R_{X_1}$};
\node[name=RX2, ell, below=5mm of X2]        {$R_{X_2}$};
\node[name=RW, ell, below=5mm of W]        {$R_{W_1}$};
\node[name=RY, ell, below=5mm of Y]        {$R_Y$};

\draw[->,line width=1.2pt,>=stealth]
(X1) edge (X2)
(X2) edge (W)
(W) edge (RX1)
(W) edge (RX2)
(W) edge (RY);
\draw[->,line width=1.2pt,>=stealth] (X2) to[out=30,in=150] (Y);
\draw[->,line width=1.2pt,>=stealth] (X1) to[out=40,in=140] (Y);

\end{tikzpicture}
}
}
\end{figure}

\subsubsection*{DGM (d)}

\begin{figure}[H]
\centering
\noindent
\parbox[t]{0.58\textwidth}{
\[
\begin{aligned}
X_1 &\sim \text{Bern}(0.5) \\
X_2 &\sim \text{Bern}\!\left(\text{expit}(-1 + 3X_1)\right) \\
Y &\sim \text{N}\!\left(2.6X_1 + 2X_2 + X_1 X_2,\ 1\right) \\
W_1 &\sim \text{Bern}\!\left(\text{expit}(-1 + 0.5Y)\right) \\
R_{W_1} &\sim \text{Bern}(0.5) \\
R_Y &\sim \text{Bern}\!\left(\text{expit}(-1 + 3W_1)\right)
\end{aligned}
\]
}
\hfill
\parbox[t]{0.38\textwidth}{
\vspace{-0.5em}
\centering
\scalebox{1.15}{%
  \begin{tikzpicture}
\sansmath
\tikzset{
  line width=1.2pt,
  >=stealth,
  ell/.style={draw, fill=white, ellipse, inner sep=2pt, line width=1.5pt},
  shaded/.style={draw, fill=gray!40, ellipse, inner sep=2pt, line width=1.5pt}
}

\node[name=X1, ell]                         {$X_1$};
\node[name=X2, ell, right=5mm of X1]       {$X_2$};
\node[name=W,  shaded, right=5mm of X2]    {$W_1$};
\node[name=Y,  shaded, right=5mm of W]     {$Y$};
\node[name=RW, ell, below=5mm of W]        {$R_{W_1}$};
\node[name=RY, ell, below=5mm of Y]        {$R_Y$};

\draw[->,line width=1.2pt,>=stealth]
(X1) edge (X2)
(Y) edge (W)
(W) edge (RY);
\draw[->,line width=1.2pt,>=stealth] (X2) to[out=30,in=150] (Y);
\draw[->,line width=1.2pt,>=stealth] (X1) to[out=40,in=140] (Y);

\end{tikzpicture}
}
}
\end{figure}

\subsubsection*{DGM (e)}

\begin{figure}[H]
\centering
\noindent
\parbox[t]{0.58\textwidth}{
\[
\begin{aligned}
X_1 &\sim \text{Bern}(0.5) \\
X_2 &\sim \text{Bern}\!\left(\text{expit}(-1 + 3X_1)\right) \\
Y &\sim \text{N}\!\left(2.6X_1 + 2X_2 + X_1 X_2,\ 1\right) \\
W_1 &\sim \text{Bern}\!\left(\text{expit}(-1 + 3Y)\right) \\
R_{X_2} &\sim \text{Bern}\!\left(\text{expit}(-1 + 3W_1)\right) \\
R_{W_1} &\sim \text{Bern}(0.5) \\
R_{X_1} &\sim \text{Bern}\!\left(\text{expit}(-1 + 3W_1)\right) \\
R_Y &\sim \text{Bern}\!\left(\text{expit}(-1 + 3W_1)\right)
\end{aligned}
\]
}
\hfill
\parbox[t]{0.38\textwidth}{
\vspace{-0.5em}
\centering
\scalebox{1.15}{%
  \begin{tikzpicture}
\sansmath
\tikzset{
  line width=1.2pt,
  >=stealth,
  ell/.style={draw, fill=white, ellipse, inner sep=2pt, line width=1.5pt},
  shaded/.style={draw, fill=gray!40, ellipse, inner sep=2pt, line width=1.5pt}
}

\node[name=X1, shaded]                         {$X_1$};
\node[name=X2, shaded, right=5mm of X1]       {$X_2$};
\node[name=W,  shaded, right=5mm of X2]    {$W_1$};
\node[name=Y,  shaded, right=5mm of W]     {$Y$};
\node[name=RX1, ell, below=5mm of X1]        {$R_{X_1}$};
\node[name=RX2, ell, below=5mm of X2]        {$R_{X_2}$};
\node[name=RW, ell, below=5mm of W]        {$R_{W_1}$};
\node[name=RY, ell, below=5mm of Y]        {$R_Y$};

\draw[->,line width=1.2pt,>=stealth]
(X1) edge (X2)
(Y) edge (W)
(W) edge (RX1)
(W) edge (RX2)
(W) edge (RY);
\draw[->,line width=1.2pt,>=stealth] (X2) to[out=30,in=150] (Y);
\draw[->,line width=1.2pt,>=stealth] (X1) to[out=40,in=140] (Y);

\end{tikzpicture}
}
}
\end{figure}

\subsubsection*{DGM (f)}

\begin{figure}[H]
\centering
\noindent
\parbox[t]{0.58\textwidth}{
\[
\begin{aligned}
X_1 &\sim \text{Bern}(0.5) \\
X_2 &\sim \text{Bern}\!\left(\text{expit}(-1 + 3X_1)\right) \\
W_1 &\sim \text{Bern}(0.5) \\
Y &\sim \text{N}\!\left(2.6X_1 + 2X_2 + 2W_1 + X_1 W_1 + X_1 X_2,\ 1\right) \\
R_{W_1} &\sim \text{Bern}(0.5) \\
R_Y &\sim \text{Bern}\!\left(\text{expit}(-1 + 3W_1)\right)
\end{aligned}
\]
}
\hfill
\parbox[t]{0.38\textwidth}{
\vspace{-0.5em}
\centering
\scalebox{1.15}{%
  \begin{tikzpicture}
\sansmath
\tikzset{
  line width=1.2pt,
  >=stealth,
  ell/.style={draw, fill=white, ellipse, inner sep=2pt, line width=1.5pt},
  shaded/.style={draw, fill=gray!40, ellipse, inner sep=2pt, line width=1.5pt}
}

\node[name=X1, ell]                         {$X_1$};
\node[name=X2, ell, right=5mm of X1]       {$X_2$};
\node[name=W,  shaded, right=5mm of X2]    {$W_1$};
\node[name=Y,  shaded, right=5mm of W]     {$Y$};
\node[name=RW, ell, below=5mm of W]        {$R_{W_1}$};
\node[name=RY, ell, below=5mm of Y]        {$R_Y$};

\draw[->,line width=1.2pt,>=stealth]
(X1) edge (X2)
(W) edge (Y)
(W) edge (RY);
\draw[->,line width=1.2pt,>=stealth] (X2) to[out=30,in=150] (Y);
\draw[->,line width=1.2pt,>=stealth] (X1) to[out=40,in=140] (Y);

\end{tikzpicture}
}
}
\end{figure}

\subsubsection*{DGM (g)}

\begin{figure}[H]
\centering
\noindent
\parbox[t]{0.58\textwidth}{
\[
\begin{aligned}
X_1 &\sim \text{Bern}(0.5) \\
X_2 &\sim \text{Bern}\!\left(\text{expit}(-1 + 3X_1)\right) \\
W_1 &\sim \text{Bern}(0.5) \\
Y &\sim \text{N}\!\left(2.6X_1 + 2X_2 + 2W_1 + X_1 W_1 + X_1 X_2,\ 1\right) \\
R_{X_2} &\sim \text{Bern}\!\left(\text{expit}(-1 + 3W_1)\right) \\
R_{W_1} &\sim \text{Bern}(0.5) \\
R_{X_1} &\sim \text{Bern}\!\left(\text{expit}(-1 + 3W_1)\right) \\
R_Y &\sim \text{Bern}\!\left(\text{expit}(-1 + 3W_1)\right)
\end{aligned}
\]
}
\hfill
\parbox[t]{0.38\textwidth}{
\vspace{-0.5em}
\centering
\scalebox{1.15}{%
  \begin{tikzpicture}
\sansmath
\tikzset{
  line width=1.2pt,
  >=stealth,
  ell/.style={draw, fill=white, ellipse, inner sep=2pt, line width=1.5pt},
  shaded/.style={draw, fill=gray!40, ellipse, inner sep=2pt, line width=1.5pt}
}

\node[name=X1, shaded]                         {$X_1$};
\node[name=X2, shaded, right=5mm of X1]       {$X_2$};
\node[name=W,  shaded, right=5mm of X2]    {$W_1$};
\node[name=Y,  shaded, right=5mm of W]     {$Y$};
\node[name=RX1, ell, below=5mm of X1]        {$R_{X_1}$};
\node[name=RX2, ell, below=5mm of X2]        {$R_{X_2}$};
\node[name=RW, ell, below=5mm of W]        {$R_{W_1}$};
\node[name=RY, ell, below=5mm of Y]        {$R_Y$};

\draw[->,line width=1.2pt,>=stealth]
(X1) edge (X2)
(W) edge (Y)
(W) edge (RX1)
(W) edge (RX2)
(W) edge (RY);
\draw[->,line width=1.2pt,>=stealth] (X2) to[out=30,in=150] (Y);
\draw[->,line width=1.2pt,>=stealth] (X1) to[out=40,in=140] (Y);

\end{tikzpicture}
}
}
\end{figure}

\subsubsection*{Full results for Simulation Study 1}

\newcommand{\colwidth}{1.42cm}   %
\newcommand{\dagwidth}{0.70cm}
\newcommand{\coefwidth}{0.95cm}

The following six tables stratify by sample size. 

\begin{table}[H]
\centering
\scriptsize
\setlength{\tabcolsep}{3pt}         %
\renewcommand{\arraystretch}{1.15}  %
\begin{tabular}{
    >{\centering\arraybackslash}p{\dagwidth}
    >{\centering\arraybackslash}p{\coefwidth}
    >{\centering\arraybackslash}p{\colwidth} >{\centering\arraybackslash}p{\colwidth} >{\centering\arraybackslash}p{\colwidth}
    >{\centering\arraybackslash}p{\colwidth} >{\centering\arraybackslash}p{\colwidth} >{\centering\arraybackslash}p{\colwidth}
    >{\centering\arraybackslash}p{\colwidth} >{\centering\arraybackslash}p{\colwidth} >{\centering\arraybackslash}p{\colwidth}
}
\toprule
\textbf{DGM} & \textbf{} &
\multicolumn{3}{c}{\textbf{Bias}} &
\multicolumn{3}{c}{\textbf{RMSE}} &
\multicolumn{3}{c}{\textbf{CI coverage}} \\
\cmidrule(lr){3-5} \cmidrule(lr){6-8} \cmidrule(lr){9-11}
 & & MAR (IPW) & MIA-T & MIA-I & MAR (IPW) & MIA-T & MIA-I & MAR (IPW) & MIA-T & MIA-I \\
\midrule
\multirow{2}{*}{(a)} & $\widehat \beta_{X_2}$
    & \xmark / \textbf{-0.01} & \cmark / 0.12 & \cmark / \textbf{-0.01} & 0.53 & 0.53 & 0.54 & 0.92 & 0.83 & \textbf{0.93} \\
 & $\widehat \beta_0$
    & \xmark / 0.20 & \cmark / \textbf{0.00} & \cmark / \textbf{0.00} & 0.38 & 0.35 & 0.35 & 0.90 & 0.80 & 0.92 \\
\midrule
\multirow{2}{*}{(b)} & $\widehat \beta_{X_2}$
    & \xmark / -0.64 & \qmark / -1.05 & \qmark / -0.97 & 1.10 & 1.21 & 1.17 & 0.64 & 0.44 & 0.61 \\
 & $\widehat \beta_0$
    & \xmark / 0.84 & \qmark / 1.27 & \qmark / 1.18 & 1.11 & 1.37 & 1.30 & 0.42 & 0.15 & 0.36 \\
\midrule
\multirow{2}{*}{(c)} & $\widehat \beta_{X_2}$
    & \xmark / 0.04 & \cmark / \textbf{-0.01} & \cmark / \textbf{-0.01} & 0.68 & 0.56 & 0.57 & 0.86 & 0.89 & \textbf{0.93} \\
 & $\widehat \beta_0$
    & \xmark / -0.08 & \cmark / \textbf{0.00} & \cmark / \textbf{0.00} & 0.51 & 0.42 & 0.43 & 0.79 & 0.82 & 0.90 \\
\midrule
\multirow{2}{*}{(d)} & $\widehat \beta_{X_2}$
    & \xmark / 0.04 & \cmark / 0.06 & \cmark / \textbf{0.02} & 0.53 & 0.55 & 0.56 & 0.89 & 0.72 & 0.92 \\
 & $\widehat \beta_0$
    & \xmark / 0.04 & \cmark / \textbf{-0.01} & \cmark / \textbf{-0.01} & 0.28 & 0.31 & 0.31 & 0.92 & 0.72 & 0.90 \\
\midrule
\multirow{2}{*}{(e)} & $\widehat \beta_{X_2}$
    & \xmark / -0.44 & \qmark / -0.63 & \qmark / -0.62 & 0.72 & 0.76 & 0.76 & 0.71 & 0.60 & 0.66 \\
 & $\widehat \beta_0$
    & \xmark / 0.54 & \qmark / 0.77 & \qmark / 0.76 & 0.67 & 0.81 & 0.81 & 0.37 & 0.14 & 0.20 \\
\midrule
\multirow{2}{*}{(f)} & $\widehat \beta_{X_2}$
    & \xmark / 0.09 & \cmark / 0.22 & \cmark / \textbf{-0.03} & 0.69 & 0.72 & 0.63 & 0.88 & 0.75 & \textbf{0.94} \\
 & $\widehat \beta_0$
    & \xmark / 0.28 & \cmark / \textbf{0.00} & \cmark / \textbf{0.00} & 0.42 & 0.34 & 0.33 & 0.88 & 0.86 & 0.92 \\
\midrule
\multirow{2}{*}{(g)} & $\widehat \beta_{X_2}$
    & \xmark / 0.12 & \qmark / \textbf{-0.01} & \qmark / \textbf{-0.03} & 0.98 & 0.62 & 0.66 & 0.77 & 0.88 & 0.89 \\
 & $\widehat \beta_0$
    & \xmark / 0.70 & \qmark / 0.92 & \qmark / 0.87 & 0.88 & 0.98 & 0.93 & 0.38 & 0.18 & 0.33 \\
\bottomrule
\end{tabular}
\caption{\label{tb:sim_n200_wdim1} Simulation results for $N=200$, $|W|=1$ scenarios. $\beta_{X_2}=3.22$ for DGMs (a) and (b) and $\beta_{X_2}=2$ otherwise. MAR (IPW): Sun \& Tchetgen Tchetgen inverse-probability weighting estimator. MIA-I: ICE estimator. MIA-T: TMLE estimator. \cmark indicates that the sufficient (and in the case of $\muMAR$, necessary) conditions hold for $\muMAR$ and $\muMIA$, respectively. \xmark indicates that the conditions for $\muMAR$ are violated. \qmark indicates that the sufficient (but not necessary) conditions for $\muMIA$ are violated. Boldface indicates approximately correct performance.}
\end{table}

\begin{table}[H]
\centering
\scriptsize
\setlength{\tabcolsep}{3pt}         %
\renewcommand{\arraystretch}{1.15}  %
\begin{tabular}{
    >{\centering\arraybackslash}p{\dagwidth}
    >{\centering\arraybackslash}p{\coefwidth}
    >{\centering\arraybackslash}p{\colwidth} >{\centering\arraybackslash}p{\colwidth} >{\centering\arraybackslash}p{\colwidth}
    >{\centering\arraybackslash}p{\colwidth} >{\centering\arraybackslash}p{\colwidth} >{\centering\arraybackslash}p{\colwidth}
    >{\centering\arraybackslash}p{\colwidth} >{\centering\arraybackslash}p{\colwidth} >{\centering\arraybackslash}p{\colwidth}
}
\toprule
\textbf{DGM} & \textbf{} &
\multicolumn{3}{c}{\textbf{Bias}} &
\multicolumn{3}{c}{\textbf{RMSE}} &
\multicolumn{3}{c}{\textbf{CI coverage}} \\
\cmidrule(lr){3-5} \cmidrule(lr){6-8} \cmidrule(lr){9-11}
 & & MAR (IPW) & MIA-T & MIA-I & MAR (IPW) & MIA-T & MIA-I & MAR (IPW) & MIA-T & MIA-I \\
\midrule
\multirow{2}{*}{(a)} & $\widehat \beta_{X_2}$
    & \xmark / -0.06 & \cmark / \textbf{0.02} & \cmark / \textbf{-0.01} & 0.34 & 0.34 & 0.32 & \textbf{0.94} & 0.92 & \textbf{0.94} \\
 & $\widehat \beta_0$
    & \xmark / 0.21 & \cmark / \textbf{0.01} & \cmark / \textbf{0.01} & 0.28 & 0.21 & 0.21 & 0.86 & \textbf{0.94} & \textbf{0.94} \\
\midrule
\multirow{2}{*}{(b)} & $\widehat \beta_{X_2}$
    & \xmark / -0.50 & \qmark / -0.89 & \qmark / -0.87 & 0.84 & 0.98 & 0.96 & 0.66 & 0.36 & 0.44 \\
 & $\widehat \beta_0$
    & \xmark / 0.77 & \qmark / 1.11 & \qmark / 1.08 & 0.95 & 1.17 & 1.13 & 0.48 & 0.12 & 0.21 \\
\midrule
\multirow{2}{*}{(c)} & $\widehat \beta_{X_2}$
    & \xmark / \textbf{0.02} & \cmark / \textbf{-0.01} & \cmark / \textbf{-0.01} & 0.45 & 0.36 & 0.36 & 0.91 & 0.92 & \textbf{0.94} \\
 & $\widehat \beta_0$
    & \xmark / -0.05 & \cmark / \textbf{0.01} & \cmark / \textbf{0.01} & 0.33 & 0.28 & 0.28 & 0.90 & 0.88 & 0.91 \\
\midrule
\multirow{2}{*}{(d)} & $\widehat \beta_{X_2}$
    & \xmark / \textbf{-0.01} & \cmark / \textbf{-0.01} & \cmark / \textbf{-0.01} & 0.33 & 0.34 & 0.34 & \textbf{0.93} & 0.88 & \textbf{0.94} \\
 & $\widehat \beta_0$
    & \xmark / 0.06 & \cmark / \textbf{0.00} & \cmark / \textbf{0.00} & 0.17 & 0.19 & 0.19 & \textbf{0.95} & 0.92 & \textbf{0.93} \\
\midrule
\multirow{2}{*}{(e)} & $\widehat \beta_{X_2}$
    & \xmark / -0.46 & \qmark / -0.57 & \qmark / -0.59 & 0.61 & 0.64 & 0.65 & 0.54 & 0.42 & 0.43 \\
 & $\widehat \beta_0$
    & \xmark / 0.58 & \qmark / 0.69 & \qmark / 0.70 & 0.65 & 0.72 & 0.73 & 0.26 & 0.10 & 0.11 \\
\midrule
\multirow{2}{*}{(f)} & $\widehat \beta_{X_2}$
    & \xmark / \textbf{0.00} & \cmark / \textbf{0.00} & \cmark / \textbf{0.00} & 0.41 & 0.42 & 0.41 & \textbf{0.95} & 0.88 & \textbf{0.93} \\
 & $\widehat \beta_0$
    & \xmark / 0.28 & \cmark / \textbf{-0.01} & \cmark / \textbf{-0.01} & 0.33 & 0.20 & 0.19 & 0.77 & \textbf{0.94} & \textbf{0.95} \\
\midrule
\multirow{2}{*}{(g)} & $\widehat \beta_{X_2}$
    & \xmark / 0.13 & \qmark / 0.07 & \qmark / \textbf{0.00} & 0.83 & 0.41 & 0.40 & 0.84 & 0.91 & \textbf{0.93} \\
 & $\widehat \beta_0$
    & \xmark / 0.70 & \qmark / 0.86 & \qmark / 0.83 & 0.82 & 0.89 & 0.85 & 0.46 & 0.08 & 0.11 \\
\bottomrule
\end{tabular}
\caption{\label{tb:sim_n500_wdim1} Simulation results for $N=500$, $|W|=1$ scenarios. $\beta_{X_2}=3.22$ for DGMs (a) and (b) and $\beta_{X_2}=2$ otherwise. MAR (IPW): Sun \& Tchetgen Tchetgen inverse-probability weighting estimator. MIA-I: ICE estimator. MIA-T: TMLE estimator. \cmark indicates that the sufficient (and in the case of $\muMAR$, necessary) conditions hold for $\muMAR$ and $\muMIA$, respectively. \xmark indicates that the conditions for $\muMAR$ are violated. \qmark indicates that the sufficient (but not necessary) conditions for $\muMIA$ are violated. Boldface indicates approximately correct performance.}
\end{table}

\begin{table}[H]
\centering
\scriptsize
\setlength{\tabcolsep}{3pt}         %
\renewcommand{\arraystretch}{1.15}  %
\begin{tabular}{
    >{\centering\arraybackslash}p{\dagwidth}
    >{\centering\arraybackslash}p{\coefwidth}
    >{\centering\arraybackslash}p{\colwidth} >{\centering\arraybackslash}p{\colwidth} >{\centering\arraybackslash}p{\colwidth}
    >{\centering\arraybackslash}p{\colwidth} >{\centering\arraybackslash}p{\colwidth} >{\centering\arraybackslash}p{\colwidth}
    >{\centering\arraybackslash}p{\colwidth} >{\centering\arraybackslash}p{\colwidth} >{\centering\arraybackslash}p{\colwidth}
}
\toprule
\textbf{DGM} & \textbf{} &
\multicolumn{3}{c}{\textbf{Bias}} &
\multicolumn{3}{c}{\textbf{RMSE}} &
\multicolumn{3}{c}{\textbf{CI coverage}} \\
\cmidrule(lr){3-5} \cmidrule(lr){6-8} \cmidrule(lr){9-11}
 & & MAR (IPW) & MIA-T & MIA-I & MAR (IPW) & MIA-T & MIA-I & MAR (IPW) & MIA-T & MIA-I \\
\midrule
\multirow{2}{*}{(a)} & $\widehat \beta_{X_2}$
    & \xmark / -0.06 & \cmark / \textbf{-0.01} & \cmark / \textbf{-0.01} & 0.24 & 0.24 & 0.22 & \textbf{0.95} & \textbf{0.93} & \textbf{0.95} \\
 & $\widehat \beta_0$
    & \xmark / 0.20 & \cmark / \textbf{0.00} & \cmark / \textbf{0.00} & 0.24 & 0.14 & 0.14 & 0.79 & \textbf{0.95} & \textbf{0.95} \\
\midrule
\multirow{2}{*}{(b)} & $\widehat \beta_{X_2}$
    & \xmark / -0.38 & \qmark / -0.83 & \qmark / -0.84 & 0.69 & 0.89 & 0.89 & 0.66 & 0.20 & 0.21 \\
 & $\widehat \beta_0$
    & \xmark / 0.68 & \qmark / 1.06 & \qmark / 1.06 & 0.87 & 1.09 & 1.08 & 0.55 & 0.03 & 0.04 \\
\midrule
\multirow{2}{*}{(c)} & $\widehat \beta_{X_2}$
    & \xmark / \textbf{0.00} & \cmark / \textbf{-0.01} & \cmark / \textbf{0.00} & 0.35 & 0.26 & 0.25 & 0.89 & 0.92 & \textbf{0.94} \\
 & $\widehat \beta_0$
    & \xmark / \textbf{-0.02} & \cmark / \textbf{0.01} & \cmark / \textbf{0.00} & 0.26 & 0.22 & 0.20 & 0.90 & 0.88 & 0.92 \\
\midrule
\multirow{2}{*}{(d)} & $\widehat \beta_{X_2}$
    & \xmark / \textbf{0.00} & \cmark / \textbf{0.00} & \cmark / \textbf{0.00} & 0.23 & 0.24 & 0.24 & \textbf{0.93} & \textbf{0.93} & \textbf{0.94} \\
 & $\widehat \beta_0$
    & \xmark / 0.05 & \cmark / \textbf{0.00} & \cmark / \textbf{0.00} & 0.13 & 0.13 & 0.13 & \textbf{0.94} & \textbf{0.94} & \textbf{0.94} \\
\midrule
\multirow{2}{*}{(e)} & $\widehat \beta_{X_2}$
    & \xmark / -0.45 & \qmark / -0.54 & \qmark / -0.56 & 0.55 & 0.58 & 0.60 & 0.40 & 0.23 & 0.22 \\
 & $\widehat \beta_0$
    & \xmark / 0.60 & \qmark / 0.67 & \qmark / 0.69 & 0.66 & 0.69 & 0.70 & 0.24 & 0.02 & 0.02 \\
\midrule
\multirow{2}{*}{(f)} & $\widehat \beta_{X_2}$
    & \xmark / \textbf{-0.01} & \cmark / \textbf{0.00} & \cmark / \textbf{0.00} & 0.29 & 0.27 & 0.27 & \textbf{0.95} & \textbf{0.94} & \textbf{0.94} \\
 & $\widehat \beta_0$
    & \xmark / 0.27 & \cmark / \textbf{0.00} & \cmark / \textbf{0.00} & 0.30 & 0.14 & 0.14 & 0.57 & \textbf{0.95} & \textbf{0.95} \\
\midrule
\multirow{2}{*}{(g)} & $\widehat \beta_{X_2}$
    & \xmark / 0.11 & \qmark / 0.05 & \qmark / \textbf{0.00} & 0.69 & 0.33 & 0.29 & 0.82 & 0.89 & \textbf{0.93} \\
 & $\widehat \beta_0$
    & \xmark / 0.68 & \qmark / 0.83 & \qmark / 0.83 & 0.76 & 0.85 & 0.84 & 0.49 & 0.02 & 0.01 \\
\bottomrule
\end{tabular}
\caption{\label{tb:sim_n1000_wdim1} Simulation results for $N=1{,}000$, $|W|=1$ scenarios. $\beta_{X_2}=3.22$ for DGMs (a) and (b) and $\beta_{X_2}=2$ otherwise. MAR (IPW): Sun \& Tchetgen Tchetgen inverse-probability weighting estimator. MIA-I: ICE estimator. MIA-T: TMLE estimator. \cmark indicates that the sufficient (and in the case of $\muMAR$, necessary) conditions hold for $\muMAR$ and $\muMIA$, respectively. \xmark indicates that the conditions for $\muMAR$ are violated. \qmark indicates that the sufficient (but not necessary) conditions for $\muMIA$ are violated. Boldface indicates approximately correct performance.}
\end{table}

\begin{table}[H]
\centering
\scriptsize
\setlength{\tabcolsep}{3pt}         %
\renewcommand{\arraystretch}{1.15}  %
\begin{tabular}{
    >{\centering\arraybackslash}p{\dagwidth}
    >{\centering\arraybackslash}p{\coefwidth}
    >{\centering\arraybackslash}p{\colwidth} >{\centering\arraybackslash}p{\colwidth} >{\centering\arraybackslash}p{\colwidth}
    >{\centering\arraybackslash}p{\colwidth} >{\centering\arraybackslash}p{\colwidth} >{\centering\arraybackslash}p{\colwidth}
    >{\centering\arraybackslash}p{\colwidth} >{\centering\arraybackslash}p{\colwidth} >{\centering\arraybackslash}p{\colwidth}
}
\toprule
\textbf{DGM} & \textbf{} &
\multicolumn{3}{c}{\textbf{Bias}} &
\multicolumn{3}{c}{\textbf{RMSE}} &
\multicolumn{3}{c}{\textbf{CI coverage}} \\
\cmidrule(lr){3-5} \cmidrule(lr){6-8} \cmidrule(lr){9-11}
 & & MAR (IPW) & MIA-T & MIA-I & MAR (IPW) & MIA-T & MIA-I & MAR (IPW) & MIA-T & MIA-I \\
\midrule
\multirow{2}{*}{(a)} & $\widehat \beta_{X_2}$
    & \xmark / -0.06 & \cmark / \textbf{-0.01} & \cmark / \textbf{0.00} & 0.18 & 0.16 & 0.15 & \textbf{0.94} & \textbf{0.95} & \textbf{0.95} \\
 & $\widehat \beta_0$
    & \xmark / 0.19 & \cmark / \textbf{0.00} & \cmark / \textbf{0.00} & 0.22 & 0.10 & 0.10 & 0.60 & \textbf{0.96} & \textbf{0.95} \\
\midrule
\multirow{2}{*}{(b)} & $\widehat \beta_{X_2}$
    & \xmark / -0.26 & \qmark / -0.83 & \qmark / -0.84 & 0.58 & 0.86 & 0.87 & 0.72 & 0.04 & 0.03 \\
 & $\widehat \beta_0$
    & \xmark / 0.57 & \qmark / 1.06 & \qmark / 1.06 & 0.74 & 1.08 & 1.08 & 0.60 & 0.00 & 0.00 \\
\midrule
\multirow{2}{*}{(c)} & $\widehat \beta_{X_2}$
    & \xmark / \textbf{0.01} & \cmark / \textbf{0.00} & \cmark / \textbf{0.00} & 0.27 & 0.18 & 0.18 & 0.90 & \textbf{0.94} & \textbf{0.95} \\
 & $\widehat \beta_0$
    & \xmark / \textbf{-0.03} & \cmark / \textbf{0.00} & \cmark / \textbf{0.00} & 0.21 & 0.14 & 0.14 & 0.92 & \textbf{0.94} & \textbf{0.95} \\
\midrule
\multirow{2}{*}{(d)} & $\widehat \beta_{X_2}$
    & \xmark / \textbf{0.00} & \cmark / \textbf{0.00} & \cmark / \textbf{0.00} & 0.16 & 0.16 & 0.16 & \textbf{0.94} & \textbf{0.94} & \textbf{0.94} \\
 & $\widehat \beta_0$
    & \xmark / 0.06 & \cmark / \textbf{0.00} & \cmark / \textbf{0.00} & 0.10 & 0.09 & 0.09 & 0.92 & \textbf{0.94} & \textbf{0.94} \\
\midrule
\multirow{2}{*}{(e)} & $\widehat \beta_{X_2}$
    & \xmark / -0.47 & \qmark / -0.55 & \qmark / -0.56 & 0.55 & 0.56 & 0.57 & 0.26 & 0.04 & 0.04 \\
 & $\widehat \beta_0$
    & \xmark / 0.62 & \qmark / 0.67 & \qmark / 0.67 & 0.68 & 0.68 & 0.68 & 0.20 & 0.00 & 0.00 \\
\midrule
\multirow{2}{*}{(f)} & $\widehat \beta_{X_2}$
    & \xmark / \textbf{0.00} & \cmark / \textbf{-0.01} & \cmark / \textbf{-0.01} & 0.20 & 0.20 & 0.20 & \textbf{0.95} & \textbf{0.94} & \textbf{0.93} \\
 & $\widehat \beta_0$
    & \xmark / 0.26 & \cmark / \textbf{0.00} & \cmark / \textbf{0.00} & 0.28 & 0.10 & 0.10 & 0.28 & \textbf{0.96} & \textbf{0.95} \\
\midrule
\multirow{2}{*}{(g)} & $\widehat \beta_{X_2}$
    & \xmark / \textbf{0.02} & \qmark / \textbf{0.01} & \qmark / \textbf{-0.01} & 0.51 & 0.23 & 0.20 & 0.85 & 0.91 & \textbf{0.94} \\
 & $\widehat \beta_0$
    & \xmark / 0.67 & \qmark / 0.83 & \qmark / 0.83 & 0.72 & 0.84 & 0.84 & 0.32 & 0.00 & 0.00 \\
\bottomrule
\end{tabular}
\caption{\label{tb:sim_n2000_wdim1} Simulation results for $N=2{,}000$, $|W|=1$ scenarios. $\beta_{X_2}=3.22$ for DGMs (a) and (b) and $\beta_{X_2}=2$ otherwise. MAR (IPW): Sun \& Tchetgen Tchetgen inverse-probability weighting estimator. MIA-I: ICE estimator. MIA-T: TMLE estimator. \cmark indicates that the sufficient (and in the case of $\muMAR$, necessary) conditions hold for $\muMAR$ and $\muMIA$, respectively. \xmark indicates that the conditions for $\muMAR$ are violated. \qmark indicates that the sufficient (but not necessary) conditions for $\muMIA$ are violated. Boldface indicates approximately correct performance.}
\end{table}

\begin{table}[H]
\centering
\scriptsize
\setlength{\tabcolsep}{3pt}         %
\renewcommand{\arraystretch}{1.15}  %
\begin{tabular}{
    >{\centering\arraybackslash}p{\dagwidth}
    >{\centering\arraybackslash}p{\coefwidth}
    >{\centering\arraybackslash}p{\colwidth} >{\centering\arraybackslash}p{\colwidth} >{\centering\arraybackslash}p{\colwidth}
    >{\centering\arraybackslash}p{\colwidth} >{\centering\arraybackslash}p{\colwidth} >{\centering\arraybackslash}p{\colwidth}
    >{\centering\arraybackslash}p{\colwidth} >{\centering\arraybackslash}p{\colwidth} >{\centering\arraybackslash}p{\colwidth}
}
\toprule
\textbf{DGM} & \textbf{} &
\multicolumn{3}{c}{\textbf{Bias}} &
\multicolumn{3}{c}{\textbf{RMSE}} &
\multicolumn{3}{c}{\textbf{CI coverage}} \\
\cmidrule(lr){3-5} \cmidrule(lr){6-8} \cmidrule(lr){9-11}
 & & MAR (IPW) & MIA-T & MIA-I & MAR (IPW) & MIA-T & MIA-I & MAR (IPW) & MIA-T & MIA-I \\
\midrule
\multirow{2}{*}{(a)} & $\widehat \beta_{X_2}$
    & \xmark / -0.06 & \cmark / \textbf{0.00} & \cmark / \textbf{0.00} & 0.12 & 0.10 & 0.10 & \textbf{0.93} & \textbf{0.95} & \textbf{0.95} \\
 & $\widehat \beta_0$
    & \xmark / 0.19 & \cmark / \textbf{0.00} & \cmark / \textbf{0.00} & 0.20 & 0.06 & 0.06 & 0.19 & \textbf{0.96} & \textbf{0.95} \\
\midrule
\multirow{2}{*}{(b)} & $\widehat \beta_{X_2}$
    & \xmark / -0.10 & \qmark / -0.84 & \qmark / -0.84 & 0.37 & 0.85 & 0.85 & 0.83 & 0.00 & 0.00 \\
 & $\widehat \beta_0$
    & \xmark / 0.39 & \qmark / 1.06 & \qmark / 1.06 & 0.51 & 1.06 & 1.06 & 0.63 & 0.00 & 0.00 \\
\midrule
\multirow{2}{*}{(c)} & $\widehat \beta_{X_2}$
    & \xmark / \textbf{0.03} & \cmark / \textbf{0.00} & \cmark / \textbf{0.00} & 0.19 & 0.11 & 0.11 & \textbf{0.93} & \textbf{0.96} & \textbf{0.96} \\
 & $\widehat \beta_0$
    & \xmark / \textbf{-0.03} & \cmark / \textbf{0.00} & \cmark / \textbf{0.00} & 0.16 & 0.09 & 0.09 & \textbf{0.93} & \textbf{0.94} & \textbf{0.95} \\
\midrule
\multirow{2}{*}{(d)} & $\widehat \beta_{X_2}$
    & \xmark / \textbf{0.00} & \cmark / \textbf{0.00} & \cmark / \textbf{0.00} & 0.10 & 0.10 & 0.10 & \textbf{0.94} & \textbf{0.95} & \textbf{0.94} \\
 & $\widehat \beta_0$
    & \xmark / 0.05 & \cmark / \textbf{0.00} & \cmark / \textbf{0.00} & 0.07 & 0.06 & 0.06 & 0.86 & \textbf{0.95} & \textbf{0.95} \\
\midrule
\multirow{2}{*}{(e)} & $\widehat \beta_{X_2}$
    & \xmark / -0.47 & \qmark / -0.55 & \qmark / -0.55 & 0.53 & 0.56 & 0.56 & 0.23 & 0.00 & 0.00 \\
 & $\widehat \beta_0$
    & \xmark / 0.63 & \qmark / 0.68 & \qmark / 0.68 & 0.68 & 0.68 & 0.68 & 0.18 & 0.00 & 0.00 \\
\midrule
\multirow{2}{*}{(f)} & $\widehat \beta_{X_2}$
    & \xmark / \textbf{0.00} & \cmark / \textbf{0.00} & \cmark / \textbf{0.00} & 0.13 & 0.12 & 0.12 & \textbf{0.95} & \textbf{0.96} & \textbf{0.95} \\
 & $\widehat \beta_0$
    & \xmark / 0.26 & \cmark / \textbf{0.00} & \cmark / \textbf{0.00} & 0.26 & 0.06 & 0.06 & 0.01 & \textbf{0.94} & \textbf{0.93} \\
\midrule
\multirow{2}{*}{(g)} & $\widehat \beta_{X_2}$
    & \xmark / \textbf{0.01} & \qmark / \textbf{-0.01} & \qmark / \textbf{0.00} & 0.32 & 0.14 & 0.13 & 0.91 & \textbf{0.95} & \textbf{0.95} \\
 & $\widehat \beta_0$
    & \xmark / 0.59 & \qmark / 0.83 & \qmark / 0.83 & 0.61 & 0.83 & 0.83 & 0.08 & 0.00 & 0.00 \\
\bottomrule
\end{tabular}
\caption{\label{tb:sim_n5000_wdim1} Simulation results for $N=5{,}000$, $|W|=1$ scenarios. $\beta_{X_2}=3.22$ for DGMs (a) and (b) and $\beta_{X_2}=2$ otherwise. MAR (IPW): Sun \& Tchetgen Tchetgen inverse-probability weighting estimator. MIA-I: ICE estimator. MIA-T: TMLE estimator. \cmark indicates that the sufficient (and in the case of $\muMAR$, necessary) conditions hold for $\muMAR$ and $\muMIA$, respectively. \xmark indicates that the conditions for $\muMAR$ are violated. \qmark indicates that the sufficient (but not necessary) conditions for $\muMIA$ are violated. Boldface indicates approximately correct performance.}
\end{table}

\begin{table}[H]
\centering
\scriptsize
\setlength{\tabcolsep}{3pt}         %
\renewcommand{\arraystretch}{1.15}  %
\begin{tabular}{
    >{\centering\arraybackslash}p{\dagwidth}
    >{\centering\arraybackslash}p{\coefwidth}
    >{\centering\arraybackslash}p{\colwidth} >{\centering\arraybackslash}p{\colwidth} >{\centering\arraybackslash}p{\colwidth}
    >{\centering\arraybackslash}p{\colwidth} >{\centering\arraybackslash}p{\colwidth} >{\centering\arraybackslash}p{\colwidth}
    >{\centering\arraybackslash}p{\colwidth} >{\centering\arraybackslash}p{\colwidth} >{\centering\arraybackslash}p{\colwidth}
}
\toprule
\textbf{DGM} & \textbf{} &
\multicolumn{3}{c}{\textbf{Bias}} &
\multicolumn{3}{c}{\textbf{RMSE}} &
\multicolumn{3}{c}{\textbf{CI coverage}} \\
\cmidrule(lr){3-5} \cmidrule(lr){6-8} \cmidrule(lr){9-11}
 & & MAR (IPW) & MIA-T & MIA-I & MAR (IPW) & MIA-T & MIA-I & MAR (IPW) & MIA-T & MIA-I \\
\midrule
\multirow{2}{*}{(a)} & $\widehat \beta_{X_2}$
    & \xmark / -0.06 & \cmark / \textbf{0.00} & \cmark / \textbf{0.00} & 0.10 & 0.07 & 0.07 & 0.87 & \textbf{0.95} & \textbf{0.94} \\
 & $\widehat \beta_0$
    & \xmark / 0.19 & \cmark / \textbf{0.00} & \cmark / \textbf{0.00} & 0.19 & 0.05 & 0.05 & 0.03 & \textbf{0.95} & \textbf{0.94} \\
\midrule
\multirow{2}{*}{(b)} & $\widehat \beta_{X_2}$
    & \xmark / -0.08 & \qmark / -0.85 & \qmark / -0.84 & 0.24 & 0.85 & 0.85 & 0.87 & 0.00 & 0.00 \\
 & $\widehat \beta_0$
    & \xmark / 0.36 & \qmark / 1.06 & \qmark / 1.06 & 0.41 & 1.06 & 1.06 & 0.41 & 0.00 & 0.00 \\
\midrule
\multirow{2}{*}{(c)} & $\widehat \beta_{X_2}$
    & \xmark / \textbf{0.03} & \cmark / \textbf{0.00} & \cmark / \textbf{0.00} & 0.15 & 0.08 & 0.08 & 0.92 & \textbf{0.94} & \textbf{0.95} \\
 & $\widehat \beta_0$
    & \xmark / \textbf{-0.03} & \cmark / \textbf{0.00} & \cmark / \textbf{0.00} & 0.12 & 0.06 & 0.06 & \textbf{0.93} & \textbf{0.94} & \textbf{0.94} \\
\midrule
\multirow{2}{*}{(d)} & $\widehat \beta_{X_2}$
    & \xmark / \textbf{0.00} & \cmark / \textbf{0.00} & \cmark / \textbf{0.00} & 0.07 & 0.07 & 0.07 & \textbf{0.96} & \textbf{0.96} & \textbf{0.95} \\
 & $\widehat \beta_0$
    & \xmark / 0.05 & \cmark / \textbf{0.00} & \cmark / \textbf{0.00} & 0.06 & 0.04 & 0.04 & 0.80 & \textbf{0.95} & \textbf{0.94} \\
\midrule
\multirow{2}{*}{(e)} & $\widehat \beta_{X_2}$
    & \xmark / -0.49 & \qmark / -0.56 & \qmark / -0.56 & 0.55 & 0.56 & 0.56 & 0.18 & 0.00 & 0.00 \\
 & $\widehat \beta_0$
    & \xmark / 0.65 & \qmark / 0.67 & \qmark / 0.67 & 0.70 & 0.68 & 0.68 & 0.12 & 0.00 & 0.00 \\
\midrule
\multirow{2}{*}{(f)} & $\widehat \beta_{X_2}$
    & \xmark / \textbf{0.01} & \cmark / \textbf{0.00} & \cmark / \textbf{0.00} & 0.09 & 0.08 & 0.08 & \textbf{0.96} & \textbf{0.96} & \textbf{0.96} \\
 & $\widehat \beta_0$
    & \xmark / 0.25 & \cmark / \textbf{0.00} & \cmark / \textbf{0.00} & 0.26 & 0.04 & 0.04 & 0.00 & \textbf{0.95} & \textbf{0.95} \\
\midrule
\multirow{2}{*}{(g)} & $\widehat \beta_{X_2}$
    & \xmark / \textbf{0.00} & \qmark / \textbf{0.00} & \qmark / \textbf{0.00} & 0.23 & 0.09 & 0.09 & \textbf{0.94} & \textbf{0.96} & \textbf{0.95} \\
 & $\widehat \beta_0$
    & \xmark / 0.56 & \qmark / 0.83 & \qmark / 0.83 & 0.57 & 0.83 & 0.83 & 0.00 & 0.00 & 0.00 \\
\bottomrule
\end{tabular}
\caption{\label{tb:sim_n10000_wdim1} Simulation results for $N=10{,}000$, $|W|=1$ scenarios. $\beta_{X_2}=3.22$ for DGMs (a) and (b) and $\beta_{X_2}=2$ otherwise. MAR (IPW): Sun \& Tchetgen Tchetgen inverse-probability weighting estimator. MIA-I: ICE estimator. MIA-T: TMLE estimator. \cmark indicates that the sufficient (and in the case of $\muMAR$, necessary) conditions hold for $\muMAR$ and $\muMIA$, respectively. \xmark indicates that the conditions for $\muMAR$ are violated. \qmark indicates that the sufficient (but not necessary) conditions for $\muMIA$ are violated. Boldface indicates approximately correct performance.}
\end{table}

\subsection{Study 2: $|W|=10$; Assumptions \ref{ass:positivity} and \ref{ass:no_csi} hold}

\label{sec:sims_highdim_W}

In Study 2, for all DGMs, 5 members of $W$ were complete and the other 5 were incomplete. We did not run the Sun \& Tchetgen Tchetgen (STT) estimator because, in the $|W|=10$ case, there were at least $2^{6}$ missingness patterns, leading to a combinatorial explosion of parameters to be estimated. That is, each missing pattern requires estimation of an intercept and a slope for every variable that is observed for that pattern. As described in the main text, we instead used MICE as the MAR-based comparator.

\subsubsection{Data generation}

\providecommand{\expit}{\operatorname{expit}}
\providecommand{\Bern}{\operatorname{Bern}}

For Study 2, we generalize the DGMs from Study 1 by replacing the univariate $W_1$ by a set
$W = \{ W_1, \dots, W_d \}$ of $d = 10$ correlated variables, including $5$ continuous variables (3 complete; 2 incomplete) and $5$ binary variables (2 complete; 3 incomplete). Each variable in $W$ had the same adjacent edges as did the univariate $W_1$ in Study 1, and we kept the induced distributions of $Y$ and of the
missingness indicators comparable to those in Study 1 through a calibration process (described at the end of this section). Each incomplete member of $W$ was missing completely at random with probability 0.10, rather than 0.5 as for $W_1$ in Study 1, so that observations with all of $W$ observed were not too rare (with 5 incomplete members, they comprised approximately $0.9^5 \approx 59\%$ of the sample). Thus, apart from this missingness probability, the $d = 1$ special case would recover the Study 1 DGMs above if the singleton $W$ were incomplete.

To produce correlation among the variables in $W$, we generated them from a shared latent Gaussian vector:
\[
\begin{aligned}
Z &\sim N_d \left(\gamma \operatorname{Pa}(W) \cdot \bm{1}_d,\ \Sigma \right), \\
W_j &= Z_j \quad (\text{continuous case}), \qquad
W_j = \mathbb{1}\{Z_j > \tau_j\} \quad (\text{binary case}),
\end{aligned}
\]
where $\operatorname{Pa}(W)$ are the parents of $W$ in the DAG at hand, namely $X_2$ in
DGMs (a)--(c), $Y$ in DGMs (d)--(e), and nothing ($\gamma = 0$) in
DGMs (f)--(g). We set $\gamma = 1$. The correlation
matrix $\Sigma$ was exchangeable with off-diagonal $\rho = 0.4$, which together with the shared parent induces correlation across all
pairs of components, including continuous--binary pairs. The thresholds $\tau_j$
are fixed constants chosen so each binary variable had marginal probability
$0.5$.

The set $W$ affected each of its children via two scalar composites, each an
equally-weighted function of the variables in $W$. The composite ultimately used to generate $Y$ had main effects of each variable in $W$:
\[
S_Y \;\propto\; \sum_{j=1}^{d} W_j.
\]
The composite ultimately used to generate each missingness indicator that was a child of $W$ additionally had 3 pairwise interactions:
\[
S_R \;\propto\; \sum_{j=1}^{d} W_j \;+\; \lambda W_1 W_2 \;+\; \lambda W_3 W_4 \;+\; \lambda W_5 W_6,
\]
where $\lambda := 1$ scales the interactions. As a multivariate counterpart to the univariate-$W$ DGMs in Section \ref{sec:sims_univariate_W}, each time $W_1$ appeared in the mean of $Y$, the multivariate version replaced it with $S_Y$ and similarly for each missingness indicator. 

We fixed the proportionality constants in $S_Y$ and $S_R$, and the intercept of each
missingness model so that the
multivariate-$W$ DGM matched its univariate-$W$ counterpart. Specifically, the
constants for $S_Y$ were chosen so that the contribution of $W$ to the mean of $Y$ had the same mean and variance as $W_1$'s did. The intercept and coefficients for $S_R$ were chosen so that each $W$-affected missingness
indicator had the same marginal probability as in the univariate-$W$ case. Under this calibration, $\beta_{X_2}$ was approximately 3.13 and $\beta_0$ approximately 0.60 in DGMs (a) and (b); in the remaining DGMs, both estimands were as in Study 1.

\subsubsection{Results}

\newcommand{\colwidthw}{2.15cm}   %

\begin{table}[H]
\centering
\scriptsize
\setlength{\tabcolsep}{3pt}         %
\renewcommand{\arraystretch}{1.15}  %
\begin{tabular}{
    >{\centering\arraybackslash}p{\dagwidth}
    >{\centering\arraybackslash}p{\coefwidth}
    >{\centering\arraybackslash}p{\colwidth} >{\centering\arraybackslash}p{\colwidth} >{\centering\arraybackslash}p{\colwidth}
    >{\centering\arraybackslash}p{\colwidth} >{\centering\arraybackslash}p{\colwidth} >{\centering\arraybackslash}p{\colwidth}
    >{\centering\arraybackslash}p{\colwidth} >{\centering\arraybackslash}p{\colwidth} >{\centering\arraybackslash}p{\colwidth}
}
\toprule
\textbf{DGM} & \textbf{} &
\multicolumn{3}{c}{\textbf{Bias}} &
\multicolumn{3}{c}{\textbf{RMSE}} &
\multicolumn{3}{c}{\textbf{CI coverage}} \\
\cmidrule(lr){3-5} \cmidrule(lr){6-8} \cmidrule(lr){9-11}
 & & MAR (MICE) & MIA-T & MIA-I & MAR (MICE) & MIA-T & MIA-I & MAR (MICE) & MIA-T & MIA-I \\
\midrule
\multirow{2}{*}{(a)} & $\widehat \beta_{X_2}$
    & \xmark / -0.05 & \cmark / \textbf{0.00} & \cmark / \textbf{0.00} & 0.16 & 0.22 & 0.21 & \textbf{0.94} & \textbf{0.94} & \textbf{0.95} \\
 & $\widehat \beta_0$
    & \xmark / \textbf{0.01} & \cmark / \textbf{0.00} & \cmark / \textbf{0.00} & 0.09 & 0.12 & 0.12 & \textbf{0.95} & \textbf{0.95} & \textbf{0.95} \\
\midrule
\multirow{2}{*}{(b)} & $\widehat \beta_{X_2}$
    & \xmark / -0.14 & \qmark / -0.21 & \qmark / -0.21 & 0.21 & 0.38 & 0.36 & 0.90 & 0.85 & 0.90 \\
 & $\widehat \beta_0$
    & \xmark / 0.04 & \qmark / 0.53 & \qmark / 0.53 & 0.10 & 0.58 & 0.57 & \textbf{0.94} & 0.32 & 0.43 \\
\midrule
\multirow{2}{*}{(c)} & $\widehat \beta_{X_2}$
    & \xmark / \textbf{0.01} & \cmark / \textbf{-0.01} & \cmark / \textbf{-0.01} & 0.13 & 0.28 & 0.26 & \textbf{0.96} & 0.92 & \textbf{0.97} \\
 & $\widehat \beta_0$
    & \xmark / \textbf{0.00} & \cmark / \textbf{0.00} & \cmark / \textbf{0.00} & 0.09 & 0.22 & 0.20 & \textbf{0.95} & 0.90 & \textbf{0.96} \\
\midrule
\multirow{2}{*}{(d)} & $\widehat \beta_{X_2}$
    & \xmark / \textbf{0.01} & \cmark / \textbf{0.00} & \cmark / \textbf{0.01} & 0.12 & 0.16 & 0.17 & \textbf{0.95} & \textbf{0.95} & \textbf{0.97} \\
 & $\widehat \beta_0$
    & \xmark / \textbf{0.00} & \cmark / \textbf{0.00} & \cmark / \textbf{0.00} & 0.06 & 0.08 & 0.11 & \textbf{0.96} & \textbf{0.97} & \textbf{0.98} \\
\midrule
\multirow{2}{*}{(e)} & $\widehat \beta_{X_2}$
    & \xmark / \textbf{0.02} & \qmark / 0.12 & \qmark / 0.13 & 0.12 & 0.29 & 0.49 & \textbf{0.97} & 0.90 & \textbf{0.97} \\
 & $\widehat \beta_0$
    & \xmark / \textbf{0.00} & \qmark / 0.10 & \qmark / 0.10 & 0.06 & 0.18 & 0.24 & \textbf{0.96} & 0.88 & \textbf{0.94} \\
\midrule
\multirow{2}{*}{(f)} & $\widehat \beta_{X_2}$
    & \xmark / -0.05 & \cmark / \textbf{-0.01} & \cmark / \textbf{-0.01} & 0.17 & 0.26 & 0.24 & \textbf{0.97} & \textbf{0.94} & \textbf{0.96} \\
 & $\widehat \beta_0$
    & \xmark / \textbf{0.01} & \cmark / \textbf{0.01} & \cmark / \textbf{0.01} & 0.09 & 0.13 & 0.12 & \textbf{0.95} & \textbf{0.95} & \textbf{0.95} \\
\midrule
\multirow{2}{*}{(g)} & $\widehat \beta_{X_2}$
    & \xmark / -0.16 & \qmark / \textbf{0.01} & \qmark / \textbf{-0.03} & 0.25 & 0.39 & 0.43 & 0.92 & 0.88 & \textbf{0.98} \\
 & $\widehat \beta_0$
    & \xmark / 0.04 & \qmark / 0.72 & \qmark / 0.72 & 0.11 & 0.75 & 0.74 & \textbf{0.93} & 0.02 & 0.02 \\
\bottomrule
\end{tabular}
\caption{\label{tb:sim_n1000_wdim10} Simulation results for $N=1{,}000$, $|W|=10$ scenarios (5 continuous, 5 binary auxiliary components; 5 complete, 5 incomplete). $\beta_{X_2}=3.13$ for DGMs (a) and (b) and $\beta_{X_2}=2$ otherwise. MAR (MICE): multiple imputation by chained equations. MIA-I: ICE estimator. MIA-T: TMLE estimator. \cmark indicates that the sufficient (and in the case of $\muMAR$, necessary) conditions hold for $\muMAR$ and $\muMIA$, respectively. \xmark indicates that the conditions for $\muMAR$ are violated. \qmark indicates that the sufficient (but not necessary) conditions for $\muMIA$ are violated. Boldface indicates approximately correct performance.}
\end{table}

\begin{table}[H]
\centering
\scriptsize
\setlength{\tabcolsep}{3pt}         %
\renewcommand{\arraystretch}{1.15}  %
\begin{tabular}{
    >{\centering\arraybackslash}p{\dagwidth}
    >{\centering\arraybackslash}p{\coefwidth}
    >{\centering\arraybackslash}p{\colwidth} >{\centering\arraybackslash}p{\colwidth} >{\centering\arraybackslash}p{\colwidth}
    >{\centering\arraybackslash}p{\colwidth} >{\centering\arraybackslash}p{\colwidth} >{\centering\arraybackslash}p{\colwidth}
    >{\centering\arraybackslash}p{\colwidth} >{\centering\arraybackslash}p{\colwidth} >{\centering\arraybackslash}p{\colwidth}
}
\toprule
\textbf{DGM} & \textbf{} &
\multicolumn{3}{c}{\textbf{Bias}} &
\multicolumn{3}{c}{\textbf{RMSE}} &
\multicolumn{3}{c}{\textbf{CI coverage}} \\
\cmidrule(lr){3-5} \cmidrule(lr){6-8} \cmidrule(lr){9-11}
 & & MAR (MICE) & MIA-T & MIA-I & MAR (MICE) & MIA-T & MIA-I & MAR (MICE) & MIA-T & MIA-I \\
\midrule
\multirow{2}{*}{(a)} & $\widehat \beta_{X_2}$
    & \xmark / -0.04 & \cmark / \textbf{0.00} & \cmark / \textbf{0.00} & 0.11 & 0.14 & 0.14 & \textbf{0.95} & \textbf{0.96} & \textbf{0.95} \\
 & $\widehat \beta_0$
    & \xmark / \textbf{0.00} & \cmark / \textbf{0.00} & \cmark / \textbf{0.00} & 0.06 & 0.08 & 0.08 & \textbf{0.94} & \textbf{0.96} & \textbf{0.95} \\
\midrule
\multirow{2}{*}{(b)} & $\widehat \beta_{X_2}$
    & \xmark / -0.13 & \qmark / -0.21 & \qmark / -0.21 & 0.17 & 0.30 & 0.29 & 0.85 & 0.81 & 0.80 \\
 & $\widehat \beta_0$
    & \xmark / \textbf{0.02} & \qmark / 0.54 & \qmark / 0.54 & 0.07 & 0.56 & 0.56 & \textbf{0.94} & 0.08 & 0.06 \\
\midrule
\multirow{2}{*}{(c)} & $\widehat \beta_{X_2}$
    & \xmark / \textbf{0.02} & \cmark / \textbf{0.00} & \cmark / \textbf{0.00} & 0.10 & 0.18 & 0.17 & \textbf{0.95} & \textbf{0.94} & \textbf{0.96} \\
 & $\widehat \beta_0$
    & \xmark / \textbf{-0.01} & \cmark / \textbf{0.00} & \cmark / \textbf{0.00} & 0.06 & 0.13 & 0.13 & \textbf{0.96} & \textbf{0.94} & \textbf{0.95} \\
\midrule
\multirow{2}{*}{(d)} & $\widehat \beta_{X_2}$
    & \xmark / \textbf{0.01} & \cmark / \textbf{-0.01} & \cmark / \textbf{0.00} & 0.08 & 0.11 & 0.11 & \textbf{0.95} & \textbf{0.96} & \textbf{0.95} \\
 & $\widehat \beta_0$
    & \xmark / \textbf{0.00} & \cmark / \textbf{0.00} & \cmark / \textbf{0.00} & 0.05 & 0.06 & 0.06 & \textbf{0.95} & \textbf{0.96} & \textbf{0.96} \\
\midrule
\multirow{2}{*}{(e)} & $\widehat \beta_{X_2}$
    & \xmark / \textbf{0.03} & \qmark / 0.11 & \qmark / 0.11 & 0.09 & 0.20 & 0.20 & \textbf{0.93} & 0.91 & 0.91 \\
 & $\widehat \beta_0$
    & \xmark / \textbf{0.00} & \qmark / 0.11 & \qmark / 0.11 & 0.05 & 0.16 & 0.16 & \textbf{0.94} & 0.79 & 0.81 \\
\midrule
\multirow{2}{*}{(f)} & $\widehat \beta_{X_2}$
    & \xmark / -0.05 & \cmark / \textbf{0.00} & \cmark / \textbf{0.00} & 0.13 & 0.19 & 0.18 & \textbf{0.95} & \textbf{0.93} & \textbf{0.93} \\
 & $\widehat \beta_0$
    & \xmark / \textbf{0.01} & \cmark / \textbf{0.00} & \cmark / \textbf{0.00} & 0.07 & 0.09 & 0.08 & \textbf{0.96} & \textbf{0.96} & \textbf{0.95} \\
\midrule
\multirow{2}{*}{(g)} & $\widehat \beta_{X_2}$
    & \xmark / -0.17 & \qmark / \textbf{0.00} & \qmark / \textbf{0.00} & 0.22 & 0.24 & 0.23 & 0.81 & \textbf{0.93} & \textbf{0.95} \\
 & $\widehat \beta_0$
    & \xmark / \textbf{0.02} & \qmark / 0.71 & \qmark / 0.71 & 0.08 & 0.72 & 0.72 & \textbf{0.94} & 0.00 & 0.00 \\
\bottomrule
\end{tabular}
\caption{\label{tb:sim_n2000_wdim10} Simulation results for $N=2{,}000$, $|W|=10$ scenarios (5 continuous, 5 binary auxiliary components; 5 complete, 5 incomplete). $\beta_{X_2}=3.13$ for DGMs (a) and (b) and $\beta_{X_2}=2$ otherwise. MAR (MICE): multiple imputation by chained equations. MIA-I: ICE estimator. MIA-T: TMLE estimator. \cmark indicates that the sufficient (and in the case of $\muMAR$, necessary) conditions hold for $\muMAR$ and $\muMIA$, respectively. \xmark indicates that the conditions for $\muMAR$ are violated. \qmark indicates that the sufficient (but not necessary) conditions for $\muMIA$ are violated. Boldface indicates approximately correct performance.}
\end{table}

\begin{table}[H]
\centering
\scriptsize
\setlength{\tabcolsep}{3pt}         %
\renewcommand{\arraystretch}{1.15}  %
\begin{tabular}{
    >{\centering\arraybackslash}p{\dagwidth}
    >{\centering\arraybackslash}p{\coefwidth}
    >{\centering\arraybackslash}p{\colwidth} >{\centering\arraybackslash}p{\colwidth} >{\centering\arraybackslash}p{\colwidth}
    >{\centering\arraybackslash}p{\colwidth} >{\centering\arraybackslash}p{\colwidth} >{\centering\arraybackslash}p{\colwidth}
    >{\centering\arraybackslash}p{\colwidth} >{\centering\arraybackslash}p{\colwidth} >{\centering\arraybackslash}p{\colwidth}
}
\toprule
\textbf{DGM} & \textbf{} &
\multicolumn{3}{c}{\textbf{Bias}} &
\multicolumn{3}{c}{\textbf{RMSE}} &
\multicolumn{3}{c}{\textbf{CI coverage}} \\
\cmidrule(lr){3-5} \cmidrule(lr){6-8} \cmidrule(lr){9-11}
 & & MAR (MICE) & MIA-T & MIA-I & MAR (MICE) & MIA-T & MIA-I & MAR (MICE) & MIA-T & MIA-I \\
\midrule
\multirow{2}{*}{(a)} & $\widehat \beta_{X_2}$
    & \xmark / -0.04 & \cmark / \textbf{0.00} & \cmark / \textbf{0.00} & 0.08 & 0.09 & 0.09 & 0.92 & \textbf{0.96} & \textbf{0.95} \\
 & $\widehat \beta_0$
    & \xmark / \textbf{0.00} & \cmark / \textbf{0.00} & \cmark / \textbf{0.00} & 0.04 & 0.05 & 0.05 & \textbf{0.94} & \textbf{0.95} & \textbf{0.93} \\
\midrule
\multirow{2}{*}{(b)} & $\widehat \beta_{X_2}$
    & \xmark / -0.12 & \qmark / -0.22 & \qmark / -0.22 & 0.14 & 0.25 & 0.25 & 0.65 & 0.61 & 0.58 \\
 & $\widehat \beta_0$
    & \xmark / \textbf{0.02} & \qmark / 0.54 & \qmark / 0.53 & 0.05 & 0.54 & 0.54 & \textbf{0.93} & 0.00 & 0.00 \\
\midrule
\multirow{2}{*}{(c)} & $\widehat \beta_{X_2}$
    & \xmark / \textbf{0.02} & \cmark / \textbf{0.01} & \cmark / \textbf{0.00} & 0.06 & 0.11 & 0.10 & \textbf{0.95} & \textbf{0.95} & \textbf{0.95} \\
 & $\widehat \beta_0$
    & \xmark / \textbf{-0.01} & \cmark / \textbf{0.00} & \cmark / \textbf{0.00} & 0.04 & 0.08 & 0.08 & \textbf{0.97} & \textbf{0.95} & \textbf{0.94} \\
\midrule
\multirow{2}{*}{(d)} & $\widehat \beta_{X_2}$
    & \xmark / \textbf{0.01} & \cmark / \textbf{0.00} & \cmark / \textbf{0.00} & 0.05 & 0.07 & 0.07 & \textbf{0.94} & \textbf{0.95} & \textbf{0.95} \\
 & $\widehat \beta_0$
    & \xmark / \textbf{0.00} & \cmark / \textbf{0.00} & \cmark / \textbf{0.00} & 0.03 & 0.04 & 0.04 & \textbf{0.95} & \textbf{0.94} & \textbf{0.94} \\
\midrule
\multirow{2}{*}{(e)} & $\widehat \beta_{X_2}$
    & \xmark / \textbf{0.03} & \qmark / 0.11 & \qmark / 0.11 & 0.06 & 0.15 & 0.15 & \textbf{0.93} & 0.82 & 0.82 \\
 & $\widehat \beta_0$
    & \xmark / \textbf{-0.01} & \qmark / 0.11 & \qmark / 0.11 & 0.03 & 0.13 & 0.13 & \textbf{0.94} & 0.64 & 0.62 \\
\midrule
\multirow{2}{*}{(f)} & $\widehat \beta_{X_2}$
    & \xmark / -0.05 & \cmark / \textbf{0.00} & \cmark / \textbf{0.00} & 0.08 & 0.11 & 0.10 & \textbf{0.95} & \textbf{0.97} & \textbf{0.95} \\
 & $\widehat \beta_0$
    & \xmark / \textbf{0.01} & \cmark / \textbf{0.00} & \cmark / \textbf{0.00} & 0.04 & 0.06 & 0.05 & \textbf{0.96} & \textbf{0.97} & \textbf{0.96} \\
\midrule
\multirow{2}{*}{(g)} & $\widehat \beta_{X_2}$
    & \xmark / -0.17 & \qmark / \textbf{0.00} & \qmark / \textbf{0.00} & 0.19 & 0.15 & 0.14 & 0.55 & \textbf{0.94} & \textbf{0.94} \\
 & $\widehat \beta_0$
    & \xmark / \textbf{0.02} & \qmark / 0.71 & \qmark / 0.71 & 0.05 & 0.72 & 0.72 & \textbf{0.94} & 0.00 & 0.00 \\
\bottomrule
\end{tabular}
\caption{\label{tb:sim_n5000_wdim10} Simulation results for $N=5{,}000$, $|W|=10$ scenarios (5 continuous, 5 binary auxiliary components; 5 complete, 5 incomplete). $\beta_{X_2}=3.13$ for DGMs (a) and (b) and $\beta_{X_2}=2$ otherwise. MAR (MICE): multiple imputation by chained equations. MIA-I: ICE estimator. MIA-T: TMLE estimator. \cmark indicates that the sufficient (and in the case of $\muMAR$, necessary) conditions hold for $\muMAR$ and $\muMIA$, respectively. \xmark indicates that the conditions for $\muMAR$ are violated. \qmark indicates that the sufficient (but not necessary) conditions for $\muMIA$ are violated. Boldface indicates approximately correct performance.}
\end{table}

\begin{table}[H]
\centering
\scriptsize
\setlength{\tabcolsep}{3pt}         %
\renewcommand{\arraystretch}{1.15}  %
\begin{tabular}{
    >{\centering\arraybackslash}p{\dagwidth}
    >{\centering\arraybackslash}p{\coefwidth}
    >{\centering\arraybackslash}p{\colwidth} >{\centering\arraybackslash}p{\colwidth} >{\centering\arraybackslash}p{\colwidth}
    >{\centering\arraybackslash}p{\colwidth} >{\centering\arraybackslash}p{\colwidth} >{\centering\arraybackslash}p{\colwidth}
    >{\centering\arraybackslash}p{\colwidth} >{\centering\arraybackslash}p{\colwidth} >{\centering\arraybackslash}p{\colwidth}
}
\toprule
\textbf{DGM} & \textbf{} &
\multicolumn{3}{c}{\textbf{Bias}} &
\multicolumn{3}{c}{\textbf{RMSE}} &
\multicolumn{3}{c}{\textbf{CI coverage}} \\
\cmidrule(lr){3-5} \cmidrule(lr){6-8} \cmidrule(lr){9-11}
 & & MAR (MICE) & MIA-T & MIA-I & MAR (MICE) & MIA-T & MIA-I & MAR (MICE) & MIA-T & MIA-I \\
\midrule
\multirow{2}{*}{(a)} & $\widehat \beta_{X_2}$
    & \xmark / -0.04 & \cmark / \textbf{0.00} & \cmark / \textbf{0.00} & 0.07 & 0.06 & 0.06 & 0.86 & \textbf{0.95} & \textbf{0.94} \\
 & $\widehat \beta_0$
    & \xmark / \textbf{0.00} & \cmark / \textbf{0.00} & \cmark / \textbf{0.00} & 0.03 & 0.04 & 0.04 & \textbf{0.94} & \textbf{0.96} & \textbf{0.94} \\
\midrule
\multirow{2}{*}{(b)} & $\widehat \beta_{X_2}$
    & \xmark / -0.12 & \qmark / -0.21 & \qmark / -0.21 & 0.13 & 0.23 & 0.23 & 0.36 & 0.35 & 0.33 \\
 & $\widehat \beta_0$
    & \xmark / \textbf{0.02} & \qmark / 0.53 & \qmark / 0.53 & 0.03 & 0.54 & 0.53 & 0.92 & 0.00 & 0.00 \\
\midrule
\multirow{2}{*}{(c)} & $\widehat \beta_{X_2}$
    & \xmark / \textbf{0.02} & \cmark / \textbf{0.00} & \cmark / \textbf{0.00} & 0.05 & 0.08 & 0.08 & \textbf{0.93} & \textbf{0.94} & \textbf{0.94} \\
 & $\widehat \beta_0$
    & \xmark / \textbf{-0.01} & \cmark / \textbf{0.00} & \cmark / \textbf{0.00} & 0.03 & 0.06 & 0.06 & \textbf{0.93} & \textbf{0.95} & \textbf{0.94} \\
\midrule
\multirow{2}{*}{(d)} & $\widehat \beta_{X_2}$
    & \xmark / \textbf{0.01} & \cmark / \textbf{0.00} & \cmark / \textbf{0.00} & 0.04 & 0.05 & 0.05 & \textbf{0.94} & \textbf{0.94} & \textbf{0.94} \\
 & $\widehat \beta_0$
    & \xmark / \textbf{0.00} & \cmark / \textbf{0.00} & \cmark / \textbf{0.00} & 0.02 & 0.03 & 0.03 & \textbf{0.95} & \textbf{0.96} & \textbf{0.95} \\
\midrule
\multirow{2}{*}{(e)} & $\widehat \beta_{X_2}$
    & \xmark / \textbf{0.03} & \qmark / 0.11 & \qmark / 0.11 & 0.05 & 0.13 & 0.13 & 0.84 & 0.68 & 0.68 \\
 & $\widehat \beta_0$
    & \xmark / \textbf{-0.01} & \qmark / 0.11 & \qmark / 0.11 & 0.02 & 0.12 & 0.12 & \textbf{0.94} & 0.28 & 0.29 \\
\midrule
\multirow{2}{*}{(f)} & $\widehat \beta_{X_2}$
    & \xmark / -0.05 & \cmark / \textbf{0.00} & \cmark / \textbf{0.00} & 0.07 & 0.08 & 0.07 & 0.92 & \textbf{0.95} & \textbf{0.94} \\
 & $\widehat \beta_0$
    & \xmark / \textbf{0.00} & \cmark / \textbf{0.00} & \cmark / \textbf{0.00} & 0.03 & 0.04 & 0.04 & \textbf{0.96} & \textbf{0.96} & \textbf{0.95} \\
\midrule
\multirow{2}{*}{(g)} & $\widehat \beta_{X_2}$
    & \xmark / -0.17 & \qmark / \textbf{0.00} & \qmark / \textbf{0.00} & 0.18 & 0.11 & 0.10 & 0.22 & \textbf{0.95} & \textbf{0.95} \\
 & $\widehat \beta_0$
    & \xmark / \textbf{0.01} & \qmark / 0.71 & \qmark / 0.71 & 0.04 & 0.71 & 0.71 & 0.92 & 0.00 & 0.00 \\
\bottomrule
\end{tabular}
\caption{\label{tb:sim_n10000_wdim10} Simulation results for $N=10{,}000$, $|W|=10$ scenarios (5 continuous, 5 binary auxiliary components; 5 complete, 5 incomplete). $\beta_{X_2}=3.13$ for DGMs (a) and (b) and $\beta_{X_2}=2$ otherwise. MAR (MICE): multiple imputation by chained equations. MIA-I: ICE estimator. MIA-T: TMLE estimator. \cmark indicates that the sufficient (and in the case of $\muMAR$, necessary) conditions hold for $\muMAR$ and $\muMIA$, respectively. \xmark indicates that the conditions for $\muMAR$ are violated. \qmark indicates that the sufficient (but not necessary) conditions for $\muMIA$ are violated. Boldface indicates approximately correct performance.}
\end{table}

\subsection{Study 3: $|W| \in \{1, 2\}$; Assumptions \ref{ass:positivity} and \ref{ass:no_csi} are violated or nearly violated}
\label{sec:sims_assumptions_violated}

Except where otherwise stated, simulation methods were as in Studies 1 and 2.  We simulated eight DGMs, indexed by a Roman numeral (I or II) and a letter ((a)--(d)). Each Roman numeral denoted a DAG, and each letter a distribution that is Markov wrt the DAG. All DAGs violated the sufficient conditions of MIA in ways that ensured bias in $\widehat \beta_0$, $\widehat \beta_{X_2}$, or both. DAG I had $W = W^-$, and DAG II had $V^- = \{Y\}$. Thus, in distributions satisfying Assumptions \ref{ass:positivity} and \ref{ass:no_csi} for which MNAR($V \cup W^+$) holds, Proposition \ref{prop:wneg_cant_fix_mnar} states that MNAR($V \cup W$) also holds, and Theorems \ref{thm:IPW_with_Wneg} and \ref{thm:IPW_with_incomplete_Y} respectively state that $\muMAR$ is incorrect. Hence, in distributions that do satisfy the assumptions, $\muMAR$ would often be biased. However, for each DAG, we specified DGMs (a)--(d) such that either MAR($V \cup W$), or only a weak form MNAR($V \cup W$) held due to a near-violation of Assumption \ref{ass:positivity} or \ref{ass:no_csi}:
\begin{outline}[enumerate]
\1[(a)] MAR($V \cup W$) held due to CSI
\1[(b)] Weak MNAR($V \cup W$) held due to a near-occurrence of CSI
\1[(c)] MAR($V \cup W$) held due to a positivity violation
\1[(d)] Weak MNAR($V \cup W$) held due to a near-violation of positivity
\end{outline}
Thus, we would expect $\muMAR$ to be unbiased for DGMs (a) and (c) for each DAG, and to be minimally biased for DGMs (b) and (d). We would furthermore expect $\muMIA$ to be biased for all DGMs. Recall that when discussing inconsistency and bias, we refer to the parameter vector $(\beta_0, \beta_{X_2})$ unless specifically referring to one component.

\subsubsection{Data generation}

\subsubsection*{DAG I}

\begin{figure}[H]
\centering
\begin{tikzpicture}
\sansmath
\tikzset{
  line width=1.2pt,
  >=stealth,
  ell/.style={draw, fill=white, ellipse, inner sep=2pt, line width=1.5pt},
  shaded/.style={draw, fill=gray!40, ellipse, inner sep=2pt, line width=1.5pt}
}
\node[name=X1, ell]                      {$X_1$};
\node[name=X2, shaded, right=10mm of X1]     {$X_2$};
\node[name=W,  shaded, right=10mm of X2]     {$W_1$};
\node[name=Y,  shaded, right=10mm of W]      {$Y$};
\node[name=RX2, ell, below=8mm of X2]       {$R_{X_2}$};
\node[name=RW,  ell, below=8mm of W]        {$R_{W_1}$};
\node[name=RY,  ell, below=8mm of Y]        {$R_Y$};
\draw[->,line width=1.2pt,>=stealth]
(X1) edge (X2)
(W)  edge (Y)
(RW) edge[color=red, dotted] (RY)
(W)  edge (RX2)
(W)  edge[color=blue, dashed] (RY);
\draw[->,line width=1.2pt,>=stealth] (X2) to[out=30,in=150] (Y);
\draw[->,line width=1.2pt,>=stealth] (X1) to[out=40,in=140] (Y);
\end{tikzpicture}
\caption{Blue dashed edge: CSI or near-CSI. Red edge: Determinism or near-determinism.}
\end{figure}

For all DGMs I(a)--I(d):
\begin{align*}
X_1 &\sim \text{Bern}(0.5) \\
X_2 &\sim \text{Bern}\!\left(\text{expit}(-1 + 3X_1)\right) \\
W_1 &\sim \text{Bern}(0.5) \\
Y &\sim \text{N}\!\left(2.6X_1 + 2X_2 + 3W_1 + X_1 X_2 + X_2 W_1,\ 1\right) \\
R_{W_1} &\sim \text{Bern}(0.5) \\
R_{X_2} &\sim \text{Bern}(\varsigma) \\
R_Y &\sim \text{Bern}(\varsigma).
\end{align*}

In DGM I(a):
\begingroup
\addtolength{\jot}{0.5em}
\begin{align*}
\varsigma &= \begin{cases}
\text{expit}(-0.5 + 3W_1) & \text{if } R_{W_1} = 1 \\
0.5 & \text{if } R_{W_1} = 0,
\end{cases}
\end{align*}
\endgroup
In DGM I(b):
\begingroup
\addtolength{\jot}{0.5em}
\begin{align*}
\varsigma &= \begin{cases}
\text{expit}(-0.5 + 3W_1) & \text{if } R_{W_1} = 1 \\
\text{expit}(-0.5 + 0.75W_1) & \text{if } R_{W_1} = 0
\end{cases}
\end{align*}
\endgroup
In DGM I(c):
\begingroup
\addtolength{\jot}{0.5em}
\begin{align*}
\varsigma &= \begin{cases}
\text{expit}(-0.5 + 3W_1) & \text{if } R_{W_1} = 1 \\
0 & \text{if } R_{W_1} = 0
\end{cases}
\end{align*}
\endgroup

In DGM I(d):
\begingroup
\addtolength{\jot}{0.5em}
\begin{align*}
\varsigma &= \begin{cases}
\text{expit}(-0.5 + 3W_1) & \text{if } R_{W_1} = 1 \\
\text{expit}(-4.5 + 3W_1) & \text{if } R_{W_1} = 0
\end{cases}
\end{align*}
\endgroup

\subsubsection*{DAG II}

\begin{figure}[H]
\centering
\begin{tikzpicture}
\sansmath
\tikzset{
  line width=1.2pt,
  >=stealth,
  ell/.style={draw, fill=white, ellipse, inner sep=2pt, line width=1.5pt},
  shaded/.style={draw, fill=gray!40, ellipse, inner sep=2pt, line width=1.5pt}
}
\node[name=X1, ell]                      {$X_1$};
\node[name=X2, ell, right=10mm of X1]     {$X_2$};
\node[name=W, ell, right=10mm of X2]     {$W_1$};
\node[name=Y,  shaded, right=10mm of W]      {$Y$};
\node[name=W2, shaded, right=10mm of Y]      {$W_2$};
\node[name=RW, ell, below=8mm of W2]         {$R_{W_2}$};
\node[name=RY, ell, below=8mm of Y]         {$R_Y$};
\draw[->,line width=1.2pt,>=stealth]
(X1) edge (X2)
(W)  edge (Y)
(RW) edge[color=red, dotted] (RY)
(W2) edge (Y)
(W2) edge[color=blue, dashed] (RY);
\draw[->,line width=1.2pt,>=stealth] (X2) to[out=30,in=150] (Y);
\draw[->,line width=1.2pt,>=stealth] (X1) to[out=40,in=140] (Y);
\end{tikzpicture}
\caption{Blue dashed edge: CSI or near-CSI. Red edge: Determinism violating positivity or near-determinism.}
\end{figure}

In all DGMs II(a)--II(d):
\begin{align*}
X_1 &\sim \text{Bern}(0.5) \\
X_2 &\sim \text{Bern}\!\left(\text{expit}(-1 + 3X_1)\right) \\
W_1 &\sim \text{Bern}(0.5) \\
W_2 &\sim \text{Bern}(0.5) \\
Y &\sim \text{N}\!\left(2.6X_1 + 2X_2 + 2W_1 + 2W_2 + X_1 X_2,\ 1\right) \\
R_{W_2} &\sim \text{Bern}\!\left(\text{expit}(-1 + 3W_1)\right) \\
R_Y &\sim \text{Bern}(\varsigma),
\end{align*}

In DGM II(a):
\begingroup
\addtolength{\jot}{0.5em}
\begin{align*}
\varsigma &= \begin{cases}
\text{expit}(-1 + 2W_2) & \text{if } R_{W_2} = 1 \\
0.5 & \text{if } R_{W_2} = 0,
\end{cases}
\end{align*}
\endgroup
and in DGM II(b):
\begingroup
\addtolength{\jot}{0.5em}
\begin{align*}
\varsigma &= \begin{cases}
\text{expit}(-1 + 2W_2) & \text{if } R_{W_2} = 1 \\
\text{expit}(-1 + 0.5W_2) & \text{if } R_{W_2} = 0
\end{cases}
\end{align*}
\endgroup
In DGM II(c):
\begingroup
\addtolength{\jot}{0.5em}
\begin{align*}
\varsigma &= \begin{cases}
\text{expit}(-1 + 2W_2) & \text{if } R_{W_2} = 1 \\
0 & \text{if } R_{W_2} = 0
\end{cases}
\end{align*}
\endgroup

In DGM II(d):
\begingroup
\addtolength{\jot}{0.5em}
\begin{align*}
\varsigma &= \begin{cases}
\text{expit}(-1 + 2W_2) & \text{if } R_{W_2} = 1 \\
\text{expit}(-3.6 + 2W_2) & \text{if } R_{W_2} = 0
\end{cases}
\end{align*}
\endgroup

\subsubsection{Results}

\newcommand{\colwidthiii}{1.42cm}   %
\newcommand{\dagwidthiii}{0.95cm}
\newcommand{\coefwidthiii}{0.95cm}

Tables \ref{tb:sim3_N200}--\ref{tb:sim3_N10000} present results for Study 3. As expected, when MAR($V \cup W$) held (DGMs (a) and (c)), the MAR estimator was approximately unbiased at larger sample sizes; at $N = 10{,}000$, its bias was at most 0.01 in absolute value for both estimands and for both DAGs, apart from a bias of $-0.03$ for $\beta_0$ in DGM II(c). When only a weak form of MNAR($V \cup W$) held (DGMs (b) and (d)), the bias of the MAR estimator was modest but did not diminish with sample size, ranging from 0.02 to 0.14 at $N = 10{,}000$. At the smallest sample sizes ($N \le 500$), most estimators were biased, as was also the case in Study 1. By design, the MIA estimators were biased for $\beta_0$ throughout, with bias of approximately 0.63 for DAG I and 0.53 for DAG II at $N = 10{,}000$; for DAG I, they were also biased for $\beta_{X_2}$ (approximately 0.21). For DAG II, however, the MIA estimators were approximately unbiased for $\beta_{X_2}$ despite the violation of MIA's sufficient conditions, illustrating that those conditions are not necessary.

\providecommand{\cmark}{\ding{51}}
\providecommand{\xmark}{\ding{55}}
\providecommand{\qmark}{\textbf{?}}

\begin{table}[H]
\centering
\footnotesize
\setlength{\tabcolsep}{3pt}
\begin{tabular}{@{}p{\dagwidthiii}p{\coefwidthiii}>{\centering\arraybackslash}p{\colwidthiii}>{\centering\arraybackslash}p{\colwidthiii}>{\centering\arraybackslash}p{\colwidthiii}>{\centering\arraybackslash}p{\colwidthiii}>{\centering\arraybackslash}p{\colwidthiii}>{\centering\arraybackslash}p{\colwidthiii}>{\centering\arraybackslash}p{\colwidthiii}>{\centering\arraybackslash}p{\colwidthiii}>{\centering\arraybackslash}p{\colwidthiii}@{}}
\toprule
\multicolumn{2}{@{}l}{\textbf{DGM}} & \multicolumn{3}{c}{\textbf{Bias}} & \multicolumn{3}{c}{\textbf{RMSE}} & \multicolumn{3}{c}{\textbf{CI coverage}} \\
\cmidrule(lr){3-5} \cmidrule(lr){6-8} \cmidrule(lr){9-11}
& & MAR (IPW) & MIA-T & MIA-I & MAR (IPW) & MIA-T & MIA-I & MAR (IPW) & MIA-T & MIA-I \\
\midrule
I(a) & $\widehat \beta_{X_2}$ & \cmark / 0.51 & \qmark / 0.55 & \qmark / 0.23 & 1.52 & 1.11 & 0.94 & 0.67 & 0.69 & 0.90 \\
 & $\widehat \beta_0$ & \cmark / 0.26 & \qmark / 0.66 & \qmark / 0.61 & 0.60 & 0.79 & 0.73 & 0.88 & 0.58 & 0.65 \\
I(b) & $\widehat \beta_{X_2}$ & \xmark / 0.57 & \qmark / 0.61 & \qmark / 0.27 & 1.53 & 1.17 & 0.96 & 0.64 & 0.64 & 0.89 \\
 & $\widehat \beta_0$ & \xmark / 0.32 & \qmark / 0.65 & \qmark / 0.61 & 0.62 & 0.79 & 0.73 & 0.87 & 0.58 & 0.66 \\
I(c) & $\widehat \beta_{X_2}$ & \cmark / 0.47 & \qmark / 0.55 & \qmark / 0.22 & 1.48 & 1.10 & 0.93 & 0.67 & 0.69 & 0.90 \\
 & $\widehat \beta_0$ & \cmark / 0.29 & \qmark / 0.68 & \qmark / 0.63 & 0.63 & 0.81 & 0.75 & 0.86 & 0.55 & 0.63 \\
I(d) & $\widehat \beta_{X_2}$ & \xmark / 0.57 & \qmark / 0.56 & \qmark / 0.18 & 1.48 & 1.11 & 0.90 & 0.67 & 0.68 & 0.91 \\
 & $\widehat \beta_0$ & \xmark / 0.33 & \qmark / 0.66 & \qmark / 0.62 & 0.64 & 0.80 & 0.74 & 0.86 & 0.57 & 0.65 \\
\midrule
II(a) & $\widehat \beta_{X_2}$ & \cmark / 0.22 & \qmark / 0.17 & \qmark / \textbf{0.01} & 1.00 & 0.82 & 0.78 & 0.86 & 0.77 & \textbf{0.93} \\
 & $\widehat \beta_0$ & \cmark / 0.18 & \qmark / 0.53 & \qmark / 0.53 & 0.45 & 0.65 & 0.65 & 0.89 & 0.57 & 0.69 \\
II(b) & $\widehat \beta_{X_2}$ & \xmark / 0.23 & \qmark / 0.16 & \qmark / \textbf{0.01} & 0.97 & 0.83 & 0.78 & 0.84 & 0.74 & 0.92 \\
 & $\widehat \beta_0$ & \xmark / 0.21 & \qmark / 0.53 & \qmark / 0.53 & 0.48 & 0.67 & 0.66 & 0.85 & 0.56 & 0.68 \\
II(c) & $\widehat \beta_{X_2}$ & \cmark / 0.20 & \qmark / 0.17 & \qmark / \textbf{0.01} & 1.00 & 0.83 & 0.75 & 0.86 & 0.76 & 0.92 \\
 & $\widehat \beta_0$ & \cmark / 0.16 & \qmark / 0.51 & \qmark / 0.51 & 0.49 & 0.65 & 0.65 & 0.85 & 0.56 & 0.67 \\
II(d) & $\widehat \beta_{X_2}$ & \xmark / 0.20 & \qmark / 0.18 & \qmark / 0.04 & 0.98 & 0.80 & 0.78 & 0.84 & 0.75 & 0.91 \\
 & $\widehat \beta_0$ & \xmark / 0.19 & \qmark / 0.51 & \qmark / 0.51 & 0.47 & 0.64 & 0.63 & 0.86 & 0.58 & 0.70 \\
\bottomrule
\end{tabular}
\caption{\label{tb:sim3_N200}Simulation results for Study 3, $N = 200$. For DAG I, $\beta_{X_2} = 2.5$ and $\beta_0 = 1.5$; for DAG II, $\beta_{X_2} = 2$ and $\beta_0 = 2$. MAR (IPW): Sun \& Tchetgen Tchetgen inverse-probability weighting estimator. MIA-I: ICE estimator. MIA-T: TMLE estimator. \cmark\ indicates that the sufficient (and, in the case of $\muMAR$, necessary) conditions for $\muMAR$ and $\muMIA$, respectively, hold. \xmark\ indicates that the conditions for $\muMAR$ are violated; in DGMs (b) and (d), this violation is weak by construction. \qmark\ indicates that the sufficient (but not necessary) conditions for $\muMIA$ are violated. Boldface indicates approximately correct performance.}
\end{table}

\begin{table}[H]
\centering
\footnotesize
\setlength{\tabcolsep}{3pt}
\begin{tabular}{@{}p{\dagwidthiii}p{\coefwidthiii}>{\centering\arraybackslash}p{\colwidthiii}>{\centering\arraybackslash}p{\colwidthiii}>{\centering\arraybackslash}p{\colwidthiii}>{\centering\arraybackslash}p{\colwidthiii}>{\centering\arraybackslash}p{\colwidthiii}>{\centering\arraybackslash}p{\colwidthiii}>{\centering\arraybackslash}p{\colwidthiii}>{\centering\arraybackslash}p{\colwidthiii}>{\centering\arraybackslash}p{\colwidthiii}@{}}
\toprule
\multicolumn{2}{@{}l}{\textbf{DGM}} & \multicolumn{3}{c}{\textbf{Bias}} & \multicolumn{3}{c}{\textbf{RMSE}} & \multicolumn{3}{c}{\textbf{CI coverage}} \\
\cmidrule(lr){3-5} \cmidrule(lr){6-8} \cmidrule(lr){9-11}
& & MAR (IPW) & MIA-T & MIA-I & MAR (IPW) & MIA-T & MIA-I & MAR (IPW) & MIA-T & MIA-I \\
\midrule
I(a) & $\widehat \beta_{X_2}$ & \cmark / 0.20 & \qmark / 0.26 & \qmark / 0.22 & 0.94 & 0.67 & 0.59 & 0.88 & 0.86 & 0.91 \\
 & $\widehat \beta_0$ & \cmark / 0.12 & \qmark / 0.63 & \qmark / 0.63 & 0.34 & 0.68 & 0.68 & \textbf{0.96} & 0.30 & 0.33 \\
I(b) & $\widehat \beta_{X_2}$ & \xmark / 0.29 & \qmark / 0.27 & \qmark / 0.21 & 0.90 & 0.65 & 0.56 & 0.87 & 0.86 & 0.92 \\
 & $\widehat \beta_0$ & \xmark / 0.22 & \qmark / 0.63 & \qmark / 0.63 & 0.36 & 0.68 & 0.68 & \textbf{0.94} & 0.30 & 0.34 \\
I(c) & $\widehat \beta_{X_2}$ & \cmark / 0.22 & \qmark / 0.27 & \qmark / 0.20 & 0.95 & 0.69 & 0.58 & 0.86 & 0.83 & 0.92 \\
 & $\widehat \beta_0$ & \cmark / 0.13 & \qmark / 0.65 & \qmark / 0.65 & 0.34 & 0.69 & 0.69 & \textbf{0.96} & 0.28 & 0.31 \\
I(d) & $\widehat \beta_{X_2}$ & \xmark / 0.25 & \qmark / 0.23 & \qmark / 0.17 & 0.92 & 0.65 & 0.56 & 0.87 & 0.86 & \textbf{0.93} \\
 & $\widehat \beta_0$ & \xmark / 0.21 & \qmark / 0.64 & \qmark / 0.64 & 0.37 & 0.69 & 0.68 & \textbf{0.94} & 0.26 & 0.30 \\
\midrule
II(a) & $\widehat \beta_{X_2}$ & \cmark / 0.13 & \qmark / \textbf{0.02} & \qmark / \textbf{0.01} & 0.62 & 0.46 & 0.46 & 0.92 & 0.89 & \textbf{0.93} \\
 & $\widehat \beta_0$ & \cmark / 0.11 & \qmark / 0.53 & \qmark / 0.53 & 0.30 & 0.58 & 0.58 & \textbf{0.93} & 0.39 & 0.43 \\
II(b) & $\widehat \beta_{X_2}$ & \xmark / 0.05 & \qmark / \textbf{-0.02} & \qmark / \textbf{-0.03} & 0.60 & 0.48 & 0.47 & 0.91 & 0.89 & \textbf{0.94} \\
 & $\widehat \beta_0$ & \xmark / 0.16 & \qmark / 0.55 & \qmark / 0.55 & 0.32 & 0.59 & 0.59 & 0.90 & 0.36 & 0.40 \\
II(c) & $\widehat \beta_{X_2}$ & \cmark / 0.09 & \qmark / \textbf{0.00} & \qmark / \textbf{-0.01} & 0.67 & 0.47 & 0.47 & 0.89 & 0.90 & \textbf{0.93} \\
 & $\widehat \beta_0$ & \cmark / 0.07 & \qmark / 0.53 & \qmark / 0.53 & 0.32 & 0.58 & 0.58 & 0.92 & 0.36 & 0.40 \\
II(d) & $\widehat \beta_{X_2}$ & \xmark / 0.06 & \qmark / \textbf{-0.01} & \qmark / \textbf{-0.01} & 0.60 & 0.45 & 0.45 & 0.91 & 0.91 & \textbf{0.95} \\
 & $\widehat \beta_0$ & \xmark / 0.17 & \qmark / 0.53 & \qmark / 0.53 & 0.31 & 0.58 & 0.58 & 0.90 & 0.38 & 0.43 \\
\bottomrule
\end{tabular}
\caption{\label{tb:sim3_N500}Simulation results for Study 3, $N = 500$. For DAG I, $\beta_{X_2} = 2.5$ and $\beta_0 = 1.5$; for DAG II, $\beta_{X_2} = 2$ and $\beta_0 = 2$. MAR (IPW): Sun \& Tchetgen Tchetgen inverse-probability weighting estimator. MIA-I: ICE estimator. MIA-T: TMLE estimator. \cmark\ indicates that the sufficient (and, in the case of $\muMAR$, necessary) conditions for $\muMAR$ and $\muMIA$, respectively, hold. \xmark\ indicates that the conditions for $\muMAR$ are violated; in DGMs (b) and (d), this violation is weak by construction. \qmark\ indicates that the sufficient (but not necessary) conditions for $\muMIA$ are violated. Boldface indicates approximately correct performance.}
\end{table}

\begin{table}[H]
\centering
\footnotesize
\setlength{\tabcolsep}{3pt}
\begin{tabular}{@{}p{\dagwidthiii}p{\coefwidthiii}>{\centering\arraybackslash}p{\colwidthiii}>{\centering\arraybackslash}p{\colwidthiii}>{\centering\arraybackslash}p{\colwidthiii}>{\centering\arraybackslash}p{\colwidthiii}>{\centering\arraybackslash}p{\colwidthiii}>{\centering\arraybackslash}p{\colwidthiii}>{\centering\arraybackslash}p{\colwidthiii}>{\centering\arraybackslash}p{\colwidthiii}>{\centering\arraybackslash}p{\colwidthiii}@{}}
\toprule
\multicolumn{2}{@{}l}{\textbf{DGM}} & \multicolumn{3}{c}{\textbf{Bias}} & \multicolumn{3}{c}{\textbf{RMSE}} & \multicolumn{3}{c}{\textbf{CI coverage}} \\
\cmidrule(lr){3-5} \cmidrule(lr){6-8} \cmidrule(lr){9-11}
& & MAR (IPW) & MIA-T & MIA-I & MAR (IPW) & MIA-T & MIA-I & MAR (IPW) & MIA-T & MIA-I \\
\midrule
I(a) & $\widehat \beta_{X_2}$ & \cmark / 0.07 & \qmark / 0.21 & \qmark / 0.22 & 0.57 & 0.44 & 0.44 & \textbf{0.95} & 0.90 & 0.90 \\
 & $\widehat \beta_0$ & \cmark / 0.07 & \qmark / 0.63 & \qmark / 0.63 & 0.21 & 0.65 & 0.65 & \textbf{0.98} & 0.07 & 0.08 \\
I(b) & $\widehat \beta_{X_2}$ & \xmark / 0.12 & \qmark / 0.19 & \qmark / 0.19 & 0.57 & 0.43 & 0.43 & \textbf{0.95} & 0.90 & 0.89 \\
 & $\widehat \beta_0$ & \xmark / 0.18 & \qmark / 0.64 & \qmark / 0.64 & 0.26 & 0.66 & 0.66 & \textbf{0.95} & 0.05 & 0.06 \\
I(c) & $\widehat \beta_{X_2}$ & \cmark / 0.07 & \qmark / 0.22 & \qmark / 0.22 & 0.58 & 0.43 & 0.43 & \textbf{0.95} & 0.90 & 0.92 \\
 & $\widehat \beta_0$ & \cmark / 0.06 & \qmark / 0.62 & \qmark / 0.62 & 0.21 & 0.65 & 0.65 & \textbf{0.98} & 0.08 & 0.08 \\
I(d) & $\widehat \beta_{X_2}$ & \xmark / 0.18 & \qmark / 0.21 & \qmark / 0.20 & 0.58 & 0.43 & 0.41 & \textbf{0.96} & 0.89 & 0.91 \\
 & $\widehat \beta_0$ & \xmark / 0.15 & \qmark / 0.63 & \qmark / 0.63 & 0.25 & 0.65 & 0.65 & \textbf{0.96} & 0.05 & 0.06 \\
\midrule
II(a) & $\widehat \beta_{X_2}$ & \cmark / 0.07 & \qmark / \textbf{0.01} & \qmark / \textbf{0.01} & 0.44 & 0.31 & 0.31 & \textbf{0.94} & \textbf{0.93} & \textbf{0.94} \\
 & $\widehat \beta_0$ & \cmark / 0.07 & \qmark / 0.53 & \qmark / 0.53 & 0.20 & 0.56 & 0.56 & \textbf{0.96} & 0.12 & 0.12 \\
II(b) & $\widehat \beta_{X_2}$ & \xmark / 0.06 & \qmark / \textbf{0.01} & \qmark / \textbf{0.01} & 0.43 & 0.32 & 0.32 & \textbf{0.93} & \textbf{0.94} & \textbf{0.93} \\
 & $\widehat \beta_0$ & \xmark / 0.11 & \qmark / 0.53 & \qmark / 0.53 & 0.22 & 0.56 & 0.56 & \textbf{0.94} & 0.10 & 0.12 \\
II(c) & $\widehat \beta_{X_2}$ & \cmark / 0.05 & \qmark / \textbf{-0.01} & \qmark / \textbf{-0.01} & 0.47 & 0.31 & 0.31 & 0.91 & \textbf{0.94} & \textbf{0.95} \\
 & $\widehat \beta_0$ & \cmark / \textbf{0.02} & \qmark / 0.53 & \qmark / 0.53 & 0.21 & 0.56 & 0.56 & \textbf{0.96} & 0.11 & 0.11 \\
II(d) & $\widehat \beta_{X_2}$ & \xmark / \textbf{0.03} & \qmark / \textbf{0.00} & \qmark / \textbf{0.00} & 0.41 & 0.31 & 0.31 & \textbf{0.94} & \textbf{0.94} & \textbf{0.95} \\
 & $\widehat \beta_0$ & \xmark / 0.15 & \qmark / 0.54 & \qmark / 0.54 & 0.24 & 0.56 & 0.56 & 0.91 & 0.10 & 0.11 \\
\bottomrule
\end{tabular}
\caption{\label{tb:sim3_N1000}Simulation results for Study 3, $N = 1{,}000$. For DAG I, $\beta_{X_2} = 2.5$ and $\beta_0 = 1.5$; for DAG II, $\beta_{X_2} = 2$ and $\beta_0 = 2$. MAR (IPW): Sun \& Tchetgen Tchetgen inverse-probability weighting estimator. MIA-I: ICE estimator. MIA-T: TMLE estimator. \cmark\ indicates that the sufficient (and, in the case of $\muMAR$, necessary) conditions for $\muMAR$ and $\muMIA$, respectively, hold. \xmark\ indicates that the conditions for $\muMAR$ are violated; in DGMs (b) and (d), this violation is weak by construction. \qmark\ indicates that the sufficient (but not necessary) conditions for $\muMIA$ are violated. Boldface indicates approximately correct performance.}
\end{table}

\begin{table}[H]
\centering
\footnotesize
\setlength{\tabcolsep}{3pt}
\begin{tabular}{@{}p{\dagwidthiii}p{\coefwidthiii}>{\centering\arraybackslash}p{\colwidthiii}>{\centering\arraybackslash}p{\colwidthiii}>{\centering\arraybackslash}p{\colwidthiii}>{\centering\arraybackslash}p{\colwidthiii}>{\centering\arraybackslash}p{\colwidthiii}>{\centering\arraybackslash}p{\colwidthiii}>{\centering\arraybackslash}p{\colwidthiii}>{\centering\arraybackslash}p{\colwidthiii}>{\centering\arraybackslash}p{\colwidthiii}@{}}
\toprule
\multicolumn{2}{@{}l}{\textbf{DGM}} & \multicolumn{3}{c}{\textbf{Bias}} & \multicolumn{3}{c}{\textbf{RMSE}} & \multicolumn{3}{c}{\textbf{CI coverage}} \\
\cmidrule(lr){3-5} \cmidrule(lr){6-8} \cmidrule(lr){9-11}
& & MAR (IPW) & MIA-T & MIA-I & MAR (IPW) & MIA-T & MIA-I & MAR (IPW) & MIA-T & MIA-I \\
\midrule
I(a) & $\widehat \beta_{X_2}$ & \cmark / \textbf{-0.01} & \qmark / 0.21 & \qmark / 0.21 & 0.24 & 0.27 & 0.27 & \textbf{0.97} & 0.76 & 0.78 \\
 & $\widehat \beta_0$ & \cmark / \textbf{0.02} & \qmark / 0.63 & \qmark / 0.63 & 0.09 & 0.64 & 0.64 & \textbf{1.00} & 0.00 & 0.00 \\
I(b) & $\widehat \beta_{X_2}$ & \xmark / 0.10 & \qmark / 0.20 & \qmark / 0.20 & 0.26 & 0.26 & 0.26 & \textbf{0.96} & 0.76 & 0.79 \\
 & $\widehat \beta_0$ & \xmark / 0.12 & \qmark / 0.63 & \qmark / 0.63 & 0.15 & 0.64 & 0.64 & 0.89 & 0.00 & 0.00 \\
I(c) & $\widehat \beta_{X_2}$ & \cmark / \textbf{0.00} & \qmark / 0.21 & \qmark / 0.21 & 0.25 & 0.27 & 0.27 & \textbf{0.97} & 0.76 & 0.78 \\
 & $\widehat \beta_0$ & \cmark / \textbf{0.01} & \qmark / 0.63 & \qmark / 0.63 & 0.09 & 0.63 & 0.63 & \textbf{0.99} & 0.00 & 0.00 \\
I(d) & $\widehat \beta_{X_2}$ & \xmark / 0.14 & \qmark / 0.21 & \qmark / 0.21 & 0.27 & 0.27 & 0.27 & \textbf{0.95} & 0.74 & 0.76 \\
 & $\widehat \beta_0$ & \xmark / 0.10 & \qmark / 0.63 & \qmark / 0.63 & 0.13 & 0.64 & 0.64 & \textbf{0.94} & 0.00 & 0.00 \\
\midrule
II(a) & $\widehat \beta_{X_2}$ & \cmark / \textbf{0.01} & \qmark / \textbf{0.00} & \qmark / \textbf{0.00} & 0.19 & 0.14 & 0.14 & \textbf{0.97} & \textbf{0.94} & \textbf{0.95} \\
 & $\widehat \beta_0$ & \cmark / \textbf{0.01} & \qmark / 0.53 & \qmark / 0.53 & 0.08 & 0.54 & 0.54 & \textbf{1.00} & 0.00 & 0.00 \\
II(b) & $\widehat \beta_{X_2}$ & \xmark / \textbf{0.02} & \qmark / \textbf{0.00} & \qmark / \textbf{0.00} & 0.18 & 0.13 & 0.13 & \textbf{0.96} & \textbf{0.96} & \textbf{0.96} \\
 & $\widehat \beta_0$ & \xmark / 0.06 & \qmark / 0.53 & \qmark / 0.53 & 0.10 & 0.54 & 0.54 & \textbf{0.97} & 0.00 & 0.00 \\
II(c) & $\widehat \beta_{X_2}$ & \cmark / \textbf{0.01} & \qmark / \textbf{0.00} & \qmark / \textbf{0.00} & 0.20 & 0.14 & 0.14 & \textbf{0.97} & \textbf{0.95} & \textbf{0.95} \\
 & $\widehat \beta_0$ & \cmark / \textbf{-0.02} & \qmark / 0.54 & \qmark / 0.54 & 0.09 & 0.54 & 0.54 & \textbf{0.99} & 0.00 & 0.00 \\
II(d) & $\widehat \beta_{X_2}$ & \xmark / \textbf{0.02} & \qmark / \textbf{0.00} & \qmark / \textbf{0.00} & 0.18 & 0.14 & 0.14 & \textbf{0.96} & \textbf{0.95} & \textbf{0.94} \\
 & $\widehat \beta_0$ & \xmark / 0.11 & \qmark / 0.53 & \qmark / 0.53 & 0.14 & 0.54 & 0.54 & 0.83 & 0.00 & 0.00 \\
\bottomrule
\end{tabular}
\caption{\label{tb:sim3_N5000}Simulation results for Study 3, $N = 5{,}000$. For DAG I, $\beta_{X_2} = 2.5$ and $\beta_0 = 1.5$; for DAG II, $\beta_{X_2} = 2$ and $\beta_0 = 2$. MAR (IPW): Sun \& Tchetgen Tchetgen inverse-probability weighting estimator. MIA-I: ICE estimator. MIA-T: TMLE estimator. \cmark\ indicates that the sufficient (and, in the case of $\muMAR$, necessary) conditions for $\muMAR$ and $\muMIA$, respectively, hold. \xmark\ indicates that the conditions for $\muMAR$ are violated; in DGMs (b) and (d), this violation is weak by construction. \qmark\ indicates that the sufficient (but not necessary) conditions for $\muMIA$ are violated. Boldface indicates approximately correct performance.}
\end{table}

\begin{table}[H]
\centering
\footnotesize
\setlength{\tabcolsep}{3pt}
\begin{tabular}{@{}p{\dagwidthiii}p{\coefwidthiii}>{\centering\arraybackslash}p{\colwidthiii}>{\centering\arraybackslash}p{\colwidthiii}>{\centering\arraybackslash}p{\colwidthiii}>{\centering\arraybackslash}p{\colwidthiii}>{\centering\arraybackslash}p{\colwidthiii}>{\centering\arraybackslash}p{\colwidthiii}>{\centering\arraybackslash}p{\colwidthiii}>{\centering\arraybackslash}p{\colwidthiii}>{\centering\arraybackslash}p{\colwidthiii}@{}}
\toprule
\multicolumn{2}{@{}l}{\textbf{DGM}} & \multicolumn{3}{c}{\textbf{Bias}} & \multicolumn{3}{c}{\textbf{RMSE}} & \multicolumn{3}{c}{\textbf{CI coverage}} \\
\cmidrule(lr){3-5} \cmidrule(lr){6-8} \cmidrule(lr){9-11}
& & MAR (IPW) & MIA-T & MIA-I & MAR (IPW) & MIA-T & MIA-I & MAR (IPW) & MIA-T & MIA-I \\
\midrule
I(a) & $\widehat \beta_{X_2}$ & \cmark / \textbf{0.00} & \qmark / 0.21 & \qmark / 0.21 & 0.17 & 0.24 & 0.24 & \textbf{0.98} & 0.57 & 0.59 \\
 & $\widehat \beta_0$ & \cmark / \textbf{0.01} & \qmark / 0.63 & \qmark / 0.63 & 0.06 & 0.63 & 0.63 & \textbf{0.99} & 0.00 & 0.00 \\
I(b) & $\widehat \beta_{X_2}$ & \xmark / 0.12 & \qmark / 0.22 & \qmark / 0.22 & 0.20 & 0.25 & 0.25 & \textbf{0.94} & 0.54 & 0.57 \\
 & $\widehat \beta_0$ & \xmark / 0.11 & \qmark / 0.63 & \qmark / 0.63 & 0.12 & 0.63 & 0.63 & 0.81 & 0.00 & 0.00 \\
I(c) & $\widehat \beta_{X_2}$ & \cmark / \textbf{0.01} & \qmark / 0.21 & \qmark / 0.21 & 0.16 & 0.24 & 0.24 & \textbf{0.98} & 0.56 & 0.59 \\
 & $\widehat \beta_0$ & \cmark / \textbf{0.01} & \qmark / 0.63 & \qmark / 0.63 & 0.06 & 0.63 & 0.63 & \textbf{0.99} & 0.00 & 0.00 \\
I(d) & $\widehat \beta_{X_2}$ & \xmark / 0.14 & \qmark / 0.21 & \qmark / 0.21 & 0.22 & 0.24 & 0.24 & 0.92 & 0.56 & 0.58 \\
 & $\widehat \beta_0$ & \xmark / 0.09 & \qmark / 0.63 & \qmark / 0.63 & 0.11 & 0.63 & 0.63 & 0.88 & 0.00 & 0.00 \\
\midrule
II(a) & $\widehat \beta_{X_2}$ & \cmark / \textbf{0.00} & \qmark / \textbf{0.00} & \qmark / \textbf{0.00} & 0.13 & 0.10 & 0.10 & \textbf{0.98} & \textbf{0.94} & \textbf{0.94} \\
 & $\widehat \beta_0$ & \cmark / \textbf{0.01} & \qmark / 0.53 & \qmark / 0.53 & 0.06 & 0.53 & 0.53 & \textbf{0.99} & 0.00 & 0.00 \\
II(b) & $\widehat \beta_{X_2}$ & \xmark / \textbf{0.02} & \qmark / \textbf{0.00} & \qmark / \textbf{0.00} & 0.13 & 0.10 & 0.10 & \textbf{0.97} & \textbf{0.95} & \textbf{0.95} \\
 & $\widehat \beta_0$ & \xmark / 0.05 & \qmark / 0.53 & \qmark / 0.53 & 0.08 & 0.54 & 0.54 & \textbf{0.95} & 0.00 & 0.00 \\
II(c) & $\widehat \beta_{X_2}$ & \cmark / \textbf{0.00} & \qmark / \textbf{0.01} & \qmark / \textbf{0.01} & 0.14 & 0.10 & 0.10 & \textbf{0.97} & \textbf{0.95} & \textbf{0.95} \\
 & $\widehat \beta_0$ & \cmark / \textbf{-0.03} & \qmark / 0.53 & \qmark / 0.53 & 0.07 & 0.53 & 0.53 & \textbf{0.98} & 0.00 & 0.00 \\
II(d) & $\widehat \beta_{X_2}$ & \xmark / \textbf{0.02} & \qmark / \textbf{0.00} & \qmark / \textbf{0.00} & 0.13 & 0.10 & 0.10 & \textbf{0.96} & \textbf{0.94} & \textbf{0.94} \\
 & $\widehat \beta_0$ & \xmark / 0.11 & \qmark / 0.53 & \qmark / 0.53 & 0.13 & 0.54 & 0.54 & 0.70 & 0.00 & 0.00 \\
\bottomrule
\end{tabular}
\caption{\label{tb:sim3_N10000}Simulation results for Study 3, $N = 10{,}000$. For DAG I, $\beta_{X_2} = 2.5$ and $\beta_0 = 1.5$; for DAG II, $\beta_{X_2} = 2$ and $\beta_0 = 2$. MAR (IPW): Sun \& Tchetgen Tchetgen inverse-probability weighting estimator. MIA-I: ICE estimator. MIA-T: TMLE estimator. \cmark\ indicates that the sufficient (and, in the case of $\muMAR$, necessary) conditions for $\muMAR$ and $\muMIA$, respectively, hold. \xmark\ indicates that the conditions for $\muMAR$ are violated; in DGMs (b) and (d), this violation is weak by construction. \qmark\ indicates that the sufficient (but not necessary) conditions for $\muMIA$ are violated. Boldface indicates approximately correct performance.}
\end{table}

\pagebreak 
\printbibliography